\documentclass[PhD,binding=0.6cm]{sapthesis}

\usepackage{microtype}
\usepackage{amsmath,amssymb}
\usepackage[utf8]{inputenx}
\usepackage{accents}
\usepackage{mathrsfs}
\usepackage{setspace}
\usepackage[fontsize=12]{fontsize}

\usepackage{hyperref}

\newcommand{\MYhref}[3][black]{{{\href{#2}{\color{#1}{#3}}}}}

\hypersetup{pdftitle={Probing new physics on the horizon of black holes with gravitational waves}}

\usepackage{lipsum}
\usepackage{curve2e}
\definecolor{gray}{gray}{0.4}

\title{Probing new physics on the horizon of black holes with gravitational waves}
\author{Elisa Maggio}
\IDnumber{1468241}
\course[Physics]{Fisica}
\courseorganizer{Scuola di dottorato Vito Volterra}
\cycle{XXXIV}
\submitdate{2022}
\copyyear{2022}
\advisor{Prof. Paolo Pani}
\authoremail{elisa.maggio@uniroma1.it}

\examdate{22 February 2022}
\examiner{Prof. Alfredo Urbano}
\examiner{Prof. Massimo Bianchi}
\examiner{Prof. Enrico Barausse}

\begin{document}

\frontmatter
\maketitle

\begin{abstract}
Black holes are the most compact objects in the Universe. According to general relativity, black holes have a horizon that hides a singularity where Einstein’s theory breaks down.
Recently, gravitational waves opened the possibility to probe the existence of horizons and investigate the nature of compact objects. This is of particular interest given some quantum-gravity models which predict the presence of horizonless and singularity-free compact objects.
Such exotic compact objects can emit a different gravitational-wave signal relative to the black hole case.
In this thesis, we analyze the stability of horizonless compact objects, and derive a generic framework to compute their characteristic oscillation frequencies. We provide an analytical, physically-motivated template to search for the gravitational-wave echoes emitted by these objects in the late-time postmerger signal.
Finally, we infer how extreme mass-ratio inspirals observable by future gravitational-wave detectors will allow for model-independent tests of the black hole paradigm.
\end{abstract}

\tableofcontents

\chapter{Preface}

The work presented in this thesis has been carried out mainly at the Physics Department of Sapienza University of Rome in the research group of gravity theory and gravitational wave phenomenology. Part of this work was carried out at the Consortium for Fundamental Physics, School of Mathematics and Statistics, University of Sheffield, United Kingdom, and at the Centro de Astrof\'{i}sica e Gravitaç\~{a}o (CENTRA), Instituto Superior T\'{e}cnico, Universidade de Lisboa, Portugal. I thank these institutions for their kind hospitality.

\section*{List of publications}

The work in this thesis was accomplished with different scientific collaborations, whose members I kindly acknowledge.

\begin{itemize}
    \item Chapter~\ref{chapter2} is the outcome of a collaboration with Paolo Pani and Guilherme Raposo based on:
    \begin{itemize}
        \item[] E. Maggio, P. Pani, G. Raposo, ``Testing the nature of dark compact objects with gravitational waves,'' Invited chapter for C. Bambi, S. Katsanevas, K.D. Kokkotas (editors), Handbook of Gravitational Wave Astronomy, Springer, Singapore (2021),
        \MYhref{https://arxiv.org/abs/2105.06410}{\texttt{arXiv:2105.06410}},
        \MYhref{https://doi.org/10.1007/978-981-15-4702-7_29-1}{\texttt{https://doi.org/10.1007/978-981-15-4702-7$\_$29-1}}.
    \end{itemize}
    \item Chapter~\ref{chapter3} is the outcome of a collaboration with Luca Buoninfante, Anupam Mazumdar and Paolo Pani based on:
    \begin{itemize}
        \item[] E. Maggio, L. Buoninfante, A. Mazumdar, P. Pani, ``How does a dark compact object ringdown?,'' Phys. Rev. D \textbf{102}, 064053 (2020),\\ \MYhref{https://arxiv.org/abs/2006.14628}{\texttt{arXiv:2006.14628}}.
    \end{itemize}
    \item Chapter~\ref{chapter4} is the outcome of a collaboration with Vitor Cardoso, Sam Dolan, Valeria Ferrari and Paolo Pani based on:
    \begin{itemize}
        \item[] E. Maggio, P. Pani, and V. Ferrari, ``Exotic compact objects and how to quench their ergoregion instability,'' Phys. Rev. D \textbf{96}, 104047 (2017), \MYhref{https://arxiv.org/abs/1703.03696}{\texttt{arXiv:1703.03696}}.
        \item[] E. Maggio, V. Cardoso, S. Dolan, and P. Pani, ``Ergoregion instability of exotic compact objects: electromagnetic and gravitational perturbations and the role of absorption,'' Phys. Rev. D \textbf{99}, 064007 (2019), \MYhref{https://arxiv.org/abs/1807.08840}{\texttt{arXiv:1807.08840}};
    \end{itemize}
    \item Chapter~\ref{chapter5} is the outcome of a collaboration with Swetha Bhagwat, Paolo Pani and Adriano Testa based on:
    \begin{itemize}
        \item[] E. Maggio, A. Testa, S. Bhagwat, and P. Pani, ``Analytical model for gravitational-wave echoes from spinning remnants,'' Phys. Rev. D \textbf{100}, 064056 (2019),  \MYhref{https://arxiv.org/abs/1907.03091}{\texttt{arXiv:1907.03091}}.
    \end{itemize}
    \item Chapter~\ref{chapter6} is the outcome of a collaboration with Maarten van de Meent and Paolo Pani based on:
    \begin{itemize}
        \item[] E. Maggio, M. van de Meent, P. Pani, ``Extreme mass-ratio inspirals around a spinning horizonless compact object,'' Phys. Rev. D in press (2021),  \MYhref{https://arxiv.org/abs/2106.07195}{\texttt{arXiv:2106.07195}}.
    \end{itemize}
\end{itemize}

As a part of the activities during my PhD, I served as a member of the
LISA Consortium, being involved in the writing of the LISA Fundamental
Physics and the LISA Waveform White Papers, the LISA Figure of Merit
analysis, and the LISA Early Career Scientists (LECS) group.
Part of the outcome of these activities is in preparation or has been
submitted for publication and is not included in this thesis.

\section*{Conventions}

In this work, geometrized units, $G = c = 1$, are adopted where $G$ is the gravitational constant, and $c$ is the speed of light.

The signature of the metric adopts the $(-, +, +, +)$ convention. The Greek letters run over the four-dimensional spacetime indices.
The comma stands for an ordinary derivative, and the semi-colon stands for a covariant derivative. 

$\mathbf{M}^*$ is the complex conjugate of a matrix, and $\mathbf{M}^\top$ is the transpose of a matrix. $\mathfrak{R}(n)$ and $\mathfrak{I}(n)$ are the real and the imaginary part of a number, respectively.

\section*{Abbreviations}

\begin{tabular}{ l l }
 BH & Black Hole \\ 
 ECO & Exotic Compact Object \\  
 EMRI & Extreme Mass-Ratio Inspiral \\
 ISCO & Innermost Stable Circular Orbit \\
 LIGO & Large Interferomenter Gravitational-wave Observatory \\
 LISA & Laser Interferometer Space Antenna \\
 GR & General Relativity \\
 GW & Gravitational Wave \\
 NS & Neutron Star \\
 PN & Post-Newtonian \\
 QNM & Quasi-Normal Mode \\
 SNR & Signal-to-Noise Ratio \\
 TH & Tidal Heating \\
 ZAMO & Zero Angular Momentum Observer
\end{tabular}

\chapter{Introduction}

The landmark detection of gravitational waves (GWs) provides the unique opportunity to test gravity in the strong-field regime and infer the nature of astrophysical sources. So far, the ground-based detectors LIGO and Virgo have detected 
ninety GW events from the coalescence of compact binaries~\cite{LIGOScientific:2018mvr,LIGOScientific:2020ibl,LIGOScientific:2021usb,LIGOScientific:2021djp}.
These detections allowed us to observe for the first time the coalescence of two black holes (BHs) and revealed that their masses can be heavier than the ones observed in the electromagnetic spectrum~\cite{LIGOScientific:2016lio,LIGOScientific:2020ibl}. Recent important discoveries include the first multi-messenger observation of a binary neutron star (NS) merger~\cite{LIGOScientific:2017vwq,LIGOScientific:2017ync} and the observation of the formation of an intermediate-mass BH~\cite{LIGOScientific:2020iuh}.

Furthermore, GWs provide a new channel for probing Einstein's theory of gravity in a regime inaccessible to traditional astronomical observations, namely the strong-field and highly dynamical one.
Several consistency tests of the GW data with the predictions of general relativity (GR) have been performed. No evidence for new physics has been reported within current measurement accuracies~\cite{LIGOScientific:2016lio,LIGOScientific:2019fpa,LIGOScientific:2020tif,LIGOScientific:2021sio}.

The GW signal emitted by compact binary coalescences is characterized by three main stages: the \emph{inspiral} phase, when the two bodies orbit around each other and the emission of GWs makes the orbit shrink, the \emph{merger} phase, when the two bodies coalesce, and the \emph{ringdown} when the final remnant relaxes to an equilibrium solution. The study of the different stages of the GW signal allows us to infer the properties of the compact objects and understand their nature. 

Several extensions of GR predict the existence of regular and horizonless compact objects, also known as exotic compact objects (ECOs)~\cite{Giudice:2016zpa,Cardoso:2019rvt}.
Indeed, the presence of the horizon poses some theoretical problems, the most notable ones being the existence of a singularity in the black-hole interior and the Hawking information loss paradox~\cite{Mathur:2009hf}. 

ECOs can mimic the features of BHs through electromagnetic observations since they can be as compact as BHs~\cite{Abramowicz:2002vt}. Indeed, the supermassive object at the center of the M87 galaxy observed by the Event Horizon Telescope poorly constrains few models of ECOs~\cite{EventHorizonTelescope:2019pgp}.
Furthermore, current GW observations do not exclude ECOs that could potentially explain events in the mass gap between NSs and BHs and due to pair-instability supernova processes~\cite{LIGOScientific:2020iuh,Bustillo:2020syj,LIGOScientific:2020zkf}.

One way to distinguish ECOs from BHs is by analyzing the ringdown stage of a compact binary coalescence. The ringdown is dominated by the complex characteristic frequencies -- the so-called quasi-normal modes (QNMs) -- of the remnant, that differ dramatically if the latter is a BH or an ECO~\cite{Cardoso:2016rao}. 
By inferring the QNMs of the remnant, we can test whether they are compatible with the predicted spectrum for a BH. 

Current observations of the fundamental QNM in the ringdown of binary coalescences are compatible with remnant BHs as predicted by GR~\cite{LIGOScientific:2019fpa,LIGOScientific:2020tif,LIGOScientific:2021sio}; however, the characterization of the remnant is still an open problem.
The no-hair theorems establish that BHs in GR are determined uniquely by two parameters, i.e., their mass and angular momentum~\cite{Carter:1971zc,Robinson:1975bv}. Therefore, the measurement of one complex QNM allows us only to estimate the parameters of the BH. 
A test of the BH paradigm would require the identification of at least two complex QNM frequencies.
Louder GW events, to be collected as detector sensitivity improves, and more sophisticated parametrized waveforms will allow us to extract more information about the remnant.

If the remnant of a merger is an ECO that is almost as compact as a BH, the prompt ringdown signal would be nearly indistinguishable from that of a BH~\cite{Cardoso:2016rao}. A characteristic fingerprint of ECOs would be the appearance of a modulated train of GW echoes at late times due to the absence of the horizon~\cite{Cardoso:2016rao,Maggio:2019zyv}. 
Tentative evidence for GW echoes in LIGO/Virgo data has been reported in the last few years~\cite{Abedi:2016hgu}, but recent independent searches argued that the statistical significance for GW echoes is consistent with noise~\cite{Westerweck:2017hus,LIGOScientific:2020tif,LIGOScientific:2021sio}.

Besides ECO fingerprints in the GW emission, in this thesis we analyze the astrophysical viability of ECOs as BH alternatives.
Indeed, spinning horizonless compact objects are prone to the so-called ergoregion instability when spinning sufficiently fast~\cite{Friedman:1978wla,10.1093/mnras/282.2.580,Kokkotas:2002sf}. The endpoint of the instability could be a slowly-spinning ECO~\cite{Cardoso:2014sna,Brito:2015oca} or dissipation within the object could lead to a stable remnant~\cite{Maggio:2017ivp,Maggio:2018ivz,Wang:2019rcf}. If confirmed, the ergoregion instability could provide a strong theoretical argument in favor of the BH paradigm for which rapidly spinning compact objects must have a horizon.

The prospect for detectability of new physics will improve in the future with the next-generation detectors like the ground-based observatories Einstein Telescope~\cite{Punturo:2010zz} and Cosmic Explorer~\cite{Reitze:2019iox}, and the space-based Laser Interferometer Space Antenna (LISA)~\cite{LISA:2017pwj}.
In particular, LISA is an extremely promising observatory of fundamental physics. Planned for launch in 2034, LISA will detect GWs in a lower frequency band than ground-based detectors. LISA will observe a plethora of astrophysical sources, particularly extreme mass-ratio inspirals (EMRIs) in which a stellar-mass object orbits around the supermassive object at the center of a galaxy~\cite{Gair:2017ynp}.

EMRIs are unique probes of the nature of supermassive compact objects. 
Since LISA will observe inspirals that can last for years, the phase shift of the waveform will be tracked with high precision and deviations from GR will be measured accurately. During the inspiral around a horizonless supermassive object, extra resonances would be excited leaving a characteristic imprint in the waveform~\cite{Cardoso:2019nis,Maggio:2021uge}. Any evidence of partial reflectivity at the surface of the object would also indicate a departure from the classical BH picture~\cite{Datta:2019epe,Maggio:2021uge}.

Within this broad and flourishing context, this thesis aims to investigate the nature of compact objects and test the existence of horizons with GWs.
This work is organized as follows.

Chapter~\ref{chapter1} is dedicated to the tests of the BH paradigm that have been currently performed using GWs. A review of the recent GW observations is presented, and the stages of the gravitational waveform are analyzed. In particular, the consistency tests of GR and the parametrized tests of deviations from GR are described. Finally, the prospects of detecting deviations from GR with next-generation detectors are assessed.

Chapter~\ref{chapter2} illustrates the theoretical motivations for studying horizonless compact objects. A parametrized classification of horizonless compact objects is presented depending on their deviations from the standard BH picture. Some remarkable models of ECOs are reviewed, and their phenomenology is compared to the BH case.

Chapter~\ref{chapter3} derives the GW signatures of static horizonless compact objects, particularly their QNM spectrum in the ringdown. The system is described by perturbation theory. The imposition of the boundary conditions that describe the response of the compact object to perturbations requires careful analysis. A numerical procedure for the derivation of the QNMs is illustrated. 
The QNMs of horizonless compact objects are derived both for remnants almost as compact as BHs and with smaller compactness. In the former case, the presence of characteristic low frequencies in the QNM spectrum is highlighted. In the latter case, an extended version of the BH membrane paradigm is applied to derive model-independent deviations from the BH QNM spectrum. Finally, current constraints on horizonless compact objects and prospects of detectability are assessed.

Chapter~\ref{chapter4} analyzes spinning horizonless compact objects that are prone to the ergoregion instability above a critical threshold of the spin. The QNMs of spinning horizonless compact objects are derived. From the analysis of the imaginary part of the QNMs, the conditions for which horizonless compact objects are unstable are identified. An analytical description of the QNMs in terms of superradiance is detailed, and ways of quenching the ergoregion instability are provided.

Chapter~\ref{chapter5} describes the GW echoes that would be emitted after the prompt ringdown in the case of a horizonless merger remnant. An analytical, physically-motivated template for GW echoes is provided that can be implemented to perform a matched-filter-based search for echoes. Finally, the properties of GW echoes are analyzed, and the prospects of detection with current and future detectors are assessed.

Chapter~\ref{chapter6} is devoted to the analysis of EMRIs with a central horizonless compact object. During the inspiral, extra resonances are excited when the orbital frequency matches the characteristic frequencies of the central ECO. The impact of the resonances in the GW dephasing with respect to the BH case is assessed. This analysis shows that LISA will be able to probe the reflectivity of compact objects with unprecedented accuracy.

\mainmatter

\chapter{Tests of the black hole paradigm} \label{chapter1}

\begin{flushright}
    \emph{Sì, ma dapprincipio non lo si sapeva, - precisò Qfwfq, - ossia, uno poteva anche prevederlo, ma così, un po' a naso, tirando a indovinare.
    Io, non per vantarmi, fin da principio scommisi che l’universo ci sarebbe stato, e l’azzeccai, e anche sul come sarebbe stato vinsi parecchie scommesse, col Decano (k)yK.
    }\\
    \vspace{0.1cm}
    Italo Calvino, Le Cosmicomiche
\end{flushright}
\vspace{0.5cm}

Astrophysical BHs are the end result of gravitational collapse and hierarchical mergers. 
The no-hair theorems in GR establish that rotating BHs are well described by the Kerr geometry~\cite{Carter:1971zc,Robinson:1975bv}. 
Kerr BHs are determined uniquely by two parameters, i.e., their mass $M$ and angular momentum $J$ defined through the dimensionless spin parameter $\chi \equiv J/M^2$. 
As such, every property of BHs is characterized only by two parameters, i.e., the mass and the spin.  Observations of deviations from the properties of Kerr BHs would be an indication of departure from GR.

In this chapter, the tests of the BH paradigm that have been performed with current GW observations are reviewed. Moreover, the prospects of detection of deviations from GR with next-generation detectors are assessed.

\section{Gravitational waves from compact binary coalescences}

\subsection{Review of current observations}

On September 14, 2015, the first GW emitted by the coalescence of a compact binary was detected~\cite{LIGOScientific:2016aoc}. The signal, GW150914, is compatible with the inspiral of two BHs as predicted by GR~\cite{LIGOScientific:2016lio}. The remnant BH has final mass $M=62^{+4}_{-4}M_\odot$ and spin $\chi=0.67^{+0.05}_{-0.07}$, where $\sim 3.0M_\odot c^2$ were radiated in GWs.

During the first observing run (O1), which ran from September 12, 2015 to January 19, 2016, the two Advanced LIGO detectors~\cite{Abbott:2016xvh} observed a total of three GW events from the coalescence of binary BHs~\cite{LIGOScientific:2018mvr}. The second observing run (O2), which ran from November 30, 2016 to August 25, 2017, was joined by the Advanced Virgo detector~\cite{VIRGO:2014yos} in the last month of data taking, enabling the first three-detector observations of GWs. 

On August 17, 2017, the detectors made their first observation of a binary NS inspiral~\cite{LIGOScientific:2017vwq}. The signal, GW170817, was detected with a combined signal-to-noise ratio (SNR) of $32.4$, which is the highest SNR in both the O1 and O2 datasets.
GW170817 is the most precisely localized event and allowed for the first multi-messenger observation in the electromagnetic spectrum~\cite{LIGOScientific:2017ync}. Indeed, a short $\gamma$-ray burst was associated with the NS merger, followed by a transient kilonova counterpart across the electromagnetic spectrum in the same sky location.
In addition to GW170817, during O2, a total of seven binary BH mergers were detected~\cite{LIGOScientific:2018mvr}. 

During the first half of the third observing run (O3a), which ran from April 1, 2019 to October 1, 2019, forty-four GW events were detected~\cite{LIGOScientific:2020ibl,LIGOScientific:2021usb}. For the first time, the observations include binary systems with asymmetric mass ratios~\cite{LIGOScientific:2020zkf,LIGOScientific:2020stg} and an intermediate-mass BH with $142^{+28}_{-16}M_\odot$~\cite{LIGOScientific:2020iuh}.
The event, GW190521, is consistent with the merger of two BHs whose primary component lies in the mass gap produced by pulsational pair-instability supernova processes. Indeed, it is predicted that stars with a helium core are subject to an instability which leaves the remnants with mass less than $65 M_\odot$~\cite{Barkat:1967zz}. BHs with mass larger than this value might form via hierarchical mergers of smaller BHs. Recent studies showed that the event GW190521 is also consistent with the head-on collision of two horizonless vector boson stars forming a remnant BH~\cite{Bustillo:2020syj}.
    
During O3a, the event GW190814~\cite{LIGOScientific:2020zkf} was also detected, whose secondary mass lies in the lower mass gap of $2.5 - 5M_\odot$ between known NSs and BHs~\cite{Ozel:2010su}. Some ECOs such as boson stars and gravastars can potentially support masses beyond $2.5 M_\odot$. The nature of GW190814 is unknown, and the hypothesis of an exotic secondary is not excluded.

In the second half of the third observing run (O3b), which ran from November 1, 2019 to March 27, 2020, thirty-five GW events were detected including the first observations of NS-BH coalescences~\cite{LIGOScientific:2021qlt,LIGOScientific:2021djp}.

\subsection{Stages of the waveform} \label{sec:stages}

%
\begin{figure}[t]
\centering
\includegraphics[width=0.69\textwidth]{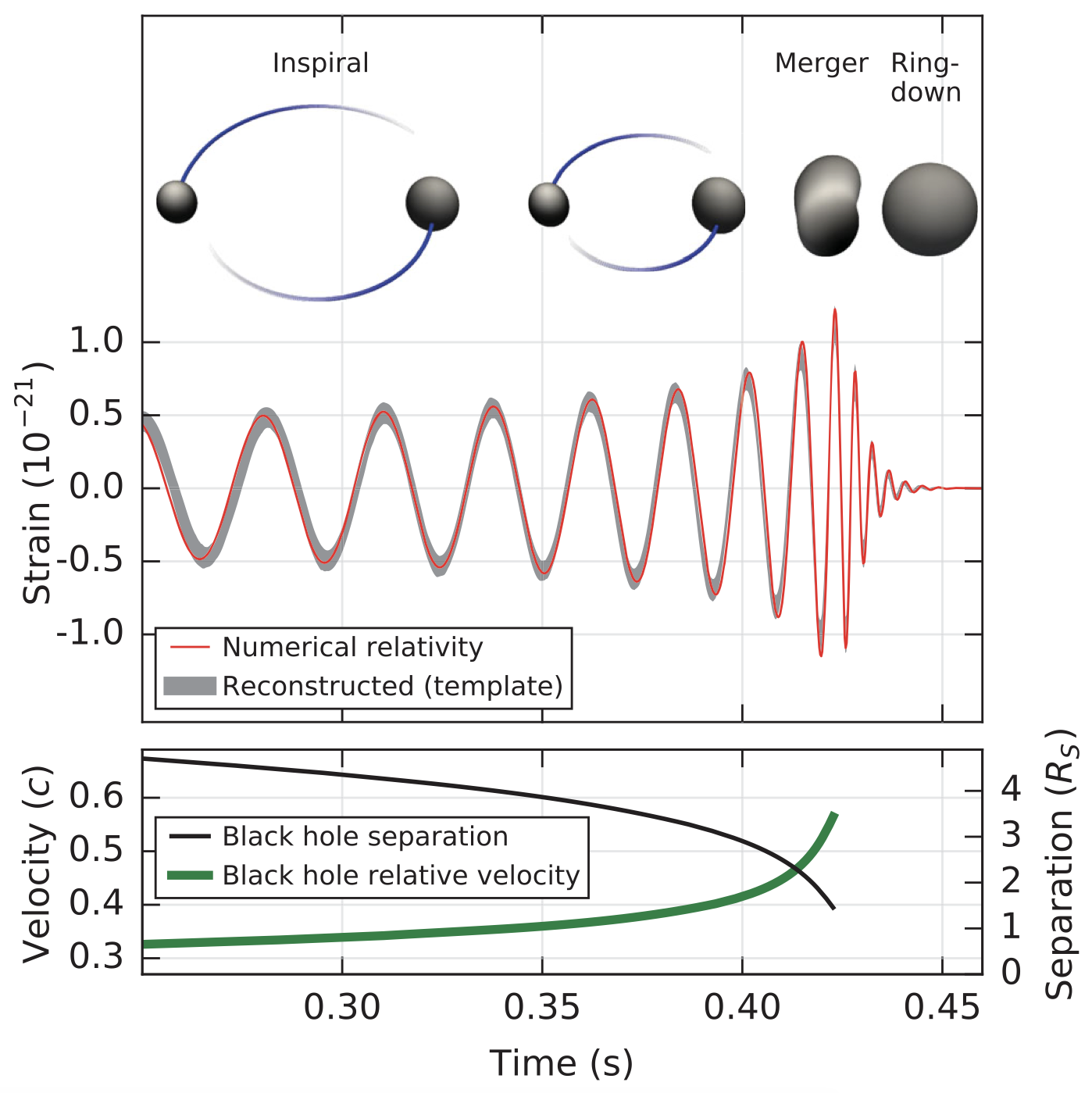}
\caption{Top: The stages of a compact binary coalescence. Estimated GW strain amplitude from the event GW150914 compared to the numerical waveform of a binary BH coalescence. Bottom: The Keplerian effective separation in units of Schwarzschild radii ($R_S = 2M$) and the effective relative velocity of the binary system.~\cite{LIGOScientific:2016aoc}} 
\label{fig:stages}
\end{figure}
The signal emitted by a compact binary coalescence is characterized by three main stages, as shown in Fig.~\ref{fig:stages}. 
\begin{itemize}
    \item[(i)] The \emph{inspiral} is a phase during which the two compact objects spiral in towards each other as they lose energy to gravitational radiation. At this stage, the two compact objects have large separations and small velocities. The gravitational waveform is well approximated by the post-Newtonian (PN) theory~\cite{Blanchet:2013haa,Blanchet:1995ez,Kidder:1995zr,Damour:2001bu,Arun:2008kb}. The latter is a perturbative approach to solve the Einstein field equations in which an expansion in terms of the velocity parameter $v/c$ is performed.
    \item[(ii)] The \emph{merger} is a rapid phase in which the two compact objects coalesce to form a final remnant. This stage can be described only by numerical simulations that take into account the nonlinearities of the dynamics.
    \item[(iii)] The \emph{ringdown} is a final phase in which the remnant settles down to its stationary state. This stage is described by perturbation theory. The ringdown is dominated by the complex characteristic frequencies of the remnant, the so-called quasi-normal modes~(QNMs), which describe the response of the compact object to a perturbation~\cite{Teukolsky:1973ha,Press:1973zz,Teukolsky:1974yv,Chandrasekhar:1975zza,Leaver:1985ax,Kokkotas:1999bd,Berti:2009kk}. The BH ringdown signal can be modeled as a superposition of exponentially damped sinusoids
    \begin{equation}
        h = \sum_{\ell mn} A_{\ell mn}(r) e^{-t/\tau_{\ell mn}} \sin(\omega_{\ell mn} t + \phi_{\ell mn}) ~_{-2}S_{\ell m}(\theta,\varphi) \,,
    \end{equation}
    where $\omega_{\ell mn}$ are the characteristic frequencies of the remnant, $\tau_{\ell mn}$ are the damping times, $A_{\ell mn}(r)\propto 1/r$ is the amplitude of the signal at a distance $r$, $\phi_{\ell mn}$ is the phase, and $~_{s}S_{\ell m}(\theta,\varphi)\propto e^{i m\varphi}$ are the spin-weighted spheroidal harmonics which depend on the location of the observer to the source. Each mode is described by three integers, namely the angular number ($\ell \geq 0$), the azimuthal number $m$ (such that $|m| \leq \ell$), and the overtone number ($n \geq 0$). From the detection of the ringdown it is possible to infer the QNMs of the remnant and understand the nature of the compact object.
\end{itemize}
%

\section{Inspiral-merger-rindown consistency test}

%
\begin{figure}[t]
\centering
\includegraphics[width=0.69\textwidth]{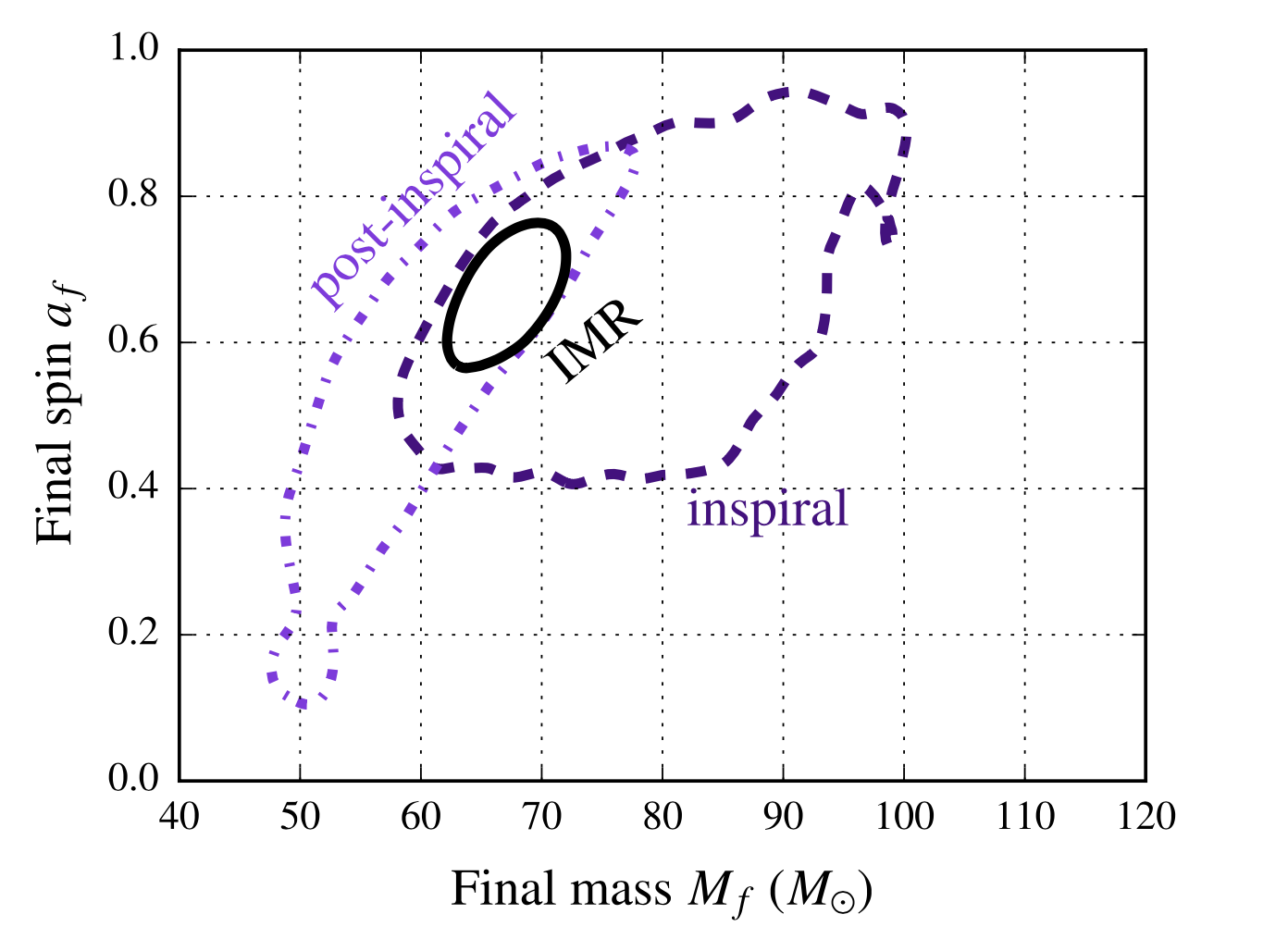}
\caption{$90\%$ credible regions in the joint posterior distributions for the mass $M_f$ and spin $a_f$ of the merger remnant as determined from the inspiral and postinspiral signals, and a full inspiral-merger-ringdown analysis. The posterior distributions have a significant region of overlap.~\cite{LIGOScientific:2016lio}} 
\label{fig:IMR}
\end{figure}
One way of testing that a gravitational waveform is consistent with the predictions of GR for a binary BH is the inspiral-merger-ringdown  test~\cite{Ghosh:2016qgn}. The test consists in comparing the estimates of the mass and the spin of the remnant obtained from the inspiral (low-frequency) and postinspiral (high-frequency) parts of the waveform.
If GR describes well both the adiabatic and the nonlinear regimes, the estimates on the parameters from the two phases are consistent with each other within the statistical uncertainties.

The inspiral-merger-ringdown consistency test was performed in the first GW event, GW150914~\cite{LIGOScientific:2016lio}. Fig.~\ref{fig:IMR} shows the $90\%$ credible contours in the estimates of the remnant mass $M_f$ and spin $a_f \equiv \chi_f M_f$ from the inspiral and the postinspiral stages. The two posterior distributions have a significant region of overlap. Moreover, they agree with the estimate performed using full inspiral-merger-ringdown waveforms.

To constrain possible departures from GR, the following parameters are defined
\begin{eqnarray}
 \frac{\Delta M_f}{M_f} &=&  2 \frac{M_f^{\rm insp} - M_f^{\rm postinsp}}{M_f^{\rm insp} + M_f^{\rm postinsp}} \,, \\
 \frac{\Delta \chi_f}{\chi_f} &=& 2 \frac{\chi_f^{\rm insp} - \chi_f^{\rm postinsp}}{\chi_f^{\rm insp} + \chi_f^{\rm postinsp}} \,,
\end{eqnarray}
that quantify the fractional difference between the estimates of the remnant mass and dimensionless spin from the inspiral and postinspiral stages. In GW150914, the joint posterior distribution of the parameters is compatible with the $(0,0)$ result expected in GR~\cite{LIGOScientific:2016lio}.

The inspiral-merger-ringdown test has also been applied to the events in the third LIGO–Virgo GW transient catalog with $\text{SNR}>6$ both in the inspiral and postinspiral regions. The measurement constraints are
%
%
%
\begin{equation}
    \frac{\Delta M_f}{M_f} = 0.03^{+0.14}_{-0.13} \,, \qquad \frac{\Delta \chi_f}{\chi_f} = -0.05^{+0.37}_{-0.38} \,,
\end{equation}
which are consistent with the expectations of GR~\cite{LIGOScientific:2021sio}.

\section{Parametrized tests}

Several parametrized tests have been performed to quantify generic deviations from GR.
These tests introduce parametrized modifications to the GR waveform to constrain the degree to which the data agree with GR predictions. The following sections analyze generic deviations from inspiral-merger-ringdown waveforms, the BH spin-induced quadrupole moment, the BH QNMs, and review some searches for GW echoes.

\subsection{Constraints on generic deviations in the waveform}

%
\begin{figure}[t]
\centering
\includegraphics[width=0.99\textwidth]{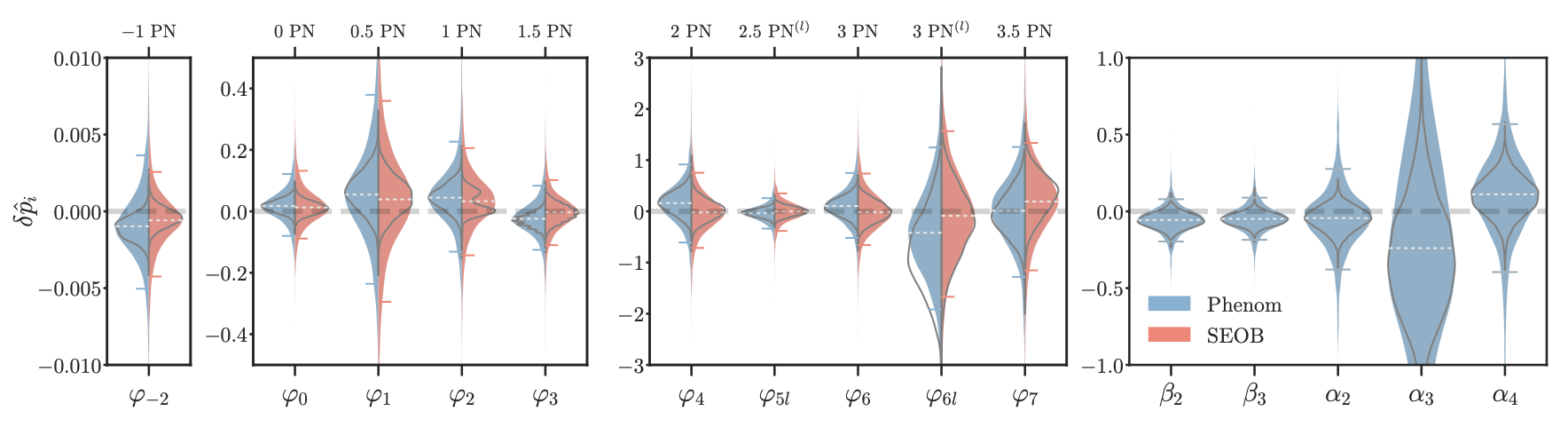}
\caption{Combined results for parametrized deviations from GR obtained from the binary BH events in the second GW transient catalog, for each deviation parameter. Phenom (SEOB) results were obtained with IMRP\textsc{henom}Pv2 (SEOBNRv4$\_$ROM) and are shown in blue (red). The error bars denote symmetric $90\%$-credible intervals, and the white dashed line marks the median. The dashed horizontal line at $\delta \hat p_i = 0$ highlights the expected GR value.~\cite{LIGOScientific:2020tif}} 
\label{fig:PN}
\end{figure}
Deviations from GR can be introduced via parametric deformations of the inspiral-merger-ringdown waveform predicted by GR, without relying on any specific alternative theory of gravity. In this framework~\cite{Li:2011cg,Agathos:2013upa}, the deviations from GR are modeled as fractional changes ${\delta \hat p_i}$ in the parameters ${p_i}$ that parametrize the GW phase as $p_i \to \left( 1 + \delta \hat p_i \right) p_i$. The fractional changes are parameters that are introduced to be constrained by the data and check the consistency with the GR values.

The parameters $p_i$ denote collectively all the inspiral $\{\varphi_i\}$ and postinspiral $\{\alpha_i, \beta_i\}$ parameters. 
In particular, the early-inspiral stage is known analytically up to the order $\left(v/c\right)^7$ and is parametrized in terms of the PN coefficients $\varphi_j$ with $j= 0, ..., 7$ and the logarithmic terms $\varphi_{jl}$ with $j=5,6$.
In addition, the coefficient at $j = -2$ is included corresponding to an effective -1PN term that, in some circumstances, can be interpreted as arising from the emission of dipolar radiation.
The transition between the inspiral and the merger-ringdown phase is parametrized in terms of the phenomenological coefficients $\beta_j$ with $j= 2, 3$. Finally, the merger-ringdown phase is parametrized in terms of the phenomenological coefficients $\alpha_j$ with $j = 2, 3, 4$. Parameters that are degenerate with either the reference time or the reference phase are not considered.

It is possible to perform two kinds of tests: a \emph{single-parameter} analysis, in which only one of the parameters is allowed to vary freely while the remaining ones are fixed to their GR value, and a \emph{multiple-parameter} analysis, in which all the parameters are allowed to vary simultaneously. The multiple-parameter analysis accounts for correlations between the parameters and provides a more conservative estimate on the agreement between a single GW event and GR.

Fig.~\ref{fig:PN} shows the combination of the parametric deviations of GR from a single-parameter analysis for the binary BH coalescences in the second GW transient catalog~\cite{LIGOScientific:2020tif}.
From left to right, the plot shows increasingly high-frequency regimes: the early-inspiral stage is from 0PN to 3.5PN, whereas the parameters $\beta_i$ and $\alpha_i$ correspond to the intermediate and merger-ringdown stages, respectively.
Phenom (SEOB) results were obtained with IMRP\textsc{henom}Pv2 (SEOBNRv4$\_$ROM) waveform model. The error bars are symmetric $90\%$-credible intervals, and the white dashed line is the median. The dashed horizontal line at $\delta \hat p_i = 0$ highlights the expected GR value.

The parameter that is constrained most tightly by the combined analysis is 
$\delta \hat \varphi_{-2} = -0.05^{+0.99}_{-1.25}\times 10^{-3}$, within $90\%$ credibility~\cite{LIGOScientific:2021sio}. The 0PN term is the second best constrained parameter with $|\delta \hat \varphi_0| \lesssim 4.4 \times 10^{-2}$~\cite{LIGOScientific:2020tif}. However, the latter constraint is weaker than the bound inferred from the double pulsar J0737-3039 by a factor $\sim 3$ due to the long observation time~\cite{Wex:2014nva}.
All the other PN orders are constrained significantly more tightly with GW observations rather than electromagnetic observations.

The results of the parametrized analysis can be used to place constraints on specific theories of gravity by building a theory-dependent mapping~\cite{Yunes:2009ke,Yunes:2016jcc}. 

\subsection{Measurement of the spin-induced quadrupole moment} \label{sec:quadrupole}

%
\begin{figure}[t]
\centering
\includegraphics[width=0.69\textwidth]{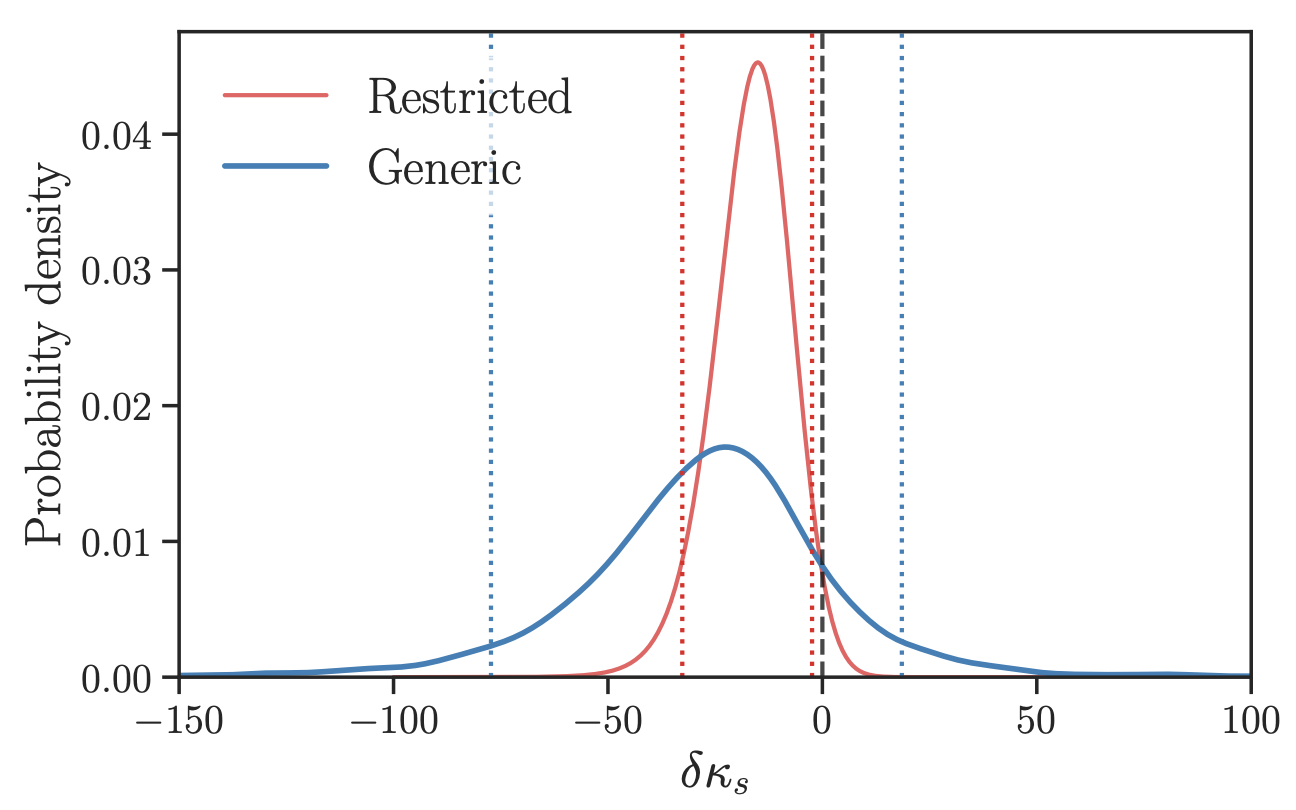}
\caption{Combined measurement of the spin-induced quadrupole moment parameter $\delta \kappa_s$ for all the events in the third GW transient catalog~\cite{LIGOScientific:2021sio}. The blue curve is the posterior obtained without assuming a unique value of $\delta \kappa_s$ in all the events, whereas the red curve is the posterior obtained by restricting $\delta \kappa_s$ to take the same value for all the events. The dotted lines bound symmetric $90\%$-credible intervals. The Kerr BH value ($\delta \kappa_s=0$) is marked by a dashed line.~\cite{LIGOScientific:2021sio}} 
\label{fig:quadrupole}
\end{figure}
The oblateness of a compact object due to its spin creates a deformation in the surrounding gravitational field, which is measured by the spin-induced quadrupole-moment~\cite{Poisson:1997ha}. 
The effect of the quadrupole moment on the orbital motion of a binary system is imprinted in the gravitational waveform at specific PN orders with a 2PN leading-order effect.

For a compact object with mass $M$ and spin $\chi$, the spin-induced quadrupole moment can be parametrized as
\begin{equation}
 Q = - \kappa \chi^2 M^3 \,,   
\end{equation}
where $\kappa$ is the spin-induced quadrupole moment coefficient that depends on the mass, spin, and internal composition of the compact object. Due to the no-hair theorems, $\kappa$ is unity for BHs in GR~\cite{Carter:1971zc,Hansen:1974zz}. For spinning NSs, $\kappa$ can vary between $\sim 2$ and $\sim 14$ depending on the equation of state~\cite{Pappas:2012ns,Harry:2018hke}, whereas for slowly spinning boson stars, $\kappa$ can vary between $\sim 10$ and $\sim 150$~\cite{Ryan:1996nk,Herdeiro:2014goa}.

The measurement of the spin-induced quadrupole moment coefficients of the binary components, $\kappa_1$ and $\kappa_2$, is challenging given the strong correlations between the binary parameters~\cite{Krishnendu:2017shb}.
For this reason, the individual deviations from unity are defined, $\delta \kappa_1$ and $\delta \kappa_2$, and the symmetric and anti-symmetric combinations of the individual deviation parameters are~\cite{LIGOScientific:2020tif}
\begin{eqnarray}
    \delta \kappa_s = \left( \delta \kappa_1 + \delta \kappa_2\right)/2 \,, \\
    \delta \kappa_a = \left( \delta \kappa_1 - \delta \kappa_2\right)/2 \,.
\end{eqnarray}
For simplicity, the analysis is restricted to $\delta \kappa_a=0$, requiring that the binary components are of the same nature. 
Nevertheless, the measurement of $\delta \kappa_s$ from individual GW events is poorly constrained. Fig.~\ref{fig:quadrupole} shows the distributions on $\delta \kappa_s$ obtained by combining all the events in the third GW transient catalog~\cite{LIGOScientific:2021sio}. The blue curve represents the posterior obtained without assuming a unique value of $\delta \kappa_s$ in all the events, whereas the red curve is the posterior obtained by restricting $\delta \kappa_s$ to take the same value for all the events.
Under the latter assumption, the following contraint is estimated $\delta \kappa_s = -16.0^{+13.6}_{-16.7}$ ~\cite{LIGOScientific:2021sio}.

\subsection{Tests of the remnant properties} \label{sec:testremnant}

%
\begin{figure}[t]
\centering
\includegraphics[width=0.59\textwidth]{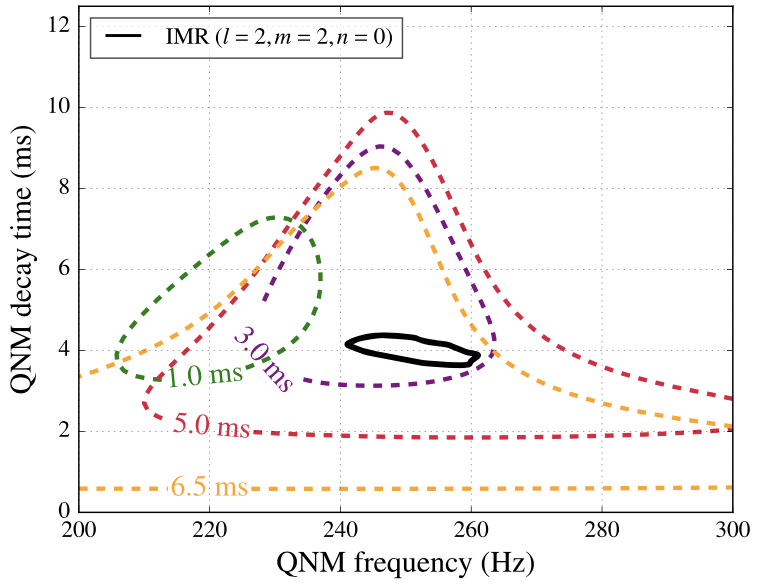}
\caption{$90\%$ credible regions in the joint posterior distributions for the QNM frequency and decay time, for several values of the starting time of the ringdown after the merger.
The solid black line shows the $90\%$ credible region for the frequency and decay time of the least-damped ($\ell=m=2, n=0$) QNM, as derived from the posterior distributions of the remnant mass and spin from inspiral-merger-ringdown waveforms.~\cite{LIGOScientific:2016lio}} 
\label{fig:fundamental_QNM}
\end{figure}
From the analysis of the postmerger signal of a compact binary coalescence, it is possible to infer the nature of the compact remnant. Due to the no-hair theorems~\cite{Carter:1971zc,Robinson:1975bv}, the QNM spectrum of a BH in GR depends only on the mass and spin of the remnant.
Therefore, the measurement of one complex QNM allows us only to infer the mass and the spin of the remnant. Conversely, the measurement of more than one QNMs would allow us to perform independent tests of the Kerr metric. This set of analyses is referred to as \emph{BH spectroscopy}~\cite{Dreyer:2003bv,Gossan:2011ha,Brito:2018rfr,Carullo:2019flw,Isi:2019aib,Bhagwat:2019dtm}. 

One test of the remnant properties consists in checking the consistency of the data with the least-damped QNM predicted for a remnant BH. The posterior estimates for the QNM frequency and decay time are a function of the unknown starting time of the ringdown after the merger.
Fig.~\ref{fig:fundamental_QNM} shows the $90\%$ credible contours for the QNM frequency and decay time as a function of the ringdown time offset for the event GW150914~\cite{LIGOScientific:2016lio}.
The solid black line shows the $90\%$ credible region of the least-damped QNM as derived from the posterior distributions of the remnant mass and spin from full inspiral-merger-ringdown waveforms.
The $90\%$ posteriors overlap with the GR prediction starting from $t_0=3 \  \text{ms}$, which is the offset time when the description of the ringdown in terms of QNMs is valid.

\begin{figure}[t]
\centering
\includegraphics[width=0.59\textwidth]{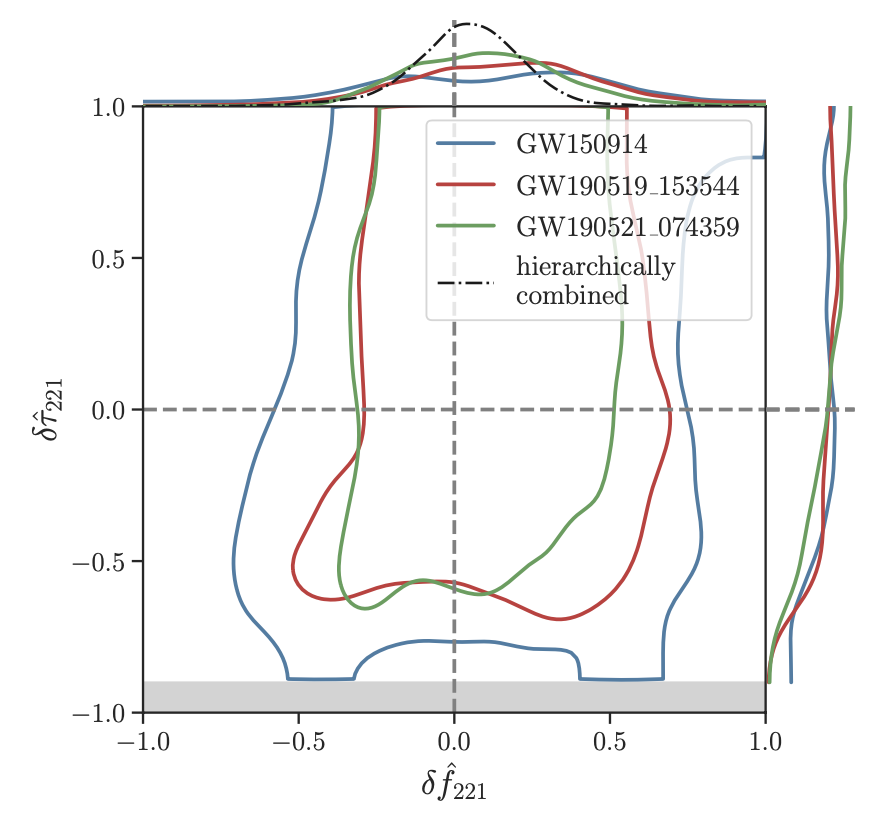}
\caption{
The $90\%$ credible region of the joint posterior distribution of
the fractional deviations of the frequency and the damping time for the $\ell=m=2, n=1$ mode. The measurement of the fractional deviation of the frequency is consistent with GR. The fractional deviation of the damping time is mostly unconstrained.~\cite{LIGOScientific:2020tif}} 
\label{fig:first_overtone}
\end{figure}
To test the BH paradigm, one would need to detect at least two QNMs in the ringdown. One test consists in incorporating the first overtone ($\ell=m=2, n=1$) in the ringdown template in time domain~\cite{Carullo:2019flw,Isi:2019aib}. The starting time of the ringdown is chosen based on an estimate of the peak of the strain from the full inspiral-merger-ringdown analyses. 
The data show evidence for the presence of overtones only for loud signals.
Fig.~\ref{fig:first_overtone} shows the joint posterior distributions for the fractional deviations in the frequency and damping time to their GR predictions for the first overtone, where
\begin{eqnarray}
 f_{221} &=& f_{221}^{\text{GR}} \left( 1 + \delta \hat f_{221}\right) \,, \\
 \tau_{221} &=& \tau_{221}^{\text{GR}} \left( 1 + \delta \hat \tau_{221}\right) \,,
\end{eqnarray}
and the “GR” superscript indicates the Kerr value corresponding to a remnant with a given mass and spin~\cite{LIGOScientific:2020tif}. A hierarchical analysis of the events in the third GW transient catalog constrains the frequency deviations to 
$\delta \hat f_{221} = 0.01^{+0.27}_{-0.28}$, whereas the damping time is essentially unconstrained~\cite{LIGOScientific:2021sio}. Recently, the observation of the $\ell=m=3$, $n=0$ mode has been claimed in the event GW190521~\cite{Capano:2021etf}.

The fractional deviations in the frequency and the damping time of the least-damped QNM are $\delta \hat f_{220} = 0.03^{+0.10}_{-0.09}$ and $\delta \hat \tau_{220} = 0.10^{+0.44}_{-0.39}$, which are obtained by combining the information from different events
using a hierarchical approach~\cite{Ghosh:2021mrv}. The event GW150914 gives the single-event most-stringent constraint with $\delta \hat f_{220} = 0.05^{+0.11}_{-0.07}$ and $\delta \hat \tau_{220} = 0.07^{+0.11}_{-0.07}$~\cite{Ghosh:2021mrv}, corresponding to a maximum allowed deviation from the least-damped QNM of a Kerr BH of $\sim 16\%$ and $\sim 33\%$ for the real and the imaginary part of the QNM, respectively. 

The bounds for the fractional deviations for the modes can be used to put constraints on possible deviations in the ringdown spectrum caused by horizonless remnant objects, as detailed in Secs.~\ref{sec:QNMmembrane} and~\ref{sec:detectabilitydoublet}.

\subsection{Searches for gravitational-wave echoes} \label{sec:echosearch}

If the remnant of a compact binary coalescence is a horizonless compact object, a train of modulated pulses --~known as GW echoes~-- is emitted in the late postmerger stage in addition to the ringdown expected from BHs~\cite{Cardoso:2019rvt,Abedi:2020ujo}. The detection of the GW echoes would be clear evidence of the existence of horizonless objects whose compactness is similar to the BH one (see Sec.~\ref{sec:echoespicture} for details).

Several matched-filtered searches have been performed to search for GW echoes. 
A waveform template in the time domain is based on the standard inspiral-merger-ringdown template in GR, $\mathcal{M}(t)$, with five additional parameters, i.e.,~\cite{Abedi:2016hgu}
\begin{equation}
    h(t) = A \sum_{i=0}^\infty (-1)^{i+1} \gamma^i \mathcal{M}(t+t_{\rm merger} + t_{\rm echo} - i \Delta t_{\rm echo}, t_0) \,,
\end{equation}
where $\mathcal{M}(t,t_0) = \Theta(t,t_0) \mathcal{M}(t)$ and $\Theta(t,t_0)$ is a smooth cut-off function. The five free parameters are: the time-interval in between successive echoes, $\Delta t_{\rm echo}$; the time of arrival of the first echo, $t_{\rm echo}$, that can be affected by the non-linear dynamics near the merger; the cut-off time, $t_0$, which quantifies the part of the GR template that produce the subsequent echoes; the damping factor of successive echoes, $\gamma$; the overall amplitude of the echo template, $A$. The $(-1)^{i}$ term represents the phase inversion of the waveform in each pulse. Extensions of the original template have been developed in Refs.~\cite{Nakano:2017fvh,Wang:2018gin}.

A phenomenological template in the time domain is based on the superposition of sine-Gaussians with several free parameters~\cite{Maselli:2017tfq}. Furthermore, some templates in the
frequency domain depend explicitly on the physical parameters of the horizonless compact object, i.e., its compactness and reflectivity~\cite{Mark:2017dnq,Testa:2018bzd,Maggio:2019zyv}.

Some unmodeled searches have also been performed. Several analyses are based on the superposition of generalized wavelets adapted from burst searches~\cite{Tsang:2018uie,Tsang:2019zra}. Moreover, searches with Fourier windows~\cite{Conklin:2017lwb,Conklin:2019fcs} use the fact that the echoes should pile up at specific frequencies.

Tentative evidence for GW echoes in LIGO/Virgo O1 and O2 events has been reported~\cite{Abedi:2016hgu,Conklin:2017lwb,Abedi:2018npz}, although independent searches argued that the statistical significance for GW echoes is low and consistent with noise~\cite{Westerweck:2017hus, Nielsen:2018lkf}. Recently, some negative searches have been performed~\cite{Uchikata:2019frs,Tsang:2019zra,Lo:2018sep}. 
Furthermore, a dedicated search for GW echoes has been performed by the LIGO/Virgo Collaboration in the events of the second and third GW transient catalogs, finding no evidence for GW echoes~\cite{LIGOScientific:2020tif,LIGOScientific:2021sio}. 

\section{Prospects with next-generation detectors}

Next-generation detectors are planned to observe GWs in a different frequency range than current detectors and with improved sensitivity, opening up the possibility of observing new GW sources~\cite{Hild:2010id}.
The ground-based observatories Einstein Telescope~\cite{Punturo:2010zz} and Cosmic Explorer~\cite{Reitze:2019iox} will observe GWs in the $5-4000 \ \text{Hz}$ band with a sensitivity of a factor of 10 better than current detectors.

The future space-based interferometer LISA~\cite{LISA:2017pwj} will detect GWs in the $10^{-4}-1 \ \text{Hz}$ frequency band from a variety of astrophysical sources. For instance, massive BHs (with masses ranging from $10^5 M_\odot$ to $10^9M_\odot$) are hosted in the center of galaxies and are expected to coalesce in bigger systems. The inspiral, merger, and ringdown phases are predicted to be in the LISA frequency band of observation with $\text{SNR} \sim 1000$~\cite{Klein:2015hvg}.

EMRIs are one of the target sources of LISA~\cite{Gair:2017ynp}. EMRIs are binary systems in which a stellar-mass object (with mass ranging $10-100M_\odot$) orbits around a supermassive object at the center of a galaxy. EMRIs occur over long timescales since the stellar-mass compact object spends $10^3-10^5$ orbits in the close vicinity of the central object. 
A large number of orbital cycles allows for precise measurements of the parameters of the binary. Moreover,  and to put GR to the most stringent tests.

EMRIs are unique probes of the nature of the central supermassive object. 
The mass quadrupole moment of the central object and possible deviations from the Kerr metric will be probed by LISA with large accuracy~\cite{Barack:2006pq,Glampedakis:2005cf}. In Sec.~\ref{chapter6}, the prospects of detection of LISA for the reflectivity of compact objects are assessed. 

\chapter{Exotic compact objects} \label{chapter2}

\begin{flushright}
    \emph{
    Per voi cadere è sbattersi giù magari dal ventesimo piano d’un grattacielo, o da un aeroplano che si guasta in volo: precipitare a testa sotto, annaspare un po’ nell’aria, ed ecco che la terra è subito lì, e ci si piglia una gran botta. 
    }\\
    \vspace{0.1cm}
    Italo Calvino, Le Cosmicomiche
\end{flushright}
\vspace{0.5cm}

\section{Motivation}

BHs are the most compact objects in the Universe. According to GR, stationary BHs have an event horizon surrounding a curvature singularity where Einstein’s theory breaks down. On the astrophysical side, the existence of BHs with masses ranging from a few to hundred solar masses has been confirmed by GW observations~\cite{LIGOScientific:2018mvr,LIGOScientific:2020ibl,LIGOScientific:2021djp}. Moreover, supermassive BHs at the center of galaxies have been observed with stellar orbits~\cite{Ghez:2008ms} and the electromagnetic emission from accretion disks~\cite{EventHorizonTelescope:2019dse}. All the observations are compatible with BHs as predicted by GR and support the \emph{Kerr hypothesis} for which any compact object heavier than a few solar masses is well described by the Kerr metric. Indeed, the Carter-Robinson uniqueness theorem establishes that the Kerr geometry is the only physically acceptable stationary solution to the Einstein vacuum field equations~\cite{Carter:1971zc,Robinson:1975bv}.

Given the observational robustness of BHs, it is natural to question the motivation for further tests of the nature of compact objects. It is worth remarking that the evidence for BHs is the observation of dark, compact, and massive objects. The Kerr geometry has been probed in the exterior spacetime approximately until the location of the light ring~\cite{EventHorizonTelescope:2019dse} which is the innermost stable circular orbit (ISCO) of photons. Investigations of the spacetime in the vicinity of the event horizon have not been performed with current measurement accuracies. For this reason, it is relevant to quantify the evidence for BHs by constraining the compactness and darkness of the objects observed so far via gravitational and electromagnetic channels.

On the theoretical side, Kerr BHs have singularities and are pathological in their interior. In particular, the existence of a curvature singularity with infinite tidal forces shows a breakdown of the Einstein equations. Moreover, the spacetime within the BH horizon can contain closed time-like curves which violate causality. Some attempts to regularize the BH solution predict that quantum fluctuations might prevent the formation of the horizon and the singularity therein~\cite{Mazur:2001fv,Mathur:2005zp}.

In the semiclassical approximation, when a massless scalar field such as that of the photon is quantized in the Schwarzschild background, one finds that the BH radiates a thermal spectrum at the Hawking temperature $T_{\rm H}= \hbar/(8 \pi k_B M)$~\cite{Hawking:1974rv}. 
The inverse dependence of the Hawking temperature on the mass implies that a BH in thermal equilibrium with its Hawking radiation has negative specific heat, hence is thermodynamically unstable~\cite{Hawking:1976de}.
Energy conservation plus the thermal radiation spectrum also imply that the BH has enormous entropy~\cite{Bekenstein:1973ur} which is far over a typical stellar progenitor.

Finally, one of the main open problems in BH physics is the information-loss paradox~\cite{Mathur:2009hf} which is related to loss of unitarity at the end of the BH evaporation due to Hawking's radiation. Several attempts to address this issue involve the formulation of a consistent quantum gravity theory that predicts modifications at the horizon relative to the classical picture (e.g., nonlocal theories~\cite{Giddings:1992hh,Giddings:2009ae,Giddings:2012bm} and string theories~\cite{Lunin:2002qf,Mazur:2004fk, Mathur:2005zp, Mathur:2008nj}) and new ways to compute the entropy~\cite{Engelhardt:2014gca,Almheiri:2019psf,Marolf:2020rpm}.

ECOs are horizonless objects that are predicted in quantum gravity extensions of GR~\cite{Nicolini:2005vd,Bena:2007kg,Giddings:2014ova,Koshelev:2017bxd,Abedi:2020ujo} and in the context of GR in the presence of exotic matter fields~\cite{Liebling:2012fv,Brito:2015pxa,Giudice:2016zpa}. These ideas have inspired a plethora of models including gravastars~\cite{Mazur:2001fv,Mazur:2004fk}, boson stars~\cite{Feinblum:1968nwc,Kaup:1968zz,Ruffini:1969qy,Seidel:1991zh}, wormholes~\cite{Einstein:1935tc,Morris:1988cz,Damour:2007ap}, fuzzballs~\cite{Mathur:2005zp,Mathur:2008nj}, and others~\cite{Bowers:1974tgi,Gimon:2007ur,Brustein:2016msz,Holdom:2016nek,Buoninfante:2019swn}. 
Some models are solutions to consistent field theories coupled to gravity~\cite{Liebling:2012fv,Mazur:2004fk}, whereas some phenomenological models do not arise from specific theories and are simple toy models to test deviations from the classical BH picture~\cite{Damour:2007ap,Pani:2010jz}. 

ECOs without a classical horizon can nonetheless mimic the features of BHs through electromagnetic observations since they can be as compact as BHs~\cite{Abramowicz:2002vt}. For this reason, ECOs are also generically called ``BH mimickers''~\cite{Cardoso:2007az,Lemos:2008cv}.
In most models, the dynamical formation of ECOs has not been explored consistently, with some notable exceptions~\cite{Seidel:1993zk,Liebling:2012fv}.

From a more phenomenological standpoint, BHs and NSs might be just two species of a larger zoo of compact objects. New species can be used to devise precision tests on the nature of compact objects. In particular, GW events that fall in the mass gap forbidden by standard stellar evolution (i.e., GW190814~\cite{LIGOScientific:2020zkf} and GW190521~\cite{LIGOScientific:2020iuh,LIGOScientific:2020ufj}) could be interpreted as mergers of exotic objects~\cite{Bustillo:2020syj}.

In summary, ECOs are a tool that allows us to quantify the observational evidence for BHs and search for signatures of alternative proposals in GW and electromagnetic data.

\section{A parametrized classification}

%
\begin{figure}[t]
\centering
\includegraphics[width=0.75\textwidth]{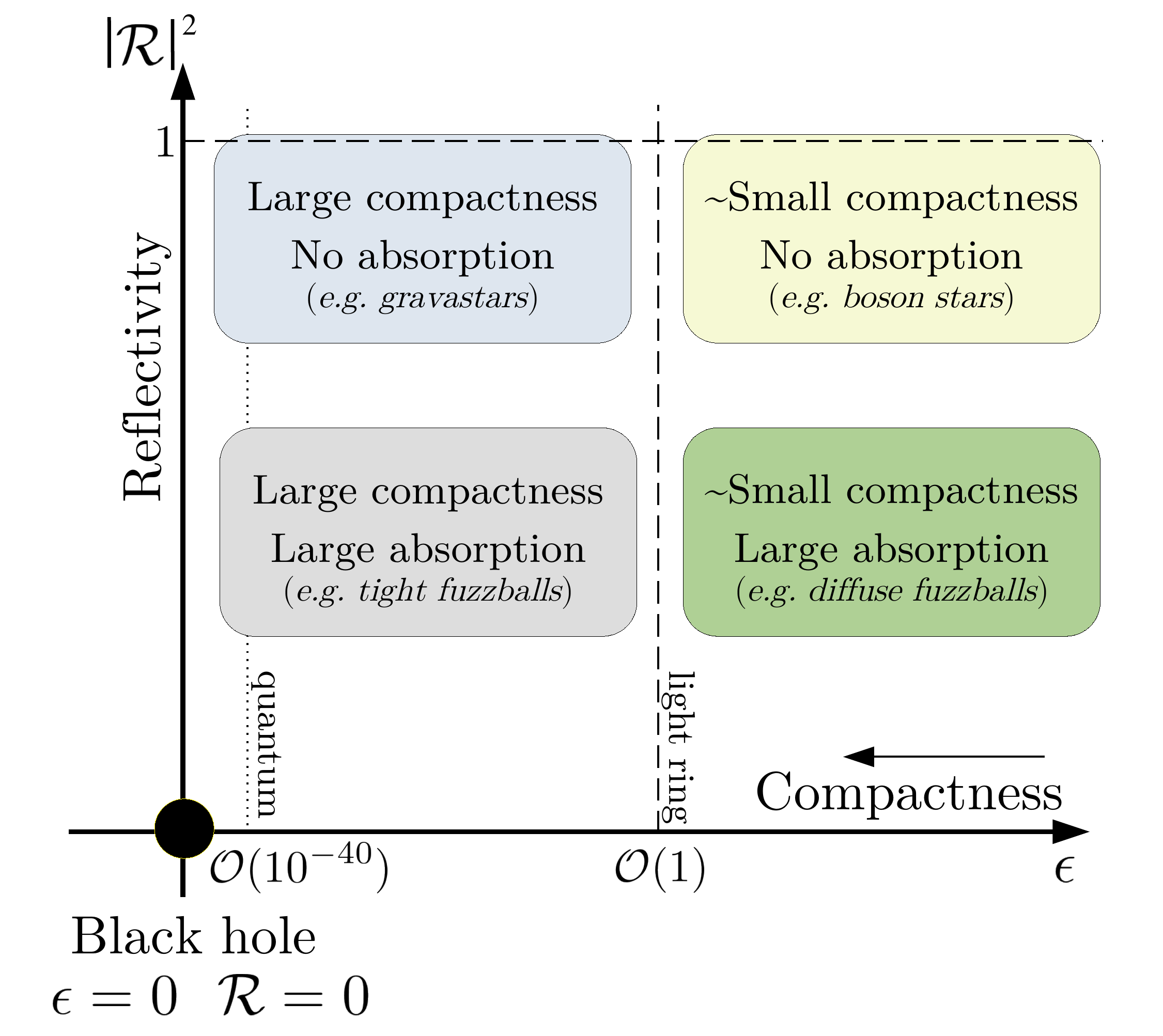}
\caption{Schematic representation of the parameter space of ECOs. The closeness parameter $\epsilon$ (horizontal axis) is related to the effective radius of the object by $r_0=r_+(1+\epsilon)$, where $r_+$ is the horizon of a Kerr BH with the same mass and spin. The reflectivity $\mathcal{R}$ (vertical axis) is related to the object's interior and is generically complex and frequency-dependent.~\cite{Maggio:2021ans}} 
\label{fig:paramspace}
\end{figure}
Horizonless compact objects deviate from BHs for two parameters (see Fig.~\ref{fig:paramspace}):
\begin{itemize}
    \item their {\it compactness}, i.e., the inverse of their effective radius in units of the total mass $M$. It is customary to define a closeness parameter $\epsilon$ from the horizon of a BH such that the effective radius of the horizonless compact object is located at 
    \begin{equation} \label{radius}
        r_0=r_+(1+\epsilon) \,,
    \end{equation}
    where $r_+ = M+\sqrt{M^2 -a^2}$, and $a$ is the spin of the compact object.
    The closeness parameter is related to the compactness of the horizonless object via
    \begin{equation} \label{compactnessECO}
        \mathcal{C} \equiv \frac{M}{r_0} = \frac{M}{r_+(1+\epsilon)} \,,
    \end{equation}
    where $0.5 \leq \mathcal{C} \leq 1$ for a Kerr BH with $0 \leq a \leq M$.
    In this framework, the BH limit corresponds to $\epsilon\to0$;
    \item their {\it reflectivity} ${\cal R}$ at the effective radius of the compact object. 
    The properties of the interior of a horizonless compact object can be parametrized in terms of its surface reflectivity, which is generically complex and frequency-dependent. In this framework, the $\mathcal{R} = 0$ case describes a totally absorbing object that reduces to the standard BH case when $\epsilon \to 0$. The $|\mathcal{R}|^2 = 1$ case describes a perfectly reflecting object for perturbations moving towards the compact object. This is the case, for example, of NSs where the most efficient absorption mechanism is due to viscosity. However, the absorption of the incoming radiation is negligible as detailed in Sec.~\ref{sec:quenchergoregion}, therefore the radiation passes unperturbed across the NS from a three-dimensional perspective. 
    Intermediate values of $\mathcal{R}$ describe partially absorbing compact objects through viscosity, dissipation, fluid mode excitations, nonlinear effects, etc.
\end{itemize}

Depending on their compactness, two categories of ECOs are~\cite{Cardoso:2017cqb}: horizonless objects with small compactness, whose effective radius is comparable with the light ring of BHs with $\epsilon=\mathcal{O}(0.1-1)$; and ultracompact objects, where Planckian corrections at the horizon as $r_0 \approx r_+ + l_{\rm Planck}$ correspond to $\epsilon=\mathcal{O}(10^{-50}-10^{-40})$ for supermassive to stellar objects depending on their mass. 
Several models of ultracompact horizonless objects are conceived by assuming that quantum fluctuations might prevent the formation of the horizon~\cite{Mazur:2001fv}. 
These models are so compact that the round-trip time of the light between the light ring and the radius of the object is longer than the instability timescale of the photon orbits.

If the remnant of a merger is a compact object with small compactness, the ringdown signal differs from the BH ringdown at early stages, as detailed in Sec.~\ref{sec:modpromptringdown}.
Conversely, if the remnant of a merger is an ultracompact horizonless object, the prompt ringdown is nearly indistinguishable from that of a BH since it is excited at the light ring which is at the same location both in the BH and the horizonless case~\cite{Cardoso:2016rao}. The details of the object's interior appear at late times in the form of a modulated train of GW echoes, as detailed in Sec.~\ref{sec:echoespicture}.

Another mention-worthy property used to classify ECOs is their \emph{softness}, which is associated with the spacetime curvature at their surface. When the underlying theory of an ECO model has a new length scale ${\cal L}\ll M$, the curvature (e.g., the Kretschmann scalar ${\cal K}$) at the surface can be much larger than the curvature at the horizon, i.e., ${\cal K}\gg 1/M^4$. On the other hand, models of ECOs that are not motivated by length scales other than $M$ cannot sustain large curvatures at their surfaces. The former case is referred to as ``hard'' ECOs, whereas the latter case is denoted by ``soft'' ECOs~\cite{Raposo:2018xkf}.

\subsection{The Buchdhal theorem}

A useful compass to navigate through the ECO atlas is provided by the Buchdhal theorem~\cite{Buchdahl:1959zz} which states that, under certain assumptions, the maximum compactness of a self-gravitating object is 
$M/r_0=4/9$ (i.e., $\epsilon\geq 1/8$). This theorem prevents the existence of ECOs with compactness arbitrarily close to that of a BH. 
In particular, the assumptions of the Buchdhal theorem are:
\begin{itemize}
    \item GR is the theory of gravity;
    \item the solution is spherically symmetric;
    \item the interior matter is a single perfect fluid;
    \item the fluid is at most mildly anisotropic, i.e., the radial pressure is larger than the tangential one, $p_r \gtrsim p_t$;
    \item the radial pressure and the energy density are positive, i.e., $p_r \geq 0$ and $\rho \geq 0$;
    \item the energy density decreases by moving outwards, i.e., $d\rho/dr<0$.
\end{itemize}
Relaxing some of these assumptions provides a way to circumvent the theorem and suggests a route to classify ECOs~\cite{Urbano:2018nrs,Cardoso:2019rvt}.
For example, the assumption of GR is violated in any modified-gravity theory, i.e., in fuzzballs in string theory~\cite{Mathur:2005zp,Mathur:2008nj} and nonlocal stars in infinite derivative gravity~\cite{Koshelev:2017bxd,Buoninfante:2019swn}.

A common property of ECOs is the presence of an anisotropic pressure. Indeed, strong tangential stresses are necessary to support compact self-gravitating configurations. This is the case, for instance, of
ultracompact anisotropic stars~\cite{1974ApJ...188..657B,Raposo:2018rjn}. 

Several models of ECOs are spherically symmetric solutions of GR supported by an exotic distribution of matter violating the energy conditions. This is the case of gravastars~\cite{Mazur:2001fv,Mazur:2004fk,Mottola:2006ew} with a dark energy interior, and wormholes~\cite{Morris:1988cz,Visserbook,Damour:2007ap} with a thin shell of exotic matter at the throat.

\section{Review of some remarkable models} \label{sec:modelECOs}

\subsection{Boson stars}

Boson stars are self-gravitating compact solutions formed by massive bosonic fields which are coupled minimally to GR~\cite{Jetzer:1991jr,Schunck:2003kk,Liebling:2012fv}.
The action of the Einstein-Klein-Gordon theory is~\cite{Wald:106274}
\begin{equation} \label{actionBS}
    \mathcal{S} = \int \left( \frac{R}{16 \pi} + \mathcal{L}_{\mathcal{M}} \right) \sqrt{-g} \ d^4 x \,,
\end{equation}
where $R$ is the Ricci scalar of the spacetime with metric $g_{\mu \nu}$ and determinant $\sqrt{-g}$, and the term $\mathcal{L}_{\mathcal{M}}$ describes the matter of the scalar field $\Phi$,
\begin{equation}
    \mathcal{L}_{\mathcal{M}} = - \frac{1}{2} \left[ g^{\mu \nu} \nabla_{\mu} \Phi^* \nabla_{\nu} \Phi + V\left(|\Phi|^2\right) \right] \,,
\end{equation}
where $\Phi^*$ is the complex conjugate of the scalar field and $V\left(|\Phi|^2\right)$ is the bosonic potential.
By varying the action in Eq.~\eqref{actionBS} with respect to the metric $g^{\mu \nu}$, the Einstein field equations are obtained; whereas by varying the action with respect to the scalar field $\Phi$, the Klein-Gordon equation is derived.

If the scalar field is complex, the boson star is a static and spherically symmetric geometry with an oscillating field~\cite{Kaup:1968zz,Ruffini:1969qy}
\begin{equation}
    \Phi(r, t) = \Phi_0(r) e^{i \omega t}\,, 
\end{equation}
where $\Phi_0(r)$ is the profile of the star and $\omega$ is the angular frequency of the phase of the field.
On the other hand, real scalar fields give rise to long-term stable oscillating geometries with a non-trivial time-dependent stress-energy tensor, called oscillatons~\cite{Seidel:1991zh}.
Both solutions arise naturally as the end-state of the gravitational collapse in the presence of bosonic fields~\cite{Seidel:1993zk,Okawa:2013jba}.

Boson stars are the most robust model of ECOs since their formation, stability, and binary coalescence have been analyzed in detail numerically~\cite{Palenzuela:2007dm,Palenzuela:2017kcg,Sanchis-Gual:2018oui}. 
Boson stars are natural candidates for dark matter. They are not meant to replace all the BHs in the Universe since their compactness is lower than the BH one. Indeed, boson stars have properties similar to NSs, e.g., having a maximum mass above which they are unstable against gravitational collapse.

There are several models of boson stars depending on the bosonic potential and the classes of self-interactions, namely:
\begin{itemize}
    \item \emph{mini boson stars} which are characterized by a non-interacting scalar field 
    where~\cite{Lee:1988av}
    \begin{equation}
        V\left( |\Phi|^2\right) = \mu^2 |\Phi|^2 \,,
    \end{equation}
    where $\mu$ is the bare mass of the field theory.
    The maximum boson star mass is $M_{\rm max} \sim 0.633 M_{\rm Planck}^2/\mu$.
    Their mass-radius diagram is qualitatively similar to the one of static NSs;
    \item \emph{massive boson stars} which are characterized by a scalar field with a quartic self-interaction potential~\cite{Colpi:1986ye}
    \begin{equation}
        V\left( |\Phi|^2\right) = \mu^2 |\Phi|^2 + \frac{\lambda}{2} |\Phi|^4 \,, 
    \end{equation}
    where $\lambda$ is a dimensionless coupling constant. 
    The maximum mass can be of the order of the Chandrasekhar mass or larger, 
    $M_{\rm max} \sim 0.062 \ \lambda^{1/2} M_{\rm Planck}^3/\mu^2$.
    This effect is caused by the self-interaction of the potential that provides an extra pressure against the gravitational collapse;
    \item \emph{solitonic boson stars} which are characterized by a potential with an attractive term~\cite{Friedberg:1986tq,Lee:1986ts}
    \begin{equation}
        V\left( |\Phi|^2\right) = \mu^2 |\Phi|^2 \left( 1-2\frac{|\Phi|^2}{\sigma_0^2} \right) \,, 
    \end{equation}
    where $\sigma_0$ is a constant that is generically assumed to be of the same order as $\mu$. The maximum mass of the boson star is $M_{\rm max} \sim M_{\rm Planck}^4/(\mu \sigma_0)^2$.
    Stationary, soliton-like configurations have also been found for complex and massive Proca fields~\cite{Brito:2015pxa}.
\end{itemize}

\subsection{Gravastars}

Gravitational-vacuum stars or gravastars are configurations supported by a negative pressure in their interior~\cite{Mazur:2004fk,Visser:2003ge,Mazur:2015kia}. 
The region with negative pressure forces the gravastar to violate some energy conditions and evade the Buchdhal theorem.
The model is singularity-free, thermodynamically stable and has no information paradox. 

Gravastars can have an arbitrary compactness depending on the model to describe the pressure. The original formulation of the gravastar has a five-layer construction with a de Sitter core, a thin shell connecting to a perfect-fluid region, and another thin shell connecting to the Schwarzschild exterior. A simpler model is the thin-shell gravastar~\cite{Visser:2003ge} that is constructed with a de Sitter core connected to the Schwarzschild exterior by a thin shell of perfect-fluid matter. 

The formation of a gravastar can occur at the endpoint of a gravitational collapse when quantum gravitational vacuum phase transition could intervene before the event horizon can form~\cite{Mazur:2001fv}. However, the dynamical formation of a non-singular gravastar is still an open issue.

\subsection{Wormholes}

Wormholes were introduced originally by Einstein and Rosen~\cite{Einstein:1935tc} in the attempt to build a geometrical model of an elementary particle in GR. Wormholes are constructed by taking two copies of a static and spherically symmetric metric with an asymptotically flat region. The two regions are connected by a wormhole whose throat occurs at the radius $r_0$. This procedure is called Schwarzschild surgery~\cite{Morris:1988cz,Visserbook}. 

The spacetime is everywhere vacuum except at the throat, where the surgery requires a thin shell of matter. The Einstein field equations yield an exotic distribution of matter that has a negative energy density and violates the weak and the dominant energy conditions.

Wormholes can be constructed with arbitrary mass and compactness, therefore they can mimic the observational features of BHs~\cite{Damour:2007ap}. 
Wormholes solutions have also been constructed in more generic gravity theories, some of which do not violate energy conditions~\cite{Kanti:2011jz}.
Nevertheless, their formation mechanism is not well understood, and wormholes are unstable under linear perturbations~\cite{Gonzalez:2008wd,Bronnikov:2012ch}. 

\subsection{Fuzzballs}

The fuzzball models are conceived in string theory to solve the loss of unitarity in the BH evaporation and the huge Bekenstein-Hawking entropy of BHs~\cite{Lunin:2001jy,Lunin:2002qf,Mathur:2005zp,Mathur:2008nj}. A classical BH is interpreted as an ensemble of regular, horizonless geometries that describes its quantum microstates~\cite{Bena:2007kg,Balasubramanian:2008da,Bena:2013dka}. These geometries are solutions to string theory and have the same mass and charge of the corresponding BH. 
In this description, quantum gravity effects are not confined close to the BH horizon, but the BH interior is formed by fluctuating geometries. For this reason, this picture is referred to as the “fuzzball” description of BHs.

The construction of the microstates has been achieved only under specific assumptions, i.e., in higher-dimensional or in non-asymptotically-flat spacetimes. None of the geometries that can be constructed in four-dimensional spacetimes could represent astrophysical BHs since  these solutions are typically non-rotating, charged, and extremal.
A general class of extremal and charged solutions in four dimensions  is described by the metric~\cite{Bena:2007kg,Gibbons:2013tqa,Bates:2003vx}
\begin{equation}
    ds^2 = -e^{2U} (dt + w)^2 + e^{-2U} \sum_{i=1}^{3} dx_i^2 \,,
\end{equation}
where $U$ is a function of eight harmonic functions associated with the electric and the magnetic charges~\cite{Bianchi:2020bxa,Bianchi:2020miz}. The fuzzballs are constructed by distributing the charges of the eight harmonic functions among $N$ centers. The geometry is regular and characterized by the absence of horizons and closed timelike curves.

\subsection{Anisotropic stars}

Anisotropic stars are compact objects which are supported by large anisotropic stresses~\cite{Bayin:1982vw,Dev:2000gt,Mak:2001eb,Andreasson:2007ck} that arise at high densities, in superfluidity, solid cores, etc.
Anisotropic stars have been studied in GR, mostly in the context of static and spherically symmetric solutions~\cite{Dev:2003qd,Herrera:2004xc,Doneva:2012rd,Silva:2014fca,Yagi:2015hda}.
Depending on the anisotropy scale, the compactness of anisotropic stars can be arbitrarily close to the BH one~\cite{Raposo:2018rjn}. Furthermore, anisotropic stars can cover a wide range of masses, hence they can mimic both stellar BHs and the supermassive BHs at the center of galaxies.

\subsection{Firewalls, nonlocal stars, and superspinars} \label{sec:ECOmodels}

Firewalls are horizonless compact objects with a BH exterior spacetime and some “hard” structure localized close to the horizon due to quantum origin~\cite{Almheiri:2012rt,Kaplan:2018dqx}. Furthermore, a classical BH with modified dispersion relations for the graviton could effectively appear as having a hard surface~\cite{Zhang:2017jze,Oshita:2018fqu}.

Nonlocal stars emerge in theories with infinite derivatives in which the nonlocality of the gravitational interaction can smear out the curvature singularity and avoid the presence of a horizon~\cite{Frolov:2015bta,Koshelev:2017bxd,Buoninfante:2018xif,Buoninfante:2019swn}. A nonlocal star is a self-gravitational bound system of gravitons interacting nonlocally. Outside the nonlocal star, the spacetime is well described by the Schwarzschild metric, whereas inside there is a nonvacuum spacetime that is conformally flat at the origin.

Superspinars are string-inspired Kerr geometries spinning above the Kerr bound~\cite{Gimon:2007ur,Pani:2010jz,Piovano:2020ooe}. Indeed, in GR the angular momentum $J$ of a Kerr BH is bounded by $J \leq M^2$. When the Kerr bound is violated, the geometry does not possess an event horizon.
Furthermore, some unknown quantum effects need to be invoked to create an effective surface to avoid naked singularities and closed timelike curves.

\section{Phenomenology of exotic compact objects}

\subsection{Tests of the multipolar structure}

Uniqueness theorems in GR predict that the outcome of the gravitational collapse is a Kerr BH which is determined uniquely by two parameters, i.e., its mass $M$ and angular momentum $J$~\cite{Carter:1971zc,Robinson:1975bv}. The multipolar structure of Kerr BHs can be written as~\cite{Hansen:1974zz}
\begin{equation}
    \mathcal{M}_\ell^{\rm BH} + i \mathcal{S}_\ell^{\rm BH} = M^{\ell +1} \left( i \chi \right)^{\ell +1} \,,
\end{equation}
where $\mathcal{M}_\ell$ and $\mathcal{S}_\ell$ are the mass and current multiple moments, respectively, $M=\mathcal{M}_0$ is the mass, $\chi=J/M^2$ is the dimensionless spin, and $J=\mathcal{S}_1$ is the angular momentum. In addition, Kerr BHs have vanishing mass (current) multiple moments when $\ell$ is odd (even) since the metric is axially and equatorially symmetric. The BH multipole moments do not depend on the azimuthal number $m$ given the axisymmetry of the metric. 

For ECOs, the tower of multipole moments is, in general, richer due to the presence of moments that break the equatorial symmetry or the axisymmetry, as in the case of multipolar boson stars~\cite{Herdeiro:2020kvf} and fuzzball microstate geometries~\cite{Bena:2020see,Bianchi:2020bxa,Bianchi:2020miz,Bena:2020uup}. The deformation of the multipoles depends on the specific ECO model and vanishes in the high-compactness limit approaching the Kerr value~\cite{Pani:2015tga,Raposo:2018xkf,Raposo:2020yjy}. The multipole moments of an ECO can be parametrized as
\begin{equation}
    \mathcal{M}_{\ell m}^{\rm ECO} = \mathcal{M}_{\ell}^{\rm BH} + \delta \mathcal{M}_{\ell m} \,, \qquad \mathcal{S}_{\ell m}^{\rm ECO} = \mathcal{S}_{\ell}^{\rm BH} + \delta \mathcal{S}_{\ell m} \,,
\end{equation}
where $\delta \mathcal{M}_{\ell m}$ and $\delta \mathcal{S}_{\ell m}$ are  model-dependent corrections to the mass and current multipole moments.

``Soft'' ECOs motivated by new physics effects whose length scale is comparable to the mass cannot have arbitrarily large deviations from the BH multipole moments. In the BH limit, the multipole moment deviations must vanish  sufficiently fast. For axisymmetric spacetimes, spin-induced moments must vanish logarithmically (or faster), whereas non-spin induced moments vanish linearly (or faster)~\cite{Raposo:2018xkf}, i.e.,
\begin{equation} \label{multipoles}
 \frac{\delta \mathcal{M}_\ell}{\mathcal{M}^{\ell +1}} \sim a_\ell \frac{\chi^\ell}{\log \epsilon} + b_\ell \epsilon \,,   
\end{equation}
and equivalently for the current multipole moments, where $a_\ell$ and $b_\ell$ are constants.

The multipolar structure of an object leaves a footprint in the GW signal emitted by a compact binary coalescence, modifying the PN structure of the  waveform at different orders. The lowest order contribution
is the quadrupole moment which enters at 2PN order~\cite{Kastha:2018bcr} as detailed in Sec.~\ref{sec:quadrupole}.
Current constraints on the parametrized PN deviations with GW observations~\cite{LIGOScientific:2019fpa,LIGOScientific:2020tif} can be mapped into constraints on $\delta \mathcal{M}_{20}$. However, such tests are challenging given the correlations between the binary component spins and the quadrupole moment where the former have not been measured accurately. 

EMRIs are expected to put stronger bounds on the multipolar structure of the central supermassive object, due to a large number of cycles before the merger. The future space mission LISA is expected to provide accurate measurements of the spin-induced quadrupole and a large set of high-order multipole moments~\cite{Barack:2006pq,Babak:2017tow,Kastha:2019brk}. 

\subsection{Tests of the tidal heating}

If the components of a binary system are dissipative objects, energy and angular momentum are dissipated in their interior in addition to the GW emission to infinity. This is the case of BHs in which energy and angular momentum are absorbed by the horizon. This effect is known as tidal heating  (TH) and can contribute to thousands of radians of accumulated orbital phase for EMRIs in the LISA band~\cite{Hartle:1973zz,Hughes:1999bq,Hughes:2001jr}.

If at least one component of the binary system is an ECO, the dissipation in their interior would be smaller than in the BH case. Indeed, exotic matter is expected to interact weakly with GWs leading to a suppressed contribution to the GW accumulated phase from TH.
This effect would allow distinguishing binary BHs from binary systems involving ECOs~\cite{Maselli:2017cmm}. For EMRIs in the LISA band, the absence of TH could be used to put a stringent upper bound on the reflectivity of ECOs~\cite{Datta:2019epe,Maggio:2021uge}.

\subsection{Measurements of the tidal deformability}

In the coalescence of a compact binary system, the gravitational field of each component acts as a tidal field on its companion, inducing some multipolar deformation in the spacetime. Tidal effects change the orbital phase and in turn the GW emission~\cite{poisson2014gravity}. This effect can be quantified in terms of the ``tidal-induced multipole moments''. Indeed, a weak tidal field can be decomposed into electric (or polar) tidal field moments, $\mathcal{E}_{\ell m}$, and magnetic (or axial) tidal field moments, $\mathcal{B}_{\ell m}$. In the nonrotating case, the ratio between the multipole moments and the tidal field moments that induce them defines the tidal deformability of the body, i.e.,
\begin{equation}
    \lambda_E^{(\ell)} = \frac{\mathcal{M}_{\ell m}}{\mathcal{E}_{\ell m}} \,, \qquad
    \lambda_B^{(\ell)} = \frac{\mathcal{S}_{\ell m}}{\mathcal{B}_{\ell m}} \,.
\end{equation}
The dimensionless tidal Love numbers can be defined as
\begin{equation}
    k_\ell^E = \text{const} \frac{\lambda_E^{(\ell)}}{M^{2 \ell +1}} \,, \qquad
    k_\ell^B = \text{const} \frac{\lambda_B^{(\ell)}}{M^{2 \ell +1}} \,,
\end{equation}
that depend on the internal composition of the central object.

A remarkable result in GR is that the tidal Love numbers of BHs are null. This was demonstrated for nonrotating BHs~\cite{Binnington:2009bb,Damour:2009vw}, then extended to slowly rotating BHs~\cite{Poisson:2014gka,Landry:2015zfa,Pani:2015hfa} and recently to Kerr BHs~\cite{Chia:2020yla,LeTiec:2020spy,LeTiec:2020bos}.
Conversely, the tidal Love numbers of ECOs are generically different from zero and can provide a smoking-gun test of the nature of compact objects~\cite{Cardoso:2017cfl}. The tidal Love numbers were computed for several models of ECOs such as boson stars~\cite{Cardoso:2017cfl,Sennett:2017etc,Mendes:2016vdr}, gravastars~\cite{Pani:2015tga,Cardoso:2017cfl,Uchikata:2016qku} and anisotropic stars~\cite{Raposo:2018rjn}.

In the case of ``hard'' ECOs, the tidal Love numbers vanish logarithmically in the BH limit~\cite{Cardoso:2017cfl}
\begin{equation}
    k_\ell^{\rm ECO} \to \frac{c_\ell}{1+d_\ell \log \epsilon} \,, \qquad \epsilon \to 0 \,,
\end{equation}
where axial and polar Love numbers coincide in the BH limit. Conversely, ``soft'' ECOs --~such as anisotropic stars in certain regimes~-- have a polynomial vanishing behavior in the BH limit~\cite{Raposo:2018rjn}
\begin{equation}
    k_\ell^{\rm ECO} \to f_\ell \left( \frac{\epsilon}{M} \right)^n \,,
\end{equation}
where $n$ is a parameter that depends on the specific model.

The effect of  tidal deformability alters the GW signal emitted by a compact binary coalescence at 5PN order.
Current and future GW detectors will be able to measure the tidal Love numbers of compact objects to distinguish ECOs from BHs~\cite{Cardoso:2017cfl,Maselli:2017cmm,Sennett:2017etc}.

\subsection{Ringdown tests}

The postmerger phase of a compact binary coalescence is dominated by the QNMs of the remnant.
In the case of a horizonless compact remnant, the QNM spectrum deviates from the one predicted for a BH in GR. 
The estimation of the fractional deviations from the GR modes in the GW events allows us to constrain possible deviations in the spectrum due to a horizonless remnant, as detailed in Sec.~\ref{sec:testremnant}. 

The vibration spectra of ECOs have been computed in a wide class of models: boson stars~\cite{Berti:2006qt,Macedo:2013jja}, gravastars~\cite{Chirenti:2007mk,Pani:2009ss,Chirenti:2016hzd,Volkel:2017ofl}, wormholes~\cite{Cardoso:2016rao,Konoplya:2016hmd,Bueno:2017hyj}, and quantum structures~\cite{Cardoso:2005gj,Eperon:2016cdd,Cardoso:2016oxy,Barcelo:2017lnx,Brustein:2017koc,Wang:2019rcf}. Typically, the QNMs of ECOs differ from the BH QNMs due to the presence of an effective radius instead of the horizon, the excitation of the internal oscillation modes~\cite{Ferrari:2000sr,Pani:2018flj,Glampedakis:2017cgd}, and the excitation of extra degrees of freedom in modified-gravity theories~\cite{Okounkova:2017yby, Blazquez-Salcedo:2016enn, Tattersall:2018nve}.

The isospectrality of axial and polar modes of BHs in GR~\cite{Chandrasekhar:1975zza} is broken in ECOs, which are expected to emit a characteristic \emph{mode doublet}. The detection of such doublet would be an irrevocable signature of new physics, whose prospects of detection are detailed in Sec.~\ref{sec:detectabilitydoublet}.

The formation of an ECO can also be constrained by looking for GW echoes in the postmerger signal of a compact binary coalescence. GW echoes are an additional signal that would be emitted after the prompt ringdown if the remnant is an ultracompact ECO. In Sec.~\ref{sec:echosearch} we reviewed the searches for GW echoes that have been currently performed.

Third-generation detectors are expected to detect the ringdown signal of massive binaries with a large SNR, which would allow putting strong constraints on the compactness and the reflectivity of the compact objects~\cite{Testa:2018bzd, Maggio:2019zyv}.

\chapter{Spectroscopy of horizonless compact objects}\label{chapter3}

\begin{flushright}
    \emph{
    Esatto, quel tempo là ci impiega, mica meno, - disse Qfwfq, - io una volta passando feci un segno in un punto dello spazio, apposta per poterlo ritrovare duecento milioni d’anni dopo, quando saremmo ripassati di lì al prossimo giro.
    }\\
    \vspace{0.1cm}
    Italo Calvino, Le Cosmicomiche
\end{flushright}
\vspace{0.5cm}

\section{A static model}

Let us analyze a static and spherically symmetric horizonless compact object. We assume that GR is a reliable approximation outside the radius of the compact object and  some modifications appear at the horizon scale as in some quantum-gravity models. Owing to the Birkhoff theorem, the exterior spacetime of a static and spherically symmetric compact object is described by the Schwarzschild metric
\begin{equation}
    ds^2 = -f(r) dt^2 + \frac{1}{f(r)} dr^2 + r^2 (d\theta^2 + \sin^2 \theta d\varphi^2) \,, \label{schwarzschild}
\end{equation}
where $(t,r,\theta,\varphi)$ are the Boyer-Lindquist coordinates, $f(r)=1-2M/r$ and $M$ is the total mass of the object.

The radius of the horizonless compact object is as in Eq.~\eqref{radius}, where $r_+=2M$ is the would-be horizon of a Schwarzschild BH with the same mass. 
Ultracompact horizonless objects ($\epsilon \ll 1$) have a compactness that is almost the same as the one of a Schwarzschild BH, i.e., $\mathcal{C} \simeq 0.5$, whereas horizonless objects with a small compactness $\left[\epsilon=\mathcal{O}(0.1-1)\right]$ have $\mathcal{C} \simeq 0.45-0.25$.
In the following, we shall not assume a specific model for the interior of the compact object that is parametrized in terms of the reflectivity at the effective radius.

In Sec.~\ref{sec:QNMspectrum}, we shall derive the QNM spectrum of ultracompact objects, whereas in Sec.~\ref{sec:membrane}, we shall derive a model-independent framework for the QNMs of horizonless compact objects using the membrane paradigm.

\section{Ringdown spectrum of ultracompact objects}
\label{sec:QNMspectrum}

Horizonless compact objects are characterized by a completely different QNM spectrum with respect to the BH case. In this section, we derive the QNM spectrum of a static ultracompact object ($\epsilon \ll 1$) with surface reflectivity $\mathcal{R}(\omega)$. 

\subsection{Linear perturbations in the Schwarzschild background}

Let us perturb the background geometry in Eq.~\eqref{schwarzschild} with a spin-$s$ perturbation, where $s=0, \pm 1, \pm 2$ for scalar, electromagnetic and gravitational perturbations, respectively. The perturbation can be decomposed as
\begin{equation} \label{decomposition}
    \Psi_s(t,r,\theta,\varphi) =  \int d\omega e^{-i \omega t} \sum_{\ell m} e^{im \varphi} ~_{s}{S}_{\ell m}(\theta) ~_{s}\psi_{\ell m}(r) \,,
\end{equation}
where $~_{s}{S}_{\ell m}(\theta)e^{im \varphi} $ are the spin-weighted spherical harmonics, $\ell$ is the angular number ($\ell \geq 0$) and $m$ is the azimuthal number ($-\ell \leq m \leq \ell$) of the perturbation. In the following, we shall omit the $s,\ell,m$ subscripts for brevity. The radial component of the perturbation is governed by a Schr\"odinger-like equation~\cite{Regge:1957td,Zerilli:1970se}
\begin{equation}
    \frac{d^2 \psi(r)}{dr_*^2} + \left[ \omega^2 - V(r)\right] \psi(r) = 0 \,, \label{waveeq}
\end{equation}
where the tortoise coordinate is defined such that $dr_*/dr=1/f(r)$, i.e., 
\begin{equation}
  r_* = r+2M \log \left( \frac{r}{2M} -1\right)  \,.
\end{equation}
Let us notice that the tortoise coordinate allows us to explore a region in close proximity to the horizon of a BH since the tortoise coordinate is finite at the effective radius, i.e., $r_*(r_0) = \text{const} \equiv r_*^0$, and diverges at the would-be horizon, i.e., $r_*(2M) \to -\infty$.

\begin{figure}[t]
\centering
\includegraphics[width=0.7\textwidth]{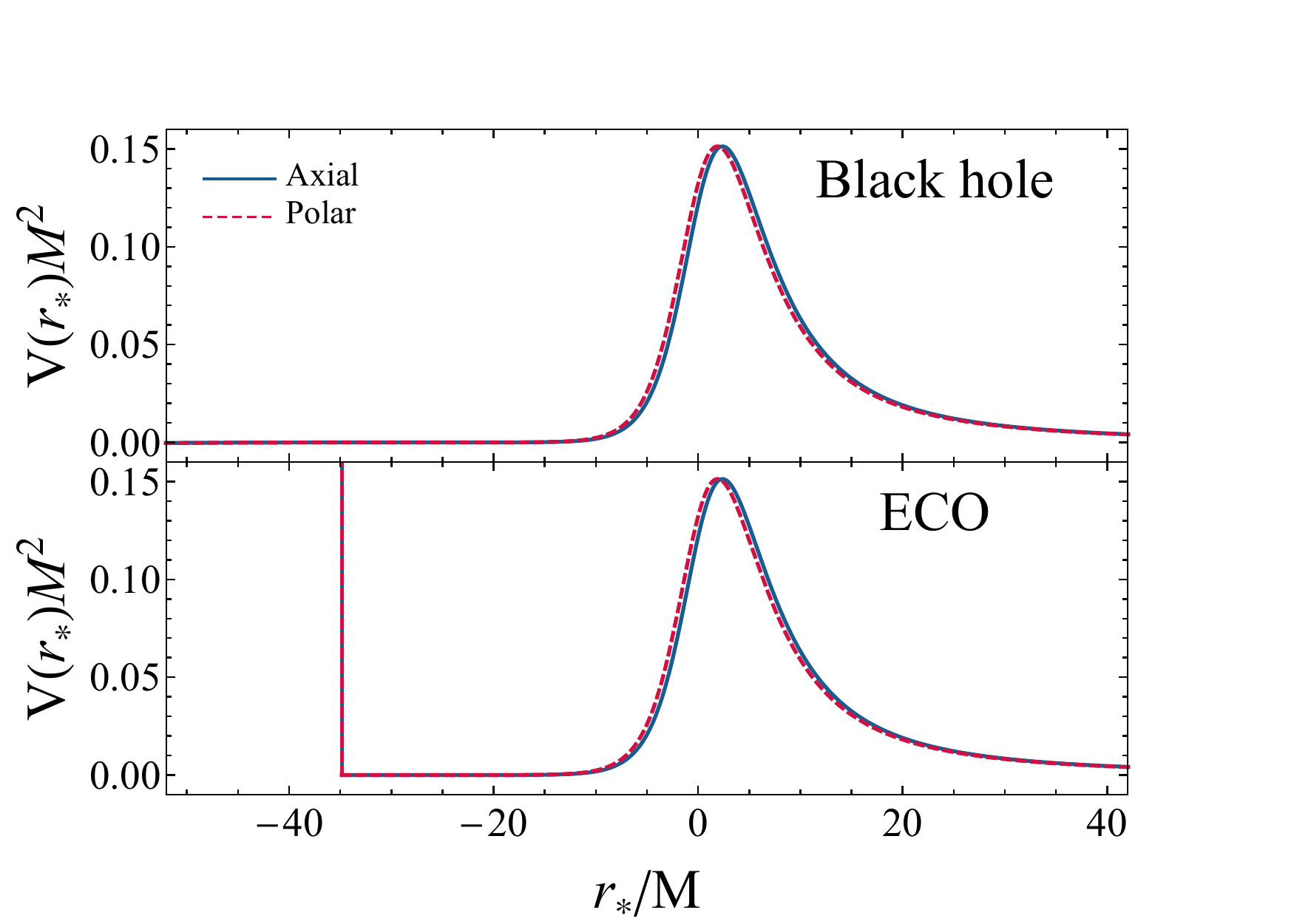}
\caption{Effective potential as a function of the tortoise coordinate of a Schwarzschild BH (top panel) and a static horizonless  compact object  with radius $r_0 = 2M(1+\epsilon)$ and $\epsilon=10^{-8}$ (bottom panel), for axial (continuous line) and polar (dashed line) $\ell=2$ gravitational perturbations. The effective potentials have a maximum approximately at the photon sphere, $r \approx 3M$. In the horizonless case, the effective potential features a characteristic cavity between the radius of the object and the photon sphere~\cite{Cardoso:2016rao,Cardoso:2016oxy,Cardoso:2019rvt,Maggio:2021ans}.} 
\label{fig:potential}
\end{figure}

The effective potential in Eq.~\eqref{waveeq} is~\cite{Regge:1957td,Zerilli:1970se}
\begin{eqnarray}
    V_{\rm axial} &=& f(r) \left[ \frac{\ell(\ell+1)}{r^2} + (1-s^2)\frac{2M}{r^3}\right] \,, \label{Vaxial} \\
    V_{\rm polar} &=& 2f(r) \left[ \frac{q^2 (q+1) r^3 + 3 q^2  M r^2 + 9 M^2 (q r + M)}{r^3 (qr+3M)^2} \right] \,, \label{Vpolar}
\end{eqnarray}
where $q=(\ell-1)(\ell+2)/2$. The potential in Eq.~\eqref{Vaxial} describes scalar, electromagnetic and axial gravitational perturbations, whereas the potential in Eq.~\eqref{Vpolar} describes polar gravitational perturbations. The tensor spherical harmonics can be classified according to their behavior under parity change, 
\begin{equation}
    \mathcal{P}\left(~_{s}{S}_{\ell m}(\theta) e^{im \varphi}\right) \to ~_{s}{S}_{\ell m}(\pi-\theta) e^{im \left(\pi + \varphi\right)} \,.
\end{equation}
In particular, we refer to axial perturbations as those with parity $(-1)^{\ell+1}$, whereas we refer to polar perturbations as those with parity $(-1)^{\ell}$. The former are described by the Regge-Wheeler wave function~\cite{Regge:1957td}, while the latter by the Zerilli wave function~\cite{Zerilli:1970se}.
Fig.~\ref{fig:potential} shows the effective potential as a function of the tortoise coordinate for axial and polar gravitational perturbations for a Schwarzschild BH (top panel) and for a static horizonless compact object (bottom panel) with $\epsilon=10^{-8}$.

Let us notice that the effective potential of a Schwarzschild BH tends to zero asymptotically both at infinity ($r_* \to +\infty$) and the horizon ($r_* \to -\infty$). As a consequence, the solution of the perturbation equation in Eq.~\eqref{waveeq} is a wave of frequency $\omega$ at the asymptotics both at infinity and the horizon. Furthermore, the effective potential displays a maximum located approximately at the photon sphere, $r \approx 3M$, that is the unstable circular orbit of photons around a Schwarzschild BH.

In the case of a horizonless ultracompact object ($\epsilon \ll 1$), the effective potential coincides with the one of a BH except for the presence of a radius at a constant $r_*^0$. The effective potential features a characteristic cavity between the radius of the object and the photon sphere. The cavity can support quasi-trapped modes that are responsible for a completely different QNM spectrum with respect to the BH case.

Let us emphasize that this description is valid when $\epsilon \ll 1$ and the effective potential is vanishing at the radius of the object, thus
the solution of Eq.~\eqref{waveeq} is a superposition of ingoing and outgoing waves at the radius of the object. Conversely, when $\epsilon \simeq 0.1-1$ (thus $r_0 \lesssim 3M$) the effective potential is not vanishing at the radius of the object and does not have an asymptotic trend, hence the solution of Eq.~\eqref{waveeq} is not a generic superposition of waves. We shall investigate the latter case in detail in Sec.~\ref{sec:membrane}.

\subsection{Boundary conditions}

The QNMs are the complex eigenvalues, $\omega = \omega_R + i \omega_I$, of the system given by Eq.~\eqref{waveeq} with two suitable boundary conditions. 
In our convention, a stable mode has $\omega_I < 0$ and corresponds to an exponentially damped sinusoidal signal with frequency $f \equiv \omega_R/(2 \pi)$ and damping time $\tau_{\rm damp} \equiv -1/\omega_I$. Conversely, an unstable mode has $\omega_I > 0$ with instability timescale $\tau_{\rm inst} \equiv 1/\omega_I$.

As a boundary condition, we impose that the perturbation is a purely outgoing wave at infinity, i.e.,
\begin{equation}
    \psi \sim e^{i \omega r_*} \,, \quad \text{as} \ r_* \to + \infty \,. \label{infBC}
\end{equation}
In the BH case, the horizon would require that the perturbation is a purely ingoing wave as $r_* \to -\infty$. In the case of a horizonless ultracompact object, the regularity at the center of the object implies the imposition of a boundary condition at the effective radius of the object. 
The perturbation can be decomposed a superposition of ingoing and outgoing waves at the radius of the object, i.e.,
\begin{equation}
    \psi \sim C_{\rm in} e^{-i \omega r_*} + C_{\rm out} e^{i \omega r_*} \,, \quad \text{as} \ r_* \to r_*^0 \,, \label{solr0}
\end{equation}
where we define the surface reflectivity of the object as~\cite{Maggio:2017ivp}
\begin{equation}
    \mathcal{R}(\omega) = \frac{C_{\rm out}}{C_{\rm in}} e^{2 i \omega r_*^0} \,. \label{R}
\end{equation}
Let us notice that, for a given wave function, $|\mathcal{R}(\omega)|^2$ defines the fraction of the reflected energy flux in units of the incident one at the radius of the object. Indeed, for $\epsilon \ll 1$ the imaginary part of the QNMs vanishes sufficiently fast that $|e^{2 i \omega r_*^0}|^2 \approx 1$ and $|\mathcal{R}(\omega)|^2 \approx |C_{\rm out}|^2/|C_{\rm in}|^2$.
The BH boundary condition is recovered for $\mathcal{R}=0$ and in the limit of $\epsilon \to 0$. Conversely, a perfectly reflecting compact object is described by $|\mathcal{R}(\omega)|^2=1$ where the outgoing energy flux at the effective radius of the object is equal to the incident one.

In the case of electromagnetic perturbations, a perfectly reflecting object can be modeled as a perfect conductor in which the electric and magnetic fields satisfy $E_\theta(r_0) = E_\varphi(r_0) = 0$ and $B_r(r_0) = 0$. The former conditions translate into
\begin{eqnarray}
    \psi(r_0) &=& 0 \quad \rm Dirichlet \ on \ axial \,, \label{dir} \\
    d\psi(r_0)/dr_* &=& 0 \quad \rm Neumann \ on \ polar \,, \label{neu}
\end{eqnarray}
where the Dirichlet boundary condition describes waves that are reflected with inverted phase ($\mathcal{R}(\omega)=-1$), whereas the Neumann boundary condition describes waves that are reflected in phase ($\mathcal{R}(\omega)=1$). The details of the derivation are given in Appendix~\ref{app:electromagneticBC}.

An analogous description of a perfectly reflecting compact object under gravitational perturbations is not available. 
We assume that the results of electromagnetic perturbations can be applied to gravitational perturbations, in which case Dirichlet and Neumann boundary conditions are imposed on axial and polar gravitational perturbations, respectively.

\subsection{Numerical procedure} \label{sec:numerics}

Equation~\eqref{waveeq} with boundary conditions at infinity in Eq.~\eqref{infBC} and at the radius of the compact object in Eq.~\eqref{dir} or~\eqref{neu} can be solved numerically with a \emph{direct integration shooting method}~\cite{Pani:2013pma}. The method starts with an analytical high-order series expansion of the solution at large distances. We use the ansatz
\begin{equation}
    \psi(r) = e^{i \omega r_*} \sum_{i=0}^{\infty} \frac{R_{\rm inf}^{(i)}}{r^i} \,, \label{BCinfDI}
\end{equation}
where the coefficients $R_{\rm inf}^{(i)}$ with $i=1, ..., \infty$ are computed by solving Eq.~\eqref{waveeq} in the large distance limit order by order, and the coefficients $R_{\rm inf}^{(i)}$ are functions of $R_{\rm inf}^{(0)}$. For simplicity, we set $R_{\rm inf}^{(0)}=1$. A high truncation order of the series expansion ($i \gtrsim 10$) is needed for the numerical stability of the solution. 

Eq.~\eqref{waveeq} is integrated with the boundary condition in Eq.~\eqref{BCinfDI} from infinity inwards up to $r = r_0$. The integration is repeated for different values of the complex frequency starting from an initial guess until the boundary condition at the radius of the object (either Eq.~\eqref{dir} or  Eq.~\eqref{neu}) is satisfied. The resulting QNM should not depend on the numerical parameters of the method, i.e., the numerical value that stands for the infinity and the truncation order of the series expansion at infinity.
The direct integration shooting method is robust when the imaginary part of the mode is sufficiently small with respect to the real part of the mode. Typically, this method allows us to compute the fundamental mode and possibly the first few overtones. 

An alternative method is based on the \emph{continued fraction technique},
where the eigenfunction is written as a series whose coefficients satisfy a finite-term recurrence relation~\cite{Leaver:1985ax}. The QNMs are the roots of $n$ implicit equations $f_n(\omega)=0$, where $n$ is the inversion index of the continued fraction. For a given $n$, the method gives some spurious roots apart from the physical QNMs. The spurious roots can be ruled out since they are not present by changing the numerical parameters of the method, i.e., the inversion index of the continued fraction. This method was derived by Leaver to compute the QNMs of Kerr BHs~\cite{Leaver:1985ax}. Appendix~\ref{app:CF} contains a generalization of the method to compact objects. The continued-fraction method is also robust for overtones with a large imaginary part of the frequency for which the direct integration fails. When they both are applicable, the two methods are in excellent agreement
within the numerical accuracy that is chosen to find the QNMs.

\subsection{Black hole vs horizonless compact object spectrum}
\label{sec:lowfrequencies}

When normalized by the mass, the QNMs of the system depend on two continuous, dimensionless parameters: the closeness parameter from the horizon of a Schwarzschild BH, $\epsilon$, and the surface reflectivity of the object, $\mathcal{R}(\omega)$. Furthermore, the QNMs depend on some integer numbers, namely the spin $s$, the angular number $\ell$ and the overtone number $n$ of the perturbation. In the following, we shall focus on the gravitational ($s=-2$) $\ell = 2$ fundamental mode ($n = 0$) that corresponds to the mode with the smallest imaginary part, i.e., with the largest damping time.

\begin{figure}[t]
\centering
\includegraphics[width=0.7\textwidth]{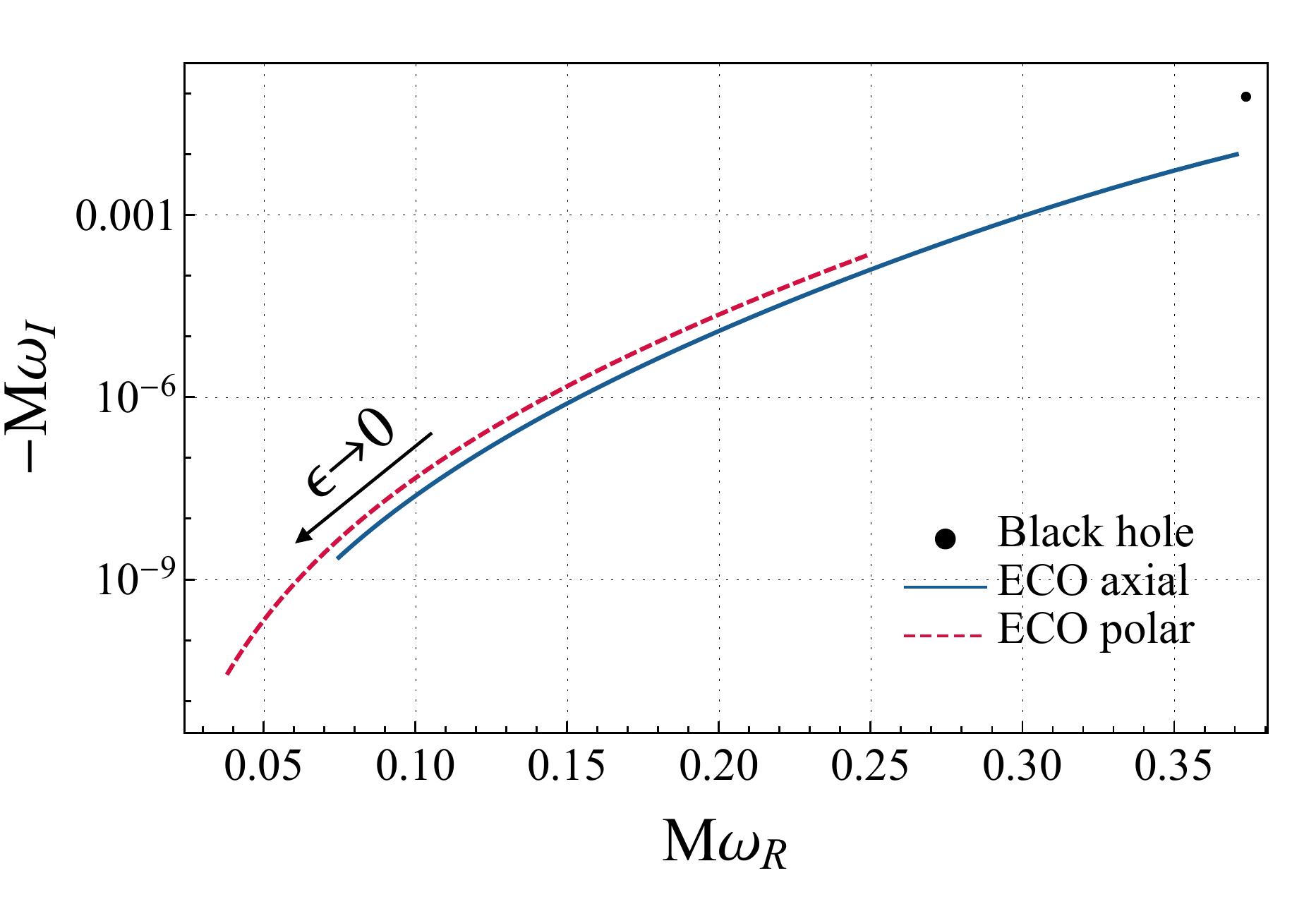}
\caption{QNM spectrum of a perfectly reflecting ECO with radius $r_0=2M(1+\epsilon)$ and $\epsilon \in (10^{-10},10^{-2})$, compared to the fundamental $\ell=2$ gravitational QNM of a Schwarzschild BH. Axial and polar modes are not isospectral at variance with the BH case. As $\epsilon \to 0$, the ECO QNMs are low-frequencies and long-lived~\cite{Cardoso:2016rao,Maggio:2021ans}.}
\label{fig:lowfrequencyQNMs}
\end{figure}
Fig.~\ref{fig:lowfrequencyQNMs} shows the QNM spectrum of a perfectly reflecting ECO compared to the fundamental $\ell=2$ QNM of a Schwarzschild BH, i.e.,
\begin{equation}
 M \omega_{\rm BH} = M (\omega_{R, \rm BH} + i \omega_{I, \rm BH}) = 0.37367 - i 0.088962 \,.   \label{BHQNM}
\end{equation}
The QNM spectrum of the ECO is derived by imposing the boundary conditions in Eqs.~\eqref{dir} and~\eqref{neu} on axial and polar perturbations, respectively. The radius of the compact object is located as in Eq.~\eqref{radius}, where $\epsilon \in (10^{-10},10^{-2})$ from the left to the right of the plot. As shown in Fig.~\ref{fig:lowfrequencyQNMs}, an important feature of ECOs is the breaking of \emph{isospectrality} between axial and polar modes unlike BHs in GR. Indeed, Schwarzschild BHs have a unique QNM spectrum despite the Regge-Wheeler potential for axial perturbations in Eq.~\eqref{Vaxial} is different from the Zerilli potential for polar perturbations in Eq.~\eqref{Vpolar}. The isospectrality can be demonstrated by showing that the Regge-Wheeler and Zerilli wave functions are related by a Darboux transformation~\cite{Chandrasekhar:1975nkd,Chandrasekhar:1975zza,Chandra,Glampedakis:2017rar}
\begin{equation} 
 \psi_{\rm RW} = A \frac{d \psi_{\rm Z}}{dr_*} + B(r) \psi_{\rm Z} \,, \label{RWfromZ}
\end{equation}
where
\begin{eqnarray}
    A &=& -M \left[ i \omega M + \frac{1}{3} q (q+1)\right]^{-1}\,, \\
    B(r) &=& \frac{q (q+1) (qr+3M)r^2 + 9M^2(r-2M)}{r^2(qr+3M)[q(q+1)+3i\omega M]}\,.
\end{eqnarray}
At the BH horizon, both the Regge-Wheeler and the Zerilli wave functions are purely ingoing. Conversely, at the effective radius of the horizonless compact object, the boundary conditions are mapped differently from Eq.~\eqref{RWfromZ} since the Regge-Wheeler and the Zerilli wave functions are a superposition of waves as in Eq.~\eqref{solr0}.

Fig.~\ref{fig:lowfrequencyQNMs} also shows that in the BH limit ($\epsilon \to 0$) the deviations from the BH QNM are arbitrarily large and the QNMs are low frequencies, i.e., $M \omega_R \ll M \omega_{R, \rm BH}$, and long-lived, i.e., $\tau_{\rm damp} \gg 1$~\cite{Cardoso:2016rao}. For example, for $\epsilon=10^{-10}$ the fundamental $\ell=2$ QNMs of a perfectly reflecting ECO are
\begin{eqnarray}
M \omega_{\rm axial} = 0.074698 - i 2.2992 \times 10^{-9} \,, \\
M \omega_{\rm polar} = 0.037914 - i 2.7385 \times 10^{-11} \,.
\end{eqnarray}
Low-frequency QNMs are a peculiar feature of horizonless compact objects whose compactness is similar to the BH one. These modes can be understood in terms of quasi-trapped modes between the effective radius of the object and the photon sphere barrier, as shown in Fig.~\ref{fig:potential}. 
The real part of the QNMs scales as the width of the cavity of the effective potential, i.e., $\omega_R \sim 1/r_*^0$; whereas the imaginary part of the QNMs is given by the modes that tunnel through the potential barrier and reach infinity, i.e., $\omega_I \sim |\mathcal{A}|^2/r_*^0$ where $|\mathcal{A}|^2$ is the tunneling probability. 
For $\epsilon \ll 1$, the QNMs can be derived analytically in the low-frequency regime as~\cite{Maggio:2018ivz,Cardoso:2019rvt}
\begin{eqnarray}
 \omega_R &\simeq& - \frac{\pi}{2 |r_*^0|} \left[ p + \frac{s(s+1)}{2}\right] \sim |\log \epsilon|^{-1} \,, \label{MomegaRa0} \\
 \omega_I &\simeq& - \frac{\beta_{s \ell}}{|r_*^0|} \left( 2 M \omega_R\right)^{2 \ell +2} \sim -|\log \epsilon|^{-(2\ell +3)} \,, \label{MomegaIa0}
\end{eqnarray}
where $\sqrt{\beta_{s \ell}} = \frac{(\ell-s)! (\ell+s)!}{(2\ell)! (2\ell+1)!!}$ and $p$ is a positive odd (even) integer for polar (axial) modes. A detailed derivation of Eqs.~\eqref{MomegaRa0} and ~\eqref{MomegaIa0} is given in Appendix~\ref{app:analytics} for their generalization to the spinning case.

\section{Membrane paradigm for compact objects}\label{sec:membrane}

In Sec.~\ref{sec:QNMspectrum}, we derived the QNM spectrum of static ultracompact objects whose effective radius is located at $r_0 = 2M (1+\epsilon)$ with $\epsilon \ll 1$.  To derive the QNM spectrum of horizonless objects with different compactness and interior solutions, we make use of the BH membrane paradigm and generalize it to the case of horizonless objects.
The membrane paradigm allows us to describe any compact object with a Schwarzschild exterior where no specific model is assumed for the object interior. GR is assumed to work sufficiently well at the radius of the compact object. This assumption is also justified in theories of gravity with higher-curvature/high-energy corrections to GR. In this case, the corrections to the metric are suppressed by powers of $l_P /r_0 \ll 1$, where $r_0$ is the object radius, and $l_P$ is the Planck length or the scale of new physics.
The membrane paradigm allows us to derive the QNMs of gravastars, wormholes, nonlocal stars, anisotropic stars, etc., after fixing the (possibly frequency-dependent) viscosity of the fictitious membrane according to the model.

\subsection{Setup}

%
\begin{figure}[t]
\centering
\includegraphics[width=0.65\textwidth]{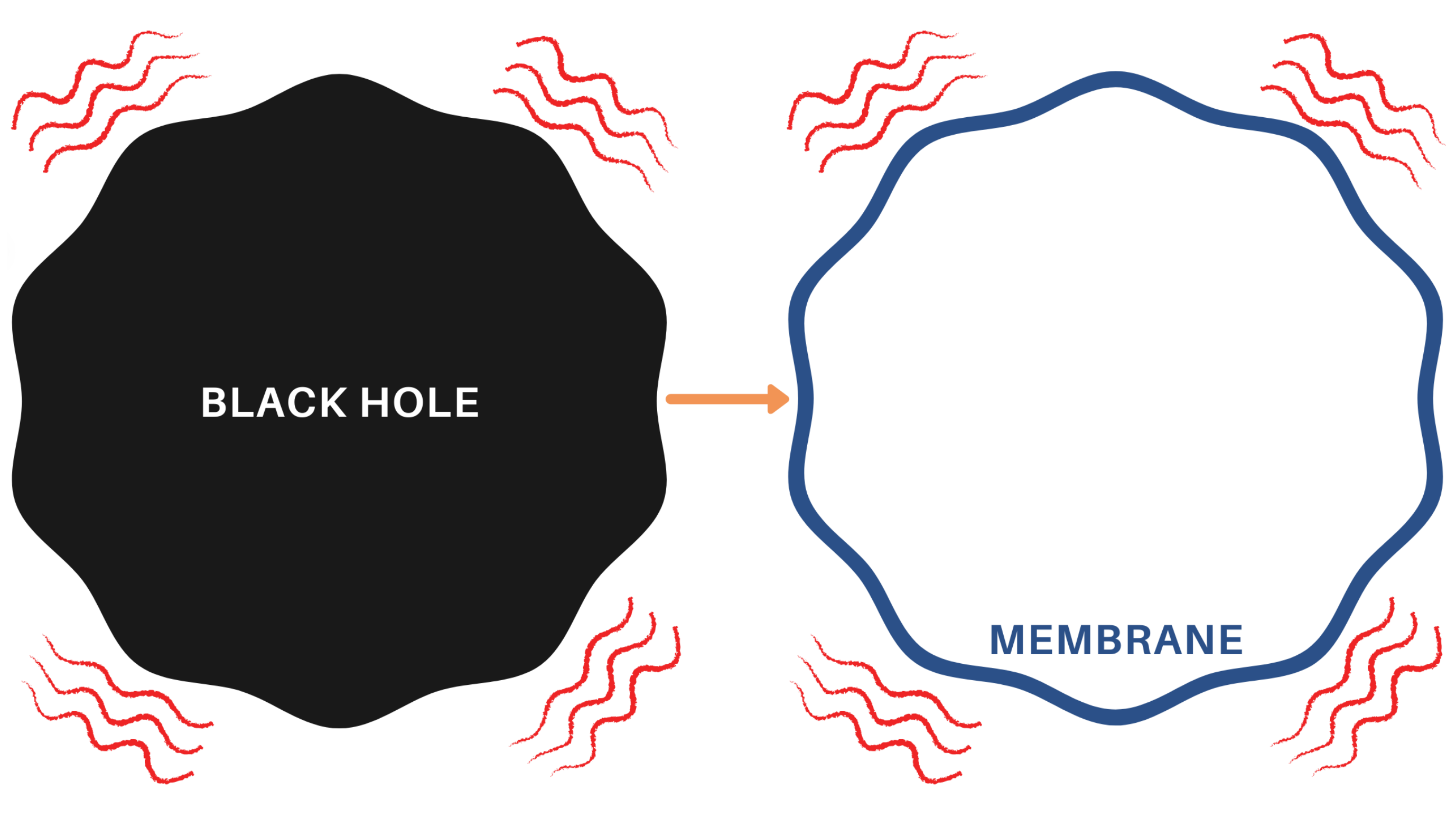}
\caption{Schematic representation of the BH membrane paradigm~\cite{Damour:1982,MembraneParadigm}. A static observer outside the horizon can replace the interior of a perturbed BH (left panel) with a perturbed fictitious membrane located at the horizon (right panel). The fictitious membrane is a viscous fluid whose properties (density, pressure, viscosity) are such that the BH phenomenology, particularly the QNM spectrum, is reproduced.~\cite{Maggio:2020jml}} 
\label{fig:membrane}
\end{figure}
According to the BH membrane paradigm, a static observer outside the horizon can replace the interior of a perturbed BH by a perturbed \emph{fictitious} membrane located at the horizon~\cite{Damour:1982,Price:1986yy,MembraneParadigm} (see Fig.~\ref{fig:membrane}). The features of the interior spacetime are mapped into the properties of the membrane that are fixed by the Israel-Darmois junction conditions~\cite{Darmois:1927,Israel:1966rt}
\begin{equation} \label{junction}
    [[K_{ab} - K h_{ab}]]=-8 \pi T_{ab} \,, \quad [[h_{ab}]]=0 \,,
\end{equation}
where $[[...]]=(...)^+ - (...)^-$ denotes the jump of a quantity across the membrane, $\mathcal{M}^+$ and $\mathcal{M}^-$ are the exterior and the interior spacetimes to the membrane, $h_{ab}$ is the induced metric on the membrane, $K_{ab}$ is the extrinsic curvature, $K=K_{ab} h^{ab}$, and $T_{ab}$ is the stress-energy tensor of the membrane. 

The fictitious membrane is such that the extrinsic curvature of the interior spacetime vanishes, i.e., $K_{ab}^-=0$~\cite{MembraneParadigm}. As a consequence, the junction conditions impose that the fictitious membrane is a viscous fluid with stress-energy tensor
\begin{equation} \label{stressenergytensor}
    T_{ab} = \rho u_a u_b + (p- \zeta \Theta) \gamma_{ab} -2 \eta \sigma_{ab} \,,
\end{equation}
where $\eta$ and $\zeta$ are the shear and bulk viscosities of the fluid, $\rho$, $p$ and $u_a$ are the density, pressure and 3-velocity of the fluid, $\Theta=u^a_{;a}$ is the expansion, $\sigma_{ab}=\frac{1}{2} \left(u_{a;c} \gamma_b^c + u_{b;c} \gamma_a^c - \Theta \gamma_{ab}\right)$ is the shear tensor, $\gamma_{ab}=h_{ab} + u_a u_b$ is the projector tensor, and the semicolon is the covariant derivative compatible with the induced metric, respectively.

The BH membrane paradigm allows us to describe the interior of a perturbed BH in terms of the shear and the bulk viscosities of a fictitious viscous fluid located at the horizon, where
\begin{equation}
    \eta_{\rm BH} = \frac{1}{16 \pi} \,, \quad \zeta_{\rm BH} = -\frac{1}{16 \pi} \,.
\end{equation}
The generalization of the BH membrane paradigm to horizonless compact objects allows us to describe several models of ECOs with different interior solutions with an exterior Schwarzschild spacetime in terms of the properties of a fictitious membrane located at the ECO radius~\cite{Abedi:2020ujo,Maggio:2020jml}. The details on the calculations are given in Appendix~\ref{app:membrane}.
The shear and the bulk viscosities of the fluid are generically complex and frequency-dependent and are related to the reflective properties of the ECO. For each model of ECO, the shear and the bulk viscosities are uniquely determined. In the following, we shall focus on the case in which $\eta$ and $\zeta$ are real and constant since the energy dissipation is absent when $\mathfrak{R}(\eta) = \mathfrak{R}(\zeta) = 0$.

\subsection{Boundary conditions}

Gravitational perturbations in the exterior Schwarzschild spacetime are governed by the Schr\"odinger-like equation in Eq.~\eqref{waveeq}, where the effective potential is in Eqs.~\eqref{Vaxial} and~\eqref{Vpolar} for axial and polar perturbations, respectively.

By imposing boundary conditions at infinity and the radius of the compact object, Eq.~\eqref{waveeq} defines the complex QNMs of the system. We  impose that the perturbation is a purely outgoing wave at infinity, whereas the condition on the inner boundary would depend on the properties of the object. We rely on the membrane paradigm to derive the boundary condition at the radius of the compact object without assuming any specific model of ECO. As detailed in Appendix~\ref{app:membrane}, the boundary conditions at the ECO radius are~\cite{Maggio:2020jml}
\begin{eqnarray}
\frac{d \psi(r_0)/dr_*}{\psi(r_0)} &=& - \frac{i \omega}{16 \pi \eta} - \frac{r_0^2 V_{\rm axial}(r_0)}{2(r_0-3M)} \,, \qquad \text{axial} \,, \label{BC-axial}\\
\frac{d \psi(r_0)/dr_*}{\psi(r_0)} &=& - 16 \pi i \eta \omega + G(r_0,\omega,\eta,\zeta) \,, \quad \text{polar} \,, \label{BC-polar}
\end{eqnarray}
where $G(r_0,\omega,\eta,\zeta)$ is a cumbersome function given in Appendix~\ref{app:membrane}. 
Let us notice that in the BH limit ($r_0 \to 2M$) the boundary conditions in Eqs.~\eqref{BC-axial} and~\eqref{BC-polar} reduce to the BH boundary condition of a purely ingoing wave at the horizon as $\eta \to \eta_{\rm BH}$. This result agrees with the standard BH membrane.

The boundary conditions in Eqs.~\eqref{BC-axial} and~\eqref{BC-polar} allow us to describe several models of ECOs in terms of the shear and bulk viscosities of the fictitious membrane located at the radius of the object.
For example, ultracompact thin-shell wormholes with Dirichlet (Neumann) boundary conditions~\cite{Cardoso:2016rao} are described by $\eta=0$ ($\eta \to \infty$). Whereas, ultracompact thin-shell gravastars~\cite{Pani:2009ss} are described by a complex and frequency-dependent shear viscosity that is expressed in terms of hypergeometric functions
\begin{eqnarray}
\eta &=& \frac{1}{16 \pi} - \frac{i \epsilon}{64 \pi M \omega} \Bigg[8 - 2 \ell^2 + 2 i M \omega + \left(1+ \ell +2iM\omega\right)
\left(2+\ell+2iM\omega\right) \nonumber \\ 
&\times& \frac{~_{2}F_1\left(\frac{1}{2}(3+\ell+2iM\omega),\frac{1}{2}(4+\ell+2iM\omega);\frac{5}{2}+\ell;1\right)}{~_{2}F_1\left(\frac{1}{2}(1+\ell+2iM\omega),\frac{1}{2}(2+\ell+2iM\omega);\frac{3}{2}+\ell;1\right)} \Bigg] + \mathcal{O}(\epsilon^2) \,.
\end{eqnarray}

In particular, the axial sector depends only on the shear viscosity of the membrane, whereas the polar sector depends also on the bulk viscosity of the fictitious fluid. In the BH limit, the dependence on the bulk viscosity disappears $\left(G(2M)=0\right)$ therefore the parameter for the bulk viscosity is not fixed by the linear perturbation analysis (see Appendix~\ref{app:membrane} for details).

\subsection{Effective reflectivity of compact objects} \label{sec:effectiverefl}

According to the membrane paradigm, the effective reflectivity of compact objects is mapped into the shear and bulk viscosities of the fictitious fluid located at the radius of the object.
To illustrate their relation, we compute the effective reflectivity of the spacetime through the scattering of a wave coming from infinity and being partially reflected after being subjected to the boundary conditions in Eqs.~\eqref{BC-axial} and~\eqref{BC-polar} at $r=r_0$, i.e., 
\begin{equation}
    \psi \sim e^{-i \omega r_*} + \mathsf{R} e^{i \omega r_*} \,, \quad r_* \to \infty \,. \label{Rscattering}
\end{equation}
Let us notice that the effective reflectivity at infinity defined in Eq.~\eqref{Rscattering} is different from the surface reflectivity defined in Eq.~\eqref{R} at the radius of ultracompact objects.

In the large-frequency limit ($M \omega \gg 1$), the potential in Eq.~\eqref{waveeq} can be neglected and the effective reflectivity reads
\begin{equation}
 |\mathsf{R}|^2 = \left( \frac{1-\eta/\eta_{\rm BH}}{1+\eta/\eta_{\rm BH}} \right)^2 \,. \label{Rhighf}
\end{equation}
Eq.~\eqref{Rhighf} shows that a compact object is a perfect absorber of high-frequency waves ($|\mathsf{R}|^2=0$) when $\eta=\eta_{\rm BH}$, whereas it is a perfect reflector of high-frequency waves ($|\mathsf{R}|^2=1$) when either $\eta=0$ or $\eta \to \infty$.

In the case of horizonless ultracompact objects with $\epsilon \ll 1$, the effective reflectivity at infinity in Eq.~\eqref{Rhighf} coincides with the surface reflectivity of the object when the latter does not have an explicit dependence on the frequency, i.e., $|\mathsf{R}|^2 = |\mathcal{R}|^2$. For $\eta=0$, the ultracompact object is perfectly reflecting ($|\mathsf{R}|^2=1$) and the boundary conditions in Eqs.~\eqref{BC-axial} and~\eqref{BC-polar} reduce to Dirichlet and Neumann boundary conditions on axial and polar modes in Eqs.~\eqref{dir} and~\eqref{neu}, respectively. 
Also for $\eta \to \infty$, the ultracompact object is perfectly reflecting.

Although $\eta$ is formally a free parameter, we  expect the most interesting range to be $\eta \in [0,\eta_{\rm BH}]$.
Indeed, from Eq.~\eqref{Rhighf} negative values of $\eta$ would correspond to $|\mathsf{R}|^2 > 1$ that would lead to superradiant instabilities~\cite{Brito:2015oca}. Similarly, for $\eta > \eta_{\rm BH}$ the effective reflectivity is a growing function of the shear viscosity, which is unphysical. For this reason, partially absorbing ultracompact objects are analyzed by considering $\eta \in (0,\eta_{\rm BH})$.

\begin{figure}[t]
\centering
\includegraphics[width=0.49\textwidth]{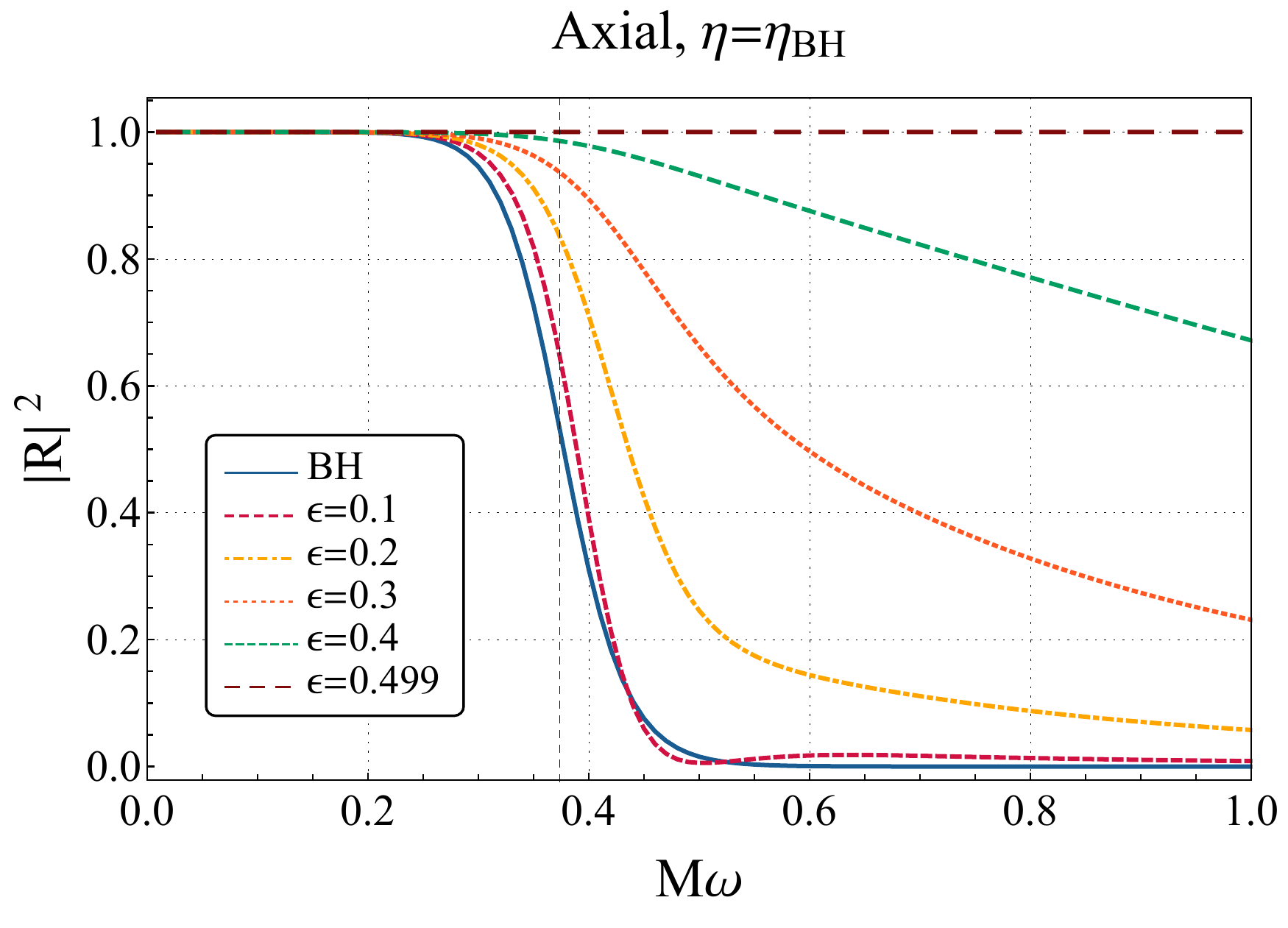}
\includegraphics[width=0.49\textwidth]{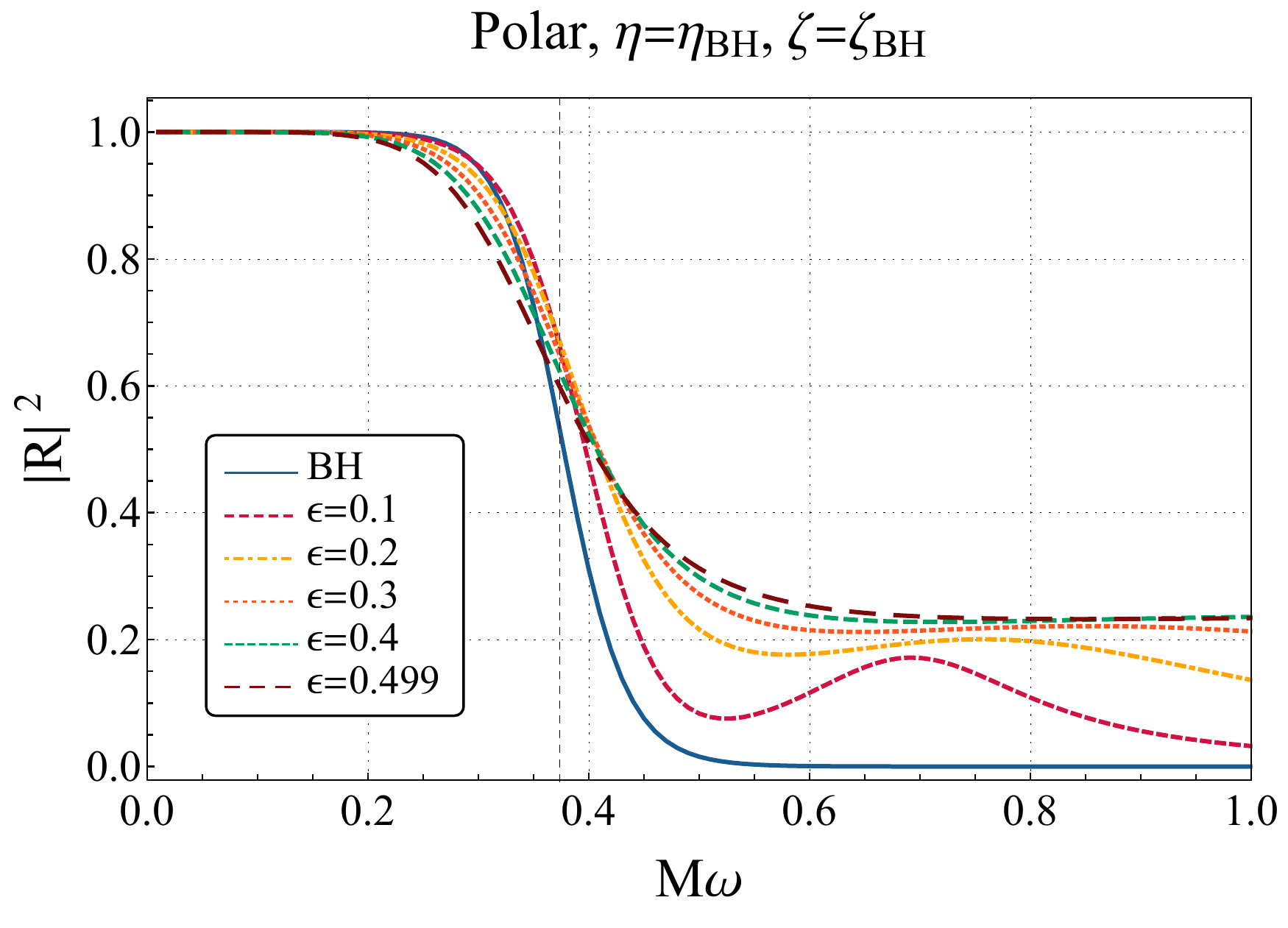}
\caption{Effective reflectivity of compact objects for axial (left panel) and polar (right panel) gravitational perturbations as a function of the frequency. The radius of the compact object is located at $r_0=2M(1+\epsilon)$, and the shear and bulk viscosities of the membrane are $\eta=\eta_{\rm BH}$ and $\zeta=\zeta_{\rm BH}$, respectively.
In both panels, the vertical dashed line corresponds to the fundamental QNM frequency of a Schwarzschild BH, i.e., $M \omega_{R,\rm{BH}} \sim 0.37367$.~\cite{Maggio:2020jml}} 
\label{fig:refl}
\end{figure}
We  compute the effective reflectivity in Eq.~\eqref{Rscattering} for generic frequencies numerically. Fig.~\ref{fig:refl} shows the effective reflectivity of compact objects with different radii compared to the BH reflectivity as a function of the frequency. The left (right) panel shows the effective reflectivity for axial (polar) gravitational perturbations with shear and bulk viscosities $\eta=\eta_{\rm BH}$ and $\zeta=\zeta_{\rm BH}$, respectively.

Interestingly, as the ECO radius approaches the photon sphere ($\epsilon \to 1/2$) the effective reflectivity tends to unity in the axial sector for any frequency. This distinctive feature can be understood by noticing that
the axial boundary condition in Eq.~\eqref{BC-axial} reduces to $\psi(r_0)=0$ as $r_0 \to 3M$ for any complex $\eta$. As a consequence, any ECO with $r_0=3M$ is a \emph{perfect reflector} of axial GWs regardless of the interior structure\footnote{The only exception is when $\eta \to -\frac{3 i \omega}{16 \pi q}(r_0-3M)$ as $r_0 \to 3M$, in
which case the divergence in Eq.~\eqref{BC-axial} cancels out. This peculiar case corresponds to thin-shell gravastars~\cite{Visser:2003ge}.}. The same universality does not occur in the polar sector.

Let us also notice that the effective reflectivity at intermediate frequencies ($M \omega = \mathcal{O}(0.1-1)$) can be larger than the BH reflectivity. For example, at the fundamental QNM frequency of a Schwarzschild BH $M \omega_{R,\rm{BH}} \sim 0.37367$, shown in Fig.~\ref{fig:refl} as a dashed vertical line, the effective reflectivity is about unity for $\epsilon=0.4$ in the axial sector. This effect has relevant consequences in the QNM spectrum of compact objects with respect to the BH QNM spectrum, as we shall discuss in Sec.~\ref{sec:QNMmembrane}.

\subsection{Quasi-normal mode spectrum} \label{sec:QNMmembrane}

%
\begin{figure}[t]
\centering
\includegraphics[width=0.49\textwidth]{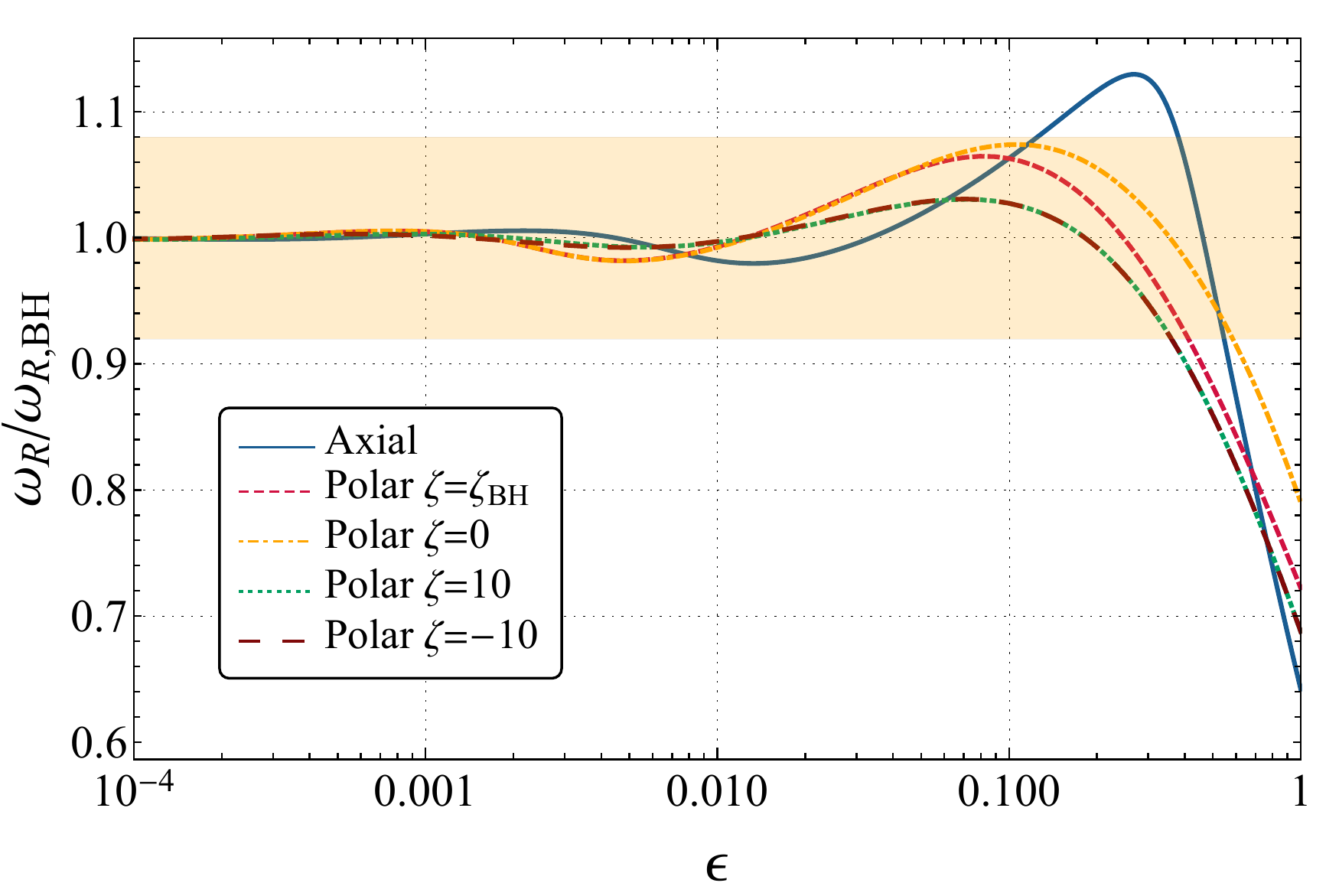}
\includegraphics[width=0.49\textwidth]{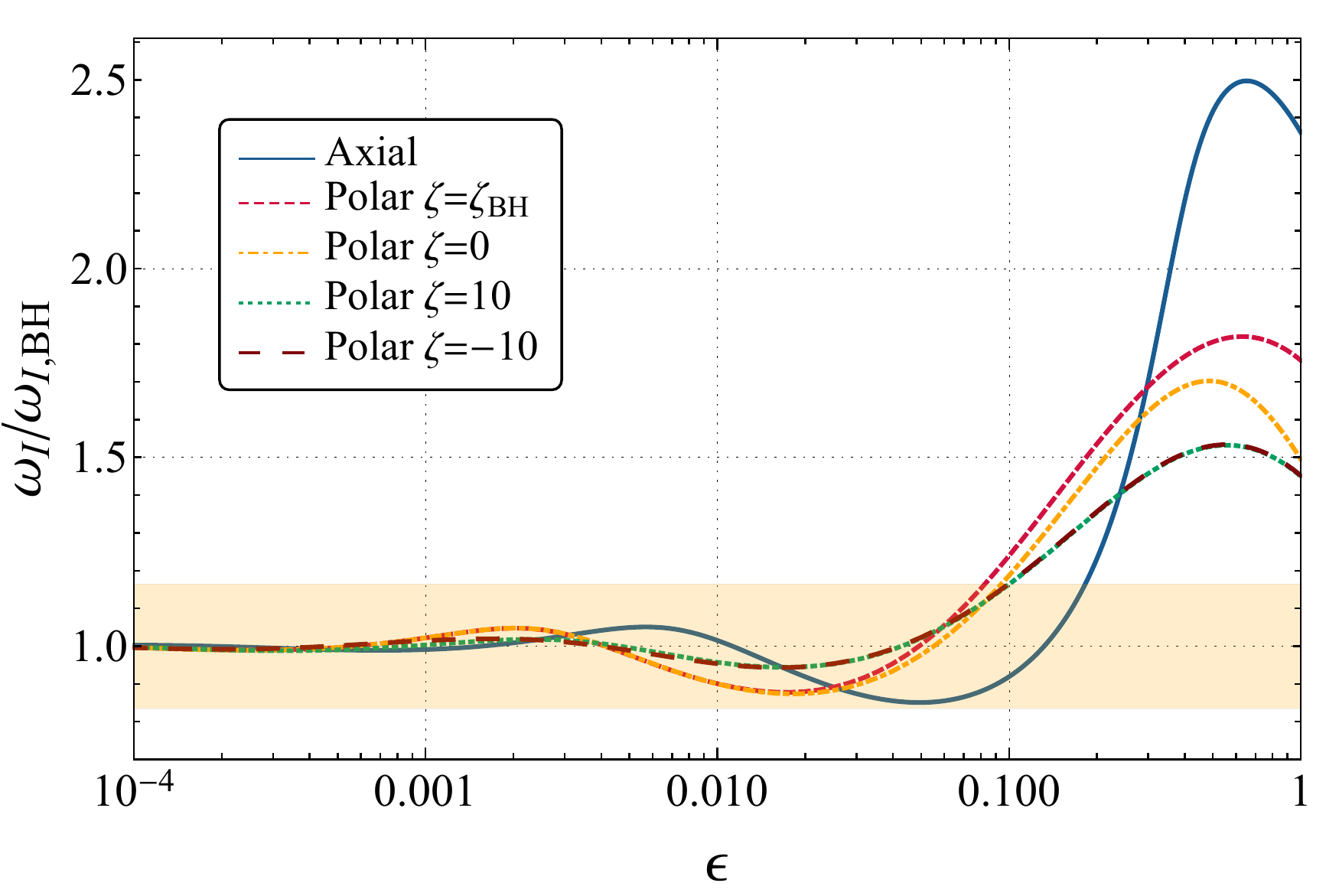}
\caption{Real (left panel) and imaginary (right panel) part of the QNMs of an ECO described by a fictitious fluid with shear viscosity $\eta=\eta_{\rm BH}$ and compared to the fundamental $\ell=2$ gravitational QNM of a Schwarzschild BH, as a function of the closeness parameter $\epsilon$ where the ECO radius is located at $r_0 = 2M(1+\epsilon$). The highlighted region corresponds to the maximum deviation (with $90\%$ credibility) for the least-damped QNM in the event GW150914 to the Kerr BH case~\cite{Ghosh:2021mrv}. Horizonless compact objects with $\epsilon \lesssim 0.1$ are compatible with current measurement accuracies.}
\label{fig:QNMsmembrane}
\end{figure}
Equation~\eqref{waveeq} with boundary conditions at infinity in Eq.~\eqref{infBC} and at the radius of the compact object in Eqs.~\eqref{BC-axial} and~\eqref{BC-polar} for the axial and polar sector, respectively, can be solved numerically to derive the QNM spectrum of a static horizonless compact object. 
When normalized by the mass, the QNMs of the object depend on three integers, i.e., the spin $s$, the angular number $\ell$, and the overtone number $n$ of the perturbation.
The QNM spectrum also depends on some continuous parameters that are related to the properties of the system, i.e., its compactness through the parameter $\epsilon$ as in Eq.~\eqref{compactnessECO}.
In the axial sector, the QNMs depend on the shear viscosity of the membrane $\eta$, whereas in the polar sector there is an additional dependence on the bulk viscosity of the membrane $\zeta$. 

We  compute the QNM spectrum with two numerical methods: a direct integration shooting method, as described in Sec.~\ref{sec:numerics}, and a method based on continued fractions, as described in Appendix~\ref{app:CF}. The continued fraction method is more robust than the direct integration for overtones with a large imaginary part of the frequency. When they both are applicable, we checked that the two methods are in excellent agreement.

Let us first analyze the QNM spectrum of a horizonless compact object with $\eta=\eta_{\rm BH}$. 
Fig.~\ref{fig:QNMsmembrane} shows the ratio of the real (left panel) and imaginary (right panel) part of the ECO QNMs to the fundamental $\ell=2$ QNM of a Schwarzschild BH, as a function of $\epsilon$. As $\epsilon \to 0$, the horizonless compact object has the same QNM spectrum of a Schwarzschild BH. This is because, as $\epsilon \to 0$, the compct object has the same reflective properties of a BH for $\eta=\eta_{\rm BH}$. 
For larger values of $\epsilon$, the compactness of the object decreases and the QNMs start deviating from the BH QNM. 
Let us notice that the isospectrality of axial and polar modes in BHs is broken for finite values of $\epsilon$. In the case of horizonless objects, the fundamental $\ell=2$ modes form a characteristic \emph{doublet}.
Polar modes show a mild dependence on the bulk viscosity of the membrane. In the large-$\zeta$ limit, the QNM spectrum is independent of the bulk viscosity, as shown by the $\zeta=\pm 10$ curves in  Fig.~\ref{fig:QNMsmembrane}.

\begin{figure}[t]
\centering
\includegraphics[width=0.7\textwidth]{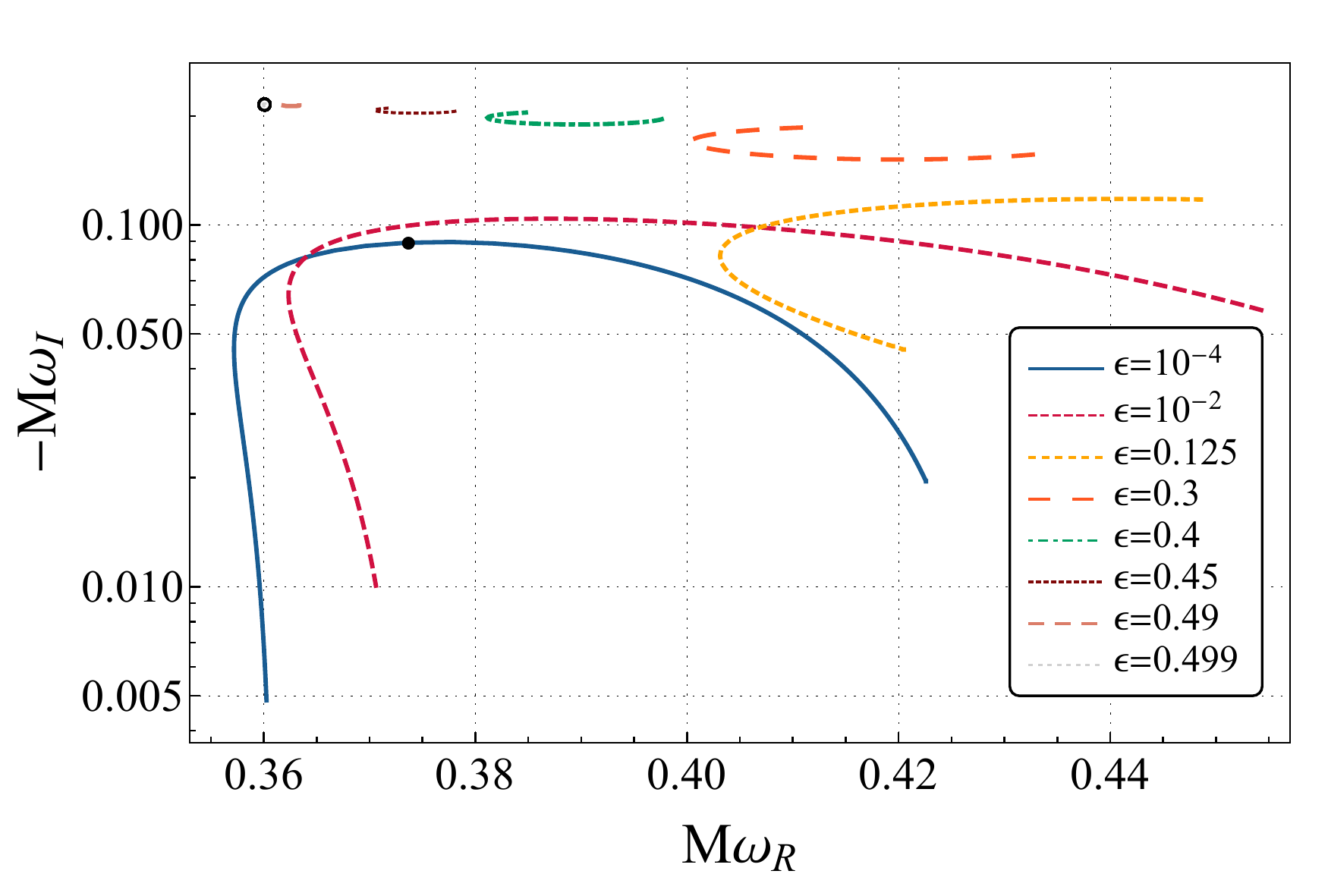}
\caption{The complex QNM plane of a horizonless compact object for axial perturbations. Each curve is the QNM spectrum of a compact object with a given radius $r_0=2M(1+\epsilon)$ parametrized by the shear viscosity of the membrane, $\eta \in [10^{-4},100]$. As the radius of the object approaches the photon sphere ($\epsilon \to 1/2$), the curves converge to a universal QNM (marked with an empty black circle) regardless of the value of $\eta$. As a reference, the fundamental QNM of a Schwarzschild BH is marked by a black dot.~\cite{Maggio:2020jml}}
\label{fig:universalQNM}
\end{figure}
The highlighted regions in Fig.~\ref{fig:QNMsmembrane} correspond to the maximum allowed deviation (with $90\%$ credibility) for the least-damped QNM in the event GW150914, and correspond to $\sim 16\%$ and $\sim 33\%$ for the real and imaginary part of the QNM, respectively~\cite{Ghosh:2021mrv}. Remarkably, Fig.~\ref{fig:QNMsmembrane} shows that horizonless compact objects with $\epsilon \lesssim 0.1$ are compatible with current measurement accuracies. Future ringdown detections would allow us to set more stringent constraints on the radius of compact objects.

Let us now change the reflective properties of the compact object via the parameter $\eta$. Fig.~\ref{fig:universalQNM} shows the complex QNM plane 
of a horizonless compact object under axial perturbations for several values of $\epsilon$. Each curve is parametrized by the shear viscosity of the membrane where $\eta \in [10^{-4},100]$. As a reference, the fundamental QNM of a Schwarzschild BH is marked by a black dot corresponding to $\eta=\eta_{\rm BH}$ and $\epsilon = 0$.
As the location of the radius of the object approaches the photon sphere ($\epsilon \to 1/2$), the axial QNMs become independent of $\eta$. Indeed, the QNMs tend to a universal mode (marked by an empty black circle in  Fig.~\ref{fig:universalQNM}) which, for $\ell = 2$, reads 
\begin{equation}
    M \omega_{\rm axial} \sim 0.3601 - i 0.2149 \,, \quad \epsilon \to 1/2 \,, \label{universalmode}
\end{equation}
As previously discussed, in this limit the object is a perfect reflector of axial GWs, regardless of the value of $\eta$. This remarkable universality does not apply to the case of polar perturbations.

\begin{figure}[t]
\centering
\includegraphics[width=0.68\textwidth]{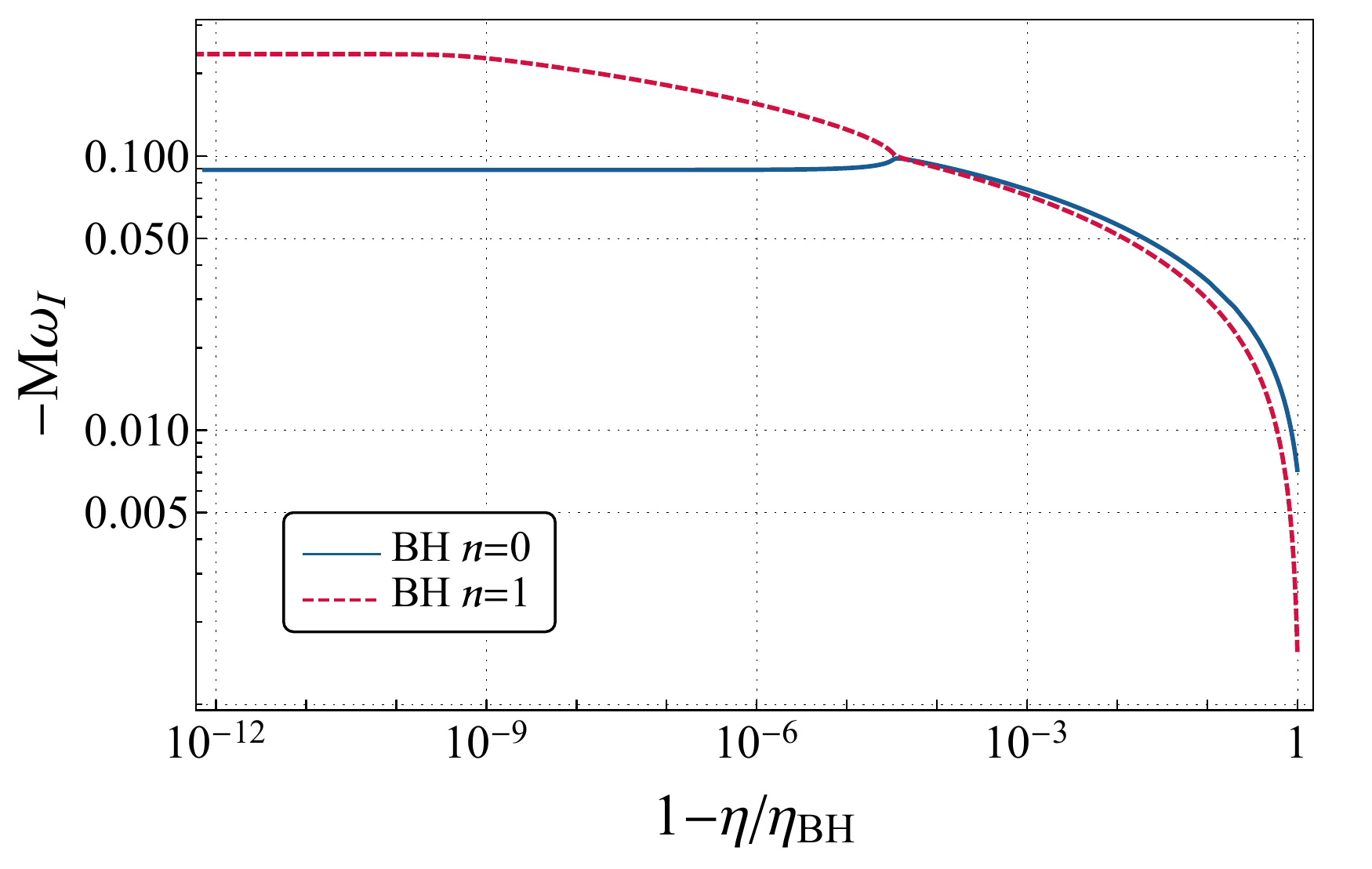}
\caption{
Imaginary part of the axial QNMs of a compact object with radius $r_0 = 2M(1+\epsilon)$ and $\epsilon=10^{-10}$ as a function of the shear viscosity of the membrane. The fundamental mode (blue curve) and the first overtone (red curve) of a Schwarzschild BH are tracked from $\eta=\eta_{\rm BH}$ (leftmost part of the plot) to the limit of a perfectly reflecting object, $\eta=0$ (rightmost part). The figure shows a crossing point after which the imaginary part of the BH overtone becomes smaller than the one of the BH fundamental mode.~\cite{Maggio:2019zyv}
}
\label{fig:overtone}
\end{figure}
Let us analyze the transition from fundamental modes to overtones as a function of the shear viscosity of the membrane for $\epsilon \ll 1$. Indeed, in the $\epsilon \to 0$ limit, the parameter $\eta$ interpolates between the BH case ($\eta=\eta_{\rm BH}$) and the perfectly reflecting case ($\eta = 0$). Fig.~\ref{fig:overtone} shows the tracking of the fundamental mode (blue curve) and the first overtone (red curve) of a Schwarzschild BH by changing the shear viscosity of the membrane. We  notice that the change in the imaginary part of the QNMs is drastic even for small variations of the shear viscosity with respect to the BH case. Fig.~\ref{fig:overtone} displays a crossing point after which the BH overtone has a smaller imaginary part than the BH fundamental mode and becomes more relevant in the ringdown stage. This trend is general and BH higher overtones become long-lived in the $\eta \to 0$ limit.
This transition could explain the presence of low-frequency QNMs in the case of perfectly reflecting objects, as discussed in Sec.~\ref{sec:lowfrequencies}. 

We notice that the tracking of higher overtones as a function of the shear viscosity is numerically challenging. Indeed, the BH QNM spectrum is unstable against small deformations of the eigenvalue problem~\cite{Jaramillo:2020tuu}. In our numerical analysis, we have seen hints of this instability (which is more severe for high-order overtones) due to finite-$\epsilon$ effects and the slightly different boundary conditions when approaching the BH limit.

\subsection{Current constraints and prospects of detectability} \label{sec:detectabilitydoublet}

%
\begin{figure}[t]
\centering
\includegraphics[width=0.49\textwidth]{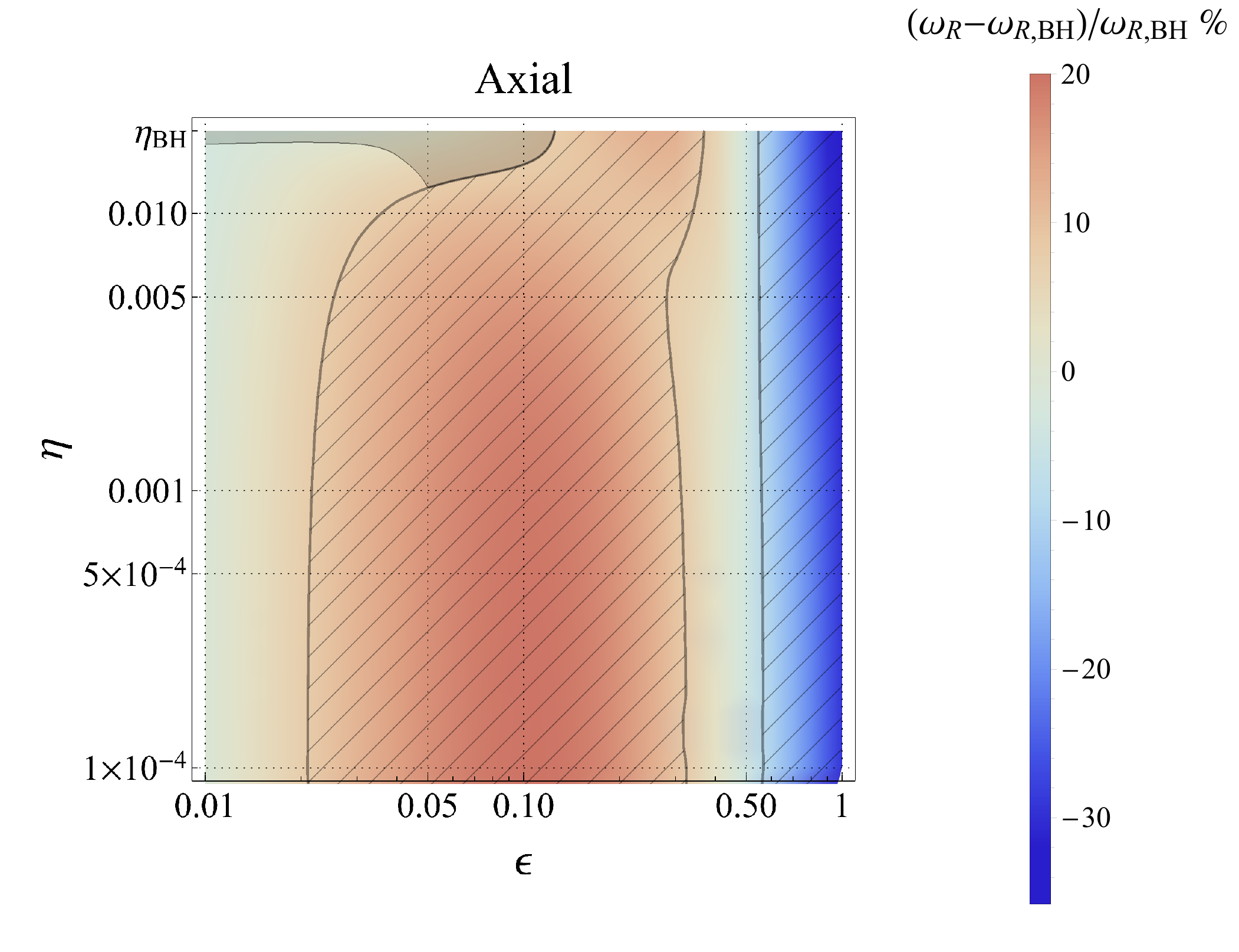}
\includegraphics[width=0.49\textwidth]{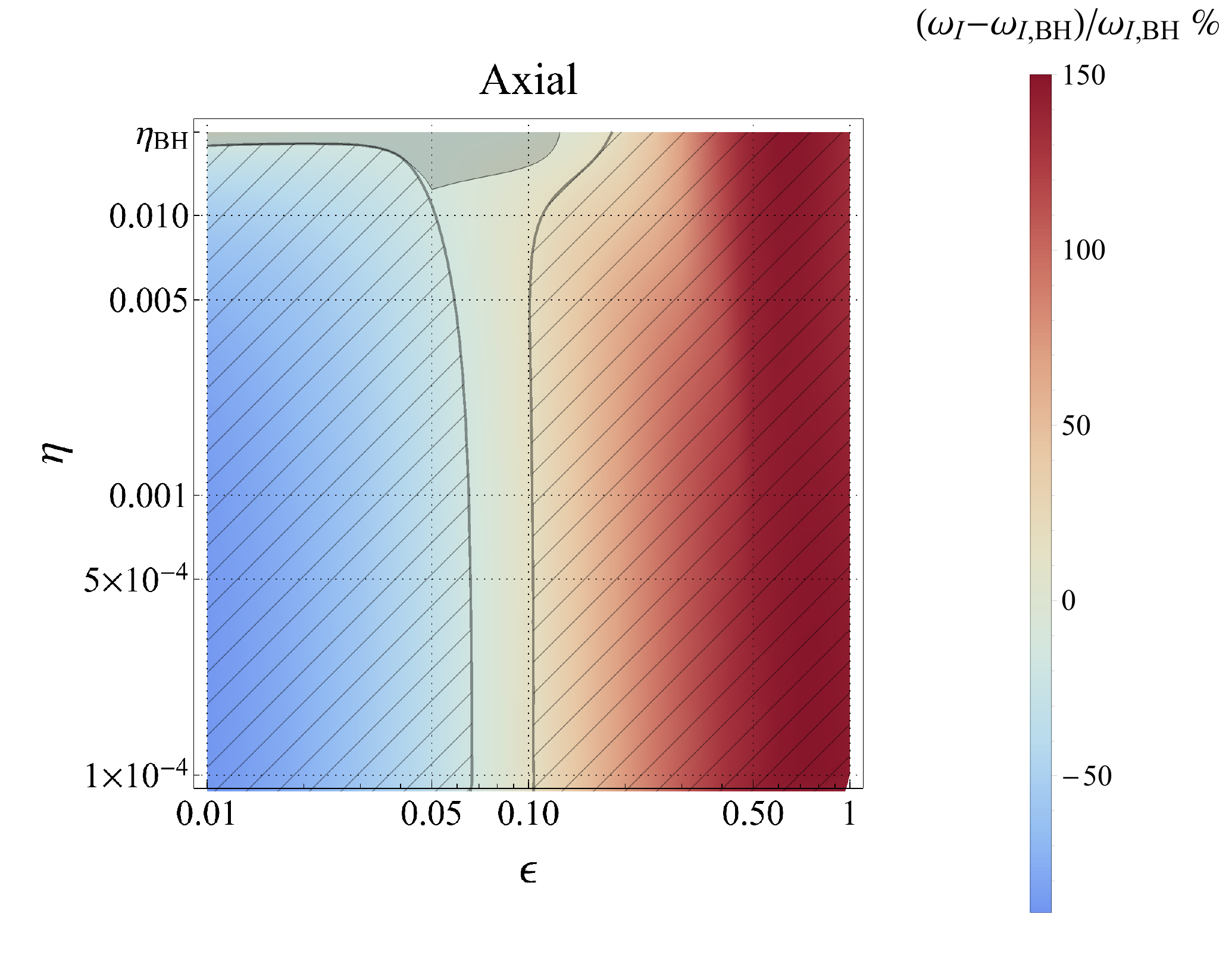}
\includegraphics[width=0.49\textwidth]{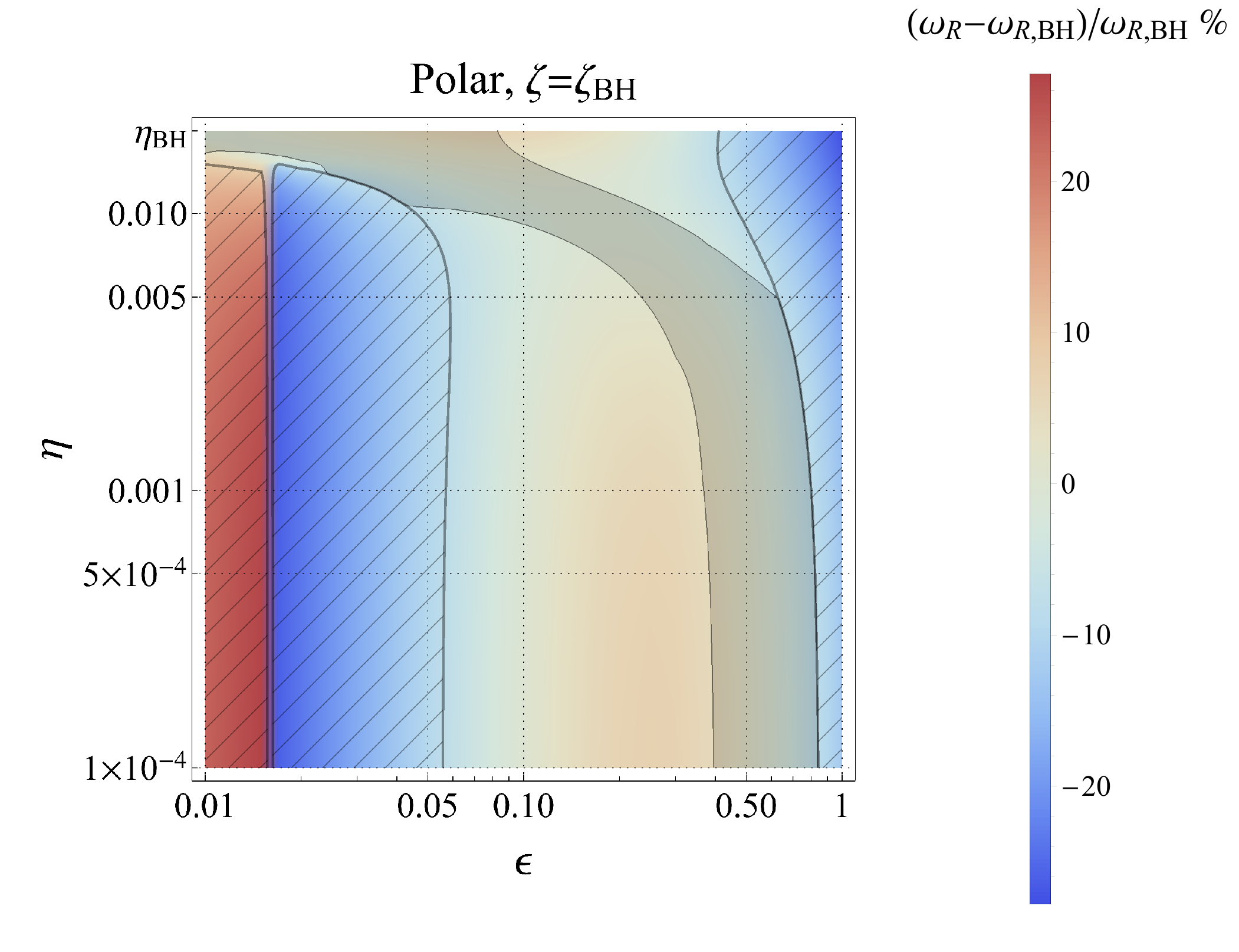}
\includegraphics[width=0.49\textwidth]{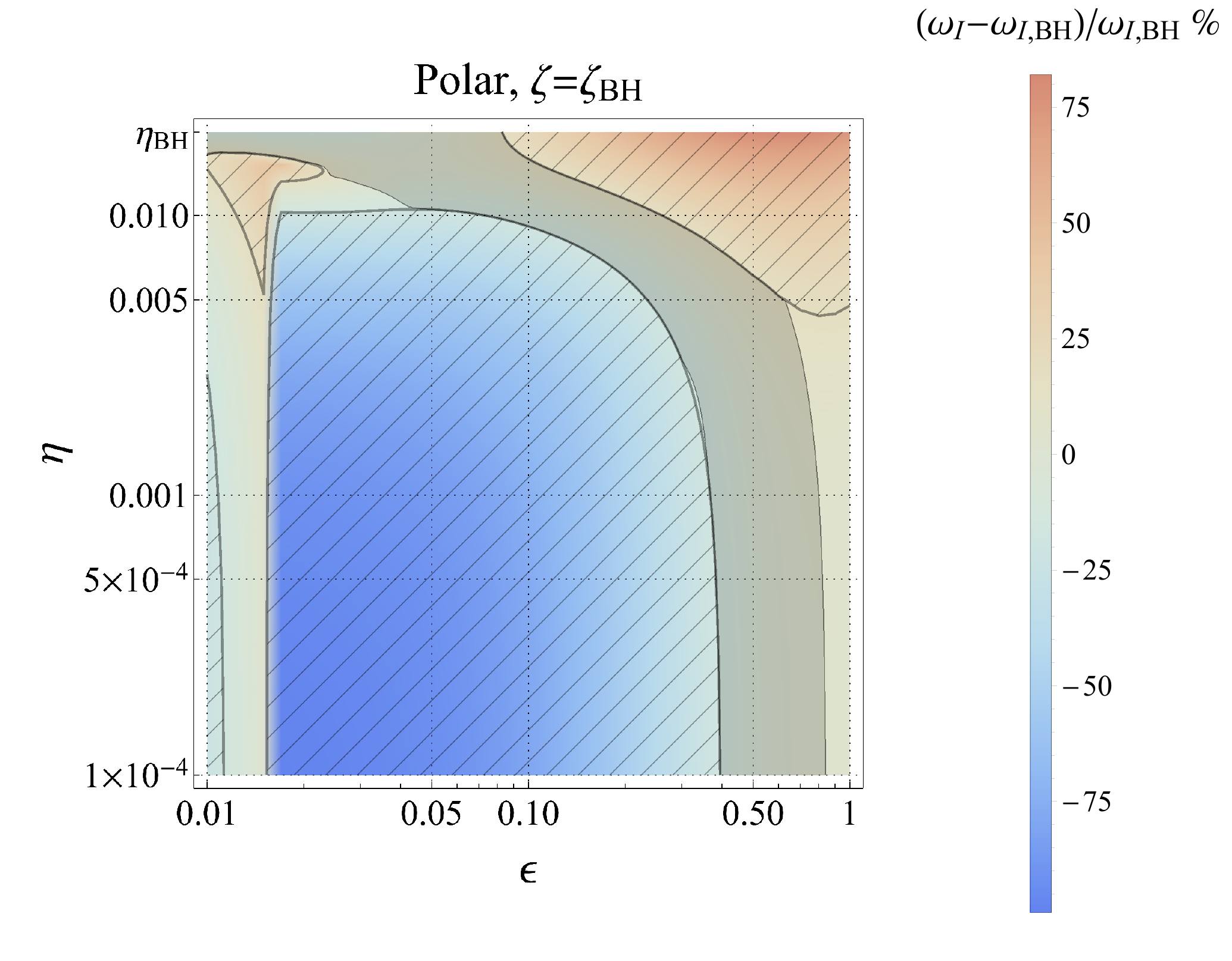}
\caption{Relative percentage difference of the real (left panels) and imaginary (right panels) part of the QNMs of a horizonless compact object to the fundamental QNM of a Schwarzschild BH under axial (top panels) and polar (bottom panels) perturbations. The QNMs of the compact object are parametrized by the compactness of the object through the parameter $\epsilon \in [0.01,1]$ and the shear viscosity $\eta \in [0,\eta_{\rm BH}]$. The dashed areas are the regions that would be excluded by individual measurements of the real (left) and imaginary (right) part of the QNMs with the same accuracy as in GW150914~\cite{Ghosh:2021mrv}. The dark shaded region is the area that would not be excluded by a simultaneous measurement of the frequency and the damping time with current measurement accuracies.~\cite{Maggio:2020jml}}
\label{fig:plot2d}
\end{figure}
\begin{figure}[t]
\centering
\includegraphics[width=0.49\textwidth]{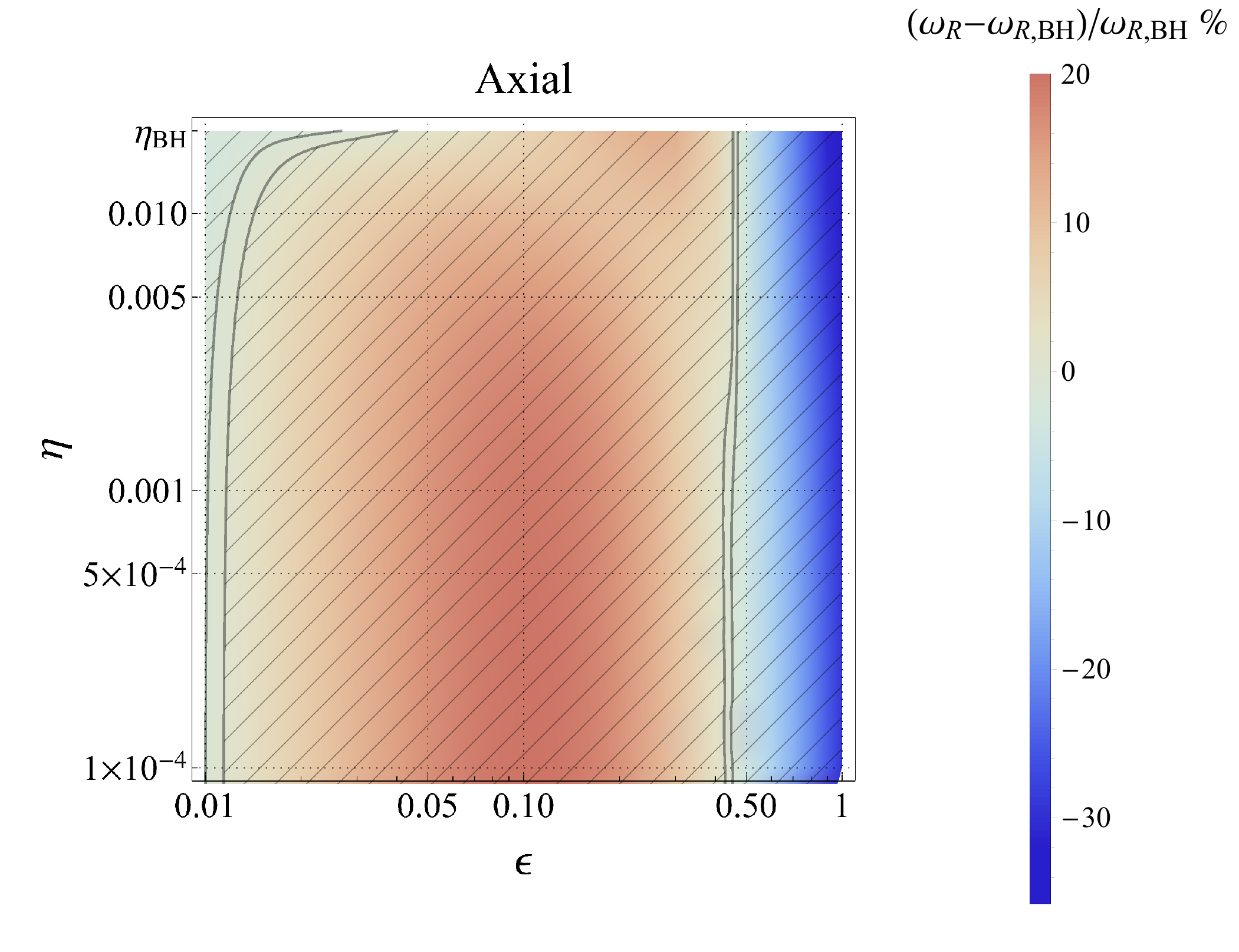}
\includegraphics[width=0.49\textwidth]{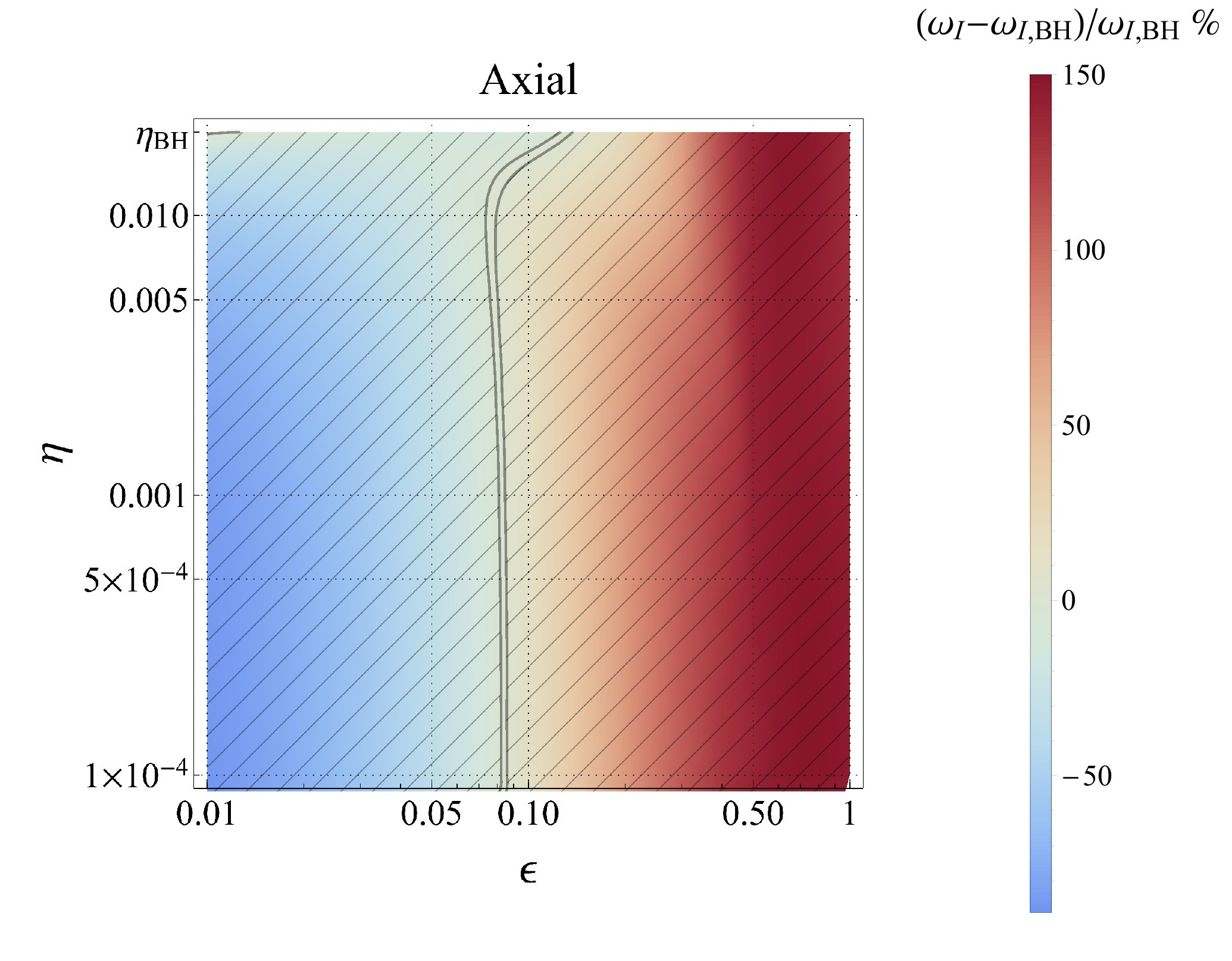}
\includegraphics[width=0.49\textwidth]{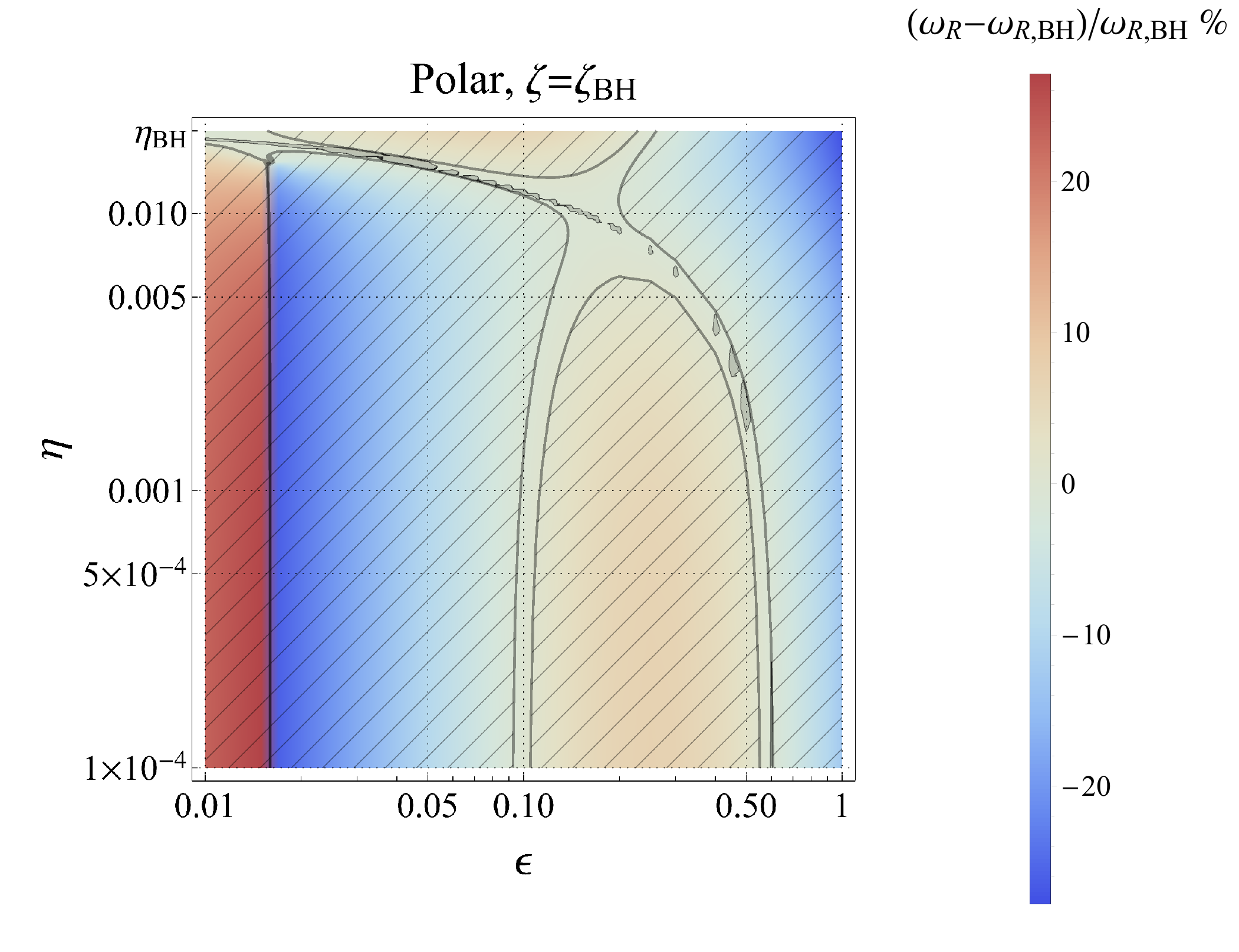}
\includegraphics[width=0.49\textwidth]{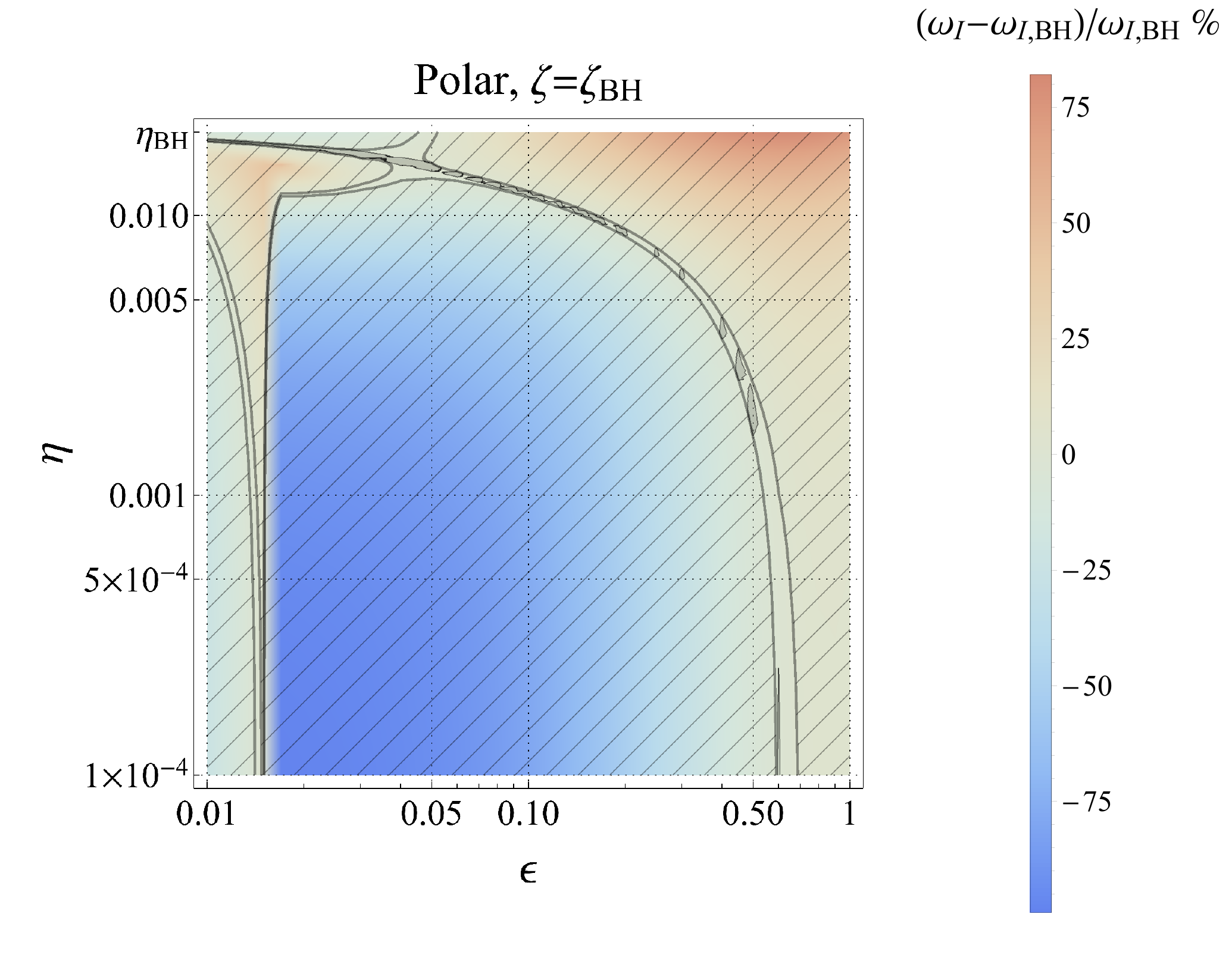}
\caption{Same as Fig.~\ref{fig:plot2d}. The dashed areas are the regions that would be excluded by individual measurements of the real and imaginary part of the QNMs by next-generation detectors, assuming an order of magnitude improvement in the ringdown measurements relative to current detectors. The dark shaded region that would be compatible with a simultaneous measurement of the frequency and the damping time is absent.
Next-generation detectors will allow us to constraint the whole region of the $(\epsilon, \eta)$ parameter space shown in the diagrams.}
\label{fig:plot2dSNR10}
\end{figure}
Current measurement accuracies impose strong constraints on the compactness and reflectivity of horizonless compact objects. Fig.~\ref{fig:plot2d} shows the relative percentage difference between the BH QNM and the QNMs of a compact object with radius $r_0 = 2M(1+\epsilon)$ as a function of the closeness parameter $\epsilon$ and the shear viscosity $\eta \in [0,\eta_{\rm BH}]$. The left (right) panels show the relative percentage difference of the real (imaginary) part of the QNMs under axial and polar perturbations in the top and bottom panels, respectively. The contour lines correspond to the accuracy within which the least-damped QNM of the remnant of GW150914 has been measured~\cite{Ghosh:2021mrv}. 
Indeed, GW150914 gives the single-event most-stringent constraints with a maximum allowed deviation from the least-damped QNM of a Kerr BH of $\sim 16\%$ and $\sim 33\%$ for the real and imaginary part of the QNM, respectively.
The dashed areas are the regions of the $(\epsilon,\eta)$ parameter space that would be excluded by the individual measurement of the real and imaginary part of the fundamental QNM. The dark shaded areas are the regions that would not be excluded by a simultaneous measurement of the real and imaginary part of the QNM in the axial (top panels) and polar (bottom panels) sectors. 

Interestingly, already with the current LIGO/Virgo accuracy we can potentially place strong constraints on the parameter space of horizonless compact objects. By combining the information from the real and imaginary
part of the QNMs, Fig.~\ref{fig:plot2d} shows that only a small region of the parameter space with $\epsilon \lesssim 0.1$ and $\eta \approx \eta_{\rm BH}$ (dark shaded area) is compatible with current constraints in the axial sector. For polar perturbations, a wider region in the parameter space with $\epsilon \gtrsim 0.1$ and $0 < \eta < \eta_{\rm BH}$ is compatible with current constraints. We assess that current measurement accuracies impose a strong lower bound on the compactness of the merger remnant, which cannot be smaller than $90\%$ the BH compactness.

Next-generation detectors, i.e., the Einstein Telescope~\cite{Hild:2010id} and LISA~\cite{LISA:2017pwj}, will have an overall improvement of the 
SNR by an order of magnitude. The sensitivity of the detectors will allow us to resolve the fundamental QNM at percent level. As shown in Fig.~\ref{fig:plot2dSNR10}, almost the whole region of the $(\epsilon, \eta)$ parameter space would be constrained.

\begin{figure}[t]
\centering
\includegraphics[width=0.65\textwidth]{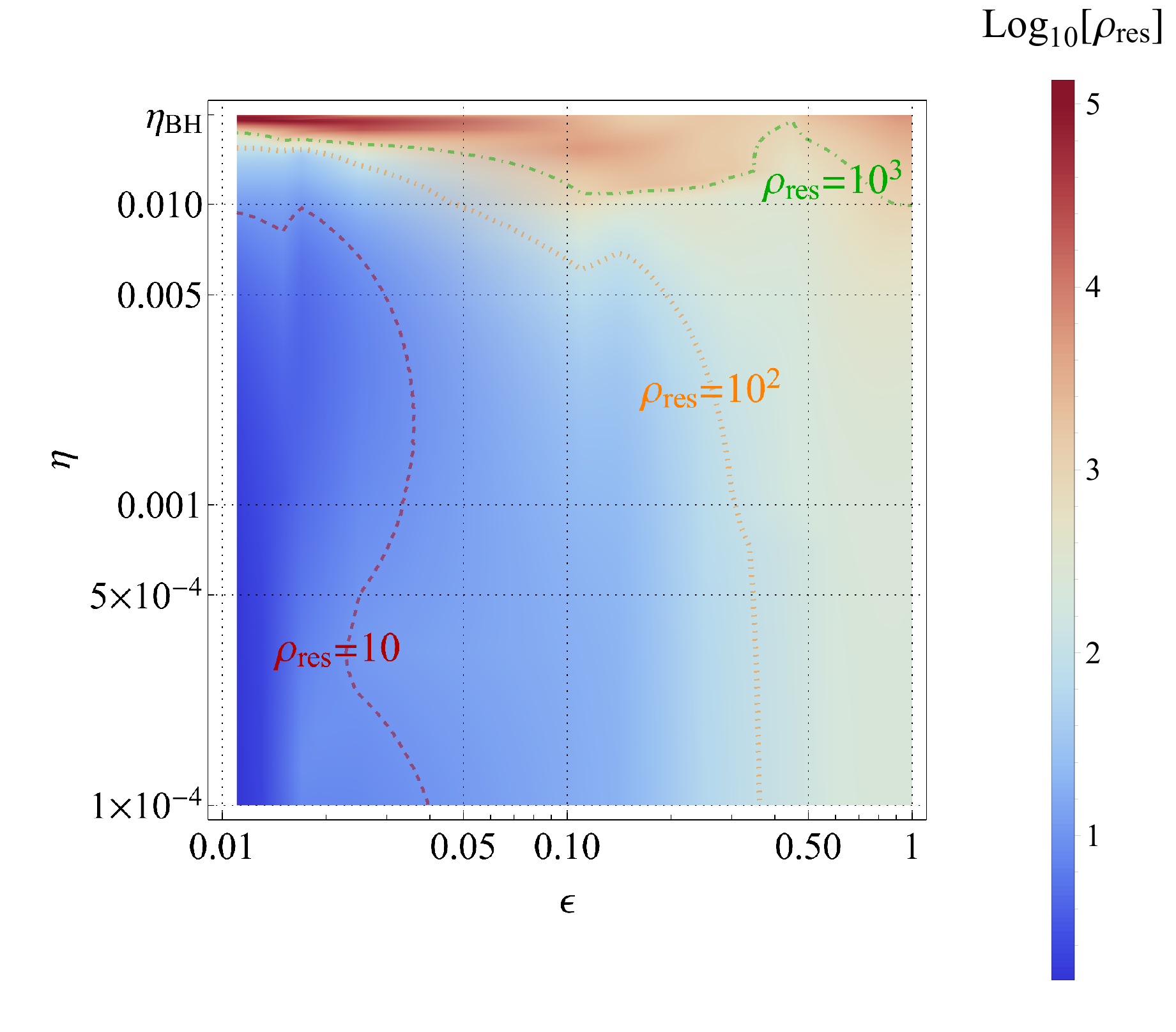}
\caption{Minimum SNR required for the resolvability of the axial-polar QNM doublet according to the Rayleigh criterion.~\cite{Maggio:2020jml}}
\label{fig:doublet}
\end{figure}
Another signature of new physics is given by the presence of the mode \emph{doublet} in the axial and polar sectors, as discussed in Sec.~\ref{sec:QNMmembrane}. A necessary condition to resolve the doublet is based on the Rayleigh resolvability criterion~\cite{Berti:2005ys}
\begin{eqnarray}
\text{max}[\sigma_{f_1},\sigma_{f_2}] &<& |f_1 - f_2| \,, \\
\text{max}[\sigma_{Q_1},\sigma_{Q_2}] &<& |Q_1 - Q_2| \,,
\end{eqnarray}
where $f_i = \omega_R^{(i)}/2 \pi$ and $Q_i = \pi f_i \tau_i$ are the frequency and the quality factor of the $i$-th mode, $\tau_i$ is the damping factor of the $i$-th mode, and $\sigma_X$ is the uncertainty associated to a quantity $X$. The uncertainties on the parameters are computed with a Fisher analysis assuming that the amplitude ratio between the axial and polar modes is $1/10$ (see Appendix~\ref{app:fisher} for details).
Fig.~\ref{fig:doublet} shows the minimum SNR required to resolve the doublet, $\rho_{\rm res}$, according to the Rayleigh resolvability criterion.  In the $\eta \approx \eta_{\rm BH}$ region, $\rho_{\rm res} > 10^3$ and it can be smaller for $\eta < \eta_{\rm BH}$. A comparison between Fig.~\ref{fig:plot2d} and Fig.~\ref{fig:doublet} shows that the resolution of the doublet requires a larger SNR than the detection of the deviations from the BH QNM.

\section{Appendix: Boundary condition for perfectly reflecting objects} \label{app:electromagneticBC}

Let us derive the boundary conditions that describe a static and perfectly reflecting ultracompact object under electromagnetic perturbations~\cite{Brito:2015oca}. 
The background geometry is the Schwarzschild metric, and the radius of the compact object is located as in Eq.~\eqref{radius} where $\epsilon \ll 1$.
The background geometry is perturbed by a test electromagnetic field that is governed by the Maxwell equations
\begin{equation}
    F^{\mu \nu}_{\ \ ;\nu} = 0 \label{maxwell}
\end{equation}
where $F_{\mu \nu} = A_{\nu, \mu} - A_{\mu, \nu}$ is the Maxwell tensor, $A_{\mu}$ is the electromagnetic four-potential, the comma stands for an ordinary derivative and the semi-colon stands for a covariant derivative. The spherical symmetry of the background allows us to expand the electromagnetic four-potential as
\begin{equation} \label{Amu}
    A_{\mu}(t,r,\theta,\varphi) = \sum_{\ell,m} \left[
    \begin{pmatrix}
         0 \\ 0 \\ a^{\ell m}(t,r) \underaccent{\bar}{S}_{\ell m}(\theta,\varphi)
    \end{pmatrix}
   + 
   \begin{pmatrix}
         f^{\ell m}(t,r)Y_{\ell m}(\theta, \varphi) \\ 
         h^{\ell m}(t,r)Y_{\ell m}(\theta, \varphi) \\ 
         k^{\ell m}(t,r) \underaccent{\bar}{Y}_{\ell m}(\theta, \varphi)
    \end{pmatrix}
   \right] \,,
\end{equation}
where the vector spherical harmonics are given by
\begin{eqnarray}
    \underaccent{\bar}{S}_{\ell m}^\top(\theta,\varphi) &=& \left( \frac{1}{\sin\theta} \partial_{\varphi} Y_{\ell m}(\theta,\varphi), - \sin \theta \partial_{\theta} Y_{\ell m}(\theta,\varphi)\right) \,, \\
    \underaccent{\bar}{Y}_{\ell m}^\top(\theta, \varphi) &=& \left(\partial_\theta Y_{\ell m}(\theta,\varphi), \partial_\varphi Y_{\ell m}(\theta,\varphi) \right) \,,
\end{eqnarray}
where $Y_{\ell m}(\theta,\varphi)$ are the scalar spherical harmonics. The first term in the right-hand side of Eq.~\eqref{Amu} has parity $(-1)^{\ell +1}$ and corresponds to axial modes, whereas the second term in the right-hand side of Eq.~\eqref{Amu} has parity $(-1)^\ell$ and corresponds to polar modes. By defining 
\begin{equation}
    \Upsilon^{\ell m}(t,r) = \frac{r^2}{\ell (\ell+1)} \left[\partial_t h^{\ell m}(t,r) - \partial_r f^{\ell m}(t,r)\right] \,,
\end{equation}
and by assuming the time dependence $a^{\ell m}, \Upsilon^{\ell m} \propto e^{-i \omega t}$, Eq.~\eqref{maxwell} translates into a a Schr\"odinger-like equation as in Eq.~\eqref{waveeq}, where 
\begin{equation}
    \psi(r) \equiv
\begin{cases}
    a^{\ell m}(r) \quad \text{for axial modes} \\
    \Upsilon^{\ell m}(r) \quad \text{for polar modes}
\end{cases} \,,
\end{equation}
and the effective potential is in Eq.~\eqref{Vaxial} with $s=-1$.

We  model the perfectly reflecting compact object as a perfect conductor at $r=r_0$, where the electric field has vanishing tangential components and the magnetic field has a vanishing parallel component
\begin{eqnarray}
    E_{\theta}(r_0) \propto F_{\theta t}(r_0) &=& 0 \,, \label{Etheta}\\
    E_{\varphi}(r_0) \propto F_{\varphi t}(r_0) &=& 0 \,, \label{Ephi} \\
    B_r(r_0) \propto F_{\varphi \theta}(r_0) &=& 0 \,. \label{Br}
\end{eqnarray}
Eqs.~\eqref{Etheta}--\eqref{Br} yield to 
\begin{eqnarray}
    a^{\ell m}(t,r_0) &=& 0 \,, \label{alm} \\
    f^{\ell m}(t,r_0) - \partial_t k^{\ell m}(t,r_0) &=& 0 \label{Ylm0} \,,
\end{eqnarray}
and the Maxwell equations~\eqref{maxwell} imply
\begin{equation}
    f^{\ell m}(t,r_0) - \partial_t k^{\ell m}(t,r_0) = - f(r_0) \partial_r \Upsilon(t,r_0) \,.
\end{equation}
Eqs.~\eqref{alm} and~\eqref{Ylm0} correspond to Dirichlet ($\psi(r_0)=0$) and Neumann ($\partial_r \psi(r_0)=0$) boundary conditions on axial and polar perturbations, respectively, as detailed in Eqs.~\eqref{dir} and~\eqref{neu}.

\section{Appendix: Continued fractions method} \label{app:CF}

The continued fraction method allows us to compute the QNMs of compact objects as roots of implicit equations~\cite{Leaver:1985ax}. The eigenfunction can be written as a series whose coefficients satisfy a finite-term recurrence relation. To optimize the recurrence relation, it is important to choose a suitable ansatz for the eigenfunction. We analyze the case of a horizonless compact object under gravitational perturbations that is governed by Eq.~\eqref{waveeq} with effective potentials in Eqs.~\eqref{Vaxial} and~\eqref{Vpolar} and $s=-2$. Let us first focus on axial perturbations, where the solution of Eq.~\eqref{waveeq} can be written as~\cite{Pani:2009ss}
\begin{equation}
    \psi(r) = (r-2M)^{2 i M \omega} e^{i \omega r_*} \phi(z) \,, \label{solCF}
\end{equation}
where $\psi(r)$ is the Regge-Wheeler wave function, $z \equiv 1-R_2/r$, and $R_2 \gtrsim r_0$ is located outside the radius of the compact object. The function $\phi(z)$ satisfies the differential equation
\begin{equation}
\left(c_0 + c_1 z + c_2 z^2 + c_3 z^3\right)\frac{d^2 \phi}{dz^2} + \left(d_0 + d_1 z + d_2 z^2\right) \frac{d \phi}{dz} + \left(e_0 + e_1 z\right) \phi = 0 \,, \label{eq-phiz}
\end{equation}
where
\begin{eqnarray}
c_0=1-\frac{2M}{R_2} \,, \ c_1=\frac{6M}{R_2}-2 \,, \ c_2=1-\frac{6M}{R_2} \,, \ c_3=\frac{2M}{R_2} \,, \\
\nonumber d_0=2i\omega R_2 + \frac{6M}{R_2} -2 \,, \ d_1=2\left(1-\frac{6M}{R_2}\right) \,, \ d_2=\frac{6M}{R_2} \,, \\
\nonumber e_0=\frac{6M}{R_2}-\ell(\ell+1) \,, \  e_1=-\frac{6M}{R_2} \,.
\end{eqnarray}
Let us perform a series expansion of $\phi(z)$ as
\begin{equation}
    \phi(z) = \sum_{n=0}^{\infty} a_n z^n \,. \label{seriesexpCF}
\end{equation}
By substituting Eq.~\eqref{seriesexpCF} in Eq.~\eqref{eq-phiz}, we  derive a four-term recurrence relation for the expansion coefficients $a_n$:
\begin{eqnarray} \label{4term}
\alpha_1 a_2 + \beta_1 a_1 + \gamma_1 a_0 = 0 \,, \quad n=1 \,, \\
\nonumber \alpha_n a_{n+1} + \beta_n a_n + \gamma_n a_{n-1} + \delta_n a_{n-2} = 0 \,, \quad n \geq 2 \,,
\end{eqnarray}
where
\begin{eqnarray}
\alpha_n &=& n(n+1)c_0 \,, \quad n\geq1 \,, \\
\nonumber \beta_n &=& (n-1)nc_1+nd_0 \,, \quad n\geq1 \,, \\
\nonumber \gamma_n &=& (n-2)(n-1)c_2+(n-1)d_1+e_0 \,, \quad n\geq1 \,, \\
\nonumber \delta_n &=& (n-3)(n-2)c_3+(n-2)d_2+e_1 \,, \quad n\geq2 \,.
\end{eqnarray}
The four-term recurrence relation~\eqref{4term} can be reduced to a three-term recurrence relation a via Gaussian elimination by defining~\cite{Leaver:1985ax}
\begin{equation}
\widehat{\alpha}_0 = -1 \,, \qquad \widehat{\beta}_0 = \frac{a_1}{a_0} \,.
\end{equation}
The term $a_1/a_0$ is determined by imposing the continuity of the solution~\eqref{solCF} and its derivative at $r=R_2$, namely
\begin{equation}
\frac{a_1}{a_0} = \frac{R_2}{\psi(R_2)} \left[ \frac{d\psi(R_2)}{dr} - \frac{i \omega}{f(R_2)}\psi(R_2) \right] \,, \label{a1a0}
\end{equation}
where the values of $\psi(R_2)$ and $d\psi(R_2)/dr$ are computed numerically by integrating Eq.~\eqref{waveeq} from $r=r_0$ up to $r=R_2$ with a suitable boundary condition at $r_0$. The remaining coefficients can be determined by recursion from Eq.~\eqref{4term}. In the case of polar perturbations, we  integrate Eq.~\eqref{waveeq} numerically with the effective potential in Eq.~\eqref{Vpolar} from $r=r_0$ up to $r=R_2$. We  obtain the values of the Zerilli function $\psi_{\rm Z}(R_2)$ and its derivative $d\psi_{\rm Z}(R_2)/dr$ from which we  derive the value of the Regge-Wheeler function in Eq.~\eqref{a1a0} using the relation in Eq.~\eqref{RWfromZ}.

Finally, by defining
\begin{eqnarray}
\widehat{\alpha}_n = \alpha_n \,, \ \widehat{\beta}_n = \beta_n \,, \ \widehat{\gamma}_n = \gamma_n \,, &n=1 \,, \\
\widehat{\alpha}_n = \alpha_n \,, \ \widehat{\beta}_n = \beta_n - \frac{\widehat{\alpha}_{n-1} \delta_n}{\widehat{\gamma}_{n-1}} \,, \ \widehat{\gamma}_n = \gamma_n - \frac{\widehat{\beta}_{n-1} \delta_n}{\widehat{\gamma}_{n-1}} \,, \ \widehat{\delta}_n=0 \,, &n\geq2 \,,
\end{eqnarray}
the four-term recurrence relation~\eqref{4term} reduces to the following three-term relation
\begin{equation}
\widehat{\alpha}_n a_{n+1} + \widehat{\beta}_n a_n + \widehat{\gamma}_n a_{n-1} = 0 \,,
\end{equation}
that can be recast in
\begin{equation}
0 = f_0(\omega) = \widehat{\beta}_0 - \frac{\widehat{\alpha}_0 \widehat{\gamma}_1}{\widehat{\beta}_1-} \frac{\widehat{\alpha}_1 \widehat{\gamma}_2}{\widehat{\beta}_2-} \frac{\widehat{\alpha}_2 \widehat{\gamma}_3}{\widehat{\beta}_3-} ... \label{CF}
\end{equation}
Using the inversion properties of the continued fractions, Eq.~\eqref{CF} can be inverted $n$ times to yield
\begin{equation} \label{CFn}
    0 = f_n(\omega) = \widehat{\beta}_n - \frac{\widehat{\alpha}_{n-1} \widehat{\gamma}_n}{\widehat{\beta}_{n-1}-} \frac{\widehat{\alpha}_{n-2} \widehat{\gamma}_{n-1}}{\widehat{\beta}_{n-2}-} ... \frac{\widehat{\alpha}_0 \widehat{\gamma}_1}{\widehat{\beta}_0-} - \frac{\widehat{\alpha}_{n} \widehat{\gamma}_{n+1}}{\widehat{\beta}_{n+1}-} \frac{\widehat{\alpha}_{n+1} \widehat{\gamma}_{n+2}}{\widehat{\beta}_{n+2}-}  \frac{\widehat{\alpha}_{n+2} \widehat{\gamma}_{n+3}}{\widehat{\beta}_{n+3}-} ...
\end{equation}
where $n=1,2,...$ The roots of Eqs.~\eqref{CF} and~\eqref{CFn} are the QNMs of the system. Since the functions $f_n(\omega)$ have different convergence properties, each of them is best suited to find the QNMs in a given region of the parameter space. Searching for roots with $n = 0$ is usually sufficient, but when the QNMs have a large imaginary part there could be stable numerical solutions for $n = 1$ and $n = 2$.

Overall, the solution is convergent if we choose $R_2$ such that $R_2 > 2M$ and $R_2/2<r_0<R_2$~\cite{Pani:2009hk}. By defining $r_0 = 2M (1+\epsilon)$ and $R_2 = 2M (1+R_{2,0})$, we derive that the integration should be performed from $r=r_0$ up to $r=R_2$ with $\epsilon<R_{2,0}<1+2 \epsilon$ to ensure numerical stability.

\section{Appendix: Membrane paradigm} \label{app:membrane}

In this Appendix, we  provide details on the derivation of the boundary conditions in Eqs.~\eqref{BC-axial} and~\eqref{BC-polar} describing horizonless compact objects with a Schwarzschild exterior and a radius located at $r_0 = 2M (1+\epsilon)$. 
The membrane paradigm allows us to map the interior of several models of ECOs in terms of the properties of a fictitious membrane located at the radius of the object. Being $\mathcal{M}$ the whole spacetime manifold, we  define $\mathcal{M}^+$ and $\mathcal{M}^-$ as the exterior and the interior regions to the $3$-dimensional membrane (or shell) that are described by the metrics $g_{\mu \nu}^+(x^+)$ and $g_{\mu \nu}^-(x^-)$, respectively, with coordinates
\begin{equation}
    x^{+ \mu} = (t^+, r^+, \theta^+, \varphi^+) \,, \quad x^{- \mu} = (t^-, r^-, \theta^-, \varphi^-) \,.
\end{equation}
We  assume that the exterior spacetime is described by the Schwarzschild metric in Boyer-Lindquist coordinates as in Eq.~\eqref{schwarzschild} and that GR works sufficiently well near the radius of the compact object, whereas the interior spacetime can be described by any theory of gravity. 

The coordinates of the membrane are $x_m^{\mu}=(t,r_0,\theta,\varphi)$ so that the intrinsic $3$-dimensional coordinates on the shell are
\begin{equation}
y^{a}=(t,\theta,\varphi)\,.
\end{equation}
The induced metric on the membrane is defined as 
\begin{equation}
h_{ab}=e^{\mu}_ae^{\nu}_bg_{\mu\nu}\,,
\label{induced metric}
\end{equation}
where the basis of three independent generators for the shell can be chosen as
\begin{equation}
e_a^{\mu}=\frac{\partial x^{\mu}}{\partial y^a}\,.
\label{basis}
\end{equation}
The extrinsic curvature on the membrane is defined as
\begin{equation}
K_{ab}=e^{\mu}_ae^{\nu}_b\nabla_{\mu}n_{\nu}\,,
\label{extrins-curv}
\end{equation}
where $n^{\mu}$ is the normal vector to the membrane and $\nabla_{\mu} \equiv \partial/\partial x^\mu$. Since the membrane is a time-like surface, we  impose that the normal vector to the membrane is space-like, i.e., $n_{\mu}n^{\mu}=1$. The trace of the extrinsic curvature is derived as $K=h^{ab}K_{ab}$.

To embed the membrane in the manifold $\mathcal{M}$ we  impose the Israel-Darmois junction conditions~\cite{Darmois:1927,Israel:1966rt}:
\begin{equation}
h^{+}_{ab}=h^{-}_{ab}\equiv h_{ab}\,,\label{israel-first}
\end{equation}
and
\begin{equation}
(K^{+}h_{ab}-K^{+}_{ab})-\left(K^{-}h_{ab}-K^{-}_{ab}\right)=8\pi T_{ab}\,,\label{israel1}
\end{equation}
where $h_{ab}^+$ ($h_{ab}^-$) and $K_{ab}^{+}$ ($K_{ab}^{-}$) are the induced metric and extrinsic curvature defined in the exterior (interior) spacetime to the shell, 
respectively, and $T_{ab}$ is the stress-energy tensor of the matter distribution located on the membrane. 

According to the original formulation of the membrane paradigm, the matter is fictitious and is such that the extrinsic curvature in the interior spacetime vanishes~\cite{MembraneParadigm}
\begin{equation}
K_{ab}^{-}=0\,,\label{assumption}
\end{equation}
so that the junction condition in Eq.~\eqref{israel1} reduces to
\begin{equation}
Kh_{ab}-K_{ab}=8\pi T_{ab}\,,\label{israel-thorne}
\end{equation}
where, for simplicity, we  define $K_{ab} \equiv K_{ab}^+$ and the coordinates of the exterior spacetime as $x^{\mu}=(t,r,\theta,\varphi)$. Let us notice that it is always possible to find a metric for the interior spacetime $g_{\mu\nu}^-(x^-)$ that satisfies the junction conditions in Eqs.~\eqref{israel-first} and~\eqref{israel1} together with the assumption in Eq.~\eqref{assumption}.

By imposing the condition in Eq.~\eqref{assumption}, the membrane paradigm allows us to map the information on the interior spacetime into the stress-energy tensor of the fictitious membrane. The junction conditions in Eqs.~\eqref{israel-first} and~\eqref{israel1} are compatible with a membrane described by the stress-energy tensor of a viscous fluid 
\begin{equation} 
    T_{ab} = \rho u_a u_b + (p- \zeta \Theta) \gamma_{ab} -2 \eta \sigma_{ab} \,, \label{stress-energy}
\end{equation}
where $\rho$ and $p$ are the density and pressure, $\eta$ and $\zeta$ are the shear and bulk viscosities, $u_a$ is the $3$-velocity of the fluid defined in 
terms of its $4$-velocity $U_{\mu}$ as $u_a=e^{\mu}_aU_{\mu}$, $\Theta=u^a_{;a}$ is the expansion, 
$\sigma_{ab}=\frac{1}{2}\left(u_{a;c}\gamma^{c}_b+u_{b;c}\gamma^{c}_a-\Theta \gamma_{ab}\right)$
is the shear tensor, $\gamma_{ab}=h_{ab}+u_au_b$ is the projector tensor, and $u_{b;a}$ is the $3$-dimensional 
covariant derivative compatible with the induced metric $h_{ab}$~\cite{Kojima:1992ie,Uchikata:2016qku}. 

\subsection*{Unperturbed background}

Let us first analyze the background spacetime as in Eq.~\eqref{schwarzschild}. In this case, the induced metric on the membrane reads
\begin{equation}
h_{tt}=-f(r_0)\,,\quad h_{\theta\theta}=\frac{h_{\varphi\varphi}}{{\rm sin}^2\theta}=r_0^2\,.
\label{3-metric backg}
\end{equation}
The normal vector to the membrane is determined by imposing the four conditions $e_{a}^\mu n_\mu=0$ and $n^\mu n_\mu=1$; its components read
\begin{equation}
n_{t}=0\,,\quad n_r=\frac{1}{\sqrt{f(r)}}\,,\quad n_{\theta}=0,\quad n_{\varphi}=0.\label{normal vector-down}
\end{equation}
The extrinsic curvature is diagonal,
\begin{eqnarray}
K_{tt}= -\frac{1}{2} \sqrt{f(r_0)} f^{\prime}(r_0)\,,\quad 
K_{\theta\theta}=\frac {K_{\varphi\varphi} }{\sin^2 \theta}= r_0\sqrt{f(r_0)}\,, \nonumber
\end{eqnarray}
where the prime denotes partial derivative with respect to the argument. The stress-energy tensor of the matter distribution reduces to the one of a perfect fluid since the expansion and the shear tensor are null. Indeed, the only nonvanishing component of the fluid $4$-velocity 
$U^\mu=(U^t,U^{r},U^\theta,U^\varphi)$ is $U^t=1/\sqrt{f(r_0)}$. This yields to the the fluid $3$-velocity  
\begin{equation}
u^a=\left(\frac{1}{\sqrt{f(r_0)}},0,0\right)\,.
\end{equation}
At the background level, the nonvanishing components of the junction condition in Eq.~\eqref{israel-thorne} are
\begin{eqnarray}
tt: \,\,&&  -\frac{2}{r_0}f^{3/2}(r_0)=8\pi f(r_0) \rho_0 \,, \nonumber\\
\theta\theta:\,\, && \frac{r_0\left[2f(r_0)+r_0f^{\prime}(r_0)\right]}{2\sqrt{f(r_0)}}=8\pi r_0^2p_0\,, \nonumber\\
\varphi\varphi:\,\, && \frac{r_0{\rm sin}^2\theta\left[2f(r_0)+r_0f^{\prime}(r_0)\right]}{2\sqrt{f(r_0)}}=8\pi r_0^2{\rm 
sin}^2\theta p_0\,.\qquad
\end{eqnarray}
The $(tt)$ component gives the unperturbed density of the membrane~\cite{Abedi:2020ujo}
\begin{equation}
\rho_0(r_0)=-\frac{\sqrt{f(r_0)}}{4\pi r_0}\,;
\label{rho0}
\end{equation}
whereas the angular components give the unperturbed pressure of the 
membrane
\begin{equation}
p_0(r_0)=\frac{2f(r_0)+r_0f^{\prime}(r_0)}{16\pi r_0\sqrt{f(r_0)}}\,.
\label{p0}
\end{equation}
Eqs.~\eqref{rho0} and~\eqref{p0} fix a barotropic equation of state $p = p(\rho)$. In the BH limit ($r_0 \to 2M$), the density vanishes and the pressure diverges as the redshift factor $\epsilon^{-1/2}$. The speed of sound is 
\begin{equation}
	c_s \equiv \sqrt{\frac{\partial p_0}{\partial \rho_0}} = \sqrt{\frac{1+2 \epsilon +4 \epsilon ^2}{8 \epsilon  
(1/2-\epsilon )}}\,,
	\label{sound-speed}
\end{equation}
that diverges both in the BH limit ($\epsilon \to 0$) and the photon-sphere limit ($\epsilon \to 1/2$), and is complex for $\epsilon>1/2$.  Let us notice that the properties of the 
fluid do not need to be physical since the membrane is fictitious.

\subsection*{Gravitational perturbations}

Let us work in the Regge-Wheeler gauge and analyze separately the axial and polar sectors of the gravitational perturbation~\cite{Regge:1957td}.
The perturbed metric can be cast in the following form
\begin{equation}
g_{\mu\nu}=g_{\mu\nu}^{0}(r)+\delta g_{\mu\nu}(r,\theta,t)\,,
\end{equation}
where, without loss of generality, the perturbation $\delta 
g_{\mu\nu}$ does not depend on the azimuthal angle $\varphi$ owing to the spherical symmetry of the background 
$g_{\mu\nu}^{0}$.

Because of the metric perturbations, the dissipative components of the stress-energy tensor are switched on, and both the density and the pressure of the membrane are perturbed as follows
\begin{eqnarray}
\rho&=&\rho_0+\delta \rho(t,\theta)\,,\\
p&=&p_0+\delta p(t,\theta)\,.
\end{eqnarray}
The location of the membrane is also affected by the perturbation, and the deviation is parametrized as
\begin{equation}
r_m(t,\theta)=r_0+\delta r(t,\theta)\,. \label{rm}
\end{equation}
The $4$-dimensional coordinates of the membrane are
\begin{equation}
x_m^{\mu}=\left(t,r_0+\delta r(t,\theta),\theta,\varphi\right)\,,
\end{equation}
and the perturbed tangential vectors $e^{\mu}_a$ introduced in Eq.~\eqref{basis} are
\begin{eqnarray}
e^{\mu}_t&=&\left(1,\partial_t \delta r(t,\theta),0,0\right)\,,\nonumber\\
e^{\mu}_\theta&=&\left(0,
\partial_\theta \delta r(t,\theta),1,0\right)\,,\nonumber\\
e^{\mu}_\varphi&=&\left(0,0,0,1\right)\,.
\label{perturbed-basis}
\end{eqnarray}

Let us notice that $\delta\rho(t,\theta)$, $\delta p(t,\theta)$ and $\delta r(t,\theta)$ are scalar quantities under rotations, therefore they are only affected by the polar perturbations and can be decomposed as
\begin{eqnarray}
\delta \rho (t,\theta)&=&\varepsilon\rho_1 P_\ell(\cos\theta) e^{-i\omega t}\,,\nonumber\\
\delta p(t,\theta)&=&\varepsilon p_1 P_\ell(\cos\theta) e^{-i\omega t}\,,\nonumber\\
\delta r(t,\theta)&=&\varepsilon \delta r_0  P_\ell(\cos\theta)  e^{-i\omega t}\,. \label{deltaR}
\end{eqnarray}
where $\rho_1$, $p_1$, $\delta r_0$ depend only on the unperturbed radius $r_0$, $P_\ell(\cos \theta)$ are the Legendre polynomials, and the parameter $\varepsilon$ is the perturbation order, so that all the contributions of order 
$\mathcal{O}(\varepsilon^2)$ are negligible. In the following, we shall analyze the axial and polar sectors separately.

\subsubsection*{Axial sector}

The nonvanishing components of the axial metric perturbations in the Regge-Wheeler gauge are~\cite{Regge:1957td}
\begin{eqnarray}
\delta g_{t\varphi}&=&\varepsilon e^{-i\omega t}h_0(r)\sin \theta  \partial_\theta P_\ell(\cos \theta)\,,\nonumber\\
\delta g_{r\varphi}&=&\varepsilon e^{-i\omega t}h_1(r)\sin \theta  \partial_\theta P_\ell(\cos \theta)\,.
\label{non-zero component}
\end{eqnarray}
It follows that the only nonvanishing component of the induced metric perturbation is
\begin{equation}
 \delta h_{t\varphi}=\varepsilon e^{-i\omega t}h_0(r_0)\sin \theta  \partial_\theta P_\ell(\cos \theta)\,.
\label{3-metric down}
\end{equation}
In the axial case, the normal vector to the membrane is given by Eq.~\eqref{normal vector-down} up to the first order in the perturbation.
As a consequence, the nonvanishing components 
of the extrinsic curvature perturbation are 
\begin{eqnarray}
\delta K_{t\varphi}&=&  \frac{1}{2}e^{-i\omega t}\varepsilon \sqrt{f}\left(i\omega 
h_1+h_0^{\prime}\right)\sin\theta \partial_\theta P_\ell(\cos\theta)\,, \nonumber \\
\delta K_{\theta\varphi}&=&-\frac{1}{2}e^{-i\omega t}\varepsilon \sqrt{f}h_1\left(-\cos\theta \partial_\theta 
+\sin\theta \partial^2_\theta \right)P_\ell(\cos\theta)\,.\nonumber
\end{eqnarray}

Concerning the fluid velocity, the components $U^t$, $U^r$ and $U^\theta$ are not affected by axial perturbations, whereas 
$\delta u^\varphi\neq 0$ and its expression can be found by solving the $t\varphi$ component of the junction condition, i.e., 
\begin{eqnarray}
\delta u^{\varphi}= \frac{\varepsilon e^{-i\omega t}\partial_\theta 
P_\ell(\cos \theta)\left[h_0f^{\prime}-f\left(i\omega h_1+ h_0^{\prime}\right)\right] 
}{r_0{\rm sin}\theta \sqrt{f}\left(2f-Rf^{\prime} \right)} \,.\qquad
\end{eqnarray}
The perturbed components of the stress-energy tensor are
\begin{eqnarray}
\delta T_{t\varphi}&=& - \varepsilon e^{-i \omega t}\rho_0 h_0 \sin\theta\partial_\theta P_\ell(\cos\theta )\nonumber\\
&&\nonumber-r_0^2 \sqrt{f} \sin^2\theta (p_0+\rho_0) \delta u^\varphi(t,\theta )\,,\\
\delta T_{\theta\varphi}&=& -\eta \, r_0^2 \sin^2\theta  \partial_\theta \delta u^\varphi(t,\theta ) \,.
\end{eqnarray}
The $\theta\varphi$ component of the junction condition then reduces to
\begin{eqnarray}
\frac{1}{2}\sqrt{f}h_1=-\frac{8\pi\eta  r_0\left[h_0f^{\prime}-f\left(i\omega 
h_1+ h_0^{\prime}\right)\right]}{\sqrt{f}\left(2f-r_0f^{\prime} \right)}\,.
\label{theta-phi}
\end{eqnarray}
In vacuum, the Regge-Wheeler functions are related to each other by~\cite{Regge:1957td}
\begin{equation}
    h_0(r)=-\frac{f(r)}{i\omega}\frac{d}{dr}\left[f(r)h_1(r)\right] \,.
\end{equation}
We  use this relation to write 
$h_0$ and $h_0^{\prime}$ in terms of $h_1$, $h_1^{\prime}$ and 
$h_1^{\prime\prime}$.
Furthermore, we  replace $h_1$ and its derivatives by introducing the Regge-Wheeler function~\cite{Regge:1957td}
\begin{equation}
\psi_{\rm RW}(r)=\frac{f(r)}{r}h_1(r) \,, \label{psih1}
\end{equation}
that satisfies Eq.~\eqref{waveeq} with the effective potential given in Eq.~\eqref{Vaxial}. Finally, Eq.~\eqref{theta-phi} yields to
\begin{equation}
\omega \psi(r_0) = 16 i \pi \eta  \left( \left.\frac{d \psi}{d r_*}\right|_{r_0} + 
\frac{r_0 V_{\rm axial}(r_0)}{2f(r_0)-r_0 f^{\prime}(r_0)} \psi(r_0) \right) \,.
\label{BC-eta-r02}
\end{equation}
that coincides with the boundary condition in Eq.~\eqref{BC-axial}.
%

\subsubsection{Polar sector}

The nonvanishing components of the polar metric perturbation are
\begin{eqnarray}
\delta g_{tt}&=&\nonumber\varepsilon e^{-i\omega t}P_\ell(\cos\theta)f(r){\cal H}(r)\,,\\
\delta g_{rr}&=&\nonumber\varepsilon e^{-i\omega 
t}P_\ell(\cos\theta)\frac{{\cal H}(r)}{f(r)}\,, \\
\delta g_{tr}&=&\nonumber\varepsilon e^{-i\omega t}P_\ell(\cos\theta){\cal H}_1(r)\,,\\
\delta g_{\theta\theta}&=&\frac{\delta 
g_{\varphi\varphi}}{{\rm sin}^2\theta}=\varepsilon e^{-i\omega t}P_\ell(\cos\theta)r^2{\cal K}(r)\,,
\label{non-zero component-polar}
\end{eqnarray}
where the location of the membrane is perturbed as in Eqs.~\eqref{rm} and~\eqref{deltaR}.
By projecting on the $3$-dimensional membrane, the nonvanishing components of the induced metric perturbation are 
\begin{eqnarray}
\delta h_{tt}&=&\varepsilon \left(f {\cal H}-f^{\prime}\delta r_0\right) P_\ell(\cos\theta) e^{-i\omega t}\,,\nonumber\\
\delta h_{\theta\theta}&=&\frac{h_{\varphi\varphi}}{{\rm sin}^2\theta}=\varepsilon \left(r_0^2{\cal K}+2r_0 \delta r_0\right)  
P_\ell(\cos\theta) e^{-i\omega t}\,.\qquad\quad  
\label{3-metric down-polar}
\end{eqnarray}
The perturbed components of the normal vector to the membrane up to the first order in the perturbation are
\begin{eqnarray}
\delta n_{t}&=&\frac{\varepsilon\,i\omega e^{-i\omega t}P_\ell(\cos\theta) \delta r_0}{\sqrt{f(r)}}\,,\nonumber\\
\delta 
n_r&=&\frac{\varepsilon e^{-i\omega t}P_\ell(\cos\theta){\cal H}(r)}{2\sqrt{f(r)}}\,,\nonumber\\
\delta n_{\theta}&=&-\frac{\varepsilon 
e^{-i\omega t} \partial_\theta P_\ell(\cos\theta)\delta r_0}{\sqrt{f(r)}}\,.\label{normal vector-down-polar}
\end{eqnarray}
Concerning the perturbed fluid velocity, in the polar sector $\delta U^t$ can be uniquely determined from the condition of unit norm
$U_\mu U^\mu=-1$, i.e., 
\begin{equation}
\delta U^t=\delta u^t=\frac{\varepsilon \left(f{\cal H}- \delta r_0f^{\prime}\right)}{2f^{3/2}}P_\ell(\cos\theta) e^{-i\omega 
t}\,.
\end{equation}
Moreover, $\delta U^{\varphi}=\delta u^\varphi=0$ and $\delta U^r=-\varepsilon U^t\,i\omega  e^{-i\omega 
t}P_\ell(\cos\theta)$, while $\delta U^\theta=\delta u^\theta$ is nonvanishing and can be determined by solving the 
$t\theta$ component of the junction condition in Eq.~\eqref{israel-thorne}, as we shown below.  

The extrinsic curvature has the following nonvanishing components up to first order in $\varepsilon$ 
\begin{eqnarray}
\delta K_{tt}&=&\frac{\varepsilon  e^{-i \omega t}  P_\ell(\cos\theta ) 
\left[\delta r_0 \left(4 \omega^2-{f^{\prime}}^2\right)+f 
\left(-2 \delta r_0 f^{\prime\prime}+3 {\cal H} f^{\prime}+4 i \omega {\cal H}_1\right)+2 f^2 {\cal H}^{\prime}\right]}
{4\sqrt{f}}\,,\nonumber\\
\delta K_{\theta\theta}&=& \frac{\varepsilon e^{-i \omega t} \left[f \left( 2 \delta r_0-r_0 {\cal H}+r_0^2 {\cal K}^\prime+2 r_0
{\cal K}\right)+\delta r_0 \left(r_0  f^\prime-2 \partial_\theta^2\right)\right]P_\ell(\cos\theta )}{2 
\sqrt{f}}\,,\nonumber\\
\delta K_{\varphi\varphi}&=& \nonumber \frac{\varepsilon\sin^2\theta  e^{-i \omega t} \left[ f \left(2 \delta r_0-r_0 {\cal H}+r_0^2 
{\cal K}^\prime+2 r_0 {\cal K}\right)+\delta r_0 \left(r_0  f^\prime-2 \cot\theta  
\partial_\theta\right)\right]P_\ell(\cos\theta )}{2 \sqrt{f}} \,,\\
\delta K_{t\theta}&=&-\frac{\varepsilon  e^{-i \omega t} \partial_\theta P_\ell(\cos\theta ) (f {\cal H}_1-2 i \delta r_0 \omega 
)}{2 \sqrt{f}}\,.
\end{eqnarray}
The nonvanishing components of the perturbation to the stress-energy tensor are 
\begin{eqnarray}
\delta T_{tt}&=&  \varepsilon e^{-i\omega t} P_\ell(\cos\theta )  \left[\rho_0 f^{\prime} \delta r_0+ f \left(\rho 
_1-\rho_0 {\cal H}\right)\right]\,, \nonumber\\
\delta T_{\theta\theta}&=& \nonumber \frac{r_0}{\sqrt{f}} \left\lbrace -\sqrt{f}r_0 \left[  (\zeta +\eta ) 
\partial_\theta \delta u^\theta +  (\zeta -\eta ) \cot \theta  \delta u^\theta \right]\right. 
\nonumber\\
&&\left.+\varepsilon e^{-i  \omega t}P_\ell(\cos\theta) \left[\sqrt{f}\left(p_0 r_0 {\cal K}+2 p_0 \delta r_0+p_1 r_0\right)+i   \omega
\zeta   (r_0 {\cal K}+2 \delta r_0)\right]\right\rbrace\,, \nonumber\\
\delta T_{\varphi\varphi}&=& \nonumber \frac{r_0\sin^2\theta }{\sqrt{f}} \left\lbrace -\sqrt{f}r_0  \left[  
(\zeta -\eta ) \partial_\theta \delta u^\theta +  (\zeta +\eta ) \cot \theta  \delta u^\theta \right]\right.\nonumber \\
&&\left.+\varepsilon e^{-i  \omega t}P_\ell(\cos\theta) \left[\sqrt{f}\left(p_0 r_0 {\cal K}+2 p_0 \delta r_0+p_1 r_0\right)+i  \omega
\zeta    (r_0 {\cal K}+2 \delta r_0)\right]\right\rbrace\,,\nonumber\\
\delta T_{t\theta}&=& - r_0^2 \sqrt{f}(\rho_0+p_0)\delta u^\theta  \,.
\end{eqnarray}
From the $tt$, $\theta\theta$, and $\varphi\varphi$ components  of the junction conditions we obtain 
analytical (albeit cumbersome) expressions for $\rho_1$ and $p_1,$ and the deviation of the membrane location
\begin{equation}
\delta r_0=\frac{16\pi \eta r_0 f {\cal H}_1}{2f-r_0f^\prime -32\pi \eta i\omega r_0} \,,
\end{equation}
whereas from the $t\theta$ component we derive
\begin{equation}
\delta u^\theta=\frac{\varepsilon e^{-i\omega t}\partial_\theta P_\ell(\cos \theta)r_0\sqrt{f}{\cal H}_1}{2f-r_0f^\prime -32\pi \eta i\omega r_0}\,.
\end{equation}

In vacuum, the metric perturbations, ${\cal H}(r)$, ${\cal H}_1(r)$ and ${\cal K}(r)$, are related by the following algebraic equation~\cite{Zerilli:1970se,Zerilli:1971wd}:
\begin{eqnarray}
{\cal H}(r)&=&\frac{1}{qr+3M}\left\lbrace \left[qr-\frac{\omega^2r^4}{r-2M}+M\frac{r-3M}{r-2M}\right]{\cal K}(r) \right.\nonumber\\
&&\left.+\left[ 
i\omega 
r^2+\frac{(q+1)M}{i\omega r} \right]{\cal H}_1(r)  \right\rbrace \,,
\label{algebraiceq}
\end{eqnarray}
where $q=(\ell-1)(\ell+2)/2$. The relation in Eq.~\eqref{algebraiceq} allows us to eliminate, say, ${\cal H}(r)$.
Moreover, we can rewrite ${\cal H}_1(r)$ and ${\cal K}(r)$ in terms of the Zerilli wave function $\psi_{\rm Z}(r)$ that satisfies Eq.~\eqref{waveeq} 
with the effective potential given in Eq.~\eqref{Vpolar}. Indeed,~\cite{Zerilli:1970se,Zerilli:1971wd}
\begin{eqnarray}
{\cal H}_1(r)&=&\omega h(r)\psi_{\rm Z}(r)+\omega k(r)\frac{d\psi_{\rm Z}(r)}{dr_*}\,,\\
{\cal K}(r)&=& g(r)\psi_{\rm Z}(r)+\frac{d\psi_Z(r)}{dr_*}\,,
\label{rel-psi-polar}
\end{eqnarray}
where
\begin{eqnarray}
h(r) &=&  i\frac{3qMr-qr^2+3M^2}{(r-2M)(qr+3M)}\,, \nonumber\\
k(r) &=& -i\frac{r^2}{r-2M} \,,\nonumber \\
g(r)&=&\frac{q(q+1)r^2+3qMr+6M^2}{r^2(qr+3M)}\,.
\label{coeff-zerilli}
\end{eqnarray}

The last condition that closes the system of equations and determines uniquely the boundary conditions 
for the polar metric perturbation is found from the barotropic equation of state $p=p(\rho)$ at the first order 
in the perturbation that gives
\begin{equation}
\delta p=c_s^2\delta \rho \,,\label{barotrpic}
\end{equation}
where the sound speed $c_s$ is given in Eq.~\eqref{sound-speed}.

By substituting the above algebraic equations, we obtain the 
following boundary condition
\begin{equation}
\frac{d\psi_{\rm Z}(r_0)/dr_*}{\psi_{\rm Z}(r_0)}=-16\pi\eta i\omega+G(r_0,\omega,\eta,\zeta) \,,\label{BC-barotr}
\end{equation}
where $G={A}/{B}$ and
\begin{eqnarray}
A&=&\nonumber (y-2) \left\lbrace 9 \left[-3-3 w^2 y^3+w^2 y^4-48 i \pi  w y^2 \zeta +y (2+96 i \pi  w \zeta 
)\right] \right. \\
&&\nonumber +q^3 y^2\left(-3 i+i y+16 \pi  w y^2 \eta \right)^2 +3 q \left[-9+3 y^2 (1+64 i \pi  w\zeta )\right. \\
&&\nonumber \left. -32 i \pi  w y^3 (3 \zeta +2 \eta ) +w y^4 (-3 w+16 i \pi  \eta 
)+w^2 
y^5 \left(1+256 \pi ^2 \eta^2\right)\right]\\
&&\nonumber+q^2 y \left[-18+6 y^2 (1+16 i \pi  w (\zeta -2 \eta ))+256 \pi ^2w^2 y^5 \eta ^2 \right.\\
&&\left.\left. +32 \pi  w y^4 \eta  (i+24 \pi 
w \eta )-i y^3 (-i+48 \pi  w (\zeta +\eta ))\right] \right\rbrace \,, \\
B&=&\nonumber y^2 (3+qy) \left\lbrace 3 \left[-3-3 w^2 y^3+w^2 y^4-48 i \pi  w y^2 \zeta +y (2+96 i\pi  w \zeta 
)\right]\right.\\
&&\nonumber +q \left[-9+9 y+w^2 y^5+y^2 (-3+192 i \pi  w \zeta )+y^3 (1-32 i \pi  w (3 \zeta -\eta )) \right. \\
&&\nonumber \left. -w y^4(3 w+16 i \pi \eta )\right]+q^2 y \left[3-3 y-16 i \pi  w y^3 (\zeta+\eta ) \right. \\
&& \left. \left. +y^2 (1+32 i \pi  w (\zeta +\eta ))\right]\right\rbrace \,,
\end{eqnarray}
and we define the dimensionless quantities $y=r_0/M$, $w=M\omega$. Let us notice that in the BH limit
$G(2M)=0$, and the BH boundary condition is recovered for $\eta=\eta_{\rm BH}$.
Our computations show that, in the BH limit, the boundary condition is independent of the bulk viscosity, as discussed in the main text.

\section{Appendix: Fisher information matrix} \label{app:fisher}

The Fisher information matrix of a template $\tilde h(f)$ in the frequency domain for a detector with 
noise spectral density $S_n(f)$ is defined as
\begin{equation}\label{fisher}
\Gamma_{i j} = \langle \partial_i \tilde h |  \partial_j \tilde h\rangle \,,
\end{equation}
where $i,j=1,...,N$, $N$ is the number of the parameters in the template, and the inner product between two waveforms ($h_1$ and $h_2$) is defined as
\begin{equation} \label{innerproduct}
\left\langle h_1|h_2\right\rangle \equiv 4\Re\,\int_{0}^{\infty} \frac{\tilde{h}_1 \tilde{h}^*_2}{S_n(f)} df\,,
\end{equation}
where the tilde stands for the Fourier transform of the waveform.
The SNR $\rho$ of a signal is defined as
\begin{equation}
    \rho^2 = \langle \tilde h | \tilde h\rangle \,. \label{SNR}
\end{equation}
The covariance matrix of the errors on the parameters of the template is defined as the inverse of the Fisher information matrix, i.e.,
\begin{equation}
    \Sigma_{ij} = \Gamma_{i j}^{-1} \,.
\end{equation}
Finally, the statistical error associated with the measurement of 
$i$-th parameter is derived as
\begin{equation}
    \sigma_{i}=\sqrt{\Sigma_{ii}} \,.
\end{equation}
%

\chapter{Ergoregion instability} \label{chapter4}

\begin{flushright}
    \emph{
    Una nuova popolazione cresceva sulla terra, nemica a noi. Ci davano addosso da tutte le parti, non ce ne andava bene una. 
    Adesso qualcuno dice che il gusto di tramontare, la passione d’essere distrutti facessero parte dello spirito di noi.
    }\\
    \vspace{0.1cm}
    Italo Calvino, Le Cosmicomiche
\end{flushright}
\vspace{0.5cm}

\section{A spinning model} \label{sec:spinningmodel}

Let us analyze a spinning horizonless compact object whose exterior spacetime is described by the Kerr metric. The absence of the Birkhoff theorem in axisymmetry implies that the vacuum region outside a spinning object can be described by geometries other than Kerr. 
However, in the case of a horizonless compact object with large compactness, any deviation from the multipolar structure of a Kerr BH dies off sufficiently fast within GR or in modified theories of gravity whose effects are confined near the radius of the compact object~\cite{Raposo:2018xkf,Bianchi:2020bxa,Bah:2021jno}. This assumption is justified for gravity theories in which putative extra degrees of freedom are heavy. In this case, the corrections to the metric and field equations are suppressed by powers of $l_P/r_0 \ll 1$, where $r_0$ is the radius of the object, and $l_P$ is the Planck length or the scale of new physics.
In Boyer-Lindquist coordinates, the exterior spacetime reads
\begin{eqnarray}
\nonumber    ds^2= &-& \left(1-\frac{2Mr}{\Sigma}\right)dt^2+\frac{\Sigma}{\Delta}dr^2-\frac{
4Mr}{\Sigma}a\sin^2\theta d\varphi dt \\
&+&{\Sigma}d\theta^2+
\left[(r^2+a^2)\sin^2\theta +\frac{2Mr}{\Sigma}a^2\sin^4\theta
\right]d\varphi^2\,, \label{kerr}
\end{eqnarray}
where $\Sigma=r^2+a^2\cos^2\theta$, $\Delta=r^2-2M r+a^2$, with $M$ and $J\equiv aM$ the total mass and angular momentum of the object, respectively.

Motivated by models of microscopical corrections at the horizon scale, the radius of the compact object is located as in Eq.~\eqref{radius} where $\epsilon \ll 1$.
The properties of the object's interior are parametrized in terms of a complex and frequency-dependent surface reflectivity $\mathcal{R}(\omega)$.

\section{The ergoregion}

The Kerr metric in Eq.~\eqref{kerr} admits two horizons at
\begin{equation}
    r_{\pm} = M \pm \sqrt{M^2-a^2} \,,
\end{equation}
where the BH spin is bounded by $-M \leq a \leq M$\footnote{Compact objects violating the Kerr bound, i.e., with $a \geq M$, are referred to as superspinars as detailed in Sec.~\ref{sec:ECOmodels}.}. The horizons are null hypersurfaces that separate regions of the spacetime where $r=const$ are timelike hypersurfaces from regions where $r=const$ are spacelike hypersurfaces.
In the Kerr spacetime, the horizons do not coincide with the infinite redshift surfaces located at
\begin{equation}
    r_{S_\pm} = M \pm \sqrt{M^2 - a^2 \cos^2 \theta} \,,
\end{equation}
which are the hypersurfaces where the Killing vector field $\xi^\mu = (1,0,0,0)$ becomes spacelike.
The horizons and the infinite redshift surfaces are such that $r_{S_-} \leq r_- < r_+ \leq r_{S_+}$: for $\theta=0, \pi$, the horizons coincide with the infinite redshift surfaces, i.e., $r_{S_\pm} = r_\pm$; whereas on the equatorial plane, $r_{S_+} = 2M > r_+$ and $r_{S_-} = 0$. Consequently, there exists a region outside the outer horizon where the Killing vector field becomes spacelike, i.e., 
\begin{equation}
    r_+ < r < r_{S_+} \,,
\end{equation}
that is called \emph{ergoregion}. Its outer boundary, i.e., $r = r_{S_+}$, is called \emph{ergosphere}.

A characteristic feature of the Kerr spacetime is that a static observer cannot exist inside the ergoregion but is forced to corotate with the compact object. A static observer is defined as a timelike curve whose tangent vector field is proportional to the Killing vector $\xi^\mu = (1,0,0,0)$ and whose $(r, \theta, \varphi)$ coordinates are constant along its worldline. Such an observer is not allowed in the ergoregion since $\xi^\mu$ is spacelike there.
We can define a zero angular momentum observer (ZAMO) 
whose angular velocity has the same sign of the angular momentum of the compact object~\cite{Wald:106274}.

When a particle starts its motion at infinity, the constant of motion coincides with the energy of the particle as measured by a static observer at infinity. The requirement that the energy is positive implies that the constant of motion is positive.
However, when a particle starts its motion in the ergoregion, its energy cannot be measured by a static observer since the latter is not allowed in the ergoregion. The particle energy can be measured, for example, by the ZAMO and does not coincide with the constant of motion. The requirement that the energy as measured by the ZAMO is positive implies that the constant of motion can be negative for counterrotating particles inside the ergoregion. This result has important consequences on the possibility of extracting energy from Kerr BHs, as described in the next section.

\section{The Penrose process}

%
\begin{figure}[t]
\centering
\includegraphics[width=0.7\textwidth]{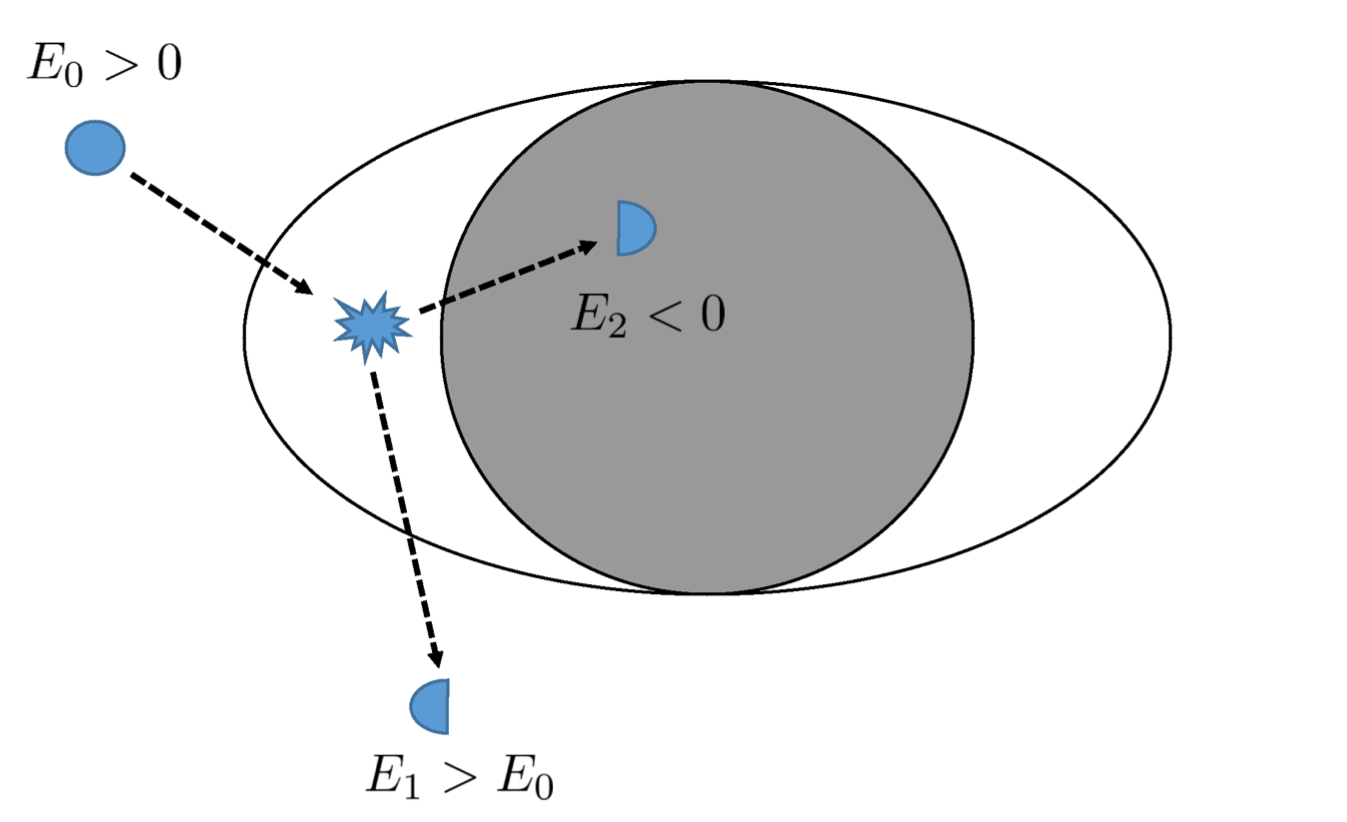}
\caption{Energy extraction from a Kerr BH via the Penrose process~\cite{Penrose:1969pc}. A particle with energy $E_0$ decays in two particles inside the ergoregion. One particle has a negative energy ($E_2 < 0$) and falls into the BH, whereas the second particle has an energy larger than the initial value ($E_1 > E_0$) and escapes at infinity.~\cite{Brito:2015oca}}
\label{fig:penrose}
\end{figure}
In Kerr spacetimes, it is possible to extract energy and angular momentum from BHs since the energy of a particle within the ergoregion as measured by an observer at infinity can be negative. This scenario was first discovered by Penrose~\cite{Penrose:1969pc} and is shown schematically in Fig.~\ref{fig:penrose}. Let us consider a particle with energy at infinity $E_0$ decaying in two particles inside the ergoregion. One of the two particles can have negative energy as measured at infinity, i.e., $E_2 < 0$, and the other particle must have energy larger than the initial value, i.e., $E_1 > E_0$. The horizon forces the negative-energy particle to fall into the BH, whereas the positive-energy particle can escape at infinity and extract energy from the BH. Indeed, the final mass of the BH is smaller than its initial value, i.e., $M_{\rm fin} = M + E_2 < M$. The same scenario holds for the BH angular momentum. Consequently, the ergoregion allows particles to acquire energy and angular momentum at the expense of the BH.

If the compact object has an ergoregion but does not have a horizon, the physical process is different. In the case of a perfectly reflecting horizonless compact object, the negative-energy states cannot be absorbed by the object and remain in orbital motion inside the ergoregion. It is therefore energetically favorable to cascade towards more negative-energy states leading to a runaway instability. This process is called \emph{ergoregion instability}~\cite{Friedman:1978wla}. The only way to prevent such an infinite cascade is by absorbing the negative-energy states efficiently. Indeed, Kerr BHs are stable against the ergoregion instability since they can absorb the radiation at the horizon. 
In the following section, we shall assess the impact of the ergoregion instability on the astrophysical viability of horizonless compact objects.

\section{Ergoregion instability in horizonless compact objects}

The ergoregion instability was proved by Friedmann in ultracompact stars under scalar and electromagnetic perturbations~\cite{Friedman:1978wla} and analyzed in uniform-density stars~\cite{CominsSchutz,10.1093/mnras/282.2.580,Kokkotas:2002sf,Brito:2015oca,Moschidis:2016zjy}, gravastars~\cite{Chirenti:2008pf}, boson stars~\cite{Cardoso:2007az}, and superspinars~\cite{Pani:2010jz}. 
In Kerr-like horizonless objects, the ergoregion instability is explained by the existence of long-lived modes in the potential cavity between the effective radius of the object and the photon sphere. As shown in Fig.~\ref{fig:lowfrequencyQNMs}, the imaginary part of the QNMs of a static horizonless object tends to zero in the limit of large compactness, i.e., $\epsilon \to 0$. In the rotating case, these modes can turn unstable\footnote{We remind the reader that, in our convention, a stable mode has a negative imaginary part of the frequency and corresponds to an exponentially damped sinusoidal signal, whereas an unstable mode has a positive imaginary part of the frequency and corresponds to an exponential growth.} due to the Zeeman splitting of the frequencies as a function of the azimuthal number of the perturbation. Indeed, in the small-spin limit, the QNMs can be expanded as~\cite{Pani:2012bp}
\begin{equation}
    \omega_{R,I} = \omega_{R,I}^{(0)} + m \chi \omega_{R,I}^{(1)} + \mathcal{O}(\chi^2) \,, \label{zeeman}
\end{equation}
where $\chi \equiv a/M$ is the dimensionless spin, $\omega_{R,I}^{(0)}$ are the real and imaginary parts of the QNMs in the static case, and $\omega_{R,I}^{(1)}$ are the first order corrections to the QNMs in the spin. In the large compactness limit, $\omega_{I}^{(0)} \to 0$ and the first order term in Eq.~\eqref{zeeman} can turn the sign of the imaginary part of the frequency to be positive depending on the azimuthal number. 
The ergoregion instability affects horizonless compact objects regardless of the azimuthal number of the perturbation due to the symmetries of the Teukolsky wave function
\begin{equation}
    m \to -m \,, \quad \omega \to -\omega^* \,, \quad ~_{s}A_{\ell m} \to ~_{s}A_{\ell -m}^* \,,
    \label{symmetry}
\end{equation}
as detailed in Sec.~\ref{sec:pertKerr}.
Let us derive the QNM spectrum of a spinning horizonless object and analyze the region of the parameter space in which the object is affected by the ergoregion instability.

\subsection{Linear perturbations in the Kerr background} \label{sec:pertKerr}

Let us perturb the background geometry in Eq.~\eqref{kerr} with a spin-$s$ perturbation, where $s=0, \pm 1, \pm 2$ for scalar, electromagnetic and gravitational perturbations, respectively. The perturbation can be decomposed as in Eq.~\eqref{decomposition} where the radial and the angular functions are governed by the Teukolsky master equations~\cite{PhysRevLett.29.1114,1973ApJ...185..635T,Teukolsky:1974yv}
\begin{eqnarray}
&&\quad \Delta^{-s} \frac{d}{dr}\left(\Delta^{s+1} \frac{d_{s}R_{\ell m}}{dr}\right) \nonumber \\
&& + \left[\frac{K^{2}-2 i s (r-M) K}{\Delta}+4 i s \omega r -\lambda\right]~_{s}R_{\ell m}=0\,, \label{wave_eq} \\
&&\quad \left[\left(1-x^2\right)~_{s}S_{\ell m,x}\right]_{,x}+ \bigg[(a\omega x)^2-2a\omega s x + s \nonumber \\
&& +~_{s}A_{\ell m}-\frac{(m+sx)^2}{1-x^2}\bigg]~_{s}S_{\ell m}=0\,, \label{angular}
\end{eqnarray}
where $~_sS_{\ell m}(\theta) e^{i m \varphi}$ are the spin-weighted spheroidal harmonics, $x\equiv\cos\theta$, $K=(r^2+a^2)\omega-am$, and the separation constants $\lambda$ and $~_{s}A_{\ell m}$ are related by $\lambda \equiv~_{s}A_{\ell m}+a^2\omega^2-2am\omega$.
When $a=0$, the angular eigenvalues are $\lambda=(\ell-s)(\ell+s+1)$, whereas when $a\neq0$ the angular eigenvalues can be computed either numerically or with approximated analytical expansions (see Sec.~\ref{sec:numerical_procedure_spin} for more details). 

It is convenient to introduce the 
Detweiler function~\cite{1977RSPSA.352..381D}
\begin{equation}
 _{s}X_{\ell m} = \Delta^{s/2} \left(r^2+a^2\right)^{1/2} \left[\alpha \
_{s}R_{\ell m}+\beta \Delta^{s+1} \frac{d_{s}R_{\ell m}}{dr}\right]\,,\label{DetweilerX}
\end{equation}
where $\alpha$ and $\beta$ are radial functions that are given in Appendix~\ref{app:detweiler}. By defining the
tortoise coordinate such that
$dr_*/dr=(r^2+a^2)/\Delta$, Eq.~\eqref{wave_eq} becomes a Schrödinger-like equation
\begin{equation}
 \frac{d^2_{s}X_{\ell m}}{dr_*^2}- V(r,\omega) \,_{s}X_{\ell m}=0\,, \label{final}
\end{equation}
where the effective potential is
\begin{equation}
 V(r,\omega)=\frac{U\Delta }{\left(r^2+a^2\right)^2}+G^2+\frac{dG}{dr_*}\,, \label{pot_detweiler}
\end{equation}
where
\begin{eqnarray}
G &=& \frac{s(r-M)}{r^2+a^2}+\frac{r \Delta}{(r^2+a^2)^2} \,, \\
U &=& V_S+\frac{2\alpha' + (\beta' \Delta^{s+1})'}{\beta \Delta^s} \,, \\
V_S &=& -\frac{1}{\Delta}\left[K^2-is\Delta'K+\Delta(2isK'-\lambda)\right] \,,
\end{eqnarray}
and the prime denotes a derivative with respect to $r$. In the following, we shall define $R_s \equiv~_sR_{\ell m}$, $X_s \equiv~_sX_{\ell m}$ and omit the $\ell, m$ subscripts for brevity.

The radial functions $\alpha$ and $\beta$ are such that the effective potential in Eq.~\eqref{pot_detweiler} is purely
real (see Appendix~\ref{app:detweiler} for a derivation). 
Let us notice that the Detweiler effective potential has the following asymptotics
\begin{equation} 
V(r,\omega) \sim 
\begin{cases}
 \displaystyle 
-\omega^2 & \text{ as } r_* \to + \infty\\ 
 \displaystyle  
 -k^2 & \text{ as } r_* \to 
- \infty
\end{cases} \,,\\
\end{equation}
where $k = \omega - m \Omega_H$ and $\Omega_H = a/(2Mr_+)$ is the angular velocity of a Kerr BH at the horizon.
Consequently, the two independent solutions of Eq.~\eqref{final} have the asymptotic behavior
\begin{equation} \label{asymptoticsplus}
X_s^+ \sim 
\begin{cases}
 \displaystyle 
e^{+i \omega r_*} & \text{ as } r_* \to + \infty\\ 
 \displaystyle  
 B_{\rm out}(\omega)e^{+i k r_*}  +  B_{\rm in}(\omega) e^{- i k r_*} & \text{ as } r_* \to 
- \infty
\end{cases} \,,\\
\end{equation}
\begin{equation} \label{asymptoticsminus}
X_s^- \sim 
\begin{cases}
 \displaystyle 
 A_{\rm out}(\omega)e^{+i \omega r_*}  +  A_{\rm in}(\omega) e^{-i \omega r_*} & \text{ as } r_* \to + \infty \\ 
 \displaystyle  
 e^{-i k r_*} & \text{ as } r_* \to - \infty \\
\end{cases} \,,
\end{equation}
where the Wronskian of the solutions is
\begin{equation} \label{wronskiandef}
    W_{\rm BH} = \frac{dX_s^+}{dr_*} X_s^- - X_s^+ \frac{dX_s^-}{dr_*}= 2 i k B_{\rm out} \,.
\end{equation}
Since the effective potential in Eq.~\eqref{pot_detweiler} is real, $X_s^\pm$ and their complex conjugates $X_s^{\pm *}$ are independent solutions to the same equation which satisfy complex conjugated boundary conditions. Via the Wronskian relationships, the asymptotic coefficients satisfy the relations~\cite{Casals:2005kr,Vilenkin:1978uc}
\begin{eqnarray} \label{wronskian}
\nonumber &&|B_{\rm out}|^2 - |B_{\rm in}|^2 = \omega/k \,, \\
\nonumber &&|A_{\rm in}|^2 - |A_{\rm out}|^2 = k/\omega \,, \\
 &&\omega A_{\rm in} = k B_{\rm out} \,, \quad  \omega A_{\rm out}^* = - k B_{\rm in} \,.
\end{eqnarray}
%
%

\subsection{Boundary conditions}

The QNMs of a Kerr-like horizonless object are derived by adding to Eq.~\eqref{final} two suitable boundary conditions.
At infinity, we  impose that the perturbation is a purely outgoing wave
\begin{equation} \label{BCinfX}
    X_s \sim e^{i \omega r_*} \,, \quad \text{as} \ r_* \to + \infty \,.
\end{equation}
The regularity at the center of the object implies the imposition of a boundary condition at the effective radius of the object.
For $\epsilon \ll 1$, the effective potential in the Detweiler equation is constant at the radius of the object, $V \approx -k^2$, so that the perturbation can be decomposed as a superposition of ingoing and outgoing waves, i.e., 
\begin{equation} \label{R_1}
    X_s \sim \mathbb{C}_{\rm in} e^{-i k r_*} + \mathbb{C}_{\rm out} e^{i k r_*} \,, \quad \text{as} \ r_* \to r_*^0 \,.
\end{equation}
where we define the surface reflectivity of the object as
\begin{equation} \label{R_2}
    \mathcal{R}(\omega) = \frac{\mathbb{C}_{\rm out}}{\mathbb{C}_{\rm in}} e^{2 i k r_*^0} \,.
\end{equation}
A perfectly reflecting object, where the outgoing energy flux at the effective radius is equal to the incident one, has $|\mathcal{R}(\omega)|^2=1$. Two notable examples of perfectly reflecting boundary conditions are
\begin{eqnarray}
    X_{s}(r_0) &=& 0 \quad \rm Dirichlet \ on \ axial \label{BCXdir} \\
    dX_{s}(r_0)/dr_* &=& 0 \quad \rm Neumann \ on \ polar \label{BCXneu}
\end{eqnarray}
where the Dirichlet boundary condition describes waves that are totally reflected with inverted phase ($\mathcal{R}(\omega)=-1$), whereas the Neumann boundary condition describes waves that are totally reflected in phase ($\mathcal{R}(\omega)=1$). In general, a partially absorbing object is described by the boundary condition
\begin{equation}
\left. \frac{dX_{s}/dr_*}{X_{s}} \right|_{r_0}= - i k \frac{1-\mathcal{R(\omega)}}{1+\mathcal{R(\omega)}} \,, \label{BC}
\end{equation}
that reduces to the BH boundary condition of a purely ingoing wave when $\mathcal{R}=0$.
In the following, we shall derive the QNMs of a perfectly reflecting ultracompact object. 

In the case of electromagnetic perturbations, a perfectly reflecting object can be modeled as a perfect conductor. In Appendix~\ref{app:electromagneticBCspin}, we show that this condition translates into Dirichlet and Neumann boundary conditions on axial and polar modes, respectively. An analogous description of a perfectly reflecting compact object under gravitational perturbations is not available. We assume that the results of electromagnetic perturbations can also be applied to gravitational perturbations.
 
\subsection{Numerical procedure} \label{sec:numerical_procedure_spin}

Equation~\eqref{final} with boundary conditions at infinity in Eq.~\eqref{BCinfX} and the radius of the object in Eqs.~\eqref{BCXdir} and~\eqref{BCXneu} can be solved numerically with a direct integration shooting method. We  start with an analytical high-order series expansion of the solution at infinity and we integrate Eq.~\eqref{final} from infinity to the radius of the object. We repeat the integration for different values of the complex frequency until the boundary condition in Eq.~\eqref{BCXdir} or~\eqref{BCXneu} is satisfied.
 
The angular eigenvalues are computed numerically using continued fractions~\cite{Berti:2005gp}.
For $a\omega\ll1$, the eigenvalues can also be expanded analytically as
\begin{equation}
    ~_{s}A_{\ell m}=\sum_{p=0}^{\infty} f_{s \ell m}^{(p)} (a\omega)^{p} \,,
\end{equation}
where $f^{(p)}_{s \ell m}$ are known expansion coefficients~\cite{Berti:2005gp}
\begin{eqnarray}
f^{(0)}_{s \ell m} &=& (\ell-s)(\ell+s+1) \,, \\
f^{(1)}_{s \ell m} &=& - \frac{2 m s^2}{\ell (\ell+1)} \,, \\
f^{(2)}_{s \ell m} &=& h(\ell+1) - h(\ell) -1 \,,
\end{eqnarray}
and
\begin{eqnarray}
h(\ell) = \frac{\left[\ell^2 - \left(\text{max}(|m|,|s|)\right)^2\right] \left[\ell^2 - \left(\frac{ms}{\text{max}(|m|,|s|)}\right)^2\right] \left(\ell^2-s^2\right)}{2 \left(\ell-\frac{1}{2}\right) \ell^3 \left(\ell+\frac{1}{2}\right)} \,.
\end{eqnarray}
We verified that the analytical approximation up to the second order is in agreement with the numerical computation: the approximation
differs from the exact eigenvalue of $\lesssim 2\%$ in the electromagnetic QNMs and $\lesssim 4\%$ in the gravitational QNMs for a compact object with high spin.

\subsection{Instability in the quasi-normal mode spectrum} \label{sec:QNMsspin}

The QNM spectrum of the system depends on two continuous dimensionless parameters, i.e., the spin of the object $\chi$ and its compactness through the parameter $\epsilon$. We  focus on perfectly reflecting compact objects that are described by the boundary conditions in Eq.~\eqref{BCXdir} and~\eqref{BCXneu}.

Furthermore, the QNMs depend on four integers, i.e., the spin $s$, the angular number $\ell$, the azimuthal number $m$, and the overtone number $n$ of the perturbation. We  focus on fundamental modes ($n = 0$) with $\ell = m = 1$ for scalar and electromagnetic perturbations, and $\ell = m = 2$ for gravitational perturbations that, in the unstable case, correspond to the modes with the largest imaginary part and thus the shortest instability timescale. The symmetries of the Teukolsky wave function guarantee that we can focus on the modes with $m \geq 0$ without loss of generality.

\begin{figure}[t]
\centering
\includegraphics[width=0.49\textwidth]{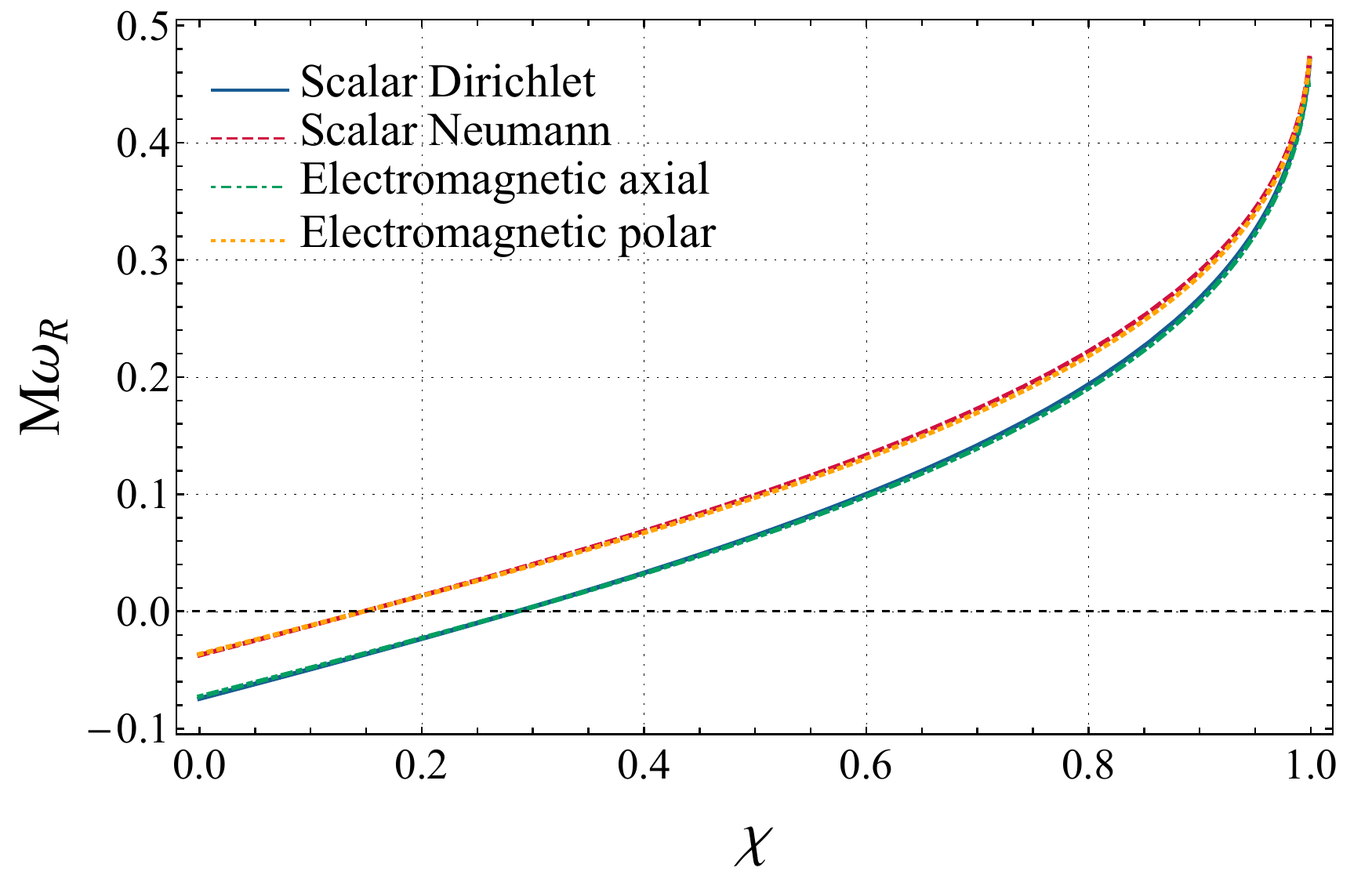}
\includegraphics[width=0.49\textwidth]{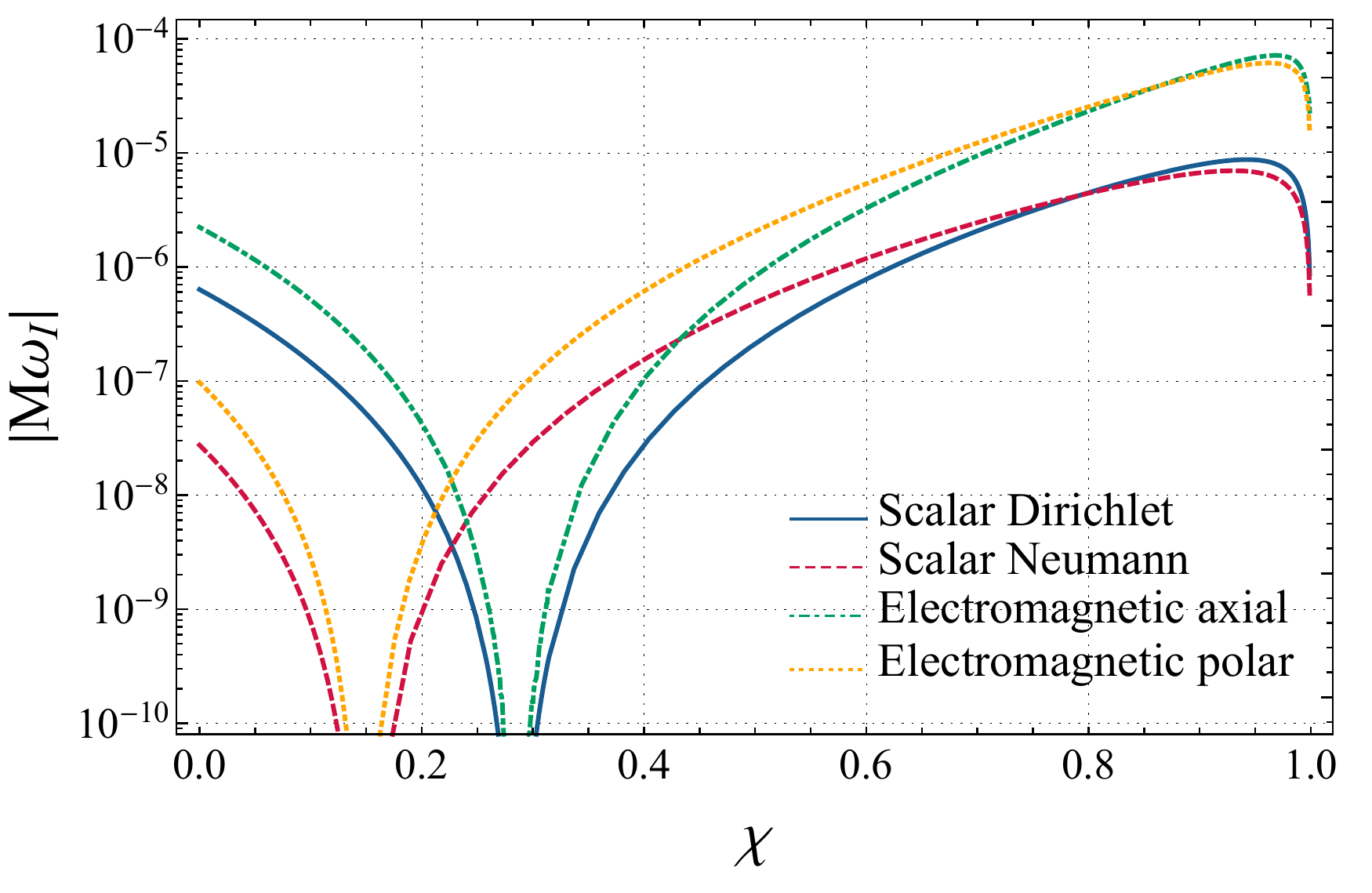}
\caption{Real (left panel) and imaginary (right panel) part of the scalar and electromagnetic QNMs ($\ell=m=1$, $n=0$) of an ECO as a function of the spin. The radius of the object is located at $r_0 = r_+ (1+\epsilon)$, where $\epsilon=10^{-10}$. The cusps in the imaginary part of the frequency correspond to the threshold of the ergoregion instability above which the QNMs are unstable. The scalar QNMs with Dirichlet (Neumann) boundary conditions are isospectral to the electromagnetic axial (polar) QNMs.~\cite{Maggio:2017ivp,Maggio:2018ivz}}
\label{fig:QNMergoregionscem}
\end{figure}
Fig.~\ref{fig:QNMergoregionscem} shows the scalar QNMs of a perfectly reflecting compact object a a function of the spin, where the radius of the object is located at $r_0 = r_+(1+\epsilon)$ and $\epsilon=10^{-10}$. The real part of the QNM has a zero crossing at a critical value of the spin that depends on Dirichlet or Neumann boundary condition. Most importantly, the imaginary part of the QNM changes sign for the same critical value of the spin, turning the object from stable to unstable. 

\begin{figure}[t]
\centering
\includegraphics[width=0.49\textwidth]{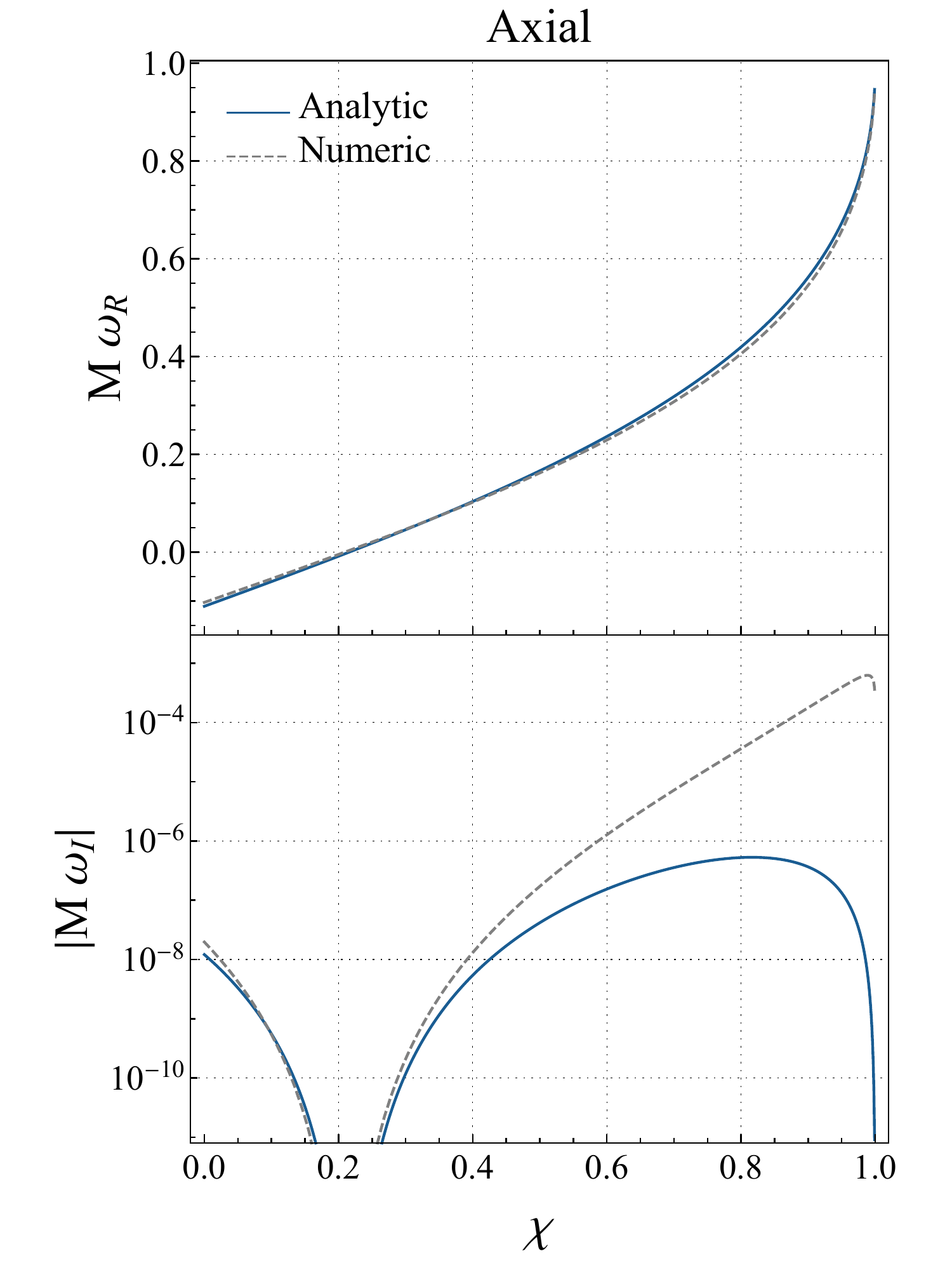}
\includegraphics[width=0.49\textwidth]{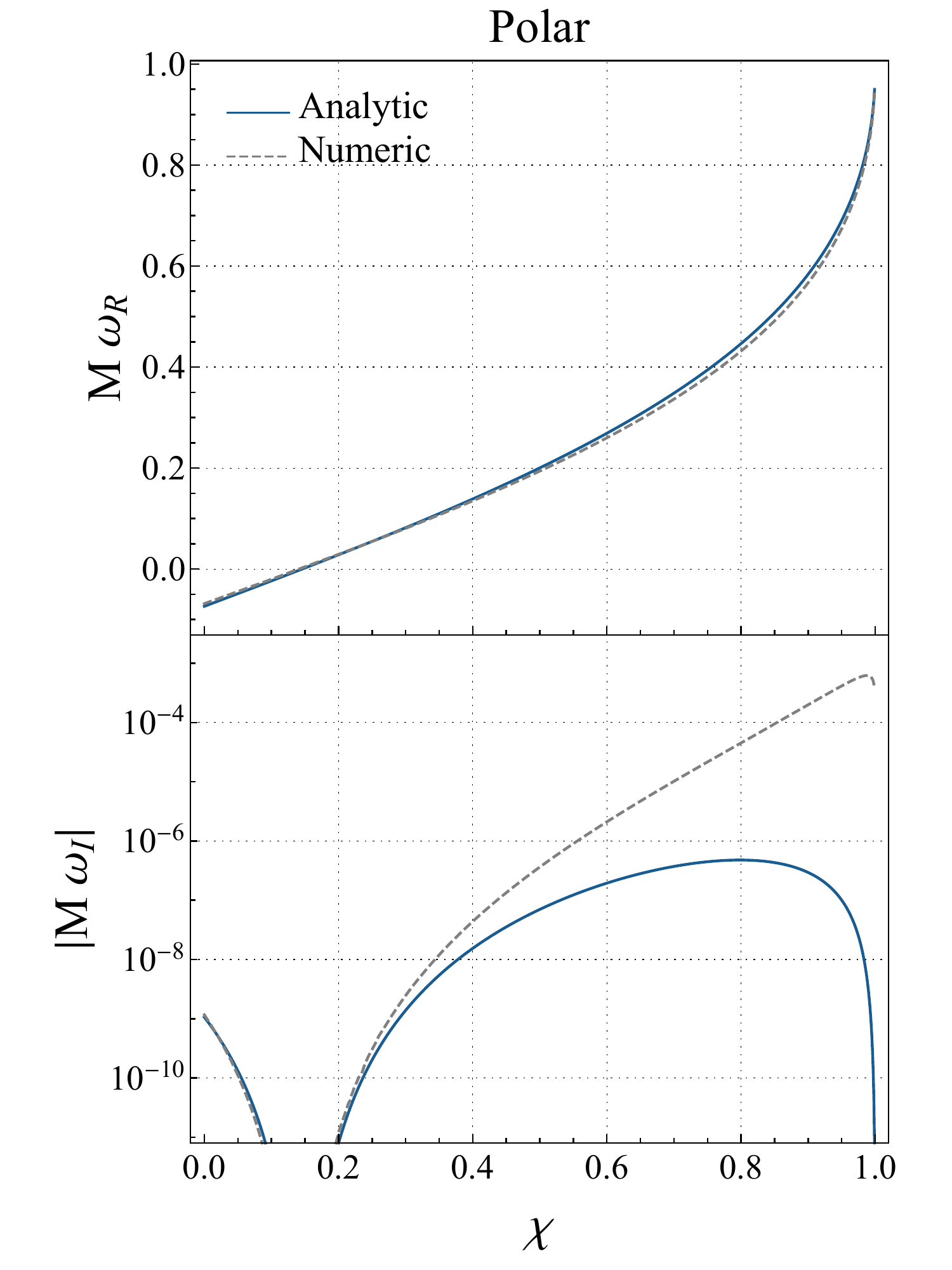}
\caption{Real (left panel) and imaginary (right panel) part of the gravitational QNMs ($\ell=m=2$, $n=0$) of an ECO as a function of the spin. The radius of the object is located at $r_0 = r_+ (1+\epsilon)$, where $\epsilon=10^{-10}$. The cusps in the imaginary part of the frequency correspond to the threshold of the ergoregion instability above which the QNMs are unstable. The QNMs computed numerically (dashed curves) agree with the QNMs computed analytically (continuous curves) using Eqs.~\eqref{MomegaR} and~\eqref{MomegaI} that are valid when $M\omega \ll 1$~\cite{Maggio:2018ivz}.}
\label{fig:QNMergoregiongrav}
\end{figure}

Above the threshold of the ergoregion instability, the imaginary part of the QNMs is positive and the modes are unstable. An interesting feature is that the threshold of the instability is the same both for scalar and electromagnetic perturbations with Dirichlet (Neumann) boundary conditions on electromagnetic axial (polar) modes. This feature is because, in the zero-frequency limit, the scalar and electromagnetic wave functions are related by a Darboux transformation~\cite{Maggio:2018ivz}, i.e.,
\begin{equation}
    R_{-1} = R_0 + \frac{i \Delta}{a m} R_0' \,,
\end{equation}
or equivalently
\begin{equation}
    R_0 = - \frac{i a m}{\ell (\ell+1)} \left( R'_{-1} + \frac{i a m}{\Delta} R_{-1}\right) \,,
\end{equation}
where the prime denotes a derivative with respect to $r$.
In the gravitational case, the threshold of the ergoregion instability is slightly shifted, as shown in Fig.~\ref{fig:QNMergoregiongrav} for axial and polar perturbations, respectively.
For $\epsilon \ll 1$, the QNMs can be derived analytically in the low-frequency regime~\cite{Maggio:2018ivz,Cardoso:2019rvt}
\begin{eqnarray}
 \omega_R &\simeq& - \frac{\pi}{2 |r_*^0|} \left[ p + \frac{s(s+1)}{2}\right] + m \Omega_H \,, \label{MomegaR} \\
 \omega_I &\simeq& - \frac{\beta_{s \ell}}{|r_*^0|} \left( \frac{2 M r_+}{r_+ - r_-}\right) \left[\omega_R (r_+ - r_-)\right]^{2 \ell+1} \left(\omega_R -m \Omega_H\right)  \,, \label{MomegaI}
\end{eqnarray}
where $\sqrt{\beta_{s \ell}} = \frac{(\ell-s)! (\ell+s)!}{(2\ell)! (2\ell+1)!!}$, $p$ is a positive odd (even) integer for Neumann (Dirichlet) boundary conditions on scalar perturbations and polar (axial) modes in electromagnetic and gravitational perturbations.
A detailed derivation of Eqs.~\eqref{MomegaRa0} and ~\eqref{MomegaIa0} is given in Appendix~\ref{app:analytics}.
As shown in Fig.~\ref{fig:QNMergoregiongrav}, the analytical QNMs agree 
with the numerical QNMs in the regime of validity of the approximation, i.e., $M \omega \ll 1$.

\begin{figure}[t]
\centering
\includegraphics[width=0.49\textwidth]{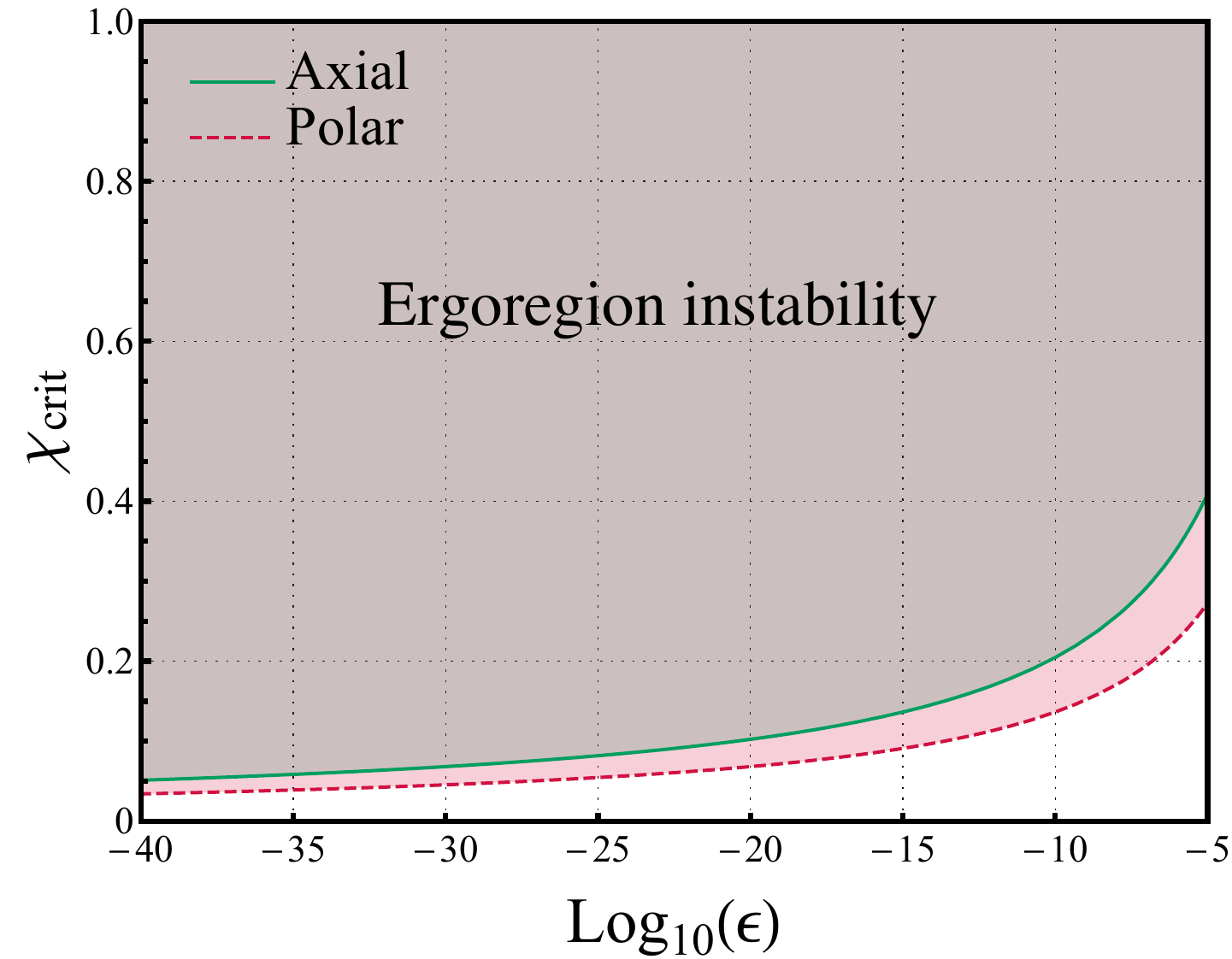}
\caption{Critical value of the spin above which a Kerr-like horizonless object is affected by the ergoregion instability as a function of the compactness of the object. The threshold is computed as in Eq.~\eqref{chi_crit} for axial and polar gravitational $\ell=m=2$ perturbations.~\cite{Maggio:2018ivz,Cardoso:2019rvt}}
\label{fig:chi_crit}
\end{figure}
Furthermore, the critical value of the spin can be computed analytically from Eqs.~\eqref{MomegaR} and~\eqref{MomegaI} that are accurate when $\omega_R \simeq \omega_I \simeq 0$. The ergoregion instability occurs for $\chi>\chi_{\rm crit}$ where~\cite{Maggio:2018ivz, Cardoso:2019rvt}
\begin{equation} \label{chi_crit}
    \chi_{\rm crit} \simeq \frac{\pi}{m |\log \epsilon|} \left[ p+\frac{s(s+1)}{2} \right] \,.
\end{equation}
Fig.~\ref{fig:chi_crit} shows the threshold of the ergoregion instability for gravitational $\ell=m=2$ perturbations as a function of the compactness of the object.
For example, an ECO with Planckian corrections at the horizon scale ($\epsilon \sim 10^{-40}$) is unstable if spinning above $\chi_{\rm crit} \simeq 0.03, 0.05$ for axial and polar perturbations, respectively. We  conclude that even slowly spinning Kerr-like horizonless objects are unstable due to the ergoregion instability. 

The timescale of the instability is defined as $\tau_{\rm inst} \equiv 1/\omega_I$. From Fig.~\ref{fig:QNMergoregiongrav}, for an ECO with $\epsilon=10^{-10}$ and spin $\chi=0.7$, the instability timescale of the $\ell=m=2$ mode is
\begin{equation}
    \tau_{\rm inst} \in (5,7) \left( \frac{M}{10 M_\odot}\right) \ \rm s \,,
\end{equation}
where the lower (upper) bound is for polar (axial) gravitational perturbations. Let us notice that the low-frequency approximation of $\omega_I$ in Eq.~\eqref{MomegaI} is not accurate for the unstable modes with large spin, as shown in Fig.~\ref{fig:QNMergoregiongrav} from the disagreement between the numerical and the analytical curves. For this reason, we  use the numerical values of the imaginary part of the QNMs for the calculation of the instability timescales.

The ergoregion instability acts on a timescale which is short compared to the accretion timescale of astrophysical BHs, i.e., $\tau_{\rm Salpeter} \sim 4 \times 10^7 \ \rm yr$. However, the instability timescale is longer than the decay time of the BH ringdown, i.e., $\tau_{\rm ringdown} \sim 0.5 \ \rm ms$ for a $10M_{\odot}$ compact object. If the remnant of a compact binary coalescence was an ECO, the ergoregion instability would spin down the remnant over a timescale $\tau_{\rm inst}$ until the condition for the stability, $\chi=\chi_{\rm crit}$, is satisfied~\cite{Brito:2015oca} via the emission of GWs. The incoherent superposition of the GW signals from the unresolved sources in the population would produce a stochastic GW background due to spin loss~\cite{Fan:2017cfw,Du:2018cmp}. 
The absence of such background in the first observing run of Advanced LIGO already puts strong constraints on perfectly reflecting ECOs that can be a small percentage of the astrophysical population~\cite{Barausse:2018vdb}.

\section{How to quench the ergoregion instability} \label{sec:quenchergoregion}

The case of a horizonless compact object with a perfectly reflecting surface is an idealization. In reality, we expect a compact object to absorb part of the ingoing radiation through viscosity, dissipation, fluid mode excitation, nonlinear effects, etc. Given that Kerr BHs can absorb radiation efficiently and are stable against the ergoregion instability, it is relevant to ask whether some absorption at the surface of Kerr-like compact objects can quench the ergoregion instability.

Let us define the reflection and transmission coefficients of a wave coming from the left of the photon-sphere barrier with unitary amplitude as
\begin{equation} \label{RBH_TBH}
    \mathcal{R}_{\text{BH}} = \frac{B_{\rm in}}{B_{\rm out}} \,, \qquad \mathcal{T}_{\text{BH}} = \frac{1}{B_{\rm out}} \,.
\end{equation}
After each bounce in the cavity between the ECO radius and the photon sphere the perturbation acquires a factor $\mathcal{R}\mathcal{R}_{\rm BH}$, where $\mathcal{R}$ is the ECO surface reflectivity and $\mathcal{R}_{\rm BH}$ is defined in Eq.~\eqref{RBH_TBH}. Due to the conservation of the Wronskian, $|\mathcal{R}_{\rm BH}| = \left| A_{\rm out} / A_{\rm in} \right|$ where $A_{\rm in}$ and $A_{\rm out}$ are the coefficients of the incident and reflected wave, respectively, at the photon sphere for a left-moving wave originating at infinity.
It follows that the amplification factor in the cavity is the same as the amplification factor of BHs which is defined as %
\begin{equation}
    Z_{s \ell m} = \left| \frac{A_{\rm out}}{A_{\rm in}} \right|^2 -1 \,.
\end{equation}
The condition for the energy in the cavity to grow indefinitely is $|\mathcal{R} \mathcal{R}_{\rm BH}|^2>1$ which implies that the object is unstable for the ergoregion instability if
\begin{equation} \label{ergoregionsuperradiance}
    |\mathcal{R}|^2 > \frac{1}{1+Z_{s \ell m}} \,.
\end{equation}
By definition, the surface reflectivity is $|\mathcal{R}|^2 \leq 1$, therefore Eq.~\eqref{ergoregionsuperradiance} implies that the ergoregion instability occurs when the real part of the QNM is in the superradiant regime, i.e.,  $Z_{s \ell m}>0$. In order to quench the ergoregion instability at any frequencies, the surface absorption, $1-|\mathcal{R}|^2$, needs to be larger than the maximum amplification factor of superradiance, namely
\begin{equation}
    1-|\mathcal{R}|^2 \gtrsim Z_{\rm max} \,.
\end{equation}
\begin{figure}[t]
\centering
\includegraphics[width=0.65\textwidth]{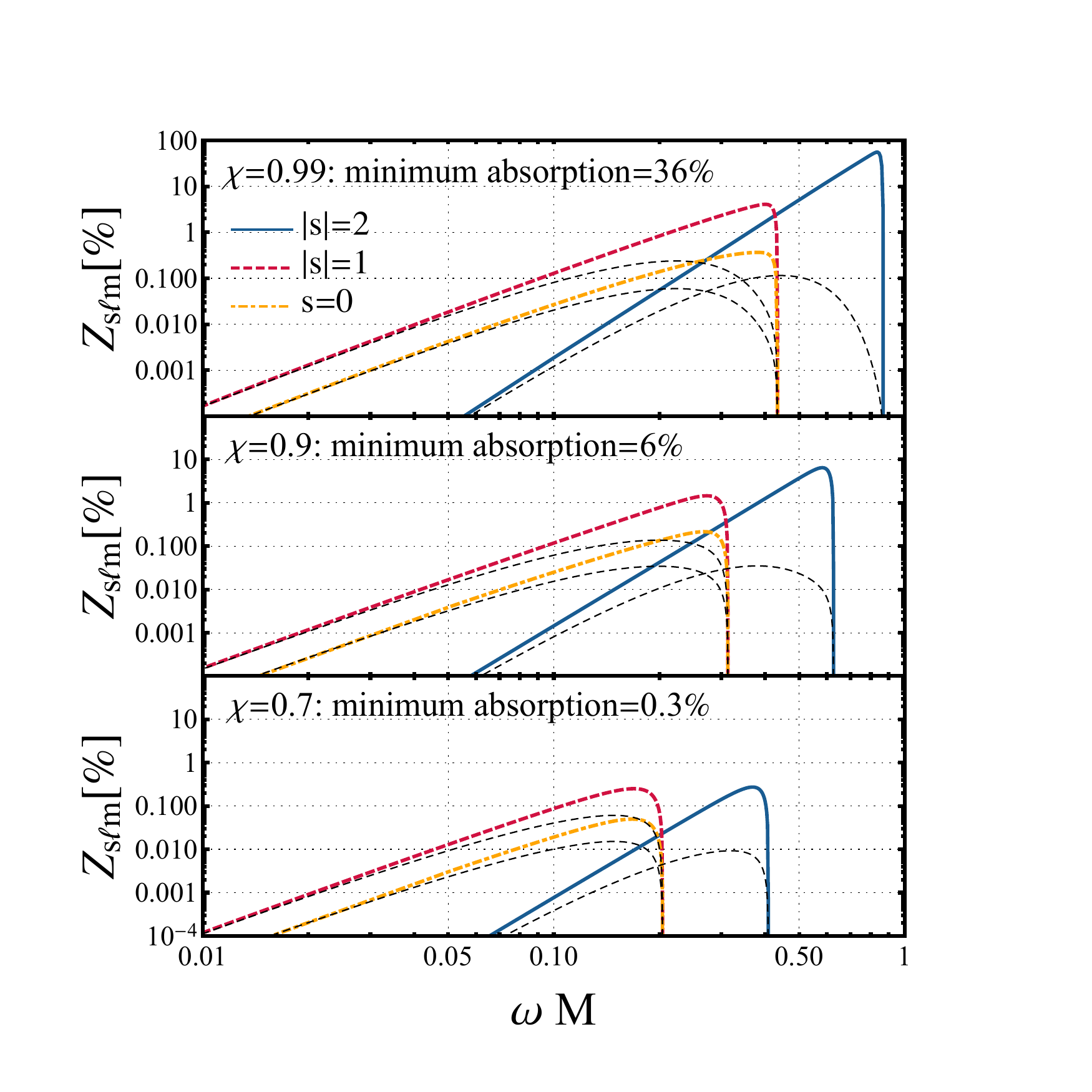}
\caption{Superradiant amplification factor of a BH as a function of the frequency for ($\ell=m=1$) scalar, electromagnetic and ($\ell=m=2$) gravitational perturbations. The analytical approximation valid at low-frequency (black dashed lines) is compared to the exact numerical result.
The minimum absorption rate to have a stable Kerr-like horizonless object for any type of perturbation is $0.3\%$ ($6\%$) for a remnant with spin $\chi=0.7$ ($\chi=0.9$).~\cite{Maggio:2018ivz}} 
\label{fig:Z}
\end{figure}
Figure~\ref{fig:Z} shows the amplification factor of a BH as a function of the frequency under scalar, electromagnetic and gravitational perturbations and for several values of the BH spin. In order to have a stable Kerr-like horizonless object under any type of perturbation, the surface absorption needs to be at least $0.3\%$ ($6\%$) for an ECO with $\chi=0.7$ ($\chi=0.9$). Let us notice that the maximum amplification factor of an extremal BH is $\approx 138 \%$ for $\ell=m=2$ gravitational perturbations~\cite{Brito:2015oca,Teukolsky:1974yv}, therefore an absorption rate of $\approx 60 \%$ would allow for stable Kerr-like horizonless objects with any spin. 

A natural question is whether this level of absorption is achievable by standard matter in compact objects. The reflective properties of compact objects depend on the specific model, but should generically be more extreme than those of an ordinary NS. For NSs, the most efficient absorption mechanism is due to viscosity. A rough estimate of the kinematic viscosity yields~\cite{1987ApJ...314..234C}
\begin{equation}
 \nu \approx 10^{-17}\left(\frac{\rho}{10^{14}\,{\rm g/cm}^3}\right)^{5/4}\left(\frac{T}{10^8\, K}\right)^{-2}\, {\rm s}\,. \label{kinviscosity}
\end{equation}
where $\rho$ and $T$ are the typical density and temperature of a NS, respectively.
As a response to some external perturbation, a viscous fluid can dissipate radiation. The fraction of gravitational energy converted into mechanical energy in a viscous and compressible fluid was estimated in Refs.~\cite{1971ApJ...165..165E,1985ApJ...292..330P}, finding that dissipation occurs through sound waves that propagate in the interior of the fluid and through shear waves that heat the surface. In the limit $\nu\omega\ll1$, which is valid in the entire parameter space of interest, and after an angle average, the fraction of absorbed energy in the flat spacetime approximation reads~\cite{1971ApJ...165..165E}
\begin{eqnarray}
 \nonumber 1-|\mathcal{R}|^2&\sim&\frac{64 \rho}{3\omega^2}(\omega\nu)^{3/2} \\
 &\approx& 0.004\, \left(\frac{M}{r_0}\right)^{27/4} \left[\frac{10^3\, K}{T}\right]^{3} \sqrt{\frac{0.01}{\omega M}}\left(\frac{20 M_\odot}{M}\right)^4\,,\label{fraction}
\end{eqnarray}
where we have normalized the physical quantities by their typical values expected for an ECO in the BH limit, namely a density similar to that of a fastly spinning Kerr BH, $r_0\sim M$, and a low temperature.
As a reference, the local temperature of an isolated gravastar is of the order of the Hawking temperature $T_{\rm H}\approx 10^{-7}\,{\rm K}$ for $M\sim 20\,{\rm M_\odot}$~\cite{Mazur:2001fv}. This temperature is negligible in astrophysical scenarios and the object would be in thermal equilibrium with the hotter environment. The temperature of the interstellar medium typically ranges between $10\,{\rm K}$ and $10^4\,{\rm K}$, so the normalization $T\approx 10^3\,{\rm K}$ adopted above is a conservative upper bound.
The estimate in Eq.~\eqref{fraction} is only indicative and shows that absorption at percent level can be naturally achieved by ECOs.

Some models of quantum BHs have a frequency-dependent reflectivity $\mathcal{R}(\omega) = e^{-|k|/(2T_{\rm H})}$ that allows for stable solutions against the ergoregion instability for any spin~\cite{Oshita:2019sat}. This model of horizonless compact object is analyzed in Sec.~\ref{sec:boltzmann}.

\section{Appendix: The Detweiler transformation} \label{app:detweiler}

In this appendix, we derive the transformation that brings the Teukolsky equation~\eqref{wave_eq} in the Schrödinger-like form in  Eq.~\eqref{final} with real effective potential.
In doing so, we  revisit and extend the original calculations by Detweiler~\cite{1977RSPSA.352..381D}. In particular, we refer the reader to the original work in Ref.~\cite{1977RSPSA.352..381D} for the explicit transformations in the electromagnetic case. In in the gravitational case, instead, we  correct several mistakes in Eqs.~(B3)--(B14) in Ref.~\cite{1977RSPSA.352..381D} and provide explicit expressions of $\alpha$ and $\beta$ extending the calculations in Ref.~\cite{10.2307/79029} to gravitational perturbations.

The Starobinsky identity for gravitational perturbations is~\cite{Teukolsky:1974yv}
\begin{equation}
\frac{1}{4} R_{2} = \mathscr{D} \mathscr{D} \mathscr{D} \mathscr{D}  R_{-2} \,, \label{staridentity}
\end{equation}
where $\mathscr{D} = \partial_r - i K/\Delta$. According to Eq.~\eqref{staridentity}, we can write
\begin{equation}
R_{2} = \mathfrak{a} R_{-2} + \frac{\mathfrak{b}}{\Delta} \frac{dR_{-2}}{dr} \,,
\end{equation}
where
\begin{eqnarray}
\mathfrak{a} &=& (a_1 + i a_2) \,, \\
\mathfrak{b} &=& i b_2 \,,
\end{eqnarray}
and
\begin{eqnarray}
\nonumber a_1 &=& 4 \left[ \frac{8 K^4}{\Delta^4} + \frac{8 K^2}{\Delta^3} \left( \frac{M^2-a^2}{\Delta} - \lambda\right) \right. - \frac{4 \omega K}{\Delta^3} (3r^2+2Mr-5a^2)\\
&&+ \left. \frac{12r^2 \omega^2 + \lambda (\lambda+2)}{\Delta^2} \right] \,, \\
a_2 &=& 4 \left[- \frac{24 \omega r K^2}{\Delta^3} + \frac{1}{\Delta^2} \left( \frac{4 \lambda (r-M)K}{\Delta}
+ 4 \omega r \lambda + 12 \omega M \right) \right] \,, \\
b_2 &=& 4 \left[\frac{8 K^3}{\Delta^2} + \frac{4 K}{\Delta} \left( \frac{2(M^2-a^2)}{\Delta} - \lambda \right) 
- \frac{8 \omega}{\Delta} (Mr-a^2) \right]\,.
\end{eqnarray}
The radial functions $\alpha$ and $\beta$ that define the Detweiler function in Eq.~\eqref{DetweilerX} are
\begin{eqnarray}
\alpha &=& \frac{\kappa \mathfrak{a} \Delta^2 + |\kappa|^2}{\sqrt{2}|\kappa| \left[a_1 \Delta^2 + \mathfrak{R} (\kappa)\right]^{1/2}} \,, \\
\beta &=& \frac{i \kappa b_2 \Delta^2}{\sqrt{2} |\kappa| \left[a_1 \Delta^2 + \mathfrak{R}(\kappa)\right]^{1/2}} \,,
\end{eqnarray}
where
\begin{eqnarray}
\nonumber \kappa &=& 4 \left[\lambda^2 (\lambda+2)^2 + 144a^2 \omega^2 (m-a\omega)^2 -  a^2 \omega^2 (40\lambda^2-48\lambda) \right.\\
&&+ \left. a\omega m (40\lambda^2+48\lambda)\right]^{1/2}+ 48 i \omega M \,, \\
\nonumber \mathfrak{R}(\kappa)  &=& 4 \left[\lambda^2 (\lambda+2)^2 + 144a^2 \omega^2 (m-a\omega)^2 
- a^2 \omega^2 (40\lambda^2-48\lambda) \right. \\
&&+ \left. a\omega m (40\lambda^2+48\lambda)\right]^{1/2} \,, \\
\nonumber |\kappa| &=& \left\{16 \left[\lambda^2 (\lambda+2)^2 + 144a^2 \omega^2 (m-a\omega)^2 \right. \right.
- a^2 \omega^2 (40\lambda^2-48\lambda) \\
&& \left. \left. +a\omega m (40\lambda^2+48\lambda)\right] 
+ \left(48 \omega M\right)^2 \right\}^{1/2}  \,.
\end{eqnarray}
With this choice of the parameters, $\alpha$ and $\beta$ satisfy the following relation
\begin{equation}
\alpha^2 - \alpha' \beta \Delta^{s+1} + \alpha \beta' \Delta^{s+1} - \beta^2 \Delta^{2s+1} V_s = \kappa \,,
\end{equation}
which guarantees that the Detweiler function defined in Eq.~\eqref{DetweilerX} satisfies the Schrödinger-like equation in Eq.~\eqref{final}. 
Eq.~\eqref{pot_detweiler} gives the following expression of the effective potential
\begin{equation}
V(r,\omega) = \frac{-K^2 + \Delta \lambda}{(r^2+a^2)^2} + \frac{\Delta (b_2 p' \Delta)'}{(r^2+a^2)^2 b_2 p} + G^2 + \frac{dG}{dr_*} \,, \label{pot_detw_final}
\end{equation}
where
\begin{equation}
p = |\kappa| \left\{ 2 \left[ a_1 \Delta^2 + \mathfrak{R}(\kappa) \right]\right\}^{-1/2} \,.
\end{equation}
The effective potential in Eq.~\eqref{pot_detw_final} is purely real and has the following asymptotics: 
at infinity $V(r\to+\infty,\omega) \to - \omega^2$, and at the horizon $V(r\to r_+,\omega) \to - k^2$.

Finally, the conserved energy flux is the same if computed by the two independent solutions of the Teukolsky equation [Eq.~\eqref{wave_eq}] or the two independent solutions of the Detweiler equation [Eq.~\eqref{final}]~\cite{10.2307/79029}. This is an important 
consistency check since the energy flux is a measurable quantity and cannot depend on the transformation of the perturbation variable.

\section{Appendix: Boundary condition for spinning perflectly reflecting objects} \label{app:electromagneticBCspin} 

In this Appendix, we  derive the boundary conditions that describe a perflectly reflecting Kerr-like object under electromagnetic perturbations~\cite{Brito:2015oca}. The horizonless object can be modeled as a perfect conductor, whose electric and magnetic fields satisfy $E_\theta(r_0) = E_\varphi(r_0) =0$ and $B_r(r_0)=0$. In the Newman-Penrose formalism, the previous conditions can be written in terms of the three complex scalars of the electromagnetic field $\phi_0, \phi_1, \phi_2$ in the frame of a ZAMO~\cite{King:1977}
\begin{eqnarray}
\nonumber E_\theta &=& \left[ \frac{\Delta^{1/2} (r^2+a^2)}{\sqrt{2} \hat\rho^* A^{1/2} (r^2 + a^2 \cos^2 \theta)} \left( \frac{\phi_0}{2} - \frac{\phi_2}{\hat\rho^2 \Delta} \right) + \text{c.c.} \right] - \frac{2 a \Delta^{1/2}}{A^{1/2}} \sin \theta \mathfrak{I}(\phi_1) \,, \\ \\
E_\varphi &=& \left[ -\frac{i \Delta^{1/2} \hat\rho}{\sqrt{2}} \left( \frac{\phi_0}{2 } + \frac{\phi_2}{\hat\rho^2 \Delta} \right) + \text{c.c.} \right] \,, \\
B_r &=& \left[ \frac{a \sin \theta}{\sqrt{2} \hat\rho A^{1/2}} \left( \phi_2 - \Delta \hat\rho^2 \frac{\phi_0}{2} \right) + \text{c.c.} \right] + 2 \frac{r^2+a^2}{A^{1/2}} \mathfrak{I}(\phi_1) \,,
\end{eqnarray}
where $\hat\rho=-(r-i a \cos \theta)^{-1}$, $A = (r^2+a^2)^2-a^2 \Delta \sin^2 \theta$, and $\text{c.c.}$ stands for the complex conjugate of the previous term. The conditions of a perfect conductor translate into
\begin{equation}
    |\Phi_0|^2 = \frac{|\Phi_2|^2}{\Delta^2} \,, \quad \mathfrak{I}(\phi_1) = 0 \,, \label{BCperfectconductor}
\end{equation}
where $\Phi_0 \equiv \phi_0$ and $\Phi_2 \equiv 2 \hat\rho^{-2} \phi_2$. We  use the decomposition
\begin{eqnarray}
    \Phi_0 &=& \int d\omega e^{-i \omega t} \sum_{\ell m} e^{i m \varphi} ~_sS_{\ell m}(\theta) ~_sR_{\ell m}(r) \,, \label{Phi0} \\
    \Phi_2 &=& \int d\omega e^{-i \omega t} \sum_{\ell m} e^{i m \varphi} ~_{-s}S_{\ell m}(\theta) ~_{-s}R_{\ell m}(r) \label{Phi2} \,,
\end{eqnarray}
and omit the $\ell, m$ subscripts for brevity. The radial and the angular functions are related by the Starobinky identities~\cite{Teukolsky:1974yv,Starobinskij2}
\begin{equation}
    \mathscr{D} \mathscr{D} R_{-1} = B R_{1} \,, \quad \mathscr{L}_0 \mathscr{L}_1 S_1 = B S_{-1} \,, \label{staroem}
\end{equation}
where $B = \sqrt{\lambda^2 +4 m a \omega -4 a^2 \omega^2}$,  $\mathscr{D} = \partial_r - i K/\Delta$, and $\mathscr{L}_n = \partial_\theta +m \csc \theta - a \omega \sin \theta + n \cot \theta$. By substituting Eqs.~\eqref{Phi0},~\eqref{Phi2} and~\eqref{staroem} in the condition for a perfect conductor in Eq.~\eqref{BCperfectconductor}, we  obtain the boundary condition for a perflectly reflecting compact object on the Teuksolsky wave function
\begin{equation}
    \partial_r R_{-1} = \left[ \frac{i K}{\Delta} - \frac{i}{2 K} \left( \lambda \pm B + 2 i \omega r \right) \right] R_{-1} \,, \label{BCem}
\end{equation}
where the plus and minus signs refer to polar and axial perturbations, respectively. In the following, we show that the boundary conditions in Eq.~\eqref{BCem} are equivalent to Dirichlet and Neumann boundary conditions on the Detweiler wave function for axial and polar perturbations, respectively, as in Eqs.~\eqref{BCXdir} and~\eqref{BCXneu}.

Near the radius of an ultracompact object ($\epsilon \ll 1$), the Teuksolsky wave function has the following asymptotics~\cite{Teukolsky:1974yv}
\begin{equation}
R_{-1} \sim \mathcal{A} \Delta e^{-i k r_*} + \mathcal{B} e^{+i
k r_*} \label{R-1} \,, \quad r_* \rightarrow -\infty \,,
\end{equation}
where $\mathcal{A} = \mathcal{A_\text{0}} + \eta \mathcal{A_{\text{1}}} + ...$ and $\mathcal{B} = \mathcal{B_\text{0}} + \eta \mathcal{B_\text{1}} + ...$, with $\eta \equiv r-r_+$.
Since $\Delta\sim (r_+-r_-)\eta$ near the surface, we  consider in Eq.~\eqref{R-1} $\mathcal{A} = \mathcal{A_\text{0}}$ and $\mathcal{B} =
\mathcal{B_\text{0}} + \eta \mathcal{B_\text{1}}$ where~\cite{Casals:2005kr}
\begin{eqnarray}
\mathcal{B_\text{0}} &=& -\frac{2^{1/2} (r_+^2 + a^2)^{1/2} k}{B}\,,\label{B0} \\
\mathcal{A_\text{0}} &=& - \frac{i B}{4 K_+ \mathbb{R}^*} \
\mathcal{B_\text{0}} \ B_{\rm in}/B_{\rm out} \,, \label{A0}
\end{eqnarray}
and $K_+ = K(r_+)$, $\mathbb{R} = i K_+ + (r_+ - r_-)/2$, and $B_{\rm in}$ and $B_{\rm out}$ are the asymptotic amplitudes defined in Eq.~\eqref{asymptoticsplus}. By inserting Eq.~\eqref{R-1} in the Teukolsky equation~\eqref{wave_eq}, we  find
\begin{equation}
\mathcal{B_\text{1}} = \left(\frac{i a m}{M(r_+ - r_-)} + \frac{2 \omega r_+ -
i \lambda}{4 M r_+ k} \right) \mathcal{B_\text0}\,. \label{B1}
\end{equation}
Equation~\eqref{R-1} with Eqs.~\eqref{B0},~\eqref{A0} and~\eqref{B1} defines the asymptotic expansion of the Teukolsky wave function $R_{-1}$ near the horizon at the first order in $\eta$. By inserting Eq.~\eqref{R-1} in the boundary condition \eqref{BCem}, we get the following expression
\begin{equation}
B_{\rm out} e^{i k r_{*}^{0}} \mp B_{\rm in} e^{-i k r_{*}^{0}} = 0\,,
\label{omega_vil0}
\end{equation}
for the two signs of Eq.~\eqref{BCem} that correspond to polar ($-$) and axial ($+$) modes, respectively. Eq.~\eqref{omega_vil0} takes the same form of Eqs.~\eqref{BCXdir} and~\eqref{BCXneu}, therefore the perfect-conductor boundary conditions imply
Dirichlet and Neumann boundary conditions on the Detweiler wave function for axial and polar modes, respectively. Let us emphasize that the boundary conditions are derived when the radius of the compact object is at microscopical distance from the would-be horizon, i.e., $\epsilon \ll 1$.

\section{Appendix: Analytical quasi-normal modes} \label{app:analytics}

In this appendix, we  derive the QNMs of a Kerr-like horizonless object analytically in the small-frequency regime. We  focus on an ultracompact object whose radius is located as in Eq.~\eqref{radius} with $\epsilon \ll 1$, and whose surface is perfectly reflecting. We  use a matched asymptotic expansion according to which the radial domain of the exterior spacetime is split into two regions: the near-region, i.e., $r-r_+ \ll 1/\omega$, and the far-region, i.e., $r- r_+ \gg M$. We  solve the radial Teukolsky equation~\eqref{wave_eq} in each region and we  match the inner and outer solutions in the overlapping region where $M \ll r-r_+ \ll 1/\omega$. Finally, we  impose the perfectly reflecting boundary conditions as in Eqs.~\eqref{BCXdir} and~\eqref{BCXneu} and we  derive the characteristic frequencies of the object.

In the region near the radius of the compact object, the radial wave equation~\eqref{wave_eq} reduces to \cite{Starobinskij2}
\begin{equation}
 \left[z(z+1)\right]^{1-s} \partial_z \left\{ \left[z(z+1)\right]^{s+1} \partial_z R_s \right\}+\left[ Q^2 + i Q s (1+2z) - \lambda z (z+1)\right] R_s = 0 \,, \label{wave_eq_near_hor}
\end{equation}
where $z=(r-r_+)/(r_+ - r_-)$, $Q=(r_+^2 + a^2)(m \Omega_H - \omega)/(r_+ - r_-)$, and 
$\lambda = (\ell-s)(\ell+s+1)$. Eq.~\eqref{wave_eq_near_hor} is valid when $M \omega \ll 1$ 
and it is derived by neglecting the terms proportional to $\omega$ in Eq.~\eqref{wave_eq} 
except for the ones which enter into $Q$. The general solution of 
Eq.~\eqref{wave_eq_near_hor} is a linear combination of hypergeometric functions, i.e.,
\begin{eqnarray}
\nonumber R_s &=& (1+z)^{iQ} \left[{C_1} z^{-iQ} 
~_{2}F_1\left(-\ell+s,\ell+1+s;1-\bar{Q}+s;-z\right) \right. \\
&& \left. + {C_2} z^{iQ-s} 
~_{2}F_1\left(-\ell+\bar{Q},\ell+1+\bar{Q};1+\bar{Q}-s;-z\right)\right] \,, \label{sol_nearhor}
\end{eqnarray}
where $\bar{Q} \equiv 2iQ$. The large-$r$ behavior of the solution is
\begin{eqnarray}
 \nonumber R_s &\sim& \left(\frac{r}{r_+ - r_-}\right)^{\ell-s} \Gamma(2\ell+1) \left[\frac{{C_1} \Gamma(1-\bar{Q}+s)}{\Gamma(\ell+1-\bar{Q}) \Gamma(\ell+1+s)} \right. \\
&& + \left. \frac{{C_2} \Gamma(1+\bar{Q}-s)}{\Gamma(\ell+1+\bar{Q}) \Gamma(\ell+1-s)} \right] + 
\left(\frac{r}{r_+ - r_-}\right)^{-\ell-1-s} \frac{(-1)^{\ell+1+s}}{2 \Gamma(2\ell+2)} \nonumber \\
&& \left[\frac{{C_1} \Gamma(\ell+1-s) 
\Gamma(1-\bar{Q}+s)}{\Gamma(-\ell-\bar{Q})} 
+ \frac{{C_2} \Gamma(\ell+1+s) \Gamma(1+\bar{Q}-s)}{\Gamma(-\ell+\bar{Q})} \right] \,, \nonumber \\ 
\label{Rhorizon}
\end{eqnarray}
where the ratio of the coefficients ${C_1}/{C_2}$ is fixed by the boundary condition at the radius of the compact object.

At infinity, the radial wave equation~\eqref{wave_eq} reduces to~\cite{Cardoso:2008kj}
\begin{equation}
r \partial_{r}^2 f_{s} + 2 \left(\ell+1-i \omega r\right) \partial_{r} f_{s} - 2 i \left(\ell+1-s\right) \omega f_{s} 
= 0 \,, \label{wave_eq_inf}
\end{equation}
where $f_{s} \equiv e^{i \omega r} r^{-\ell+s} R_{s}$. The general solution of 
Eq.~\eqref{wave_eq_inf} is a linear combination of a confluent hypergeometric function 
and a Laguerre polynomial, i.e., 
\begin{equation}
R_{s} = e^{-i \omega r} r^{\ell-s} \left[ C_3 U\left(\ell+1-s,2\ell+2,2i \omega r\right)
+ C_4 L_{-\ell-1+s}^{2\ell+1}\left(2 i \omega r\right) \right] \,,
\end{equation}
where $C_4 = (-1)^{\ell-s} C_3 \Gamma(-\ell+s)$ by imposing purely outgoing waves at infinity.
The small-$r$ behavior of the solution is
\begin{equation}
R_{s} \sim C_3 \left[ r^{\ell-s} \frac{(-1)^{\ell-s}\Gamma(\ell+1+s)}{2\Gamma(2 \ell+2)} 
+ r^{-\ell-1-s} \left(2 i \omega\right)^{-2 \ell-1} \frac{\Gamma(2\ell+1)}{\Gamma(\ell+1-s)} \right] \,. 
\label{Rinfinity}
\end{equation}

The matching of Eqs.~\eqref{Rhorizon} and \eqref{Rinfinity} in the intermediate region where $M \ll r-r_+ \ll 1/\omega$
yields
\begin{equation}
\frac{{C_1}}{{C_2}} = -\frac{\Gamma(\ell+1+s)}{\Gamma(\ell+1-s)} \left[\frac{R_+ + i (-1)^\ell 
\left(\omega (r_+ - r_-)\right)^{2\ell+1} L S_+}{R_- + i (-1)^\ell \left(\omega (r_+ - r_-)\right)^{2\ell+1} L S_-} 
\right] \,, \label{eq:ab1}
\end{equation}
where
\begin{eqnarray}
R_\pm &\equiv& \frac{\Gamma(1 \pm \bar{Q} \mp s)}{\Gamma(\ell + 1 \pm \bar{Q})}\,, \quad
S_\pm \equiv \frac{\Gamma(1 \pm \bar{Q} \mp s)}{\Gamma(-\ell \pm \bar{Q})}\,, \nonumber \\
L &\equiv& \frac{1}{2} \left[\frac{2^\ell \, \Gamma(\ell+1+s)\Gamma(\ell+1-s)}{\Gamma(2\ell+1) 
\Gamma(2\ell+2)} \right]^{2} \,.
\end{eqnarray}

\subsection*{Electromagnetic perturbations}\label{app:matchings1}

For $s=-1$, the ratio ${C_1} / {C_2}$ is derived by imposing the boundary conditions~\eqref{BCXdir} and~\eqref{BCXneu} in the near-horizon expansion of the solution in the near-region. At the radius of the object, we  obtain 
\begin{equation}
\frac{{C_1}}{{C_2}} = \mp B^{-1} \bar{Q} z_0^{\bar{Q}} \,. \label{eq:ab2}
\end{equation}
where $z_0 \equiv z(r_0)$, and the minus and plus signs refer to polar and axial perturbations, respectively.
By equating Eq.~(\ref{eq:ab1}) with Eq.~(\ref{eq:ab2}), we  obtain an algebraic equation for the real part of the QNM frequencies.
An approximate solution can be found in the small-spin and small-frequency regime for which $\bar{Q} \ll 1$. In this regime, Eq.~\eqref{eq:ab1} 
reduces to ${C_1} / {C_2}= \bar{Q} / \left[\ell \left(\ell+1\right)\right]$, whereas $B \approx \ell (\ell+1)$ in 
Eq.~\eqref{eq:ab2}. It follows
\begin{equation}
z_0^{-2iQ} = \mp 1 \,. \label{x0}
\end{equation}
By using the tortoise coordinate where $ \log (z_0) \sim r_*^0 (r_+ 
- r_-)/(r_+^2 + a^2)$, Eq.~\eqref{x0} yields
\begin{equation}
e^{-2 i Q r_*^0 (r_+ - r_-)/(r_+^2 + a^2)} = \mp 1 \,, \label{eqx0}
\end{equation}
which is analogous to Eq.~(A18) in Ref.~\cite{Cardoso:2008kj} in the case of scalar perturbations. 
The solution of Eq.~\eqref{eqx0} is
\begin{equation}
\omega_R = - \frac{\pi p}{2 |r_*^0|} + m \Omega_H \,, \label{omega_em}
\end{equation}
where $p$ is a positive odd (even) integer for polar (axial) modes. 
Equation~\eqref{omega_em} is also valid for scalar perturbations where $p$ is a 
positive odd (even) integer for the modes with Neumann (Dirichlet) boundary condition.

\subsection*{Gravitational perturbations} \label{app:matchings2}

For $s=-2$, the ratio ${C_1} / {C_2}$ is derived by imposing the boundary conditions~\eqref{BCXdir} and~\eqref{BCXneu} in the near-horizon expansion of the solution in the near-region. When $\bar{Q} \ll 1$, we  obtain at the radius of the object
\begin{equation}
\frac{{C_1}}{{C_2}} = \mp \frac{2}{(\ell+2)(\ell+1)\ell(\ell-1)} \bar{Q} z_0^{\bar{Q}} \,, 
\label{eq:ab2grav}
\end{equation}
where the minus and plus signs refer to polar and axial perturbations, respectively.
For $\bar{Q} \ll 1$, Eq.~\eqref{eq:ab1} reduces to ${C_1} / {C_2}= -2\bar{Q} / 
[(\ell+2)(\ell+1)\ell (\ell-1)]$. By equating the latter equation with Eq.~(\ref{eq:ab2grav}), it follows
\begin{equation}
z_0^{-2iQ} = \pm 1 \,, \label{x0grav}
\end{equation}
whose solution is
\begin{equation}
\omega_R = - \frac{\pi (p+1)}{2 |r_*^0|} + m \Omega_H \,, \label{omega_grav}
\end{equation}
where $p$ is a positive odd (even) integer for polar (axial) modes. By comparing Eq.~\eqref{omega_em} with Eq.~\eqref{omega_grav}, we  notice that the gravitational QNM frequencies have a $\pi$ phase shift with respect to the scalar and electromagnetic ones. 

The analytic expression of the real part of the QNMs for a generic spin-$s$ perturbation is given in Eq.~\eqref{MomegaR}.

\subsection*{Imaginary part of the quasi-normal mode frequencies} \label{app:matchings3}

To compute the imaginary part of the QNMs analytically, we  impose the boundary conditions in Eq.~\eqref{BCXdir} and~\eqref{BCXneu} on the near-horizon asymptotics of the Detweiler wave function in Eq.~\eqref{asymptoticsplus}. The computation yields
\begin{equation} \label{omegavilenkin}
\omega = - \frac{1}{2 |r_*^0|} \left( p \pi+ \Phi - i \ln \left| \frac{B_{\rm in}}{B_{\rm out}}\right| \right) + m \Omega_H \,, 
\end{equation}
where $\Phi$ is a phase that depends on the spin-$s$ of the perturbation and it is derived with the matching asymptotic expansion described above. 

According to the Wronskian relations in Eq.~\eqref{wronskian}, $|B_{\rm in}/B_{\rm out}| = |A_{\rm out}/A_{\rm in}|$ therefore the amplification factor of perturbations in the cavity between the radius of the object and the photon sphere is related to the amplification factor of BHs. For a perturbation of spin $s$, the amplification factor of BHs is defined as
\begin{equation}
Z_{s \ell m} = \left|\frac{A_{\rm out}}{A_{\rm in}}\right|^2  -1 \,, \label{Z}
\end{equation}
that has the following form in the low-frequency regime as computed by Starobinsky~\cite{Starobinskij2}
\begin{equation}
Z_{s \ell m} \equiv -D_{s \ell m} = 4 Q \beta_{s \ell}\prod_{k=1}^{\ell} \left(1 + \frac{4 Q^2}{k^2}\right)
\left[\omega (r_+ - r_-)\right]^{2\ell+1} \,. \label{amplfactor}
\end{equation}
In our calculations, we  impose $Z_{s \ell m} \equiv -\mathfrak{R}(D_{s \ell m})$ since $\omega_I\ll\omega_R$.
By inserting Eq.~\eqref{amplfactor} in Eq.~\eqref{Z} and using the Wroskian relations, we  derive
\begin{equation}
\left|\frac{B_{\rm in}}{B_{\rm out}}\right|^2 - 1 = - \mathfrak{R}(D_{s \ell m}) \,. \label{eq}
\end{equation}
Eqs.~\eqref{omegavilenkin} and~\eqref{eq} yield to the analytical expression of the imaginary part of the QNM frequencies in the low-frequency regime
\begin{equation}
    \omega_I \simeq -\frac{\mathfrak{R}(D_{s \ell m})}{4 |r_*^0|} \,,
\end{equation}
that coincides with Eq.~\eqref{MomegaI}. 
Let us notice that $Z_{s \ell m}>0$ (i.e., $\omega_I>0$) in the superradiant regime, where $\omega_R(\omega_R-m\Omega)<0$.
Consequently, the unstable modes of a
perfectly-reflecting Kerr-like object can be understood in terms of
waves amplified in the ergoregion and being reflected at the boundary.

\chapter{Gravitational-wave echoes} \label{chapter5}

\begin{flushright}
    \emph{
    Una notte osservavo come al solito il cielo col mio telescopio. Notai che da una galassia lontana cento milioni d’anni-luce sporgeva un cartello. C’era scritto: TI HO VISTO. 
    [...] Proprio duecento milioni d’anni prima, né un giorno di più né un giorno di meno, m’era successo qualcosa che avevo sempre cercato di nascondere.
    }\\
    \vspace{0.1cm}
    Italo Calvino, Le Cosmicomiche
\end{flushright}
\vspace{0.5cm}

\section{Schematic picture} \label{sec:echoespicture}

%
\begin{figure}[t]
\centering
\includegraphics[width=0.65\textwidth]{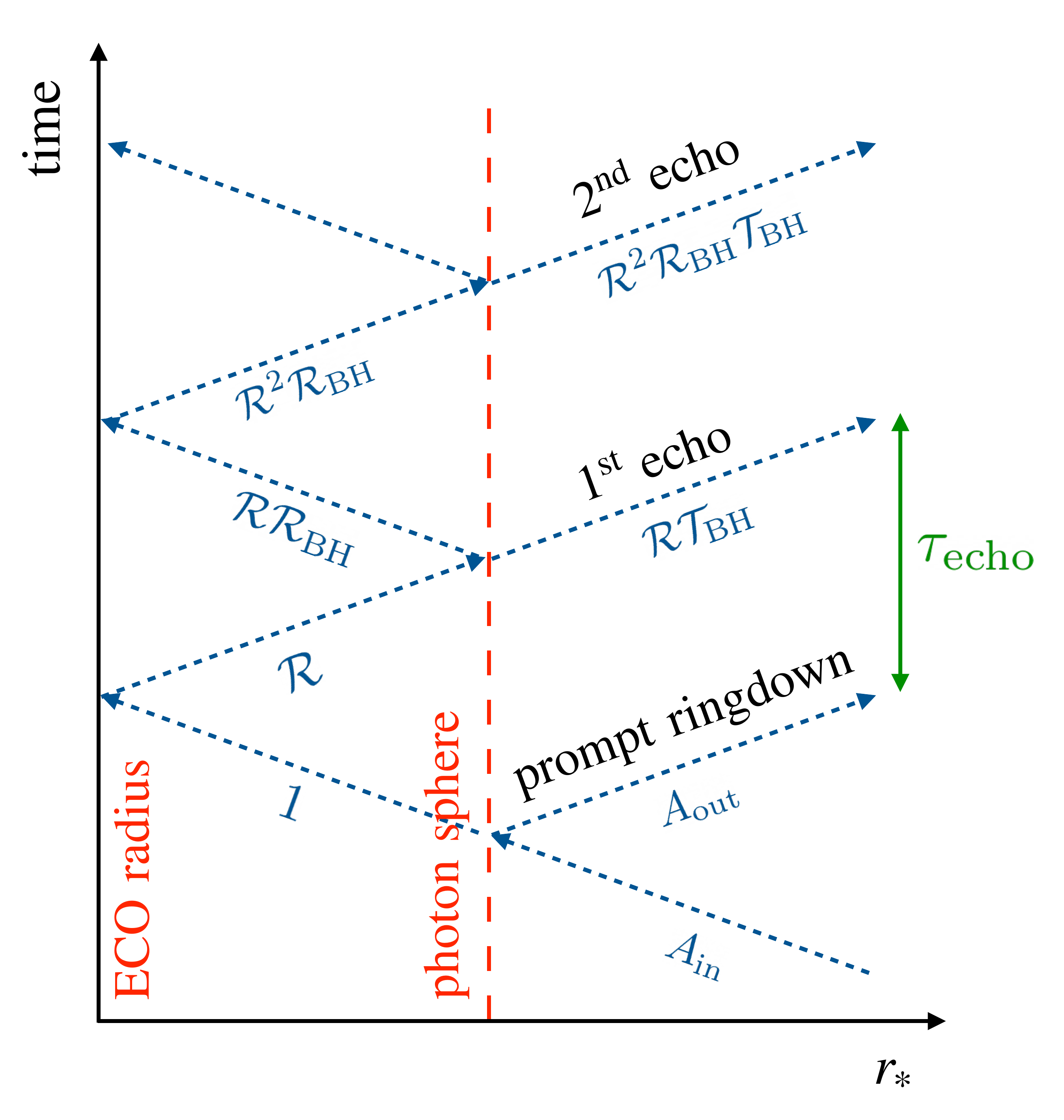}
\caption{Schematic diagram of the propagation of a perturbation in the background geometry of a horizonless compact object~\cite{Vilenkin:1978uc,Abedi:2016hgu,Cardoso:2019rvt}. When the perturbation excites the photon sphere, a prompt ringdown signal is emitted at infinity. Subsequent bounces of the perturbation in the cavity between the radius of the object and the photon sphere are responsible for the emission of GW echoes.~\cite{Maggio:2021ans}} 
\label{fig:cavityechoes}
\end{figure}
GW echoes are an additional signal that would be emitted in the postmerger phase of a compact binary coalescence when the remnant is a horizonless ultracompact object. Possible sources of GW echoes are near-horizon quantum structures~\cite{Cardoso:2016rao,Cardoso:2016oxy,Wang:2019rcf}, ultracompact NSs~\cite{Ferrari:2000sr,Pani:2018flj}, and BHs in modified theories of gravity in which the graviton reflects effectively on a hard wall~\cite{Zhang:2017jze,Oshita:2018fqu}. The key feature of the sources of GW echoes is the existence of a cavity in the effective potential of the perturbed object between the photon sphere and the effective radius of the object, as shown in Fig.~\ref{fig:cavityechoes}. If the object is sufficiently compact, the cavity can support quasi-trapped modes that leak out of the potential barrier through tunneling effects and are responsible for the emission of GW echoes.

To describe the dynamical emission of GW echoes, we analyze the scattering of a Gaussian pulse starting from infinity and going towards the compact object. As shown in Fig.~\ref{fig:cavityechoes}, when the pulse crosses the photon-sphere barrier and perturbs it, a prompt ringdown signal is emitted at infinity~\cite{Vilenkin:1978uc,Abedi:2016hgu,Cardoso:2019rvt}. The prompt ringdown emitted by an ultracompact horizonless object is almost indistinguishable from the BH ringdown since the photon sphere is approximately at the same location and has a similar shape~\cite{Cardoso:2016rao}. Afterward, the perturbation travels inside the photon-sphere barrier and is reflected by the surface of the compact object. A fraction of the radiation is absorbed by the compact object depending on its reflective properties~\cite{Maggio:2018ivz,Oshita:2019sat}. 

After each interaction with the photon sphere, a GW echo is emitted at infinity with a progressively smaller amplitude. The amplitude of the GW echoes depends on the surface reflectivity of the object $\mathcal{R}$ and the reflection ($\mathcal{R}_{\rm BH}$) and transmission ($\mathcal{T}_{\rm BH}$) coefficients of the wave coming from the left of the photon sphere. After each bounce in the cavity between the ECO radius and the photon sphere, the perturbation acquires a factor $\mathcal{R}\mathcal{R}_{\rm BH}$.

The photon-sphere barrier acts as a frequency-dependent high-pass filter. The characteristic frequencies governing the prompt ringdown are approximately the BH QNM frequencies despite the latter are not part of the QNM spectrum of horizonless compact objects. Each subsequent GW echo has a lower frequency content, and at late times the GW signal is dominated by the low-frequency QNMs of horizonless compact objects.

The delay time between subsequent GW echoes is proportional to the width of the cavity, therefore, the compactness of the object. The delay time can be computed as the round-trip time of the radiation from the photon sphere to the boundary. In the non-spinning case~\cite{Cardoso:2016rao,Cardoso:2016oxy}
\begin{equation} \label{tauecho}
    \tau_{\mathrm{echo}} = 2 M \left[1 - 2 \epsilon -2 \log (2 \epsilon) \right] \,,
\end{equation}
where the logarithmic dependence on $\epsilon$ would allow detecting even Planckian corrections ($\epsilon \sim l_{\rm Planck}/M$) at the horizon scale few $\text{ms}$ after the merger of a compact binary coalescence with a remnant of $M\sim10M_\odot$.

\section{Analytical template} \label{sec:template}

In this Section, we shall derive an analytical template for the ringdown and the GW echoes emitted by a spinning horizonless compact object. The waveform is parametrized by the standard ringdown parameters plus two quantities related to the properties of the exotic remnant. The template can be easily implemented to perform matched-filter-based searches for GW echoes and constrain models of horizonless compact objects.

\subsection{Transfer function}

We  analyze a spinning horizonless compact object whose exterior spacetime is described by the Kerr metric, as detailed in Sec.~\ref{sec:spinningmodel}.
The radius of the object is located as in Eq.~\eqref{radius}, where we  focus on ultracompact models with $\epsilon \ll 1$. We  require the location of the surface to be at a proper length $\delta \ll M$ from the would-be horizon, where
\begin{equation}
    \delta = \int_{r_+}^{r_0} dr \left. \sqrt{g_{rr}}\right|_{\theta=0} \,,
\end{equation}
and the relation between the proper length $\delta$ and $\epsilon$ is given  by 
\begin{equation}
    \epsilon \simeq \sqrt{1-\chi^2} \frac{\delta^2}{4 r_+^2} \,.
\end{equation}

Let us perturb the background geometry with a spin-$s$ perturbation. The radial component of the perturbation is governed by the inhomogeneous equation
\begin{equation} \label{detweiler_eq_inhom}
    \frac{d^2X_{s}}{dr_*^2}- V(r,\omega) X_{s}= \tilde{S}\,,
\end{equation}
where $X_s$ is the Detweiler function defined in Eq.~\eqref{DetweilerX}, $V(r,\omega)$ is the effective potential in Eq.~\eqref{pot_detweiler}, and $\tilde{S}$ is a source term. At asymptotic infinity, we  require the solution of Eq.~\eqref{detweiler_eq_inhom} to be a purely outgoing wave,  $X_s(\omega, r_* \to \infty) = \tilde{Z}^+(\omega) e^{i \omega r_*}$.

In the frequency domain, the GW signal emitted by a horizonless compact object can be written in terms of the GW signal that would be emitted by a BH and  reprocessed by a transfer function~\cite{Mark:2017dnq}, i.e.,
\begin{equation} \label{mark}
    \tilde{Z}^+ (\omega) = \tilde{Z}^+_{\rm BH} (\omega) + \mathcal{K}(\omega) \tilde{Z}^-_{\rm BH} (\omega) \,,
\end{equation}
where $\tilde{Z}^\pm_{\rm BH}$ are the responses of a Kerr BH (at infinity and near the horizon, for the plus and minus signs, respectively) to the source $\tilde{S}$, i.e.,
\begin{equation} \label{ZBHpm}
    \tilde{Z}^\pm_{\rm BH} = \frac{1}{W_{\rm BH}} \int_{-\infty}^{+\infty} dr_* \tilde{S} X_s^{\mp} \,,  
\end{equation}
where $X_s^\pm$ are the two independent solutions of the homogeneous equation associated to Eq.~\eqref{detweiler_eq_inhom} with asymptotics in Eqs.~\eqref{asymptoticsplus} and~\eqref{asymptoticsminus}, and $W_{\rm BH}$ is the Wronskian defined in Eq.~\eqref{wronskiandef}. The details of the horizonless compact object are all contained in the transfer function that is defined as~\cite{Mark:2017dnq}
\begin{equation} \label{transferfunction}
    \mathcal{K}(\omega) = \frac{\mathcal{T}_{\rm BH} \mathcal{R}(\omega) e^{-2 i k r_*^0}}{1 - \mathcal{R}_{\rm BH} \mathcal{R}(\omega) e^{-2 i k r_*^0}} \,,
\end{equation}
where $\mathcal{R}_{\rm BH}$ and $\mathcal{T}_{\rm BH}$ are the reflection and transmission coefficients of a wave coming from the left of the photon-sphere barrier defined in Eq.~\eqref{RBH_TBH}, and $\mathcal{R}(\omega)$ is the surface reflectivity of the object defined in Eq.~\eqref{R_2}.

According to Eq.~\eqref{mark}, the GW signal emitted at infinity by a horizonless compact object is the same as the one emitted by a BH at infinity with an extra GW emission that depends on the reflectivity and compactness of the object. To get an insight of the additional GW emission, let us expand the transfer function in Eq.~\eqref{transferfunction} as a geometric series~\cite{Mark:2017dnq,Correia:2018apm}
\begin{equation} \label{transferfunctionseries}
    \mathcal{K}(\omega) = \mathcal{T}_{\rm BH} \mathcal{R}(\omega) e^{-2 i k r_*^0} \sum_{j=1}^{\infty} \left[ \mathcal{R_{\rm BH}} \mathcal{R}(\omega)\right]^{j-1} e^{-2 i (j-1) k r_*^0} \,.
\end{equation}
Given Eq.~\eqref{transferfunctionseries}, the GW signal takes the form of a series of pulses where the index $j$ stands for the signal emitted by the $j$-th echo. The phase factor $2 i k r_*^0$ corresponds to the time delay between two pulses due to the round-trip time of the perturbation between the photon sphere and the radius of the object.
Subsequent echoes can have a phase inversion to each other when the factor $\mathcal{R}_{\rm BH} \mathcal{R}(\omega)$ has a negative sign.

Eq.~\eqref{mark} allows us to construct an analytical template for the GW signal emitted by a horizonless compact object. In the following sections, we  provide an analytical approximation of each term in Eq.~\eqref{mark}, namely the BH reflection coefficient and the BH responses at infinity and the horizon. In this way, the analytical template depends only on the BH ringdown parameters and the parameters of the ECO, i.e., its compactness and reflectivity. 

\subsection{Black hole reflection coefficient in the low-frequency approximation}

The low-frequency regime is the most interesting regime for GW echoes since the latter are obtained by the reprocessing of the ringdown signal, whose frequency content is initially dominated by the BH fundamental QNM and subsequently decreases in time. Hence, the low-frequency approximation becomes increasingly more accurate at late times.
The analytical approximation of the BH reflection coefficient in the small-frequency regime is computed in Appendix~\ref{app:analytics} through a matched asymptotic expansion. In particular, the BH reflection coefficient is defined in Eq.~\eqref{RBH_TBH} as the ratio of the ingoing and outgoing coefficients in the near-horizon asymptotics of the Detweiler function in Eq.~\eqref{asymptoticsplus}. The latter coefficients are related to the coefficients in the near-horizon expansion of the Teukolsky function derived in the small-frequency regime in Eq.~\eqref{eq:ab1}. For $\ell=2$, the BH reflection coefficient reads
\begin{eqnarray}
\nonumber {\cal R}_{\rm BH}^{\rm LF} &=& -8 M k e^{\frac{\zeta (\gamma-1)}{\gamma+1}} \frac{2 M k 
-i(\gamma-1)}{(\gamma-1)^2} \left[ \frac{-M (\gamma-1) \xi}{L}\right]^{\zeta (\gamma-1)} \times \\ 
&&\left[ \frac{16 k^2 
M^2}{(\gamma-1)^2}+1 \right]
\frac{\Gamma(-2+\zeta) \Gamma(-1-\zeta)}{\Gamma(-2-\zeta) \Gamma(3-\zeta) } \times \nonumber \\
&&\frac{1800 i \Gamma(-2-\zeta) + \left(\omega M (\gamma-1) \xi\right)^5 
\Gamma(3-\zeta)}{1800 i \Gamma(-2+\zeta) + \left(\omega M (\gamma-1) 
\xi\right)^5 \Gamma(3+\zeta)} \,,
\end{eqnarray}
where $\gamma=r_-/r_+$, $\xi = 1+\sqrt{1-\chi^2}$, $\zeta=i(2 \omega M - m \sqrt{\gamma})(\gamma+1)\xi/(\gamma-1)$, and 
$L$ is an arbitrary constant with the dimension of a length. The low-frequency expression of ${\cal R}_{\rm BH}$ for generic values of the angular number $\ell$ is 
provided in a publicly available {\scshape 
Mathematica}\textsuperscript{\textregistered} notebook~\cite{webpage}.

In the high-frequency regime, ${\cal R}_{\rm BH}\sim e^{-2\pi\omega/\kappa_H}$, where 
$\kappa_H=\frac{1}{2} (r_+ -r_-)/(r_+^2+a^2)$ is the surface gravity of a Kerr BH~\cite{Harmark:2007jy}. 
We  use a Fermi-Dirac interpolating function to connect the two regimes in frequency domain:
\begin{equation}
 {\cal R}_{\rm BH}={\cal R}_{\rm BH}^{\rm 
LF} \frac{\exp{\left(\frac{-2\pi\omega_{R}}{\kappa_H}\right)}+1}{\exp{\left(\frac{2\pi(|\omega|-\omega_{R})}
{\kappa_H}\right)}+1}\,, \label{RBH}
\end{equation}
where $\omega_{R}$ is the real part of the fundamental QNM of a 
Kerr BH with spin $\chi$. For $|\omega| \ll \omega_R$ the reflection coefficient reduces to ${\cal R}_{\rm 
BH}^{\rm LF}$, whereas it is exponentially suppressed when $|\omega|\gg\omega_R$.

\subsection{Modeling the black hole response at infinity}

We  model the BH response at infinity using the fundamental $\ell=m=2$ QNM. We  consider a generic linear combination of two independent polarizations, namely~\cite{Berti:2005ys,Buonanno:2006ui}
\begin{equation}
 Z_{\rm BH}^{+} (t)\sim \Theta(t - t_0) \left[\mathcal{A}_+ \cos(\omega_R 
t+\phi_+)+i\mathcal{A}_\times \sin(\omega_R t+\phi_\times)\right] e^{-t/\tau}\,, \label{ZBHplus}
\end{equation}
where $\mathfrak{R}(Z_{\rm BH}^{+})$ and $\mathfrak{I}(Z_{\rm BH}^{+})$ are the two ringdown polarizations $h_+(t)$ and $h_\times(t)$, 
respectively. In the above relation, $\Theta(t)$ is the Heaviside function, $t_0$ is the starting time of the ringdown, ${\cal A}_{+,\times}\in\mathbb{R}$ and $\phi_{+,\times}\in\mathbb{R}$ are the amplitudes and the phases of the two polarizations, respectively, and $\tau=-1/\omega_I$ is the damping time. 
Let us notice that Eq.~\eqref{ZBHplus} is the most generic expression for the ringdown with the fundamental $\ell=m=2$ mode and assumes that ${\cal A}_{+,\times}$ and $\phi_{+,\times}$ are four independent parameters. The most relevant case of a binary BH ringdown is that of circularly polarized waves~\cite{Buonanno:2006ui},
that can be obtained from Eq.~\eqref{ZBHplus} by setting ${\cal A}_{+}={\cal A}_\times$ and $\phi_+=\phi_\times$. 

Given the BH response in the time domain, the waveform in the frequency-domain is obtained through a Fourier transform
\begin{equation}
\tilde Z_{\rm BH}^{\pm}(\omega) = \int_{- \infty}^{+ \infty} \frac{dt}{\sqrt{2 
\pi}} Z_{\rm BH}^{\pm}(t) e^{i \omega t},
\end{equation}
where the response at infinity is
\begin{equation}\label{eq:bhtemplateINF}
 \tilde Z_{\rm BH}^{+}(\omega) \sim \frac{e^{i \omega t_0}}{2\sqrt{2\pi}}  \left( 
\frac{\alpha_{1+} {\cal A}_+ -\alpha_{1\times} {\cal A}_\times}{\omega - \omega_{\rm QNM}}+  \frac{\alpha_{2+} {\cal 
A}_+ -\alpha_{2\times} {\cal 
A}_\times}{\omega + \omega_{\rm QNM}^*} 
\right)\,,
\end{equation}
where $\omega_{\rm QNM}=\omega_R+i\omega_I$, $\alpha_{1+,\times}=ie^{-i(\phi_{+,\times} + t_0\omega_{\rm QNM})}$, and 
$\alpha_{2+,\times}=-\alpha_{1+,\times}^*$. 

\subsection{Modeling the black hole response at the horizon}

The BH response at the horizon, defined in Eq.~\eqref{ZBHpm}, has the same poles in the complex frequency plane as the BH response at infinity. Therefore, the near-horizon response at intermediate times can be written 
as in Eq.~\eqref{eq:bhtemplateINF} with different amplitudes and phases.
Let us assume that the source has support in the interior of the object, i.e., on the left of the effective potential barrier, where $V(r,\omega) \approx -k^2$. 
This is a reasonable assumption since the source in the exterior can hardly perturb the spacetime within the cavity~\cite{Wang:2019rcf,Oshita:2019sat}.
In this case, it is easy to show that
\begin{equation}
 \tilde Z_{\rm BH}^-=\frac{{\cal R}_{\rm BH}}{{\cal T}_{\rm BH}}\tilde Z_{\rm BH}^+ +\frac{1}{{\cal T}_{\rm BH}W_{\rm 
BH}} 
\int_{-\infty}^{+\infty} dr_*\,{\tilde S}e^{ikr_*}\,.
\end{equation}
Using Eq.~\eqref{ZBHpm} and the fact that ${\tilde S}$ has support only where $V(r,\omega) \approx -k^2$, the 
above equation can be written as
\begin{equation}
 \tilde Z_{\rm BH}^-=\frac{{\cal R}_{\rm BH}\tilde Z_{\rm BH}^+ + \tilde {\cal Z}_{\rm BH}^+}{{\cal T}_{\rm BH}}\,,
\label{ZBHminusTOT}
\end{equation}
where $\tilde {\cal Z}_{\rm BH}^+$ is the BH response at infinity to an effective source ${\tilde {\cal 
S}}(\omega,x)={\tilde S}(\omega,x) e^{2ikx}$ within the cavity. As such, the ringdown part of $\tilde {\cal Z}_{\rm 
BH}^+$ can also be generically written as in Eq.~\eqref{eq:bhtemplateINF} with different amplitudes and phases.
Two interesting features of Eq.~\eqref{ZBHminusTOT} are noteworthy. First, the final response of the horizonless compact object (in Eq.~\eqref{mark}) does not depend on the BH transmission coefficient, since the term ${\cal T}_{\rm BH}$ in the denominator of Eq.~\eqref{ZBHminusTOT} cancels out with that in the transfer function in Eq.~\eqref{transferfunction}.
Second, the first term on the right-hand side of Eq.~\eqref{ZBHminusTOT} can be computed analytically using 
the low-frequency approximation of the BH reflection coefficient in Eq.~\eqref{RBH} and the BH response at infinity in Eq.~\eqref{ZBHplus}. For this reason, in the following, we focus only on the first term of the right-hand side of Eq.~\eqref{ZBHminusTOT}. A discussion on the expressions of $\tilde {\cal Z}_{\rm BH}^+$ for several sources is given in Appendix~\ref{app:sources}.

\subsection{Parameters and validity of the template} \label{sec:validity}

%
\begin{table*}[t]
\begin{center}
 \begin{tabular}{ll}
 \hline
 \hline
  $\epsilon$ & compactness of the ECO \\
  ${\cal R}(\omega)$ & reflection coefficient at the surface of the ECO \\ 
  \hline
  $M$ & total mass \\
  $\chi$ & spin \\
  ${\cal A}_{+,\times}$ & amplitudes of the polarizations of the BH ringdown \\
  $\phi_{+,\times}$	& phases of the polarizations of the BH ringdown \\
  $t_0$	& starting time of the BH ringdown \\
  \hline
  \hline
 \end{tabular}
 \end{center}
 \caption{Parameters of the ringdown$+$echo template in Ref.~\cite{Maggio:2019zyv}. 
 The parameters $\epsilon$ and ${\cal R}(\omega)$ characterize 
 the horizonless compact object, whereas the remaining $7$ parameters characterize the BH ringdown.
 } \label{tab:template}
\end{table*}
In the frequency domain, the ringdown+echo template is given by~\cite{Maggio:2019zyv}
\begin{equation} \label{template}
    \tilde{Z}^+ (\omega) = \tilde{Z}^+_{\rm BH} (\omega) \left(1+ \frac{{\cal R}_{\rm BH} \mathcal{R}(\omega) e^{-2 i k r_*^0}}{1 - \mathcal{R}_{\rm BH} \mathcal{R}(\omega) e^{-2 i k r_*^0}}\right) \,,
\end{equation}
where the BH response at infinity is in Eq.~\eqref{eq:bhtemplateINF} and the low-frequency approximation of the BH reflection coefficient is in Eq.~\eqref{RBH}. For $\mathcal{R} = 0$, we  recover the template of a BH ringdown emitted at infinity. The additional term in Eq.~\eqref{template} is associated with the GW echoes that are emitted in the case of horizonless compact objects. Overall, the final template depends on the $7$ parameters of a standard BH ringdown (i.e., mass and spin of the remnant, amplitudes and phases of the plus and cross polarizations of the signal, and starting time of the ringdown) plus two parameters that are related to the properties of the horizonless compact object (i.e., its compactness and reflectivity), see Table~\ref{tab:template}.  For circularly polarized waves (${\cal A}_{+}={\cal A}_{\times}$ and $\phi_{+}=\phi_{\times}$) or linearly polarized waves 
($A_\times=0$), the number of the BH ringdown parameters reduces to $5$.

\begin{figure}[t]
\centering
\includegraphics[width=0.49\textwidth]{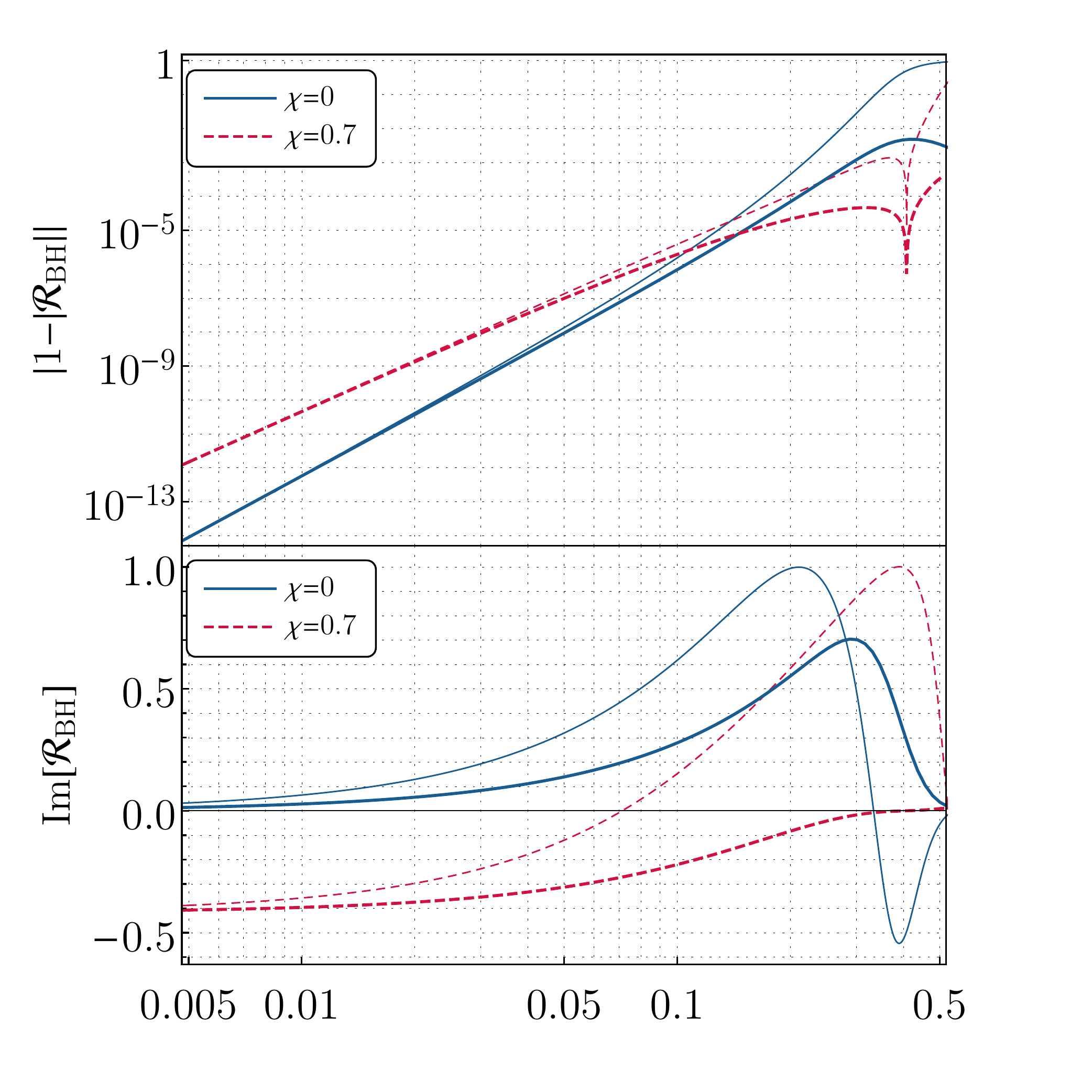}
\includegraphics[width=0.49\textwidth]{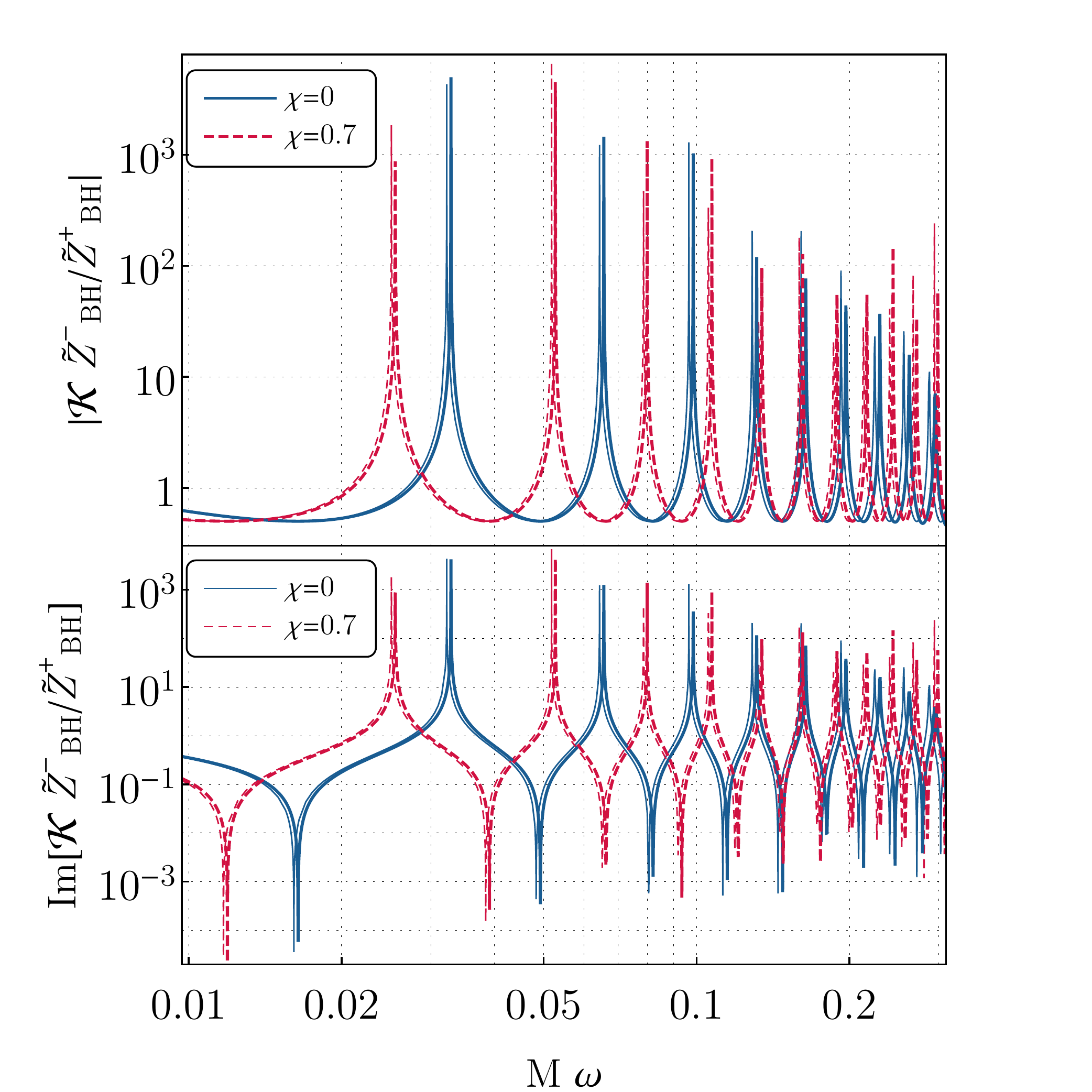}
\caption{Comparison between the analytical template (thick curves) and the result of a numerical integration of the Teukolsky
equation (thin curves) for $\chi = 0$ and $\chi = 0.7$. Left panels: the complex BH reflection coefficient. Right panels: the absolute value
(top) and the imaginary part (bottom) of the response of GW echoes $\mathcal{K} \tilde{Z}^-_{\rm BH}/\tilde{Z}^+_{\rm BH}$ as a function of the frequency. In all the panels, $\ell=m=2$ and, in the right panels, $\delta = 10^{-10} M$ and $\mathcal{R}=1$.~\cite{Maggio:2019zyv}} 
\label{fig:template}
\end{figure}
Fig.~\ref{fig:template} shows the agreement at low frequency of the analytical template with the result of a numerical integration of the Teukolsky equations. The left panels show the validity of the analytical approximation of the BH reflection coefficient in the low-frequency regime, both for a non-spinning and a spinning remnant. The right panels of Fig.~\ref{fig:template} show the quantity that is responsible for the emission of GW echoes as a function of the frequency, i.e., $\mathcal{K} \tilde{Z}^-_{\rm BH}$ normalized by the BH response $\tilde{Z}^+_{\rm BH}$. The peaks in the GW response are due to the excitation of the low-frequency QNMs that characterize horizonless compact objects (see a related discussion in Sec.~\ref{sec:lowfrequencies}).
The agreement between the analytical template and the numerical integration is very good at low frequencies (both in the absolute value and imaginary part), whereas deviations are present in the transition region where $M \omega \sim 0.1$. Crucially, the 
low-frequency resonances --~which dominate the response at infinity~-- are properly reproduced.

To quantify the validity of the template, we compute the overlap
\begin{equation}
 {\mathit O}=  \frac{|\langle \tilde h_A | \tilde h_N\rangle|}{\sqrt{|\langle \tilde h_N | \tilde h_N\rangle| |\langle 
\tilde 
h_A | \tilde h_A\rangle}|} \label{overlap}
\end{equation}
between the analytical signal $\tilde{h}_A$ and the numerical one $\tilde{h}_N$ in frequency domain, where the inner product is defined in Eq.~\eqref{innerproduct} in Appendix~\ref{app:fisher}. 
When $|{\cal R}|\sim 1$, the presence of very 
high and narrow resonances makes a quantitative comparison challenging since a slight displacement of the resonances (due for instance to finite-$\omega$ truncation errors) deteriorates the overlap. 
For example, for the representative case shown in Fig.~\ref{fig:template} ($\delta=10^{-10}M$, $\chi=0.7$, and ${\cal R}=1$), the overlap is excellent (${\mathit O}\gtrsim0.999$) when the integration is performed before the first resonance, however it reduces quickly to zero after that. 
To overcome this issue, we  compute the overlap in the case in which the resonances are less pronounced, i.e., $|{\cal R}|^2<1$. Let us consider 
a remnant with $M=30\,M_\odot$, $\chi=0.7$, $\delta=10^{-10}M$, and the aLIGO noise spectral density~\cite{zerodet}.
For ${\cal R}=0.9$ and in the frequency range $f\in(20,100)\,{\rm Hz}$ (whose upper end corresponds to the threshold $\omega M\sim0.1$ beyond which the low-frequency approximation is not accurate), the overlap is ${\mathit O}=0.48$. 
This small value is mostly due to a small displacement of the resonances. Indeed, by shifting the mass of the analytical waveform by only $1.6\%$, the overlap increases significantly, ${\mathit O}=0.995$.
For ${\cal R}=0.8$ and in the same conditions, the overlap is ${\mathit O}\approx0.8$ without the mass shift and ${\mathit O}\gtrsim0.999$ with the same mass shift as above. 
Let us notice that the agreement between the analytical template and the numerics improves as $\delta \to 0$. In this limit, the ECO QNMs appear at lower frequency and the resonant frequencies are better reproduced by the analytical framework.

\section{Properties of gravitational-wave echoes}

%
 \begin{figure*}[t]
 \centering
  \includegraphics[width=1\textwidth]{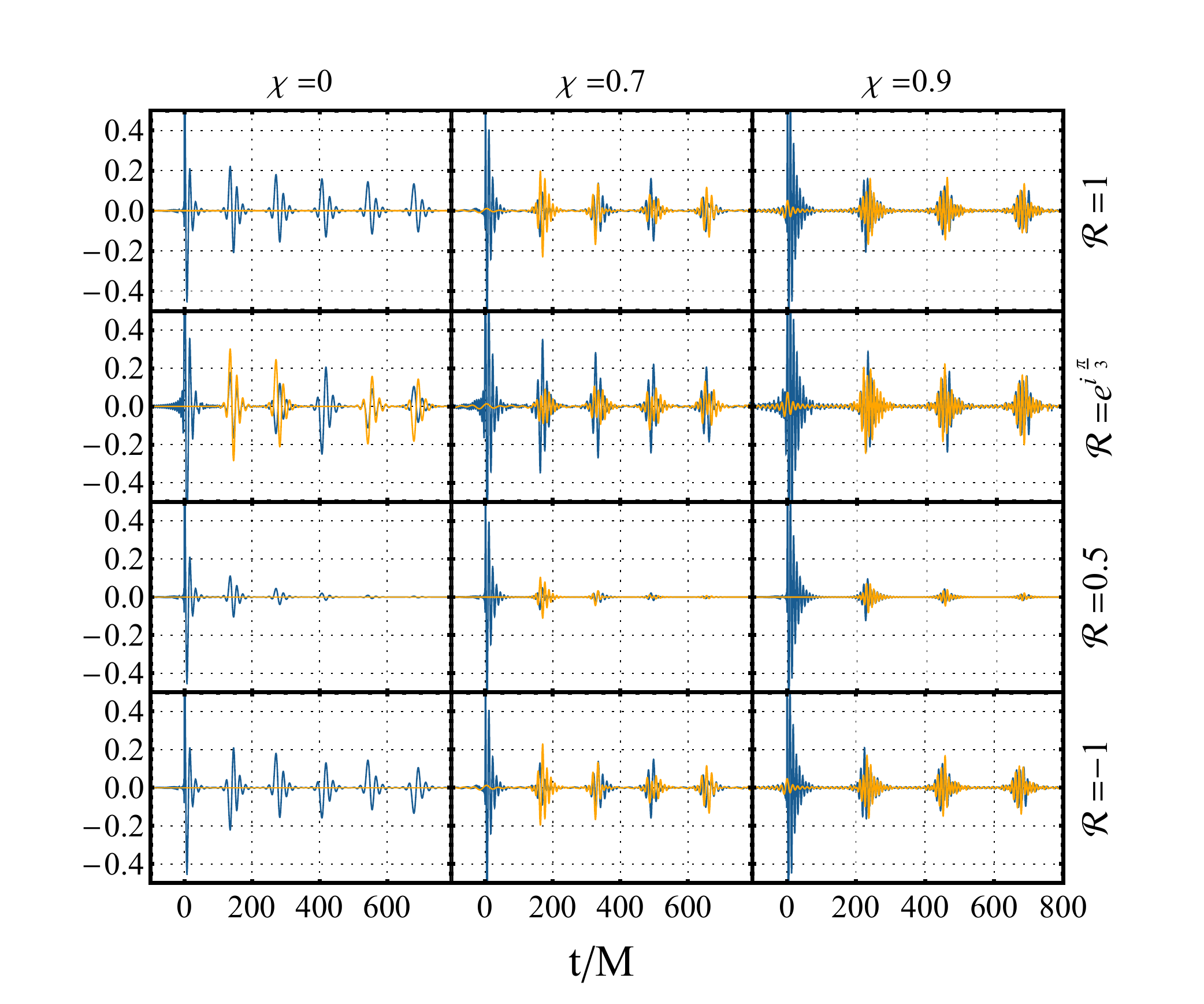}
 \caption{
 Examples of the gravitational ringdown$+$echo template in the time domain for a horizonless compact object with $\delta/M=10^{-7}$ and
 different values of the surface reflectivity ${\cal R}(\omega)={\rm const}$ and the spin $\chi$.
 The real (blue curve) and the imaginary (orange curve) part of the  waveform are the plus and 
 cross polarization of the signal, respectively. For simplicity, the ringdown signal is purely plus-polarized. 
 Each waveform is normalized to the peak of $|\mathfrak{R} [h(t)]|$ in the 
 ringdown. Additional waveforms are provided online~\cite{webpage,Maggio:2019zyv}.
 }
 \label{fig:template_time}
 \end{figure*}
The GW signal emitted by a horizonless compact object in the time domain is computed through an inverse Fourier transform of the analytical template in Eq.~\eqref{template}, i.e.,
\begin{equation}
h(t) = \frac{1}{\sqrt{2 \pi}}\int_{- \infty}^{+ \infty} d\omega 
\tilde{Z}^{+}(\omega) e^{-i \omega t}\,, \label{inverseFT}
\end{equation}
where $\mathfrak{R}\left[h(t)\right]$ and $\mathfrak{I}\left[h(t)\right]$ are the plus and cross polarizations of the GW signal, respectively.
Fig.~\ref{fig:template_time} shows the ringdown+echo waveform as a function of time for a remnant with $\delta=10^{-7}M$ and several values of the surface reflectivity and spin. 
For simplicity, we  focus on a purely plus-polarized ringdown signal, i.e., $A_\times = 0$, and each waveform is normalized to the peak of $|\mathfrak{R}\left[h(t)\right]|$ in the ringdown (the peak is not shown in the range of the $y$ axis to visualize the GW echoes better).

The time delay between subsequent echoes is constant and depends on the compactness and spin of the object. The time delay is computed as the round-trip time of the perturbation from the photon sphere to the radius of the object. In the spinning case~\cite{Abedi:2016hgu,Cardoso:2017njb}
\begin{equation} \label{tau_echo_spin}
    \tau_{\rm echo} \sim 2M \left[ 1+ (1-\chi^2)^{-1/2}\right] |\log \epsilon| \,.
\end{equation}
The logarithmic dependence on the compactness of the object allows the GW echoes to appear on a short timescale after the ringdown even for Planckian corrections at the horizon scale ($\epsilon \sim l_{\rm {Planck}}/M$).

The amplitude of the absolute value of the signal decreases monotonically and it is proportional to the product $\mathcal{R}\mathcal{R}_{\rm BH}$, i.e., it depends on the combined action of the reflection at the ECO surface and the photon-sphere barrier.
Let us notice that the spin of the object and the phase of the surface reflectivity  introduce novel effects compared to previous studies~\cite{Mark:2017dnq,Testa:2018bzd} such 
as a nontrivial amplitude modulation of subsequent echoes in each polarization of the GW signal.
This is evident, for example, in the panels of 
Fig.~\ref{fig:template_time} corresponding to $\chi=0.7$, ${\cal R}=1$ and $\chi=0$, ${\cal R}=e^{i\pi/3}$. 

\subsection{Mixing of polarizations}

An interesting feature of the GW echoes is that the signal can contain both  the plus and  cross polarizations even if 
the initial ringdown is purely plus polarized (i.e., ${\cal A}_\times=0$). This feature occurs in the cases of a spinning remnant or a complex surface reflectivity, as shown in Fig.~\ref{fig:template_time}.
This property can be explained as follows.
In the non-spinning case, and provided
\begin{equation}\label{eq:Rbarreal}
{\cal R}_{\chi=0}(\omega) = {\cal R}^*_{\chi=0}(-\omega^*) \,, 
\end{equation}
the transfer function satisfies the symmetry property
\begin{equation}\label{eq:kreal}
{\cal K}_{\chi=0}(\omega) = {\cal K}^*_{\chi=0}(-\omega^*)\,.
\end{equation}
The time domain 
echo waveforms are real (imaginary) if the ringdown waveform is real (imaginary).
Therefore, in the non-spinning case, the echo signal contains the same polarization of the BH ringdown and the two polarizations do not mix.
Remarkably, this property is broken in the following cases:
\begin{enumerate}
 \item when ${\cal R}$ is complex and does not satisfy Eq.~\eqref{eq:Rbarreal}, as shown in the second row of 
Fig.~\ref{fig:template_time};
 \item in the spinning case, even when ${\cal R}$ is real or satisfies Eq.~\eqref{eq:Rbarreal}, as shown in the second and third columns of Fig.~\ref{fig:template_time}.
\end{enumerate}
In either cases, a mixing of the polarizations occurs. In particular,
in the spinning case and when ${\cal R}$ is real, the transfer function satisfies an extended version of Eq.~\eqref{eq:kreal}
\begin{equation}\label{eq:krealspin}
{\cal K}(\omega,m) = {\cal K}^*(-\omega^*,-m) \,,
\end{equation}
that does not prevent the mixing of the polarizations due to the $m\to-m$ transformation.

As shown in Fig.~\ref{fig:template_time}, if the BH ringdown is a purely
plus-polarized wave, it can acquire a cross-polarization component upon reflection by the 
photon-sphere barrier (when $\chi\neq0$) or by the surface (when ${\cal R}$ is complex and does not satisfy 
Eq.~\eqref{eq:Rbarreal}).
The mixing of the polarizations can explain the involved echo pattern shown
in some panels of 
Fig.~\ref{fig:template_time}. For example, for $\chi=0$ and ${\cal R}=e^{i\pi/3}$ each echo is multiplied by a factor
$e^{i\pi/3}$ relative to the previous one. As a consequence, every three echoes the imaginary part of the signal (i.e., 
the cross polarization) is null.

\subsection{Phase inversion}

The phase of each subsequent echo depends on the term ${\cal R}{\cal R}_{\rm BH}$, i.e., on the combined action of the reflection at the surface of the object and the photon-sphere barrier. The
phase inversion occurs whenever ${\cal R}{\cal R}_{\rm BH}\approx -1$ for low frequencies. In Fig.~\ref{fig:template_time}, the first, the second, and fourth row all correspond to 
perfect reflectivity, $|{\cal R}|=1$, however their echo structure is different. This is because a phase term in the surface reflectivity introduces a nontrivial echo pattern.

It is worth mentioning that there exist several definitions of the radial wave function describing the perturbations of a 
Kerr metric; these are all related to each other by a linear transformation similar to Eq.~\eqref{DetweilerX}. The BH 
reflection coefficients that are defined for each function differ by a phase, while the quantity $|{\cal R}_{\rm 
BH}|^2$ (that is related to the energy damping/amplification) is invariant~\cite{Chandra}.

The transfer function in Eq.~\eqref{transferfunction} contains both the absolute value and the phase of ${\cal R}_{\rm BH}$. 
Therefore, one might wonder whether the ambiguity in the phase could affect the ECO response. For a given 
model, it should be noted that the surface reflectivity ${\cal R}$ is  affected by the same 
phase ambiguity, in accordance with the perturbation variable chosen to describe the problem. Since the transfer 
function depends on the combination ${\cal R}{\cal R}_{\rm BH}$, the phase 
ambiguity in ${\cal R}$ cancels out with that in ${\cal R}_{\rm BH}$. 
This ensures that the transfer function is invariant under the choice of the radial perturbation function, as expected 
for any measurable quantity. 

For example, at small frequencies the BH reflection coefficient derived from the 
asymptotics of the Regge-Wheeler function at $x\to-\infty$ has a phase difference of $\pi$ compared to the BH 
reflection coefficient derived from the Detweiler function with $\chi=0$. Consistently, the surface reflectivity 
associated to the former differs by a phase $\pi$ to the latter, 
i.e., if $\mathcal{R} = 1$ for the Regge-Wheeler function then $\mathcal{R} = -1$ for the Detweiler function in the same model.
All the choices of the radial wave functions are 
equivalent and --~in the same ECO model~-- the surface reflectivity ${\cal R}$ should  be 
different for each of them. This fact is particularly important in light of the mixing of the polarizations. As shown in the second row of Fig.~\ref{fig:template_time}, a phase in ${\cal R}$ introduces a mixing of the polarizations for 
any spin that results in a more involved pattern for the GW echoes.

The phase of the surface reflectivity ${\cal R}$ depends on the specific model of horizonless compact object. In the analyses of Sec.~\ref{sec:fisher}, we  
parametrize the surface reflectivity in a model-agnostic way as
\begin{equation}
    {\cal R}=|{\cal R}|e^{i\phi} \,. \label{reflF}
\end{equation}
In principle, both the 
absolute value and the phase are generically frequency-dependent; for 
simplicity, we choose them to be constants. Hence, we parametrize the template by $|{\cal R}|$ and $\phi$, different choices of which 
correspond to different models of horizonless compact objects. 

\subsection{Energy emission and superradiant instability}

%
\begin{figure}[t]
\centering
\includegraphics[width=0.75\textwidth]{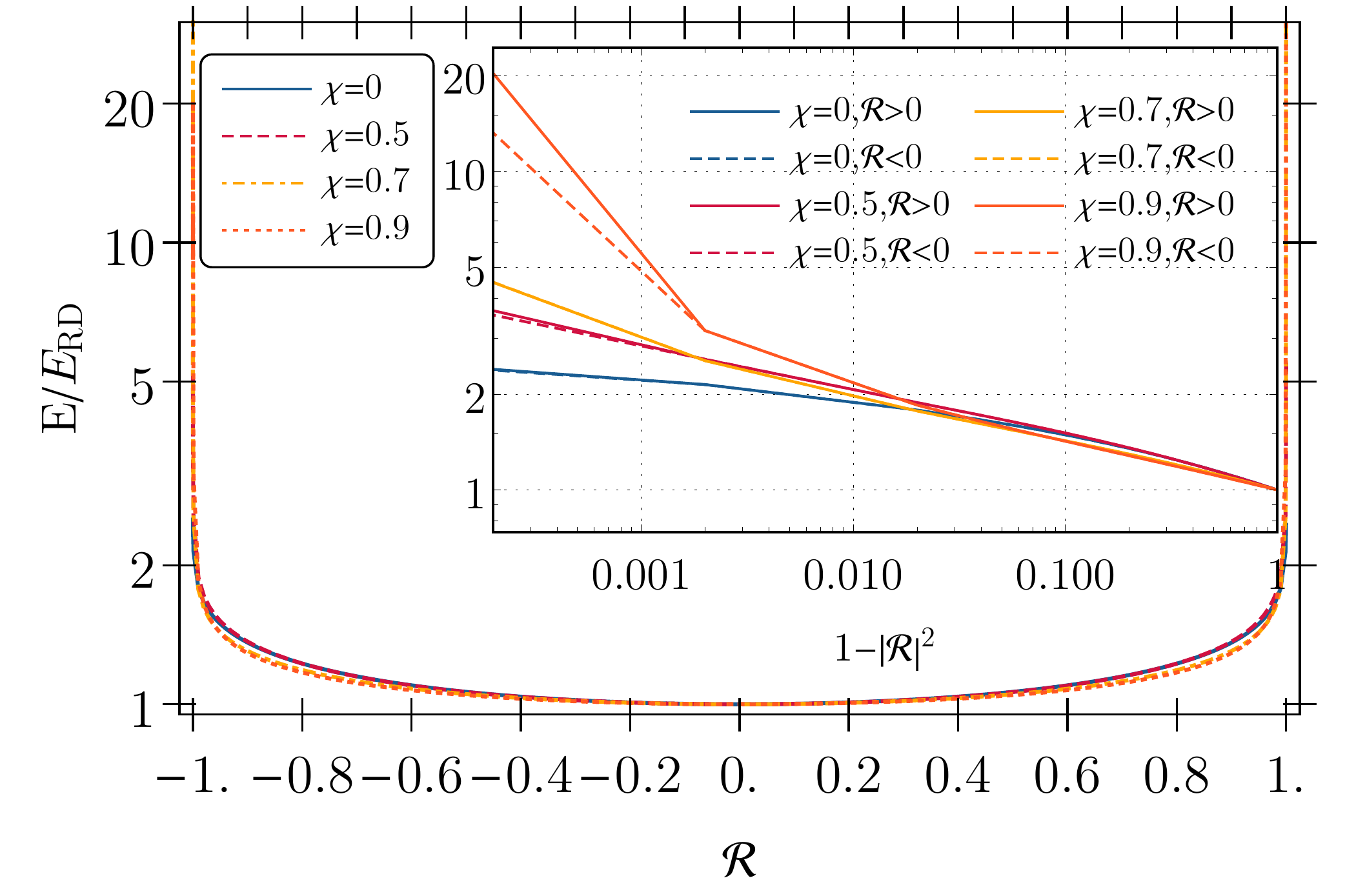}
\caption{Total energy emitted in the ringdown$+$echo signal normalized by the one of the BH ringdown as a function 
of the surface reflectivity ${\cal R}$ and for several values of the spin $\chi$. The total energy is much larger than the 
BH ringdown energy in the limit ${\cal R} \to 1$. We set $\delta=10^{-5}M$ and consider only one ringdown polarization with $\phi_+=0$; the result is independent of $\delta$ in the $\delta \ll M$ limit.~\cite{Maggio:2019zyv}} 
\label{fig:EvsR}
\end{figure}
The energy emitted in the ringdown$+$echo signal is shown in Fig.~\ref{fig:EvsR}, where the energy
\begin{equation}
 E\propto \int_{-\infty}^\infty d\omega\, \omega^2 |\hat Z^+|^2 \label{energy}
\end{equation}
is normalized by the one 
corresponding to the ringdown alone, i.e., $E_{\rm RD} \equiv E({\cal R}=0)$, and it is a function of the surface reflectivity $\mathcal{R}$.
We  use the prescription of Ref.~\cite{Flanagan:1997sx} to compute the ringdown energy, i.e., $\hat{Z}^+_{\rm BH}$ is the 
 full response in the frequency domain obtained by a Fourier transform of 
\begin{equation}
 Z_{\rm BH}^{+} (t)\sim \mathcal{A}_+\, \cos(\omega_R t+\phi_+)  e^{-|t|/\tau} \,,
\label{ZBHplusB}
\end{equation}
where the absolute value of $t$ is at variance with Eq.~\eqref{ZBHplus}.
This prescription circumvents the problem associated with the Heaviside function in Eq.~\eqref{ZBHplus} that produces a spurious 
high-frequency behavior in the energy flux, leading to an infinite energy in the ringdown signal. With the above prescription, the energy defined in 
Eq.~\eqref{energy} is finite and reduces to the energy of the BH ringdown in Ref.~\cite{Flanagan:1997sx} when ${\cal R}=0$. 

The energy emitted by a horizonless compact object can be much larger than the BH ringdown, as shown in Fig.~\ref{fig:EvsR} when ${\cal R} \approx 1$. This feature is due to the reflection and the amplification of waves in the cavity between the photon-sphere barrier and the radius of the compact object. The energy contained in the echo part 
of the signal grows fast as $|{\cal R}|\to1$ reaching a maximum value that depends on the spin and can be 
larger than the energy of the BH ringdown. This feature is due to the excitation of the resonances corresponding to the low-frequency QNMs of the  compact object.
However, it is worth noticing that the low-frequency resonances are excited only at late times, therefore the 
first few echoes contain a small fraction of the total energy of the signal. Conversely, when ${\cal R}$ is significantly 
smaller than unity, subsequent echoes are suppressed and their total 
energy is modest compared to the one of the BH ringdown.

Let us notice that when $|{\cal R}|\approx1$ the total energy is expected to diverge due to the ergoregion instability, as discussed in Sec.~\ref{chapter4}. This feature is not captured by the inverse Fourier transform of $\hat Z^+(\omega)$
since the time-domain signal is not integrable when $t\gtrsim \tau_{\rm inst}$.
Since the instability timescale is much longer than the echo delay time,  the time interval of validity of the waveform includes a large number of echoes. In particular, the ergoregion instability does not affect the first $N \sim |\log \delta/M|$ echoes~\cite{Cardoso:2019rvt}.
At late times, the signal grows when $|\mathcal{R}\mathcal{R}_{\rm BH}| > 1$, i.e. when the combined action of the reflection by the surface and the BH barrier yields an amplification factor larger than unity. When $|\mathcal{R}| \approx 1$, this requires
$|\mathcal{R}_{\rm BH}| > 1$ which occurs when the condition for superradiance, $\omega < m \Omega_H$, is satisfied. 

\subsection{Frequency content}

%
 \begin{figure}[th]
 \centering
 \includegraphics[width=0.7\textwidth]{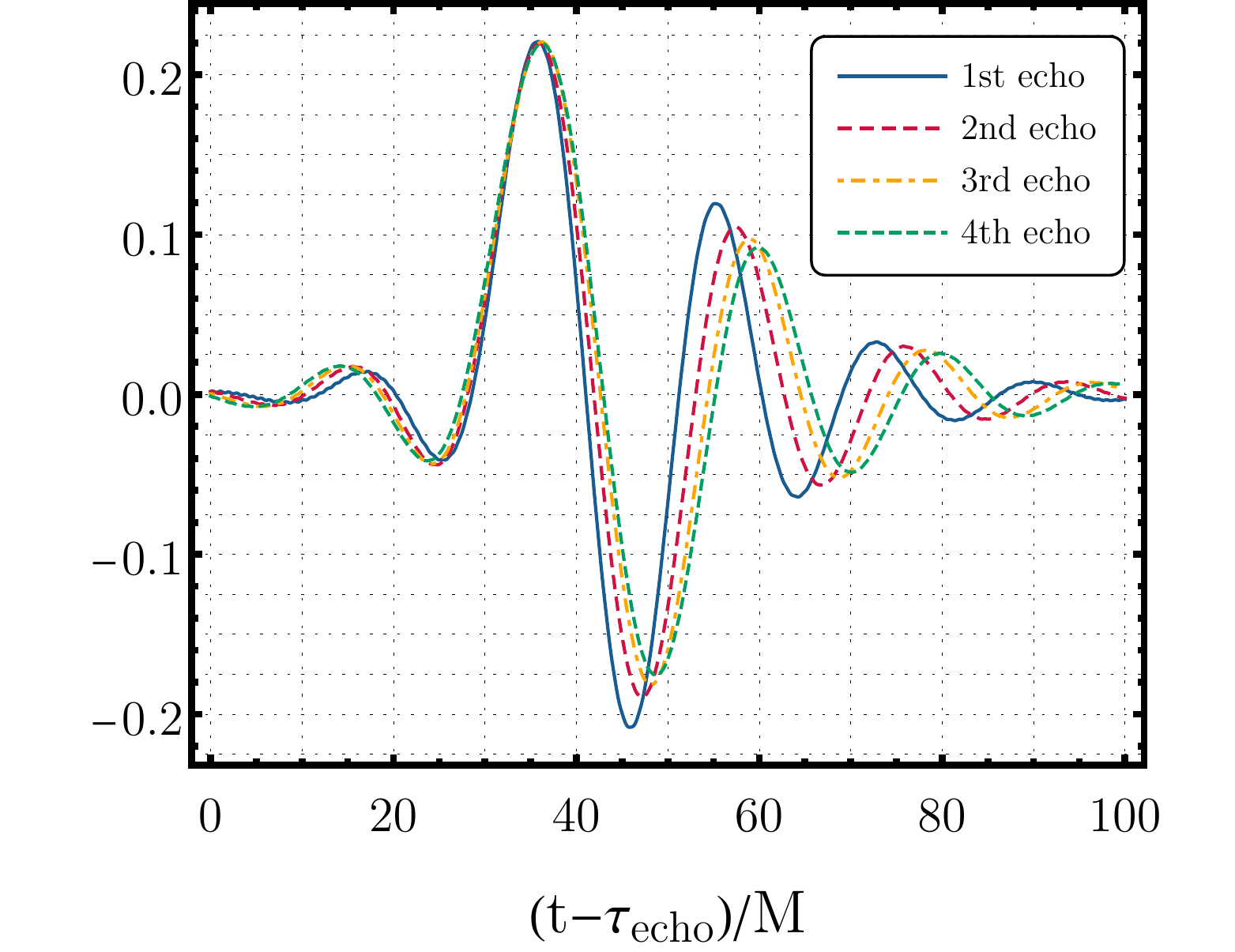}
 \caption{The first four echoes in the time-domain waveform for a horizonless compact object with $\chi=0$, ${\cal R}=1$,  
 $\delta/M=10^{-7}$. The waveform has been shifted in time and rescaled in amplitude so that the global 
 maxima of each echo are aligned. Note that each subsequent echo has a lower frequency content than the previous one.~\cite{Maggio:2019zyv}}
 \label{fig:freqcontent}
 \end{figure}
The photon-sphere barrier acts as a high-pass filter; therefore, each GW echo has a lower frequency content than the previous one. This expectation is confirmed by Fig.~\ref{fig:freqcontent}, where we  display the first four echoes that are shifted in time and rescaled in amplitude so that their global maxima are aligned, for a horizonless compact object with $\delta=10^{-7}M$, ${\cal R}=1$, and $\chi=0$. 

The frequency content of the signal starts roughly at the BH QNM frequency even if the latter is not part of the QNM spectrum of a horizonless compact object. The frequency content of each subsequent echo decreases until the signal is dominated by the low-frequency ECO QNMs at late 
times. In the example shown in Fig.~\ref{fig:freqcontent}, the frequencies of the first four echoes are approximately $M 
\omega\approx 0.34, 0.32, 0.3, 0.29$, whereas the real part of the fundamental QNM of a Schwarzschild BH is $M \omega_{R} 
\approx 0.37367$. Therefore, the frequency content between the first and the fourth echo decreases by $\approx 17 \%$. 

Let us notice that the case shown in Fig.~\ref{fig:freqcontent} is the one that provides the simplest echo pattern since $\chi=0$ and
${\cal R}$ is real. The spinning case $\chi\neq0$ or a complex choice of the surface reflectivity would provide a more involved echo pattern and polarization mixing.

Our results show that two different situations can occur: 
\begin{itemize}
 \item[A)] the reflectivity ${\cal R}$ of the 
object is small enough so that the amplitude of subsequent echoes is suppressed. In this case, most of the 
SNR is contained in the first few echoes with a frequency slightly smaller than the fundamental BH QNM. 
 \item[B)] the reflectivity ${\cal R}$ is close to unity so that subsequent echoes are relevant and contribute to the total SNR significantly. In this case, the frequency content becomes much smaller than the fundamental BH QNM at late times.
\end{itemize}
Clearly, the low-frequency approximation used to derive the analytical template is expected to be accurate in case B) and less accurate in case A).

\section{Modifications to the prompt ringdown} \label{sec:modpromptringdown}

%
\begin{figure}[t]
\centering
\includegraphics[width=0.7\textwidth]{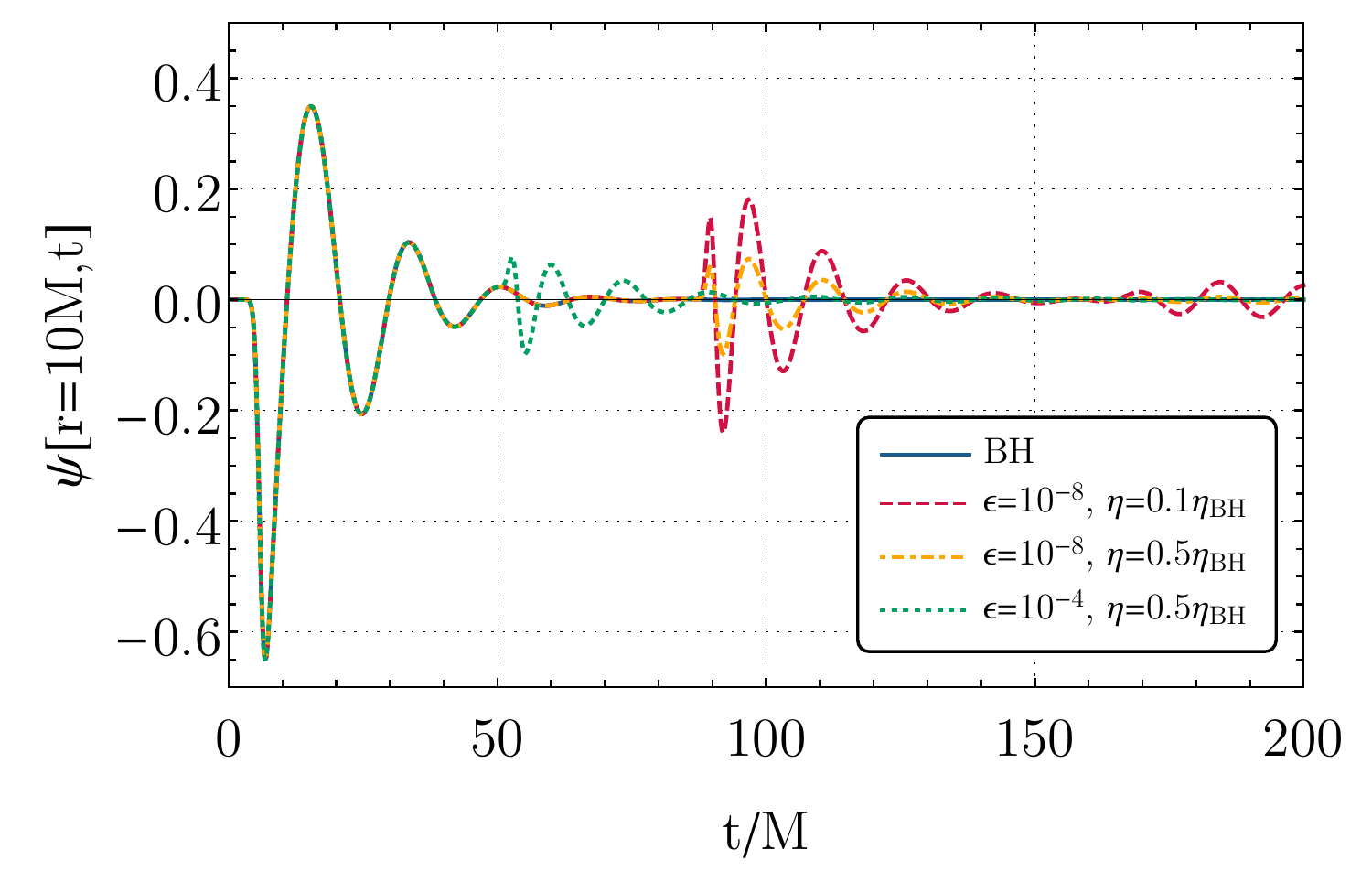}
\includegraphics[width=0.7\textwidth]{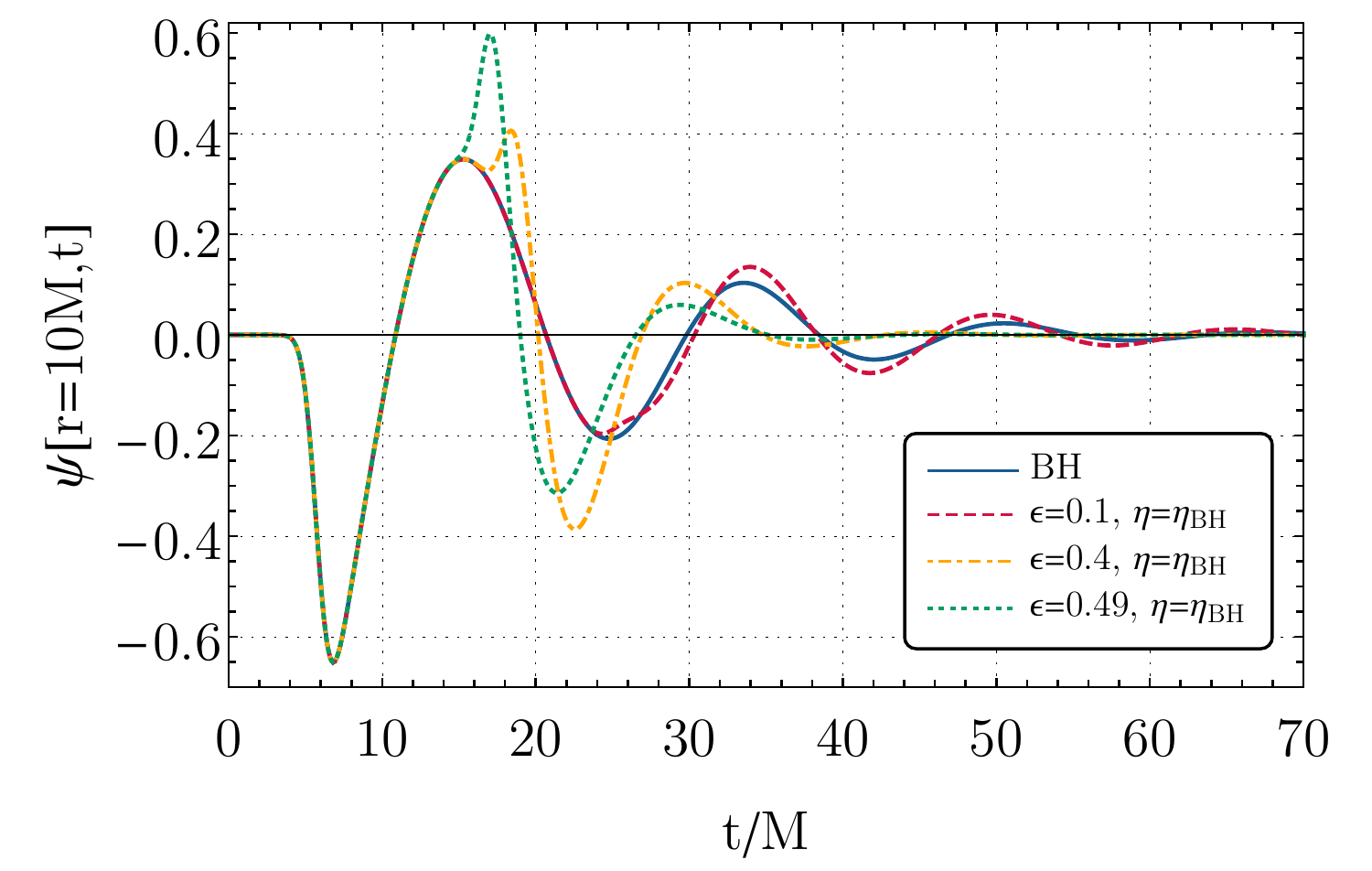}
\caption{
Ringdown of a horizonless compact object with radius $r_0=2M(1+\epsilon)$ and effective shear viscosity $\eta$. We  consider 
axial perturbations and an initial Gaussian profile. 
Top panel: when $\epsilon\ll0.01$, GW echoes appear for several values of $\eta$ related to the reflectivity of the object.
Bottom panel: a selection of ringdown waveforms for $\epsilon \gtrsim 0.01$. In this case, the prompt 
ringdown is modified and GW echoes are absent.~\cite{Maggio:2020jml}
} 
\label{fig:ringdown}
\end{figure}
The prompt ringdown is associated with the scattering of a wave packet off the photon-sphere barrier. If $\epsilon$ is sufficiently small, the following causality argument shows that the boundary conditions at the radius of the object cannot affect the prompt ringdown. The decay timescale of the prompt ringdown is associated with the instability timescale of the photon orbits at the light ring, or equivalently to the decay time of the BH fundamental QNM. Thus, the boundary condition at the radius of the compact object does not affect the prompt ringdown if the round-trip time of the radiation from the photon sphere to the boundary is much longer than the decay time of the BH fundamental QNM. In the Schwarzschild case, the round-trip time of the radiation in the cavity is in Eq.~\eqref{tauecho}, whereas the decay time of the fundamental QNM is $\tau_{\rm damp}=-1/\omega_I \approx 10 M$. Consequently, when $\epsilon\ll {\cal O}(0.01)$ the prompt ringdown is not modified and late-time GW echoes are emitted.
On the other hand, if $\epsilon\gtrsim {\cal O}(0.01)$ the object's interior should affect the prompt ringdown.

These expectations are confirmed by the ringdown waveforms shown in Fig.~\ref{fig:ringdown}, that are obtained by solving the linearized problem for axial perturbations in the Schwarzschild background.
The boundary condition at the radius of the compact object is obtained as the inverse Fourier transform of the boundary condition in Eq.~\eqref{BC-axial}. The perturbation has an initial Gaussian profile where $\psi(r_*,0)=0$ and $\partial_t \psi(r_*,0)=\exp[{-(r_*-7)^2}]$. 
The integration is performed using a fourth-order Runge-Kutta finite-difference scheme. 
The top panel of Fig.~\ref{fig:ringdown} shows the case in which $\epsilon\ll {\cal O}(0.01)$. Confirming previous 
results, the prompt ringdown is universal and the details of the object's interior appear as GW echoes after the initial ringdown. The 
time delay is given in Eq.~\eqref{tauecho}, and their phase and frequency content are modulated by the boundary conditions and the tunneling through the potential barrier. The 
amplitude of the GW echoes depends on the shear viscosity $\eta$ of the fictitious fluid located at the radius of the compact object, as described in Sec.~\ref{sec:effectiverefl}. In particular, $\eta\approx0$ corresponds to $|{\cal R}|^2\approx1$ for which the amplitudes of the subsequent echoes are only mildly damped, whereas the absorption is maximized as $\eta \approx \eta_{\rm BH}$. 
In the latter case, the linear response is identical to that of a BH, since the boundary conditions are the same in the limit $\eta\to\eta_{\rm BH}$ and $\epsilon\to0$.

The bottom panel of Fig.~\ref{fig:ringdown} focuses on the case $\epsilon\gtrsim {\cal O}(0.01)$, where the 
prompt ringdown is modified and no subsequent echoes appear. The changes to the prompt ringdown can be understood by 
considering that the part of the wave packet that initially tunnels through the barrier has enough time to be reflected 
at the radius of the object and tunnel to infinity. This process results in a superposition of the two pulses (the one directly 
reflected 
by the potential barrier and the one reflected by the object), which can interfere in an involved pattern. When the two 
pulses sum in phase, the interference can produce high peaks in the prompt ringdown.
At late time, the prompt ringdown is dominated by the fundamental QNM of the object, that is not the mode of the universal prompt ringdown in the BH case. Indeed, by fitting the time-domain waveform at late times with a damped sinusoid, we can verify that the prompt ringdown is governed by the fundamental QNM of the horizonless object shown in Fig.~\ref{fig:QNMsmembrane}.

Finally, one might wonder why there are no echoes for $\epsilon\gtrsim {\cal O}(0.01)$.
The reason is that only waves with frequency $V(r_0)<\omega^2<V_{\rm max}$ can be trapped between the radius of the object and 
the potential barrier. Therefore, when the compactness decreases, the resulting cavity is small. Furthermore, 
the transmission coefficient of the potential barrier is large when $\omega^2\lesssim V_{\rm max}$ which implies that 
these frequencies cannot be trapped efficiently. In 
practice, for $\epsilon\gtrsim{\cal O}(0.01)$ one only sees the interference between the prompt 
ringdown and the first echo, while subsequent reflections are strongly suppressed or absent, as in the bottom 
panel of Fig.~\ref{fig:ringdown}.

\section{Prospects of detection with current and future detectors} \label{sec:fisher}

In this section, we  use the template derived in Sec.~\ref{sec:template} for a preliminary error estimation of 
the ECO properties with current and future GW detectors. 

The ringdown$+$echo signal in the frequency domain displays resonances that originate from the long-lived QNMs of horizonless compact objects, as shown in the right panel of Fig.~\ref{fig:template}. The 
relative amplitude of each resonance depends on the source and the dominant modes are not necessarily the fundamental harmonics~\cite{Mark:2017dnq,Bueno:2017hyj}.
Moreover, the amplitude of the echo signal depends strongly on the surface reflectivity of the object, especially when 
$|{\cal R}|\approx1$. This suggests that the 
detectability of (or the constraints on) the echoes  depends strongly on the surface reflectivity and would be more feasible when $|{\cal R}|\approx 1$. In the following, we  quantify this expectation using a Fisher matrix analysis that is accurate at large SNR~\cite{Vallisneri:2007ev}. 
The analysis is performed as in Ref.~\cite{Testa:2018bzd}, additionally 
including the spin of the object and allowing for a 
complex reflection coefficient. 

We compute the Fisher matrix with the template $\tilde h(f) = \tilde Z^+(f)$, using the 
sensitivity curves of aLIGO with the design-sensitivity \texttt{ZERO\_DET\_high\_P}~\cite{zerodet} and two 
configurations for the third-generation instruments: Cosmic~Explorer in the narrow band variant~\cite{Evans:2016mbw,Essick:2017wyl}, and Einstein Telescope in its \texttt{ET-D} configuration~\cite{Hild:2010id}. We also consider the noise spectral density of LISA proposed in Ref.~\cite{LISA:2017pwj}.
Details on the Fisher information matrix are given in Appendix~\ref{app:fisher}.
We focus on the most relevant case of $\ell=m=2$ gravitational perturbations we consider a remnant with $M=30\,M_\odot$ 
($M=10^6\,M_\odot$) for ground (space) based detectors. 

As discussed in Sec.~\ref{sec:validity}, the most generic BH ringdown template contains $7$ parameters (mass, spin, two amplitudes, two phases, and starting time). For simplicity, we focus on a linearly polarized ringdown, and we do not include the parameters ${\cal 
A}_\times$ and $\phi_\times$. This implies that the Fisher analysis has $5$ standard ringdown parameters. 
Furthermore, the template depends on two ECO quantities, i.e., the surface reflectivity and the compactness of the ECO. We parametrize the surface reflectivity as in Eq.~\eqref{reflF}, where $|\mathcal{R}|$ and $\phi$ are assumed to be frequency independent for simplicity.
This yields three ECO dimensionless parameters: $\delta/M$, $|{\cal R}|$, and $\phi$.

We analyze two cases: (i)~a conservative case, in which the errors on the $5+3$ parameters are extracted in a Fisher matrix framework, and (ii) a more optimistic case, in which the standard ringdown parameters can be measured independently in the prompt ringdown and we are left with the 
measurement errors on the $3$ ECO parameters.

\subsection{Conservative case: 5 ringdown+3 ECO parameters}

%
 \begin{figure*}[t]
 \centering
 \includegraphics[width=0.32\textwidth]{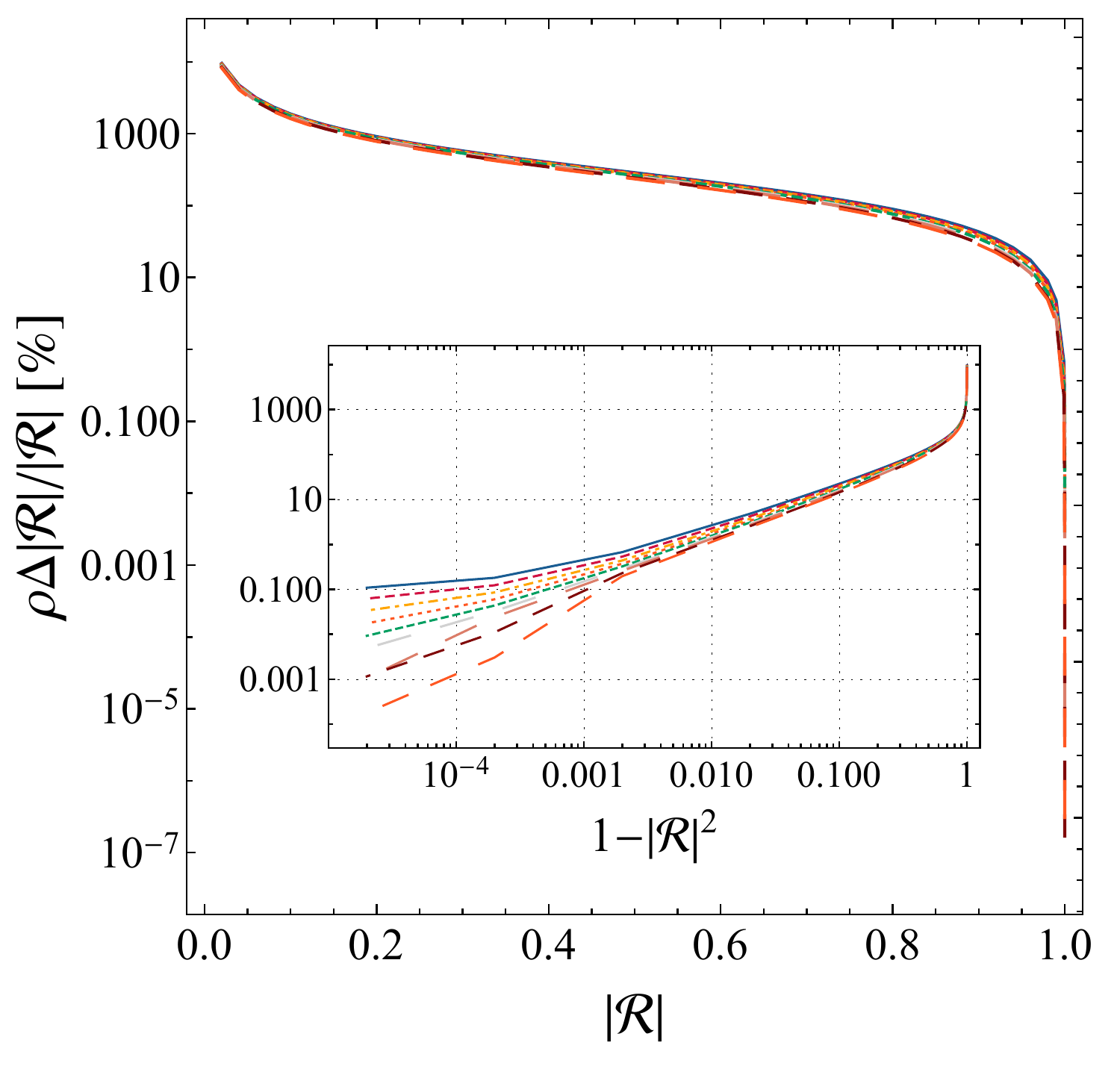}
 \includegraphics[width=0.32\textwidth]{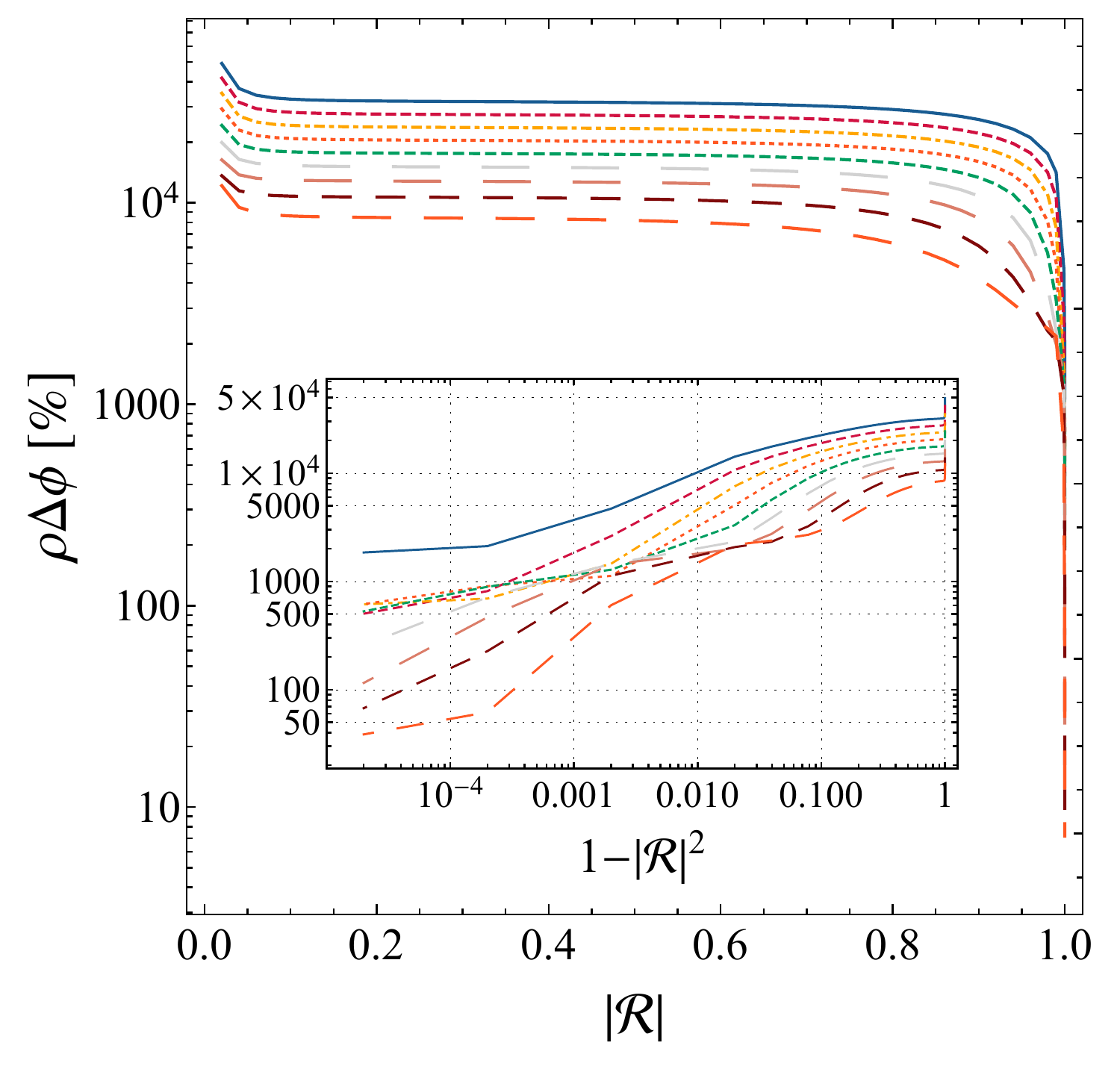}
 \includegraphics[width=0.32\textwidth]{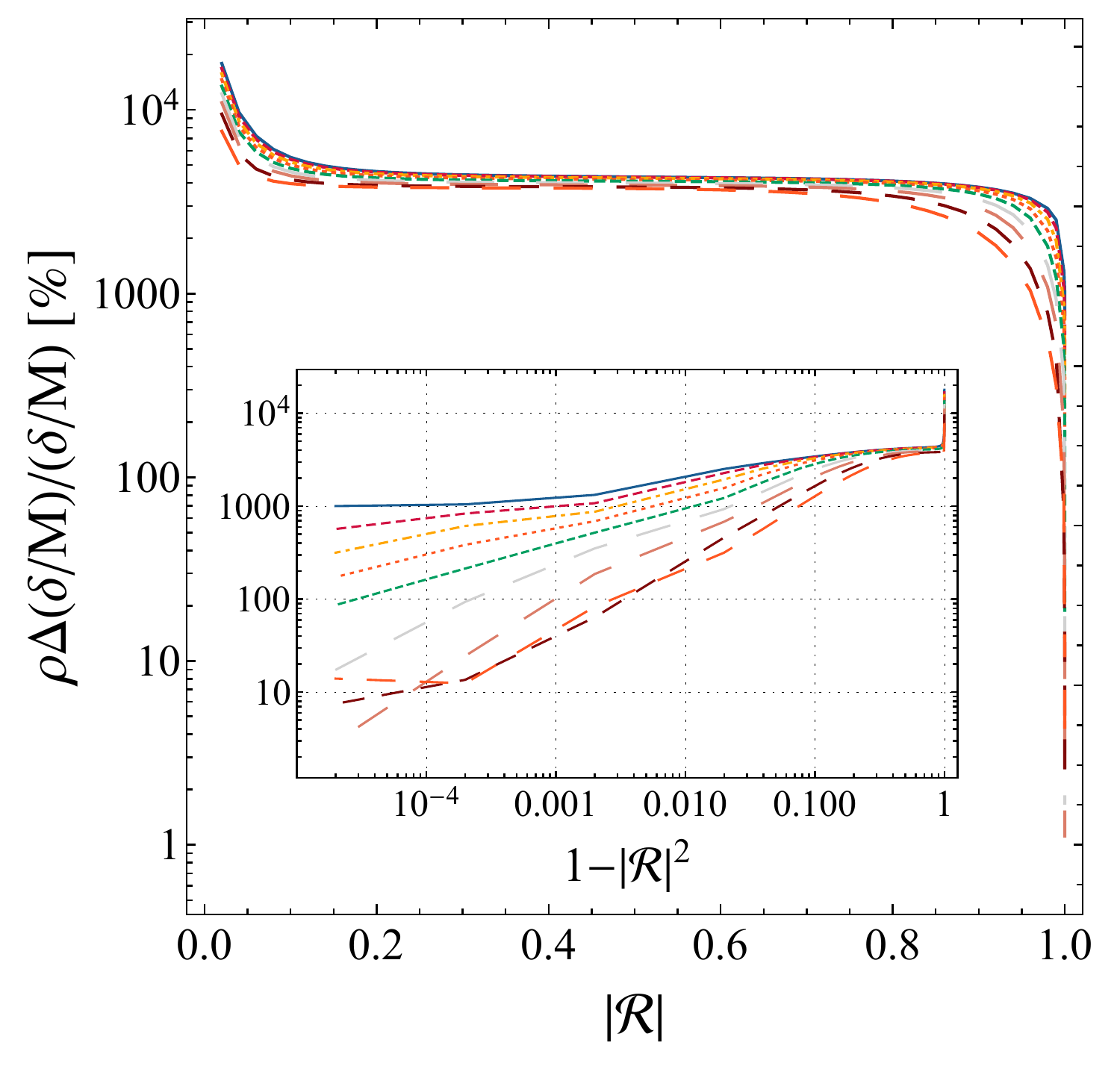}
 \caption{Left panel: relative percentage error on the reflection coefficient, $\Delta |{\cal R}| / |{\cal R}|$ multiplied by the SNR, as a function of $|{\cal R}|$ for 
different values of the spin. The inset shows the same quantity as a function of $1-|{\cal R}|^2$ in a 
 logarithmic scale. From top to bottom: $\chi=(0.9,0.8,0.7,0.6,0.5,0.4,0.3,0.2,0.1)$.
 Middle panel: same as in the left panel but for the absolute percentage error on the phase $\phi$ of the reflection coefficient, i.e.,
$\rho\Delta\phi$.
 Right panel: same as in the left panel but for the compactness parameter $\delta/M$, i.e.,
 $\Delta (\delta/M)/(\delta/M)$. 
 We assume $\delta=10^{-7} M$, where the errors are independent 
 of $\delta$ when $\delta\ll M$. We set $\phi=0$ (i.e., we
 consider a real and positive ${\cal R}$), but other choices give similar results.~\cite{Maggio:2019zyv}} 
 \label{fig:errors}
 \end{figure*}
The main results for the statistical errors on the ECO parameters are 
shown in Fig.~\ref{fig:errors}. In the large SNR limit, the errors scale as $1/\rho$, where $\rho$ is defined in Eq.~\eqref{SNR} in Appendix~\ref{app:fisher}.
Hence, Fig.~\ref{fig:errors} shows the quantities $\rho\Delta |{\cal R}| / |{\cal R}|$ (left panel), $\rho\Delta \phi 
$ (middle panel), and $\rho\Delta (\delta/M)/(\delta/M)$ (right panel) as a function of the surface reflectivity for several values of the spin.

For a given SNR, the relative errors are almost independent of the sensitivity curve of the detector, at least for the signals that are located near the minimum of each sensitivity curve.
Moreover, the statistical errors are almost independent of the compactness of the object when $\delta/M \ll 1$.
Fig.~\ref{fig:errors} shows that the statistical errors depend strongly on the surface reflectivity of the object. The reason for this can be traced back to the presence of resonances as ${\cal R} \approx 1$.
This feature confirms that it should be relatively straightforward 
to rule out or detect models with $|{\cal R}|\approx 1$, whereas it is 
increasingly more difficult to constrain models with smaller values of $|{\cal R}|$.

We  also notice that the value of the spin of the remnant affects the errors on $|{\cal R}|$ mildly, whereas it has a stronger impact on the phase of ${\cal R}$ (probably due to the aforementioned mixing of the 
polarizations) and a moderate impact on the errors on $\delta/M$.
Overall, the specific value of $\phi$ does not affect the errors significantly, although it is important to include it 
as an independent parameter to not underestimate the errors.

\subsection{Optimistic case: 3 ECO parameters}

%
 \begin{figure*}[t]
 \centering
 \includegraphics[width=0.32\textwidth]{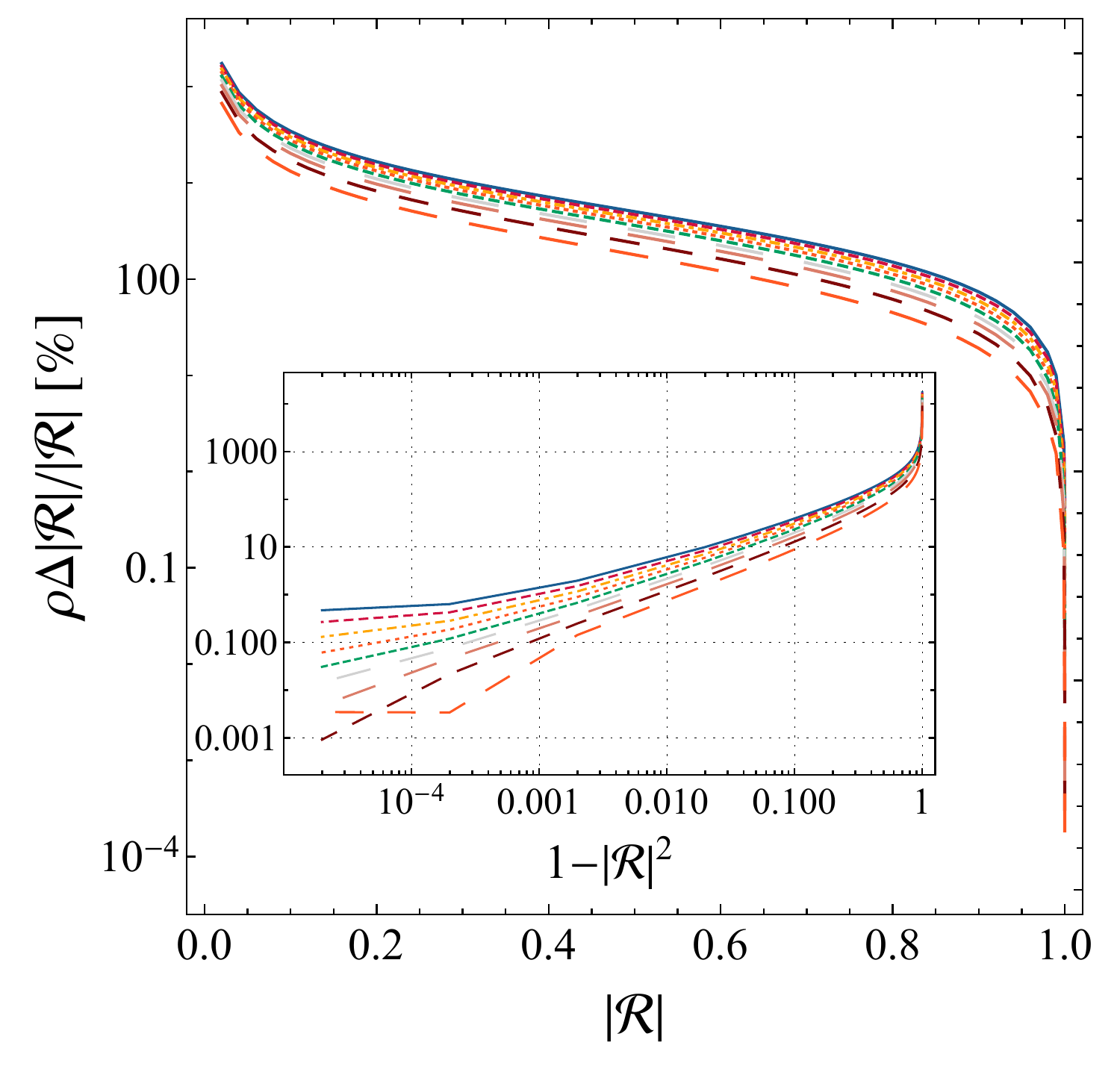}
 \includegraphics[width=0.32\textwidth]{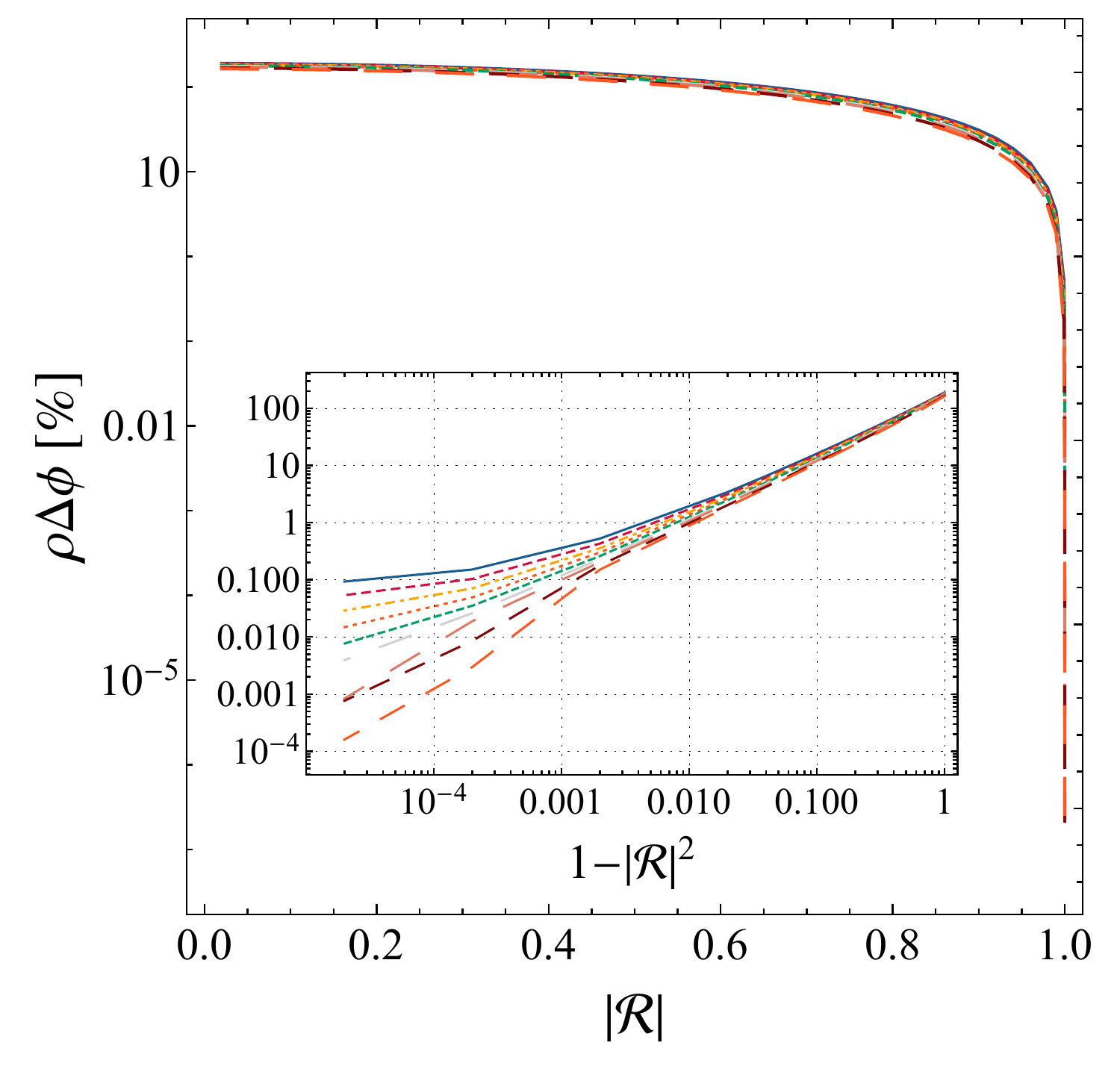}
 \includegraphics[width=0.32\textwidth]{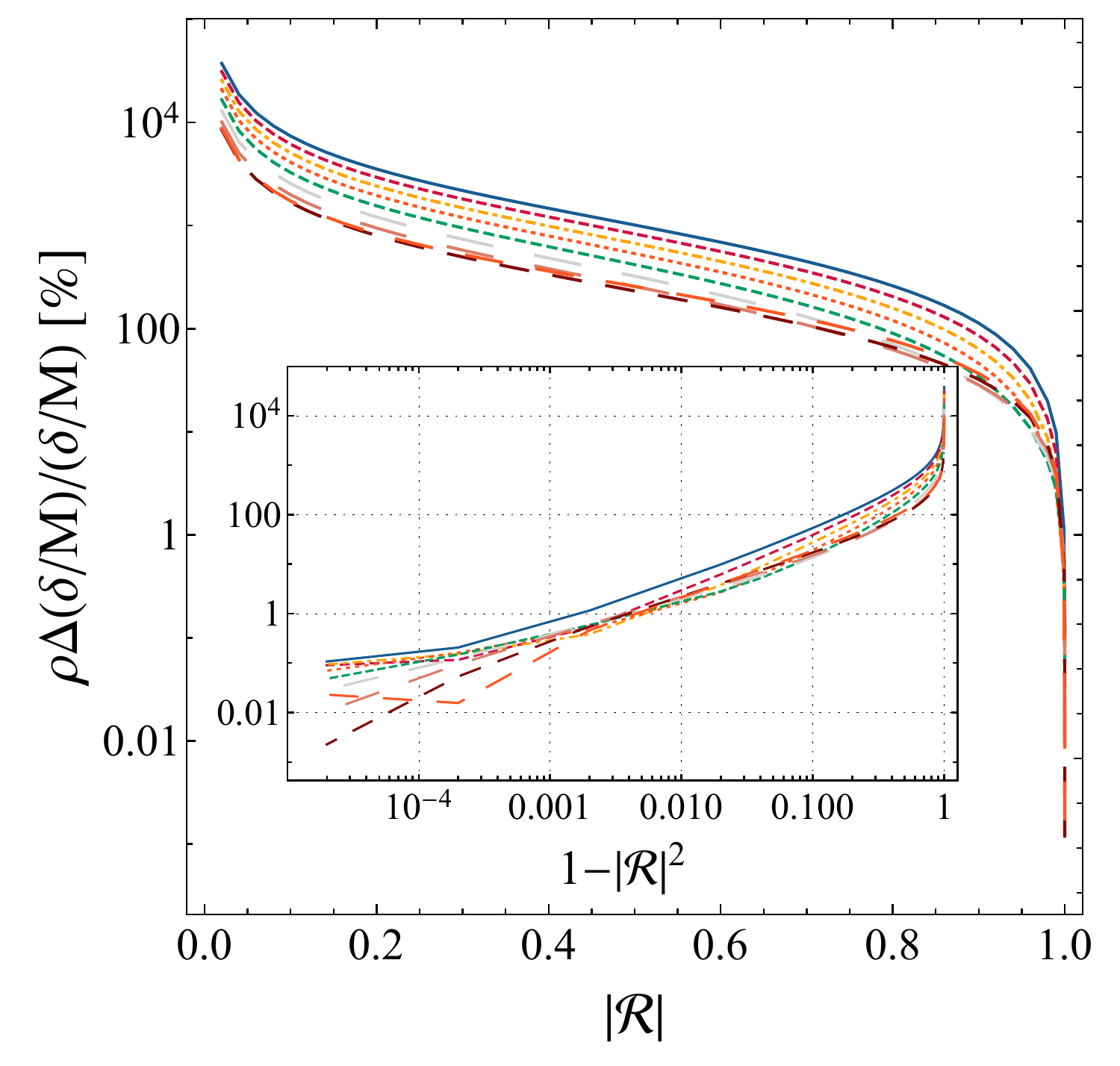}
 \caption{Same as in Fig.~\ref{fig:errors} but including only the three ECO parameters ($|{\cal R}|$, $\phi$, and 
 $\delta/M$) in the Fisher analysis.~\cite{Maggio:2019zyv}} 
 \label{fig:errors3param}
 \end{figure*}
Let us now assume that the standard ringdown parameters (mass, spin, amplitude, phase, and starting time) can be  measured independently in the prompt ringdown signal, that is identical for BHs and ECOs with $\delta/M \ll 1$. In this case, the remaining three ECO parameters ($|{\cal R}|$, $\phi$, and $\delta/M$) can be measured a posteriori, assuming that the standard ringdown parameters are known.

A representative example for this optimistic scenario is shown in Fig.~\ref{fig:errors3param}. As expected, the errors 
are significantly smaller, especially the ones on the phase of 
the reflectivity. The errors on the surface reflectivity are only mildly affected, and the projected constraints on ${\cal |R|}$ at 
different confidence levels are similar to the ones shown in Fig.~\ref{fig:RvsSNR}.

\subsection{Constraints on the reflectivity}

Let us calculate the SNR that is necessary to discriminate a 
partially-absorbing horizonless compact object from a BH on the basis of a measurement of the surface reflectivity 
at some confidence level. Clearly, if $\Delta {\cal |R|}/{\cal |R|}>100\%$, any 
measurement would be compatible with the BH case (${\cal R}=0$). On the other 
hand, relative errors $\Delta {\cal |R|}/{\cal |R|}<(4.5,0.27,0.007,0.00006)\%$ suggest that it is possible
to detect or rule out a given model at $(2,3,4,5)\sigma$ confidence level, 
respectively.
 \begin{figure}[t]
 \centering
 \includegraphics[width=0.7\textwidth]{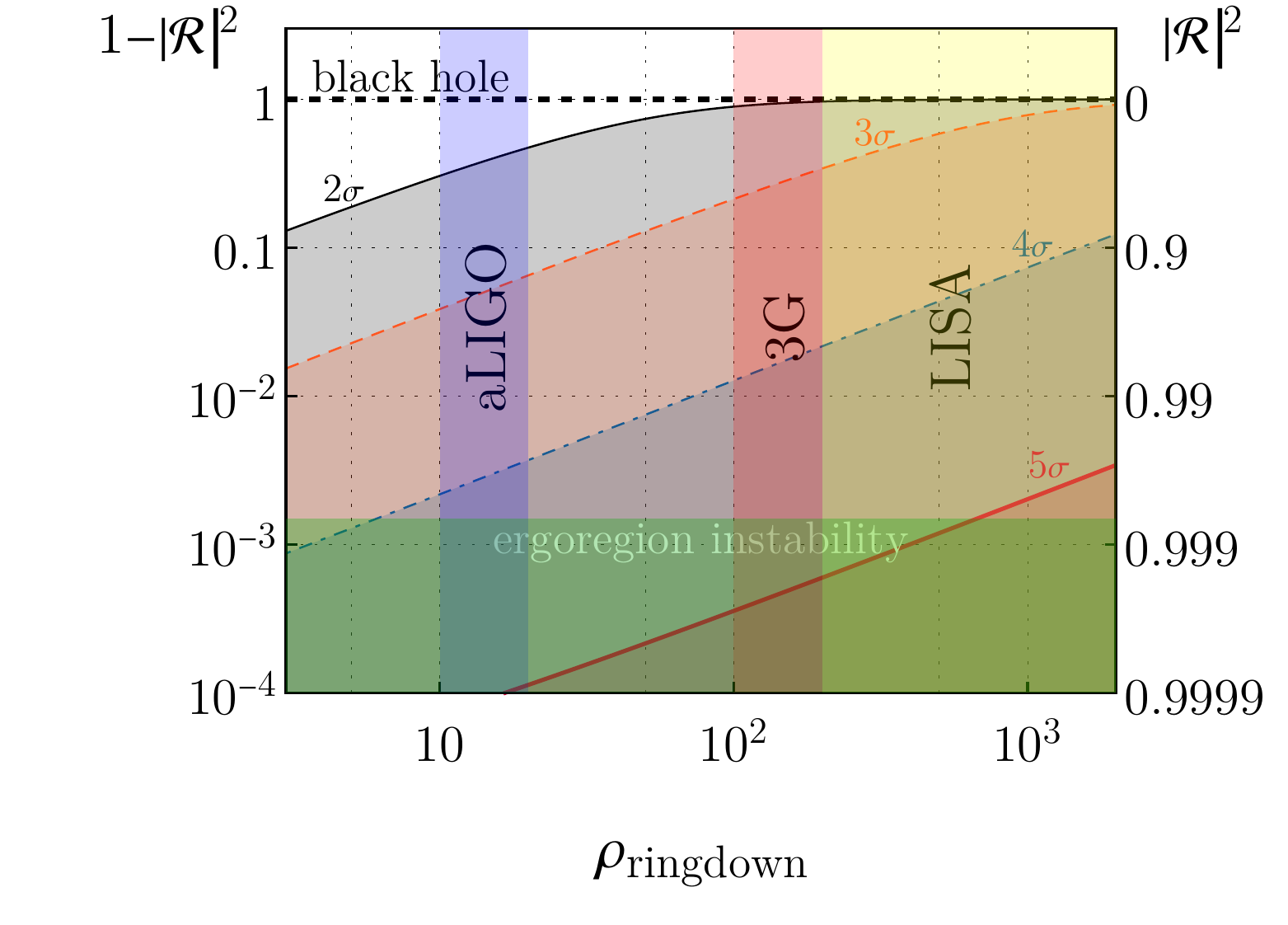}
 \caption{Projected exclusion plot for the ECO reflectivity $|{\cal R}|$ as a function of the SNR in the ringdown phase. The shaded areas are the regions that can be excluded at a 
 given confidence level ($2\sigma$, $3\sigma$, $4\sigma$, $5\sigma$). 
 Vertical bands are the typical SNRs achievable by aLIGO/Virgo, third-generation ground-based detectors, and LISA in the ringdown phase; the horizontal 
 band is the region excluded by the ergoregion instability~\cite{Maggio:2017ivp,Maggio:2018ivz}. We assume $\chi=0.7$ for the spin of the remnant, and the result depends mildly on the spin.~\cite{Maggio:2019zyv}} \label{fig:RvsSNR}
 \end{figure}
The result of this analysis is shown in Fig.~\ref{fig:RvsSNR}
that represents the exclusion plot for the parameter $|{\cal R}|$ as a function of the SNR in the ringdown phase, $\rho_{\rm ringdown}$. 
The shaded areas are the regions that can be excluded at some given confidence level. Large SNRs would allow us to 
probe values of the surface reflectivity close to the BH limit, ${\cal R}\approx 0$.

The extent of the constraints  depends strongly on the confidence 
level. For example, ${\rho_{\rm ringdown}\approx100}$ would allow us to distinguish horizonless compact objects with $|{\cal R}|^2\gtrsim 0.1$ from BHs at $2\sigma$ confidence level, but a $3\sigma$ detection would be possible if $|{\cal R}|^2\gtrsim0.8$.

Our analysis suggests that horizonless models with $|\mathcal{R}|^2 \approx 1$ can be detected or ruled out with aLIGO/Virgo (for events with $\rho_{\rm ringdown} \gtrsim 8$) at $5 \sigma$ confidence level. These events could also allow us to probe values of the reflectivity as small as $|\mathcal{R}|^2 \approx 0.8$ at $2\sigma$ confidence level.
Horizonless compact objects with $|\mathcal{R}|^2 = 1$ are already ruled out by the ergoregion instability~\cite{Maggio:2017ivp,Maggio:2018ivz}, the absence of a GW stochastic background in the LIGO O1 run due to spin loss~\cite{Barausse:2018vdb}, and the negative searches for GW echoes~\cite{LIGOScientific:2020tif}.

Excluding or detecting echoes for models with smaller values of the reflectivity (for which the ergoregion instability is absent) requires SNRs in the postmerger phase of $\mathcal{O}(100)$. This will be achievable only with third-generation detectors (Einstein Telescope and Cosmic Explorer) and with the space-based mission LISA. Our preliminary analysis confirms that very stringent constraints on (or detection of) ultracompact horizonless objects can be obtained with current (and especially future) interferometers.

\section{Appendix: Black hole response for particular sources} \label{app:sources}

In this appendix, we  provide the expressions for the BH response at the horizon for some specific toy models of the source. In the following, we assume that the source is localized within the cavity. 

The simplest case is that of a Gaussian source localized in space in which the frequency dependence can be factored out
\begin{equation}
 \tilde S(r_*,\omega) = C(\omega) \exp\left(-\frac{(r_*-r_*^s)^2}{\sigma^2}\right) \,,
\end{equation}
where $r_*^S$ is the location of the source in the tortoise coordinate, and $|r_*^s|\ll M$. It is easy to show that 
\begin{equation}
 \tilde {\cal Z}_{\rm BH}^+= e^{2ik r_*^s} \tilde Z_{\rm BH}^+ \,. \label{eq:calZlocsource}
\end{equation}
The latter equation, together with Eq.~\eqref{ZBHminusTOT}, yields
\begin{equation}
 \tilde Z_{\rm BH}^- = \left(
 \frac{e^{2ik r_*^s}+\mathcal{R}_{\rm BH}}{\mathcal{T}_{\rm BH}} \right)\tilde Z_{\rm BH}^+\,. 
\label{eq:bhtemplateHOR} 
\end{equation}
Remarkably, the above relation is independent of the width $\sigma$ of the Gaussian source  and  the function $C(\omega)$ characterizing the source. 

Inspired by Eq.~\eqref{eq:calZlocsource}, we can also parametrize the BH response $\tilde {\cal Z}_{\rm BH}^+$ in a 
model-agnostic way with a generically complex proportionality factor
\begin{equation}
 \tilde {\cal Z}_{\rm BH}^+ = \eta e^{i\nu} \tilde{Z}_{\rm BH}^{+}\,,
\end{equation}
where $\eta$ and $\nu$ are (real) parameters of the template.
Since the BH response at intermediate times is dominated by the QNM frequencies, a model in which $\tilde {\cal Z}_{\rm BH}^+ = {\cal F}(\omega) \tilde{Z}_{\rm BH}^{+}$ can be reduced effectively to $\tilde {\cal Z}_{\rm BH}^+ = {\cal F}(\omega_R) \tilde{Z}_{\rm BH}^{+}$. In such case, the term ${\cal F}(\omega_R)=\eta e^{i\nu}$ can be interpreted as a generic parametrization of a complex number. 

Finally, another possible model for the source is a plane-wave source that travels towards $r_* \to \pm\infty$, where
\begin{equation}
\tilde S(r_*, \omega) =
 \int dt e^{i \omega t} S(r_*, t)
 =\int dt e^{i \omega t} S(0, t\mp r_*) =\tilde S(0, \omega) e^{\pm i \omega r_*}\,.
\end{equation}
Using Eq.~\eqref{ZBHpm}, we  obtain
\begin{equation}\label{A7}
 \tilde Z_{\rm BH}^{+}(\omega) = \tilde Z_{\rm BH}^{-}(\omega) \frac{\int_{-\infty}^{+\infty}dr_*X_s^{-} e^{\pm i \omega 
r_*}}{\int_{-\infty}^{+\infty}dr_* X_s^{+} e^{\pm i \omega r_*}}\,,
\end{equation}
or, more explicitly,
\begin{eqnarray} \label{ZZp}
\nonumber \tilde Z_{\rm BH}^{+}(\omega) &=& \tilde Z_{\rm BH}^{-}(\omega) \left[\int_{{r_* \sim 0}}dr_*X_s^{-} e^{i \omega r_*} 
+\int^{\infty}(A_{\textrm{out}}e^{2i \omega 
		r_*} +A_{\textrm{in}})dr_* \right. \\
&+& \left. \int_{-\infty}dr_* e^{i m \Omega r_*} \right] \Big/ \left[\int_{r_* \sim 0}dr_*X_s^{+} e^{i \omega r_*}
	+\int^{\infty} e^{2i \omega r_*} dr_* \right. \nonumber \\
&+& \left. \int_{-\infty}(B_{\textrm{out}} e^{2i \omega r_*-i m \Omega 
r_*}+B_{\textrm{in}} e^{i m \Omega r_*} )dr_* \right] \,,
\end{eqnarray}
where ${r_* \sim 0}$ is the region where the potential is non-zero and we  considered the case of a plane wave traveling to $r_* \to +\infty$ for ease of notation. Since $\tilde Z_{\rm BH}^{+}(\omega)$ has poles at $\omega_{\rm QNM}=\omega_R+i\omega_I$, we also expect $\tilde Z_{\rm BH}^{-}(\omega)$ to have the same poles.
Given that $\omega_I<0$, the terms $\int^{+\infty}dr_*$ dominate the numerator and the denominator of Eq.~\eqref{ZZp} for $\omega \approx \omega_{\rm QNM}$ and yield
\begin{equation}\label{A9}
\tilde Z_{\rm BH}^{+} \approx -\left(\frac{\mathcal{R}_{\rm BH}}{\mathcal{T}_{\rm BH}}\right)^* \tilde Z_{\rm BH}^{-}\,.
\end{equation}
The case of a plane wave traveling towards $r_* \to -\infty$ gives the same relation.

\chapter{Extreme mass-ratio inspirals around a horizonless source} \label{chapter6}

\begin{flushright}
    \emph{
    Com’era il nostro gioco? È presto detto. Lo spazio essendo curvo, attorno alla sua curva facevamo correre gli atomi, come delle biglie, e chi mandava più avanti il suo atomo vinceva.
    Nel dare il colpo all’atomo bisognava calcolar bene gli effetti, le traiettorie, saper sfruttare i campi magnetici e i campi di gravitazione, se no la pallina finiva fuori pista ed era eliminata dalla gara.
    }\\
    \vspace{0.1cm}
    Italo Calvino, Le Cosmicomiche
\end{flushright}
\vspace{0.5cm}

EMRIs are binary systems in which a stellar-mass compact object orbits around a supermassive compact object at the center of a galaxy. EMRIs are one of the main target sources of the space-based interferometer LISA, and are unique probes of the nature of supermassive compact objects. The defining feature of a classical BH is to be a perfect absorber since its event horizon is a one-way hypersurface. Thus, any evidence of partial reflectivity would indicate a departure from the classical BH picture. In this chapter, we shall show that LISA would be able to probe the reflectivity of compact objects with unprecedented accuracy. 

\section{A model for the central compact object}

We  analyze a central horizonless compact object whose exterior spacetime is described by the Kerr metric, as detailed in Sec.~\ref{sec:spinningmodel}. The radius of the compact object is located as in Eq.~\eqref{radius} where $\epsilon \ll 1$. For example, if the radius of the object is at $r_0 = r_+ + l_{\rm Planck}$, then $\epsilon \sim 10^{-44}$ for a supermassive compact object with mass $M = 10^6 M_\odot$ and spin $\chi =0.9$.

The properties of the interior structure are modeled in terms of a complex and frequency-dependent reflectivity coefficient $\mathcal{R}(\omega)$ at the surface of the object.
Spinning horizonless compact objects with a perfectly reflecting surface ($|\mathcal{R}|^2=1$) are affected by an ergoregion instability when spinning sufficiently fast. In the following, we focus on stable spinning models with partial absorption ($|\mathcal{R}|^2<1$), as detailed in Sec.~\ref{sec:quenchergoregion}. We also analyze a model of quantum BH with Boltzmann reflectivity ($|\mathcal{R}(\omega)|^2 = e^{-|k|/T_{\rm H}}$, where $k$ is the corotating frequency and $T_{\rm H}$ is the Hawking temperature), that is stable against the ergoregion instability for any spin~\cite{Oshita:2019sat}.

\section{Linear perturbations from a point particle}

We  analyze a pointlike source orbiting around a central compact object which is either a Kerr BH or a Kerr-like horizonless object. The pointlike source moves along circular equatorial orbits from large distances to the ISCO.
In line with the discussion in Sec.~\ref{sec:spinningmodel}, we assume that the gravitational perturbations in the exterior spacetime are described as in the Kerr case. According to the Newman-Penrose formalism, the Weyl scalar $\Psi_4$ can be expanded as
\begin{equation} \label{weyl}
\Psi_4 = \hat\rho^4 \sum_{\ell m} \int d\omega R_{\ell m \omega}(r) {}_{-2}S_{\ell m \omega}(\theta) e^{i (m \varphi - \omega t)} \,,
\end{equation}
where $\hat\rho = (r-ia \cos \theta)^{-1}$, and the sum runs over $\ell \geq 2$ and $-\ell \leq m \leq \ell$. The radial wave function $R_{\ell m \omega}(r)$ obeys to the Teukolsky master equation~\cite{Teukolsky:1972my,Teukolsky:1973ha,Teukolsky:1974yv}
\begin{equation} \label{teukolsky_source}
\Delta^{2} \frac{d}{dr}\left(\frac{1}{\Delta} \frac{dR_{\ell m \omega}}{dr}\right) + \left[\frac{K^{2}+4 i (r-M) K}{\Delta}-8 i \omega r -\lambda \right] R_{\ell m \omega} = \mathcal{T}_{\ell m \omega} \,,
\end{equation}
and the spin-weighted spheroidal harmonics ${}_{-2}S_{\ell m \omega}(\theta) e^{i m \varphi}$ satisfy Eq.~\eqref{angular} with $s=-2$. 
The polar part of the spin-weighted spheroidal harmonics is normalized such that
\begin{equation}
\int_{-1}^{1} \left|{}_{-2}S_{\ell m \omega}(\cos \theta)\right|^2 d\cos\theta =1 \,.
\end{equation}
The source term $\mathcal{T}_{\ell m \omega}$ is constructed by projecting the stress-energy tensor $T^{\alpha \beta}$ of a pointlike source with respect to the Newman-Penrose tetrad, where~\cite{Fujita:2004rb}
\begin{equation}
T^{\alpha \beta} = \mu \frac{u^{\alpha} u^{\beta}}{\Sigma \sin \theta u^t} \delta\left(r-r(t)\right) \delta\left(\theta-\theta(t)\right) \delta\left(\varphi-\varphi(t)\right) \,,
\end{equation}
with $\mu$ being the mass of the small orbiting body, $u^{\alpha} = dz^{\alpha}/d\tau$, $z^{\alpha} = \left( t, r(t), \theta(t), \varphi(t)\right)$ is the geodesic trajectory, and $\tau$ is the particle proper time. The mass ratio of the system is defined as $q \equiv \mu/M$.

In the case of circular equatorial orbits, $\theta(t) = \pi/2$. In corotating orbits, the orbital radius is related to the orbital angular frequency by
\begin{equation}
\Omega = \sqrt{M}/(a \sqrt{M} + r^{3/2}) \,. \label{orbfreq}
\end{equation}
%

\subsection{Central black hole}

Let us first review the standard BH case. Owing to the presence of the horizon, the two independent homogeneous solutions of Eq.~\eqref{teukolsky_source} have the following asymptotic behavior
\begin{equation} \label{asymptoticsin}
R^{\rm in}_{\ell m \omega} \sim 
\begin{cases}
 \displaystyle 
B^{\rm trans}_{\ell m \omega} \Delta^2 e^{-i k r_*} & \text{ as } r_* \to - \infty\\ 
 \displaystyle  
 r^3 B^{\rm ref}_{\ell m \omega}  e^{i \omega r_*}  +  r^{-1} B^{\rm inc}_{\ell m \omega} e^{- i \omega r_*} & \text{ as } r_* \to 
+ \infty
\end{cases} \,,\\
\end{equation}
\begin{equation} \label{asymptoticsup}
R^{\rm up}_{\ell m \omega} \sim 
\begin{cases}
 \displaystyle 
 C^{\rm up}_{\ell m \omega} e^{i k r_*}  +  \Delta^2 C^{\rm ref}_{\ell m \omega} e^{-i k r_*} & \text{ as } r_* \to - \infty \\ 
 \displaystyle  
 r^3 C^{\rm trans}_{\ell m \omega} e^{i \omega r_*} & \text{ as } r_* \to + \infty \\
\end{cases} \,,
\end{equation}
where $k = \omega - m \Omega_H$, and $\Omega_H = a/(2Mr_+)$ is the angular velocity at the horizon of the Kerr BH. 
The inhomogeneous solution of the Teukolsky equation~\eqref{teukolsky_source} is constructed as~\cite{Fujita:2004rb}
\begin{eqnarray}
\nonumber R_{\ell m \omega} &=& \frac{1}{W_{\ell m \omega}} \left\{ R^{\rm up}_{\ell m \omega}(r) \int_{r_+}^r dr' \frac{\mathcal{T}_{\ell m \omega}(r') R^{\rm in}_{\ell m \omega}(r')}{\Delta^2(r')} \right. \\
&+& \left. R^{\rm in}_{\ell m \omega}(r) \int_{r}^{\infty} dr' \frac{\mathcal{T}_{\ell m \omega}(r') R^{\rm up}_{\ell m \omega}(r')}{\Delta^2(r')} \right\} \,, \label{inhomsol}
\end{eqnarray}
where $W_{\ell m \omega}$ is the Wronskian given by
\begin{eqnarray}
W_{\ell m \omega} &=& \Delta^{-1} \left( R^{\rm in}_{\ell m \omega} \frac{dR^{\rm up}_{\ell m \omega}}{dr} - R^{\rm up}_{\ell m \omega} \frac{dR^{\rm in}_{\ell m \omega}}{dr} \right) \nonumber \\
&=& 2 i \omega C^{\rm trans}_{\ell m \omega} B^{\rm inc}_{\ell m \omega} \,. 
\end{eqnarray}
The inhomogeneous solution in Eq.~\eqref{inhomsol} has the following asymptotic behavior
\begin{equation}
R_{\ell m \omega} \sim 
\begin{cases}
 \displaystyle 
 Z_{\ell m \omega}^{H} \Delta^2 e^{-i k r_*} & \text{ as } r_* \to - \infty \\ 
 \displaystyle  
 Z_{\ell m \omega}^{\infty} r^3 e^{i \omega r_*} & \text{ as } r_* \to + \infty \\
\end{cases} \,,
\end{equation}
where
\begin{eqnarray}
Z_{\ell m \omega}^{H} &=& C^{H}_{\ell m \omega} \int_{r_+}^{\infty} dr' \frac{\mathcal{T}_{\ell m \omega}(r') R^{\rm up}_{\ell m \omega}(r')}{\Delta^2(r')} \,, \\
Z_{\ell m \omega}^{\infty} &=& C^{\infty}_{\ell m \omega}  \int_{r_+}^{\infty} dr' \frac{\mathcal{T}_{\ell m \omega}(r') R^{\rm in}_{\ell m \omega}(r')}{\Delta^2(r')} \,,
\end{eqnarray}
and
\begin{equation}
C^{H}_{\ell m \omega} = \frac{B^{\rm trans}_{\ell m \omega}}{2 i \omega C^{\rm trans}_{\ell m \omega} B^{\rm inc}_{\ell m \omega}} \,, \quad C^{\infty}_{\ell m \omega} = \frac{1}{2 i \omega B^{\rm inc}_{\ell m \omega}} \,.
\end{equation}

The amplitudes $Z_{\ell m \omega}^{H}$ and $Z_{\ell m \omega}^{\infty}$ determine the gravitational energy fluxes emitted at infinity and through the horizon~\cite{Teukolsky:1974yv,Hughes:1999bq}:
\begin{eqnarray}
\dot{E}^{\infty} &=& \sum_{\ell m} \frac{|Z^{\infty}_{\ell m \omega}|^2}{4 \pi (m \Omega)^2} \,, \label{Einf}\\
\dot{E}^{H} &=& \sum_{\ell m} \frac{\alpha_{\ell m} |Z^{H}_{\ell m \omega}|^2}{4 \pi (m \Omega)^2} \label{Eh}\,,
\end{eqnarray}
where
\begin{equation}
\alpha_{\ell m} = \frac{256 (2Mr_+)^5 k (k^2 + 4 \varpi^2) (k^2 + 16 \varpi^2) (m \Omega)^3}{|c_{\ell m}|^2} \,,
\end{equation}
where $\varpi = \sqrt{M^2-a^2}/(4Mr_+)$ and

\begin{eqnarray}
|c_{\ell m}|^2 &=& [(\lambda+2)^2 + 4 m a (m\Omega) - 4a^2 (m\Omega)^2 ] \nonumber \\
& \times & [\lambda^2 + 36 m a (m\Omega) - 36 a^2 (m\Omega)^2] \nonumber \\
& + & (2 \lambda +3) [96 a^2 (m\Omega)^2 - 48 m a (m\Omega)] \nonumber \\
& + & 144 (m\Omega)^2 (M^2-a^2) \,.
\end{eqnarray}
For circular equatorial orbits, the angular momentum fluxes are related to the energy fluxes, at infinity and the horizon, by $\dot{J}^{\infty, H} = \dot{E}^{\infty, H}/\Omega$.

In the case of a central BH, the total energy flux emitted by a point particle in a  circular equatorial orbit with orbital angular frequency $\Omega$ is
\begin{equation}
 \dot E(\Omega) = \dot E^{\infty}(\Omega) +\dot E^{H}(\Omega)\,, \label{EtotBH}
\end{equation}
where $\dot E^{\infty}(\Omega)$ and $\dot E^{H}(\Omega)$ are defined in Eqs.~\eqref{Einf} and~\eqref{Eh}, respectively.

\subsection{Central horizonless compact object} \label{sec:horizonlesscase}

Let us analyze the case of a horizonless compact object whose corrections to the Kerr case are incorporated in the boundary condition at the radius of the object.
A gravitational perturbation can be written as a superposition of ingoing and outgoing waves at the radius of the object as in Eq.~\eqref{R_1}, where the surface reflectivity of the object is defined in Eq.~\eqref{R_2}.
For a generically complex and frequency-dependent reflectivity, a horizonless compact object is described by the boundary condition in Eq.~\eqref{BC} that reduces to the BH boundary condition when ${\cal R}=0$.

In the horizonless case, the solutions of the homogeneous Teukolsky equation are such that the `up' modes have the same asymptotics as in Eq.~\eqref{asymptoticsup}, whereas the `in' modes have the following asymptotics
\begin{equation} \label{RinECO}
R^{\rm in}_{\ell m \omega} \sim 
\begin{cases}
 \displaystyle 
B'^{\rm trans}_{\ell m \omega} \Delta^2 e^{-i k r_*}
+ C'^{\rm up}_{\ell m \omega} e^{i k r_*} 
& \text{ as } r_* \to r_*^0 \\
 \displaystyle  
 r^3 B'^{\rm ref}_{\ell m \omega} e^{i \omega r_*}
 +  r^{-1} B'^{\rm inc}_{\ell m \omega} e^{- i \omega r_*}
& \text{ as } r_* \to + \infty
\end{cases} \,,\\
\end{equation}
where
\begin{eqnarray}
B'^{\rm trans}_{\ell m \omega} &=& B^{\rm trans}_{\ell m \omega}+c_1 C^{\rm ref}_{\ell m \omega} \,, \\
C'^{\rm up}_{\ell m \omega} &=& c_1 C^{\rm up}_{\ell m \omega} \,, \\
B'^{\rm ref}_{\ell m \omega} &=& B^{\rm ref}_{\ell m \omega}+c_1 C^{\rm trans}_{\ell m \omega} \,, \\
B'^{\rm inc}_{\ell m \omega}  &=& B^{\rm inc}_{\ell m \omega} \,,
\end{eqnarray}
and the coefficient $c_1$ is determined by imposing the boundary condition in Eq.~\eqref{BC} where
\begin{equation}
R_{\ell m \omega} = R^{\rm in}_{\ell m \omega} + c_1 R^{\rm up}_{\ell m \omega} \,.
\end{equation}
The inhomogeneous solution of the Teukolsky function is derived as in Eq.~\eqref{inhomsol}, with $R^{\rm in}_{\ell m \omega}$ as in Eq.~\eqref{RinECO} and $R^{\rm up}_{\ell m \omega}$ as in Eq.~\eqref{asymptoticsup}, and it has the following asymptotic behavior
\begin{equation} \label{inhomECO}
R_{\ell m \omega} \sim 
\begin{cases}
 \displaystyle 
 Z_{\ell m \omega}^{H^+} \Delta^2 e^{-i k r_*} + Z_{\ell m \omega}^{H^-}  e^{i k r_*} & \text{ as } r_* \to r_*^0 \\ 
 \displaystyle  
 Z_{\ell m \omega}^{\infty} r^3 e^{i \omega r_*} & \text{ as } r_* \to + \infty \\
\end{cases} \,,
\end{equation}
where
\begin{equation}
Z_{\ell m \omega}^{H^+} = Z_{\ell m \omega}^{H} \,, \quad
Z_{\ell m \omega}^{H^-} = \frac{C'^{\rm up}_{\ell m \omega}}{B'^{\rm trans}_{\ell m \omega}} Z_{\ell m \omega}^{H} \,.
\end{equation}
To determine the energy emitted by the particle in the horizonless case, we  note that -- by assumption -- the gravitational perturbations in the neighbourhood of the particle are exactly those of a Kerr background, albeit with unusual boundary conditions. We can, therefore, determine the emitted energy by appealing to the energy balance law in the Kerr background. The energy flux to infinity is formally given by the same formula as in the BH case, Eq.~\eqref{Einf}. The energy flux to the ECO side is determined by extending $R_{\ell m \omega}$ analytically to the horizon of the Kerr background and measuring the flux there. Thus, the internal energy flux on the ECO side $\dot{E}^{\rm int}$ is given by
\begin{equation}
\dot{E}^{\rm int} = \dot{E}^{H^{+}} - \dot{E}^{H^{-}} \label{Eradius} \,,
\end{equation}
where $\dot{E}^{H^{+}}$ and $\dot{E}^{H^{-}}$ are the energy fluxes across the future and past horizon, respectively. The energy flux across the future horizon is given in Eq.~\eqref{Eh} as in the BH case, whereas the energy flux coming in across the past horizon is~\cite{Teukolsky:1974yv}
\begin{equation}
\dot{E}^{H^{-}} =\sum_{\ell m} \frac{\omega}{4 \pi k (2Mr_+)^3 (k^2 + 4 \varpi^2)} |Z^{H^{-}}_{\ell m \omega}|^2 \,.
\end{equation}
In the case of $\mathcal{R}=0$, Eq.~\eqref{Eradius} reduces to $\dot{E}^{\rm int} = \dot{E}^{H^{+}}$.
When $|\mathcal{R}(\omega)|^2=1$, the outgoing flux is equal to the ingoing flux at the radius of the object and $\dot{E}^{\rm int} = 0$, as expected from perfectly reflecting boundary conditions. 

In the ECO case, the total energy flux emitted by a point particle in a  circular equatorial orbit is
\begin{equation}
 \dot E(\Omega) = \dot E^{\infty}(\Omega) +\dot{E}^{\rm int}(\Omega)\,, \label{EtotECO}
\end{equation}
where $\dot E^{\infty}(\Omega)$ and $\dot{E}^{\rm int}(\Omega)$ are defined in Eqs.~\eqref{Einf} and~\eqref{Eradius}, respectively.

\section{Numerical procedure}

We  study the dynamics of a point particle in circular equatorial orbits around a Kerr-like horizonless object by adapting the frequency-domain Teukolsky code originally developed in Refs.~\cite{vandeMeent:2014raa, vandeMeent:2015lxa, vandeMeent:2016pee, vandeMeent:2017bcc}. In particular, the solutions to the homogeneous Teukolsky equation are calculated via the numerical Mano-Suzuki-Takasugi method~\cite{Mano:1996vt, Mano:1996gn, Fujita:2004rb,Fujita:2009us}. We have modified the boundary conditions at the radius of the compact object in terms of ${\cal R}(\omega)$ and $\epsilon$ as discussed in Sec.~\ref{sec:horizonlesscase}, and computed the energy and angular-momentum fluxes at infinity and through the object's surface. 

Our algorithm is as follows:
\begin{itemize}
\item[1.] Choose the intrinsic parameters of the binary, namely the central mass $M$, the mass ratio $q\ll1$, the primary spin $\chi$, the reflectivity $\mathcal{R}(\omega)$, and the compactness of the central object via $\epsilon$.
\item[2.] For a given $\ell=m$ mode, produce the data for a bound orbit with orbital radius $r$ and compute the energy fluxes in the cases of a central BH and a central horizonless compact object, respectively.
\item[3.] Loop on the orbital radii with an equally spaced (radial) grid starting from the ISCO radius to $r=10M$.  
\item[4.] Find the local maxima and minima in the energy fluxes at infinity for a central horizonless compact object. If present, these extrema bracket resonances in the flux that should be resolved by increasing the grid resolution. Let us notice that the initial equally-spaced grid in the orbital radii needs to be dense enough to find local maxima and minima in the energy fluxes. For this reason, we  set the initial discretization in the orbital radii to be $0.003M$.
\item[5.] Refine the grid on the orbital radii around the local maxima and minima through the bisection method until a target accuracy is reached. The refinement of the grid stops either when the difference between two subsequent orbital radii is $<10^{-5}M$ or when the difference in the energy fluxes of two subsequent points is $<10^{-5}q^2$.
\item[6.] For a given $\ell$ and each $m = \ell-1, ..., 1$ loop on the orbital radii with an equally spaced grid from the ISCO radius. 
The loop on the orbital radii stops when the total energy flux in the case of a central BH (defined in Eq.~\eqref{EtotBH}) in a given $\ell, m$ mode is $10^{-6}$ times smaller than the total energy flux in the dominant mode with $\ell=m$.
\item[7.] For a given $\ell$ and each $m = \ell-1, ..., 1$ repeat steps 4 and 5.
\item[8.] For the harmonic index $\ell=2, ..., \ell_{\rm max}=12$ repeat the steps 2 to 7.
\item[9.] For each $\ell, m$ mode, interpolate the total energy flux as a function of the orbital angular frequency.
\item[10.] Sum over the modes and perform an integration to compute the orbital phase both in the BH and in the horizonless cases. The initial condition on the orbital angular frequency is $\Omega_0 = \Omega(r=10M)$ and the integration stops at the inspiral-plunge transition frequency $\Omega(t_{\rm max}) = \Omega(r=r_{\rm ISCO} +4q^{2/3})$~\cite{Ori:2000zn}.
\end{itemize}

We  compute the gravitational waveform where for the modes with negative $m$ we make use of the following symmetries
\begin{eqnarray}
Z^{\infty}_{\ell -m \omega} &=& (-1)^{\ell} \left(Z^{\infty}_{\ell m \omega}\right)^* \,, \\
~_{-2}S_{\ell -m \omega}(\theta) &=& (-1)^{\ell} ~_{-2}S_{\ell m \omega}(\pi-\theta) \,.
\end{eqnarray}
For each $\ell, m$ mode the asymptotic amplitudes at infinity and the spin-weighted spheroidal harmonics are interpolated functions of the time-dependent orbital angular frequency. The waveform is constructed by summing over the modes with $\ell \leq 4$ and $-\ell \leq m \leq \ell$. In the case of small reflectivity ($|\mathcal{R}|^2\leq 10^{-6}$) the waveform is constructed by summing over the $\ell, m$ modes until $\ell = 5$ 
since one needs higher accuracy to keep the truncation errors smaller than the ECO corrections. We checked that the mismatch between the BH and ECO waveforms does not change quantitatively by including modes with higher $\ell $ in the waveforms.

We tested our code by reproducing the standard results for the Kerr BH case~\cite{Hughes:2001jr,Bernuzzi:2012ku,Taracchini:2013wfa,Harms:2014dqa}. Furthermore, we reproduced the results of Ref.~\cite{Datta:2019epe}, where 
the horizonless case is obtained from the BH case by artificially imposing that only a fraction $(1-|{\cal R}|^2)$ of the radiation is absorbed at the surface.

The fractional truncation error of the code in the dephasing is estimated  as 
$\Delta^{\rm tr} = 1- \Delta \phi_{\ell_{\rm max}+1}(t_f) / \Delta \phi_{\ell_{\rm max}}(t_f)$, where the energy fluxes are truncated at $\ell_{\rm max}=12$ and $t_f$ is the time in which the orbital radius reaches $r=r_{\rm ISCO} + 4 q^{2/3}$. For a reference compact object with $\chi=0.9$, $|\mathcal{R}|^2=0.9$, $\epsilon=10^{-10}$, and $q=3 \times 10^{-5}$, we  find $\Delta^{\rm tr}=2 \times 10^{-5}$. 

\section{Energy fluxes and excitation of resonances}

Horizonless compact objects contain low-frequency modes in their spectrum that are associated with long-lived states confined within the photon sphere, as described in Sec.~\ref{sec:lowfrequencies}. At variance with the BH case, these low-frequency modes can be excited during the inspiral when the orbital frequency matches the QNM frequencies, leading to resonances in the fluxes~\cite{Pani:2010em,Macedo:2013jja,Fransen:2020prl}. 
The role of the resonances in the EMRI dynamics was studied in Ref.~\cite{Cardoso:2019nis} for a perfectly reflecting and non-spinning horizonless object in the low-frequency approximation. The analysis of Ref.~\cite{Cardoso:2019nis} assessed that the impact of such excitations is negligible since the resonances are very narrow and are crossed quickly during the inspiral.
In the following, we shall extend the analysis of Ref.~\cite{Cardoso:2019nis} to the case of a partially absorbing compact object with generic spin. As we shall show, differently from the analysis in Ref.~\cite{Cardoso:2019nis}, in more generic cases the presence of resonances provides an important contribution to the EMRI dynamics. 

Another relevant feature of horizonless compact objects is the presence of partial reflectivity at the surface. In Ref.~\cite{Datta:2019epe} a phenomenological approach was adopted to parametrize the energy flux emitted by a point particle around a horizonless and partially absorbing compact object. In particular, the energy flux at infinity is modeled as the one in the BH case, whereas the energy flux on the ECO side is modeled by removing a ($|\mathcal{R}|^2$) fraction of TH from the BH energy flux through the horizon, i.e.,~\cite{Datta:2019epe}
\begin{equation} \label{EdotTH}
 \dot E (\Omega) =   E_{\rm BH}^\infty (\Omega) + \left(1-|\mathcal{R}|^2\right) E_{\rm BH}^{H} (\Omega) \,.
\end{equation}
The TH is associated with the energy and angular-momentum absorption by the compact object and results in an increase of the mass and angular momentum of the latter, unless superradiance occurs~\cite{Brito:2015oca}.
Let us notice that Eq.~\eqref{EdotTH} does not include the excitation of the low-frequency resonances in the energy fluxes both at infinity and on the ECO side. Moreover, Eq.~\eqref{EdotTH} does not fix a specific location of the radius of the compact object.
According to the analysis in Ref.~\cite{Datta:2019epe}, EMRIs could provide constraints on the reflectivity of compact objects at the level of $|\mathcal{R}|^2 \lesssim 10^{-4}$.
We shall show that, by taking a consistent model of horizonless compact object, the  bounds derived by Ref.~\cite{Datta:2019epe} can be further improved by several orders of magnitude.

\begin{figure}[t]
\centering
\includegraphics[width=0.69\textwidth]{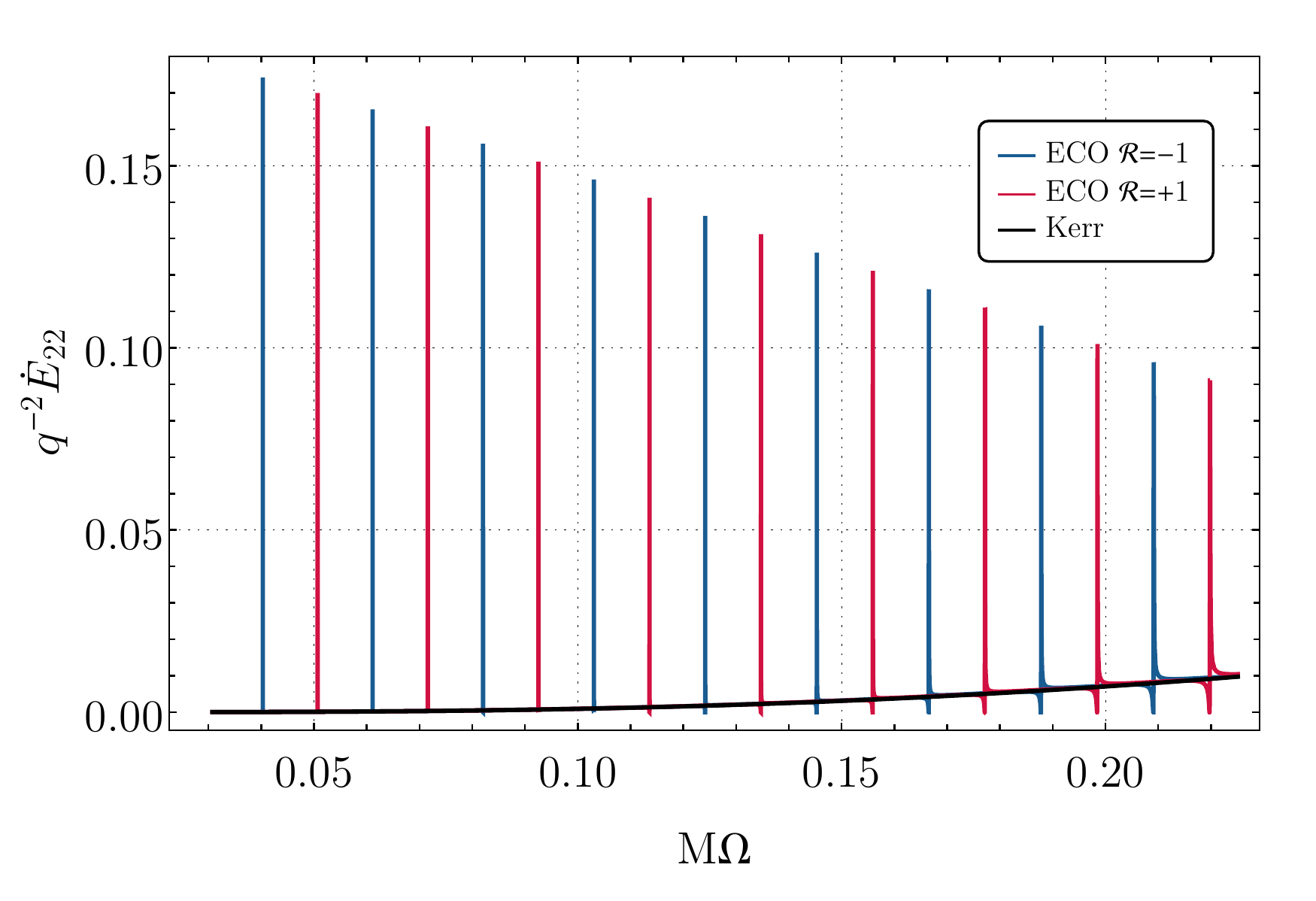}
\caption{Total energy flux of the $\ell=m=2$ mode as a function of the orbital angular frequency for a point particle in quasicircular equatorial orbits from $r=10M$ (low frequency) to $r=r_{\rm ISCO}$ (high frequency). The energy flux emitted in the case of a central BH with spin $\chi=0.9$ is compared to the case of a central horizonless compact object with a perfectly reflecting surface ($|\mathcal{R}|^2=1$), spin $\chi=0.9$, and $\epsilon=10^{-10}$. In the latter case, the energy flux is resonantly excited when the orbital frequency matches the low-frequency QNMs of the ECO~\cite{Maggio:2021uge}.} 
\label{fig:flux}
\end{figure}
Let us analyze the energy flux emitted by a point particle in quasicircular equatorial orbits around a spinning horizonless compact object, as detailed in Sec.~\ref{sec:horizonlesscase}. The energy flux is computed as in Eq.~\eqref{EtotECO} that takes into account both the excitation of the low-frequency resonances and the reflective properties of the compact object.
As a representative example, Fig.~\ref{fig:flux} shows the
$\ell=m=2$ component of the energy flux as a function of the orbital frequency for a horizonless compact object with $\epsilon=10^{-10}$, $\chi=0.9$, and two choices of perfectly reflecting boundary conditions (Dirichlet and Neumann for the lower and upper case, respectively, $\mathcal{R} = \pm 1$). 
As expected, the flux is resonantly excited when the orbital frequency matches the low-frequency QNMs of the central object, i.e.,
\begin{equation}
\Omega = \frac{\omega_R}{m} \,, 
\end{equation}
where $\omega_R$ is the real part of the QNM and $m$ is the azimuthal number of the perturbation. This is a striking difference with respect to the BH case in which the QNMs have higher frequencies and cannot be resonantly excited by quasicircular inspirals.
In the small-$\epsilon$ limit, the Dirichlet and Neumann modes are described by Eqs.~\eqref{MomegaR} and~\eqref{MomegaI} with $s=-2$.
As shown in Fig.~\ref{fig:flux}, for a compact object with a given spin and compactness, the modes are equispaced by 
\begin{equation}
\Delta \omega_R=\frac{\pi}{|r_*^0|} \sim |\log \epsilon|^{-1} \,,
\end{equation}
whereas consecutive Dirichlet and Neumann mode frequencies are separated by half this width. The difference between consecutive resonances scales as $|\log \epsilon|^{-1}$. It follows that the resonances are denser in the $\epsilon \to 0$ limit. 

Interestingly, the resonances appear at the same frequencies in all the individual fluxes: $\dot E^\infty$, $\dot E^{H^+}$, and $\dot E^{H^-}$. This is because the QNMs are associated with the poles of the Wronskian appearing in the solutions of the Teukolsky equation (as in Eq.~\eqref{inhomsol}). However, when $|{\cal R}|^2=1$, the fluxes $\dot E^{H^+}$ and $\dot E^{H^-}$ are exactly equal to each other since $\dot{E}^{\rm int}=0$. Consequently, for  perfectly reflecting compact objects the resonances appear only in the energy flux at infinity.

Equation~\eqref{MomegaI} shows that $\omega_I\ll\omega_R$, which implies that the resonances are typically very narrow and hard to resolve~\cite{Pani:2010em,Cardoso:2019nis,Fransen:2020prl}. To assess the relation between the width of the resonances and the imaginary part of the QNMs, we  make use of the harmonic oscillator model~\cite{Pons:2001xs}. According to the latter, a compact object which resonates at a QNM frequency can be modeled as a forced harmonic oscillator that satisfies~\cite{Pons:2001xs}
\begin{equation}
    \ddot{\xi} - 2 \omega_I \dot{\xi} + \left( \omega_R^2 + \omega_I^2 \right) \xi = b \omega^2 e^{-i \omega t} \,,
\end{equation}
where $\xi$ is the amplitude of the GW emitted at infinity normalized by a reference amplitude, e.g., the amplitude of the GW emitted at infinity when the central object is a BH, and the orbiting point particle acts as a driving force. The solution is $\xi(t) = \xi(\omega) e^{-i \omega t}$, where
\begin{equation}
    \xi(\omega) = \frac{-b \omega^2}{\omega^2-\omega_R^2-\omega_I^2-2i \omega_I \omega} \,.
\end{equation}
Near the resonance, the amplitude of the GW is the sum of two contributions, one due to the orbital motion and one due to the resonance to the QNM. The normalized energy flux across a single resonance is well fitted by the model~\cite{Pons:2001xs}
\begin{equation}
\frac{\dot{E}^{\rm ECO}}{\dot{E}^{\rm BH}} = \left|1-\xi(\omega)\right|^2 = \frac{\left[(1-b) \omega^2 - \omega_R^2 - \omega_I^2\right]^2+\left(2 \omega_I \omega\right)^2}{\left(\omega^2 - \omega_R^2 - \omega_I^2\right)^2 + \left(2 \omega_I \omega\right)^2} \,,
\end{equation}
where $\dot{E}$ is the total energy flux as computed in Eq.~\eqref{EtotBH} and Eq.~\eqref{EtotECO} for the BH and ECO cases, respectively, $b = 1-\left(\Omega_{\rm max}/\Omega_{\rm min}\right)^2$, and $\Omega_{\rm max}$ and $\Omega_{\rm min}$ are the orbital angular frequencies of the maximum and the minimum of each resonance.
The width of each resonance in the orbital frequency scales as $\delta \Omega \sim \omega_I$~\cite{Cardoso:2019nis}, where $\omega_I \sim \omega_R^{2\ell+2}$ from Eq.~\eqref{MomegaI}. It follows that the width of the resonances increases with the orbital angular frequency, as it is shown in Fig.~\ref{fig:flux}. 

In the non-spinning and perfectly reflecting case, we recover the results of Ref.~\cite{Cardoso:2019nis}, namely that low-frequency resonances do not contribute significantly to the GW phase due to their narrow width. 
However, for highly spinning compact objects, the ISCO frequency occurs at higher frequencies to the non-spinning case. Consequently, the resonances with larger width can be excited and contribute to a significant dephasing to the BH case (as shown in Sec.~\ref{sec:dephasing}). 

\begin{figure}[t]
\centering
\includegraphics[width=0.69\textwidth]{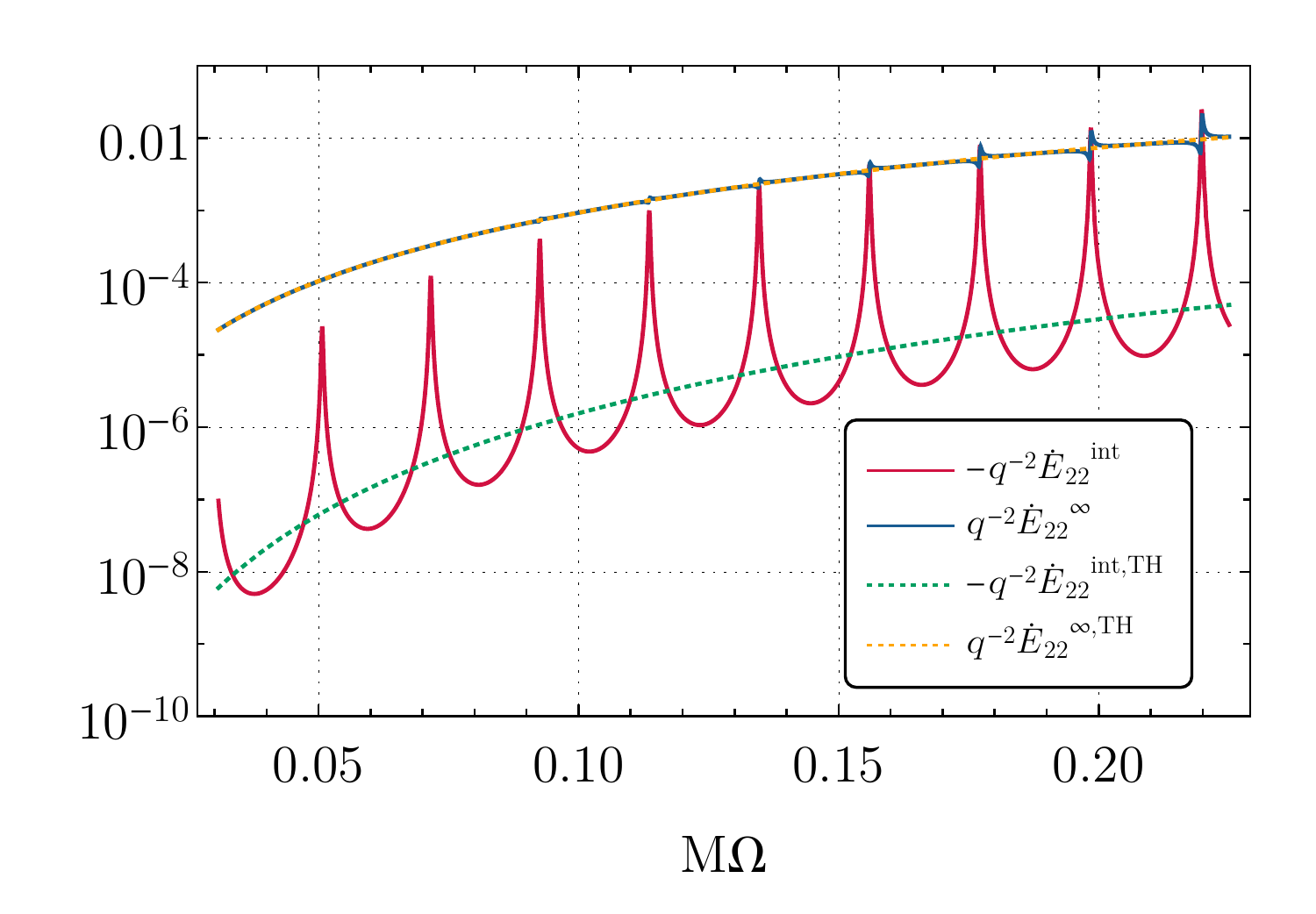}
\caption{Energy fluxes that are emitted on the ECO side and at infinity by a point particle around a central ECO with $\chi=0.9$, $\epsilon=10^{-10}$, and $\mathcal{R}=\sqrt{0.9}$ for the $\ell=m=2$ mode. The fluxes are compared with those of Ref.~\cite{Datta:2019epe} in which the effect of the ECO was accounted for by removing a fraction of the tidal heating~(TH) from the standard Kerr flux.~\cite{Maggio:2021uge}} 
\label{fig:flux2}
\end{figure}
The system shown in Fig.~\ref{fig:flux} is purely indicative since a  horizonless compact object with a perfectly reflecting surface is unstable due to the ergoregion instability and would spin down on short timescales (see Secs.~\ref{sec:QNMsspin} and~\ref{sec:quenchergoregion} for a related discussion).
Stable models of horizonless compact objects require either small values of the spin or partial absorption~\cite{Maggio:2017ivp,Maggio:2018ivz}. In both cases, the resonances are less evident, as shown in Fig.~\ref{fig:flux2} for a partially absorbing compact object with $|{\cal R}|^2=0.9$, a value that guarantees stability for a central object with spin $\chi=0.9$.

Fig.~\ref{fig:flux2} shows that, also for reflectivities smaller than unity, the resonances are excited both in the energy flux at infinity ($\dot{E}^\infty$) and on the ECO side ($\dot{E}^{\rm int}$). The resonances are less peaked than in the perfectly reflecting case but, as shown below, can still have a sufficiently large width to be efficiently excited. Overall, the energy flux on the ECO side is several orders of magnitude smaller than the energy flux at infinity.
However near the ISCO frequency, $\dot E^{\rm int}$ is comparable to $\dot E^\infty$ and contributes significantly to the GW phase.

Finally, Fig.~\ref{fig:flux2} also shows the energy fluxes at infinity and on the ECO side computed in Ref.~\cite{Datta:2019epe} by removing a fraction of TH from the BH energy flux as in Eq.~\eqref{EdotTH}. In this case, the energy flux at infinity  is similar to the exact result except for the presence of the resonances that are absent in the model of Ref.~\cite{Datta:2019epe}. On the other hand, the energy flux of the ECO side can change significantly. Due to the presence of the resonances, $\dot{E}^{\rm int}$ computed in Ref.~\cite{Datta:2019epe} is a sort of averaged value of the exact result. The latter is modulated by the presence of resonances that can be as high as the flux at infinity.

\begin{figure*}[ht!]
\centering
 \includegraphics[width=0.69\columnwidth]{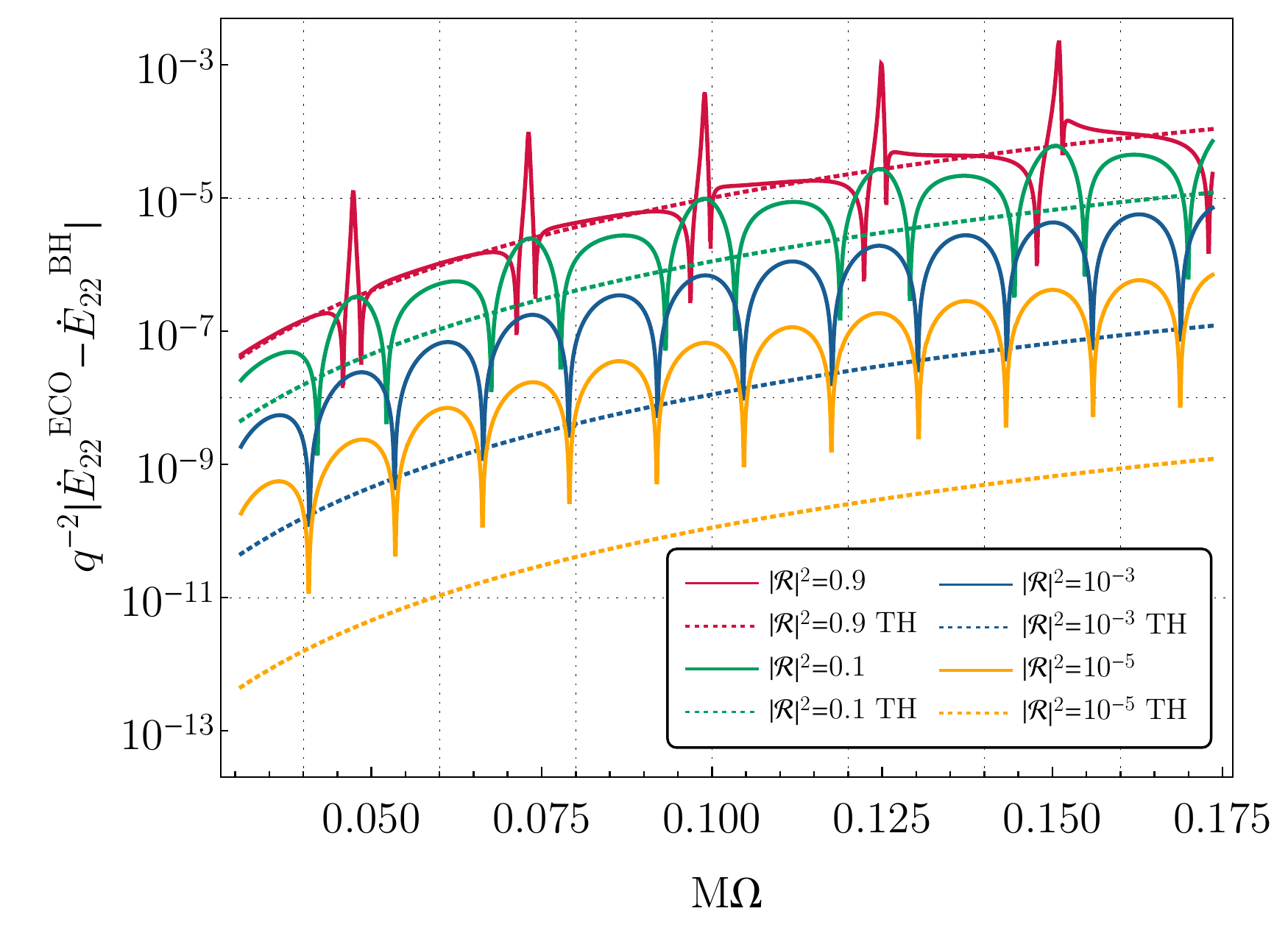}
 \includegraphics[width=0.40\columnwidth]{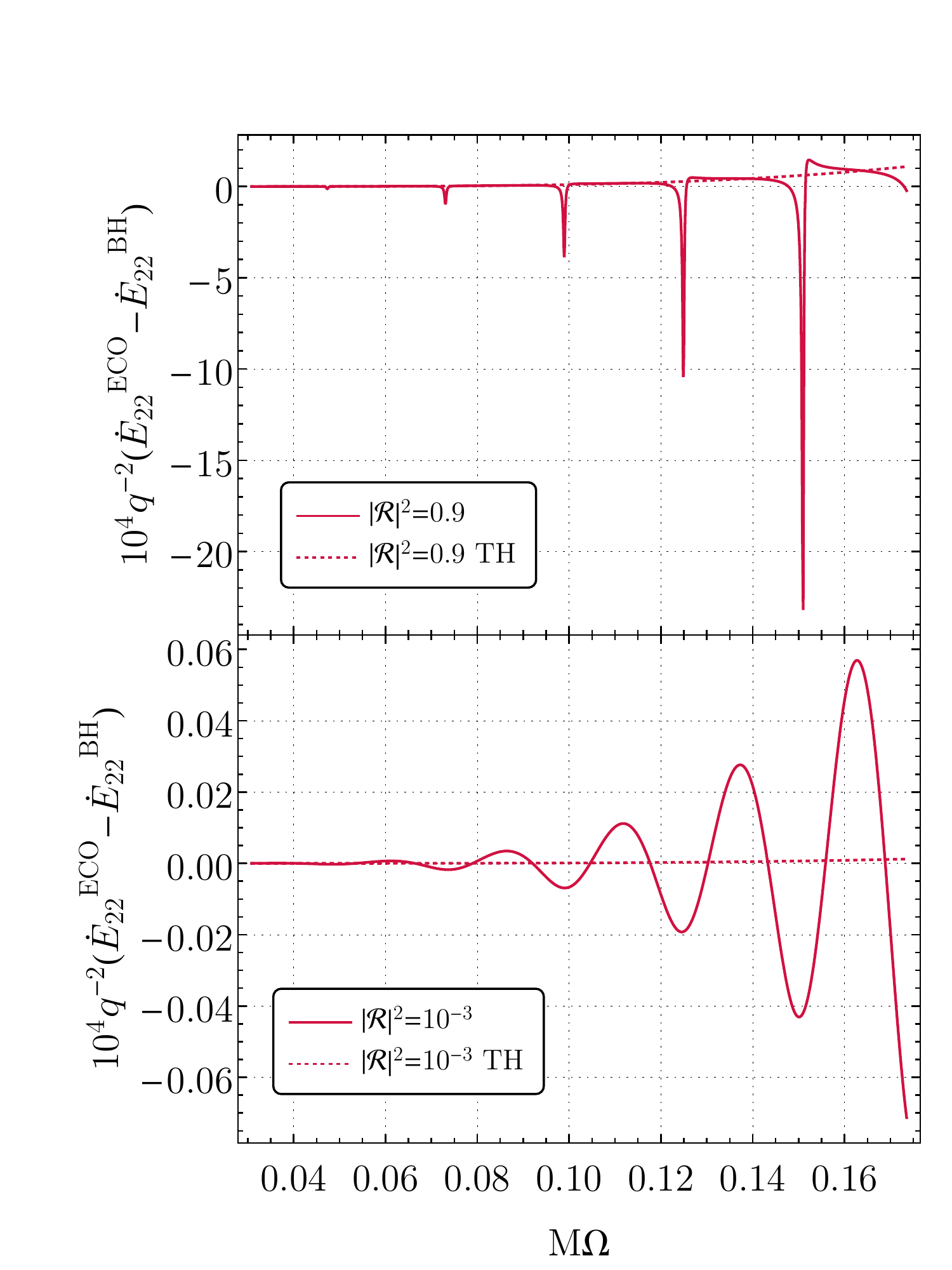}
 \includegraphics[width=0.40\columnwidth]{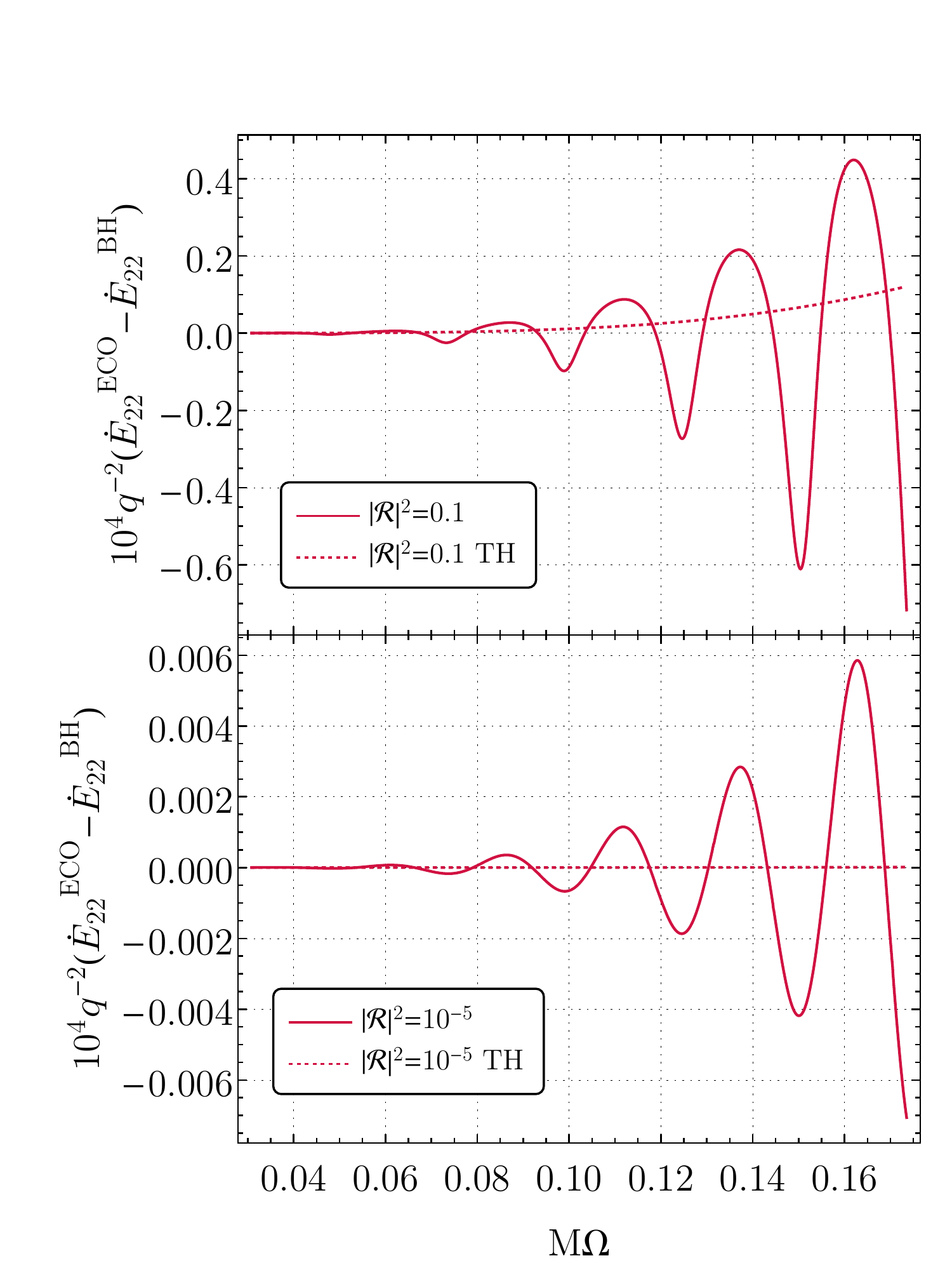}
\caption{
Difference between the total energy flux of the $\ell=m=2$ mode in the ECO case with respect to the BH case.
Top panel: absolute value of the difference for a central object with $\chi=0.8$, $\epsilon=10^{-10}$, and several values of the reflectivity. The dotted lines are the estimated differences in the total energy flux due to the absence of TH relative to the BH case, as in Ref.~\cite{Datta:2019epe}.
Bottom panels: same as in the top panel but without the absolute value and in a linear scale, to appreciate the change of sign of the oscillations associated with the resonances.~\cite{Maggio:2021uge}
} 
\label{fig:difference}
\end{figure*}
In Fig.~\ref{fig:difference} we show the difference between the total energy flux of the $\ell=m=2$ mode in the horizonless case with respect to the BH case, both in our consistent model and in the phenomenological approach of Ref.~\cite{Datta:2019epe}. In particular, the top panel shows the absolute value of the difference between the ECO and BH fluxes on a logarithmic scale, to appreciate the relatively small numbers involved. The bottom panel, instead, shows the difference between the energy fluxes on a linear scale, to appreciate the change of sign during the oscillations, for several values of the reflectivity.

For $|\mathcal{R}|^2 \approx 1$ the differences between the consistent model and the model of Ref.~\cite{Datta:2019epe} 
are due to two factors: the excitation of the resonances and the (subleading) effect that the flux computation in the consistent model accounts for the fraction of the GWs that are reflected by the object and make their way to infinity rather than being reabsorbed by the particle, as implicitly assumed in Ref.~\cite{Datta:2019epe}. 
For small values of the reflectivity, the difference between the consistent model and the phenomenological one is even more important. In this case, the resonances are suppressed in amplitude but still appear in the total energy flux with a larger width, as shown in the top panel of Fig.~\ref{fig:difference}. The bottom panel grid in Fig.~\ref{fig:difference} shows the oscillatory trend of the total energy flux in the horizonless case compared to the energy flux in Ref.~\cite{Datta:2019epe} for small reflectivities. The amplitude of the oscillations increases with the orbital angular frequency and decreases with the reflectivity. The oscillations are related to the resonances and, as we shall see in Sec.~\ref{sec:dephasing}, they can contribute significantly to the GW phase also for small values of the reflectivity.

Interestingly, when the superradiance condition is met, $\Omega<\Omega_H$, the flux on the ECO side can be negative due to the energy and angular-momentum extraction from the central object~\cite{Brito:2015oca}. Since $\dot{E}^{\rm int}$ and $\dot E^\infty$ have the opposite sign, it is interesting to check whether they can compensate each other at some given frequency, giving rise to a total zero flux and hence to ``floating'' orbits~\cite{Kapadia:2013kf,Cardoso:2011xi}. As clear from Fig.~\ref{fig:flux2}, in the case of a single mode (e.g., $\ell=m=2$) such orbits would exist near the high-frequency resonances, where $\dot{E}^{\rm int}$ (which is typically subdominant) can be as large as $\dot E^\infty$ in absolute value. When including the contribution of multipoles, we find that the total flux at infinity is larger than the flux on the ECO side because modes with different $(\ell,m)$ are resonantly excited at different frequencies. The net result is that the total flux, $\dot E^\infty+ \dot{E}^{\rm int}$, is overall positive and the orbit shrinks during the adiabatic evolution.

\section{Adiabatic evolution and dephasing} \label{sec:dephasing}

In EMRIs, the radiation-reaction timescale is much longer than the orbital period. For this reason, at the first order in the mass ratio the orbital parameters can be evolved using an adiabatic expansion~\cite{Hinderer:2008dm}. For a particle in a circular, equatorial and corotating orbit,  
the evolution of the orbital angular frequency $\Omega$ and the orbital phase $\phi$ are governed by
\begin{eqnarray}
 \dot \Omega &=& -\left(\frac{d E_b}{d \Omega}\right)^{-1} \dot E(\Omega)\,, \label{eq:adiab1}\\
 \dot \phi&=& \Omega \label{eq:adiab2}\,,
\end{eqnarray}
where $E_b$ is the binding energy of the system 
\begin{equation}
E_b = \mu \frac{1-2 v^2 + \chi v^3}{\sqrt{1-3v^2+2\chi v^3}} \,,
\end{equation}
where $v \equiv \sqrt{M/r}$, $r$ is the orbital radius that is related to the orbital angular frequency through Eq.~\eqref{orbfreq}, and $\dot E(\Omega)$ is the total energy flux defined in Eqs.~\eqref{EtotBH} and~\eqref{EtotECO} in the BH and the ECO case, respectively.

Equations~\eqref{eq:adiab1} and~\eqref{eq:adiab2} can be solved by adding two initial conditions, namely
\begin{eqnarray}
 \Omega(t=0) &=& \Omega_0 \,, \\
 \phi(t=0) &=& 0 \,,
\end{eqnarray}
without loss of generality. The GW phase of the dominant mode is related to the orbital phase by $\phi_{\rm GW} = 2 \phi$. We  compute the GW dephasing accumulated up to a certain time between the cases of a central BH and a central horizonless compact object as~\cite{Datta:2019epe}
\begin{equation}
\Delta \phi(t) = \phi_{\rm GW}^{\rm BH}(t) - \phi_{\rm GW}^{\rm ECO}(t) \,.
\end{equation}
%

\subsection{Non-spinning central object} \label{sec:dephasing_nospin}

%
\begin{figure}[t]
\centering
\includegraphics[width=0.69\textwidth]{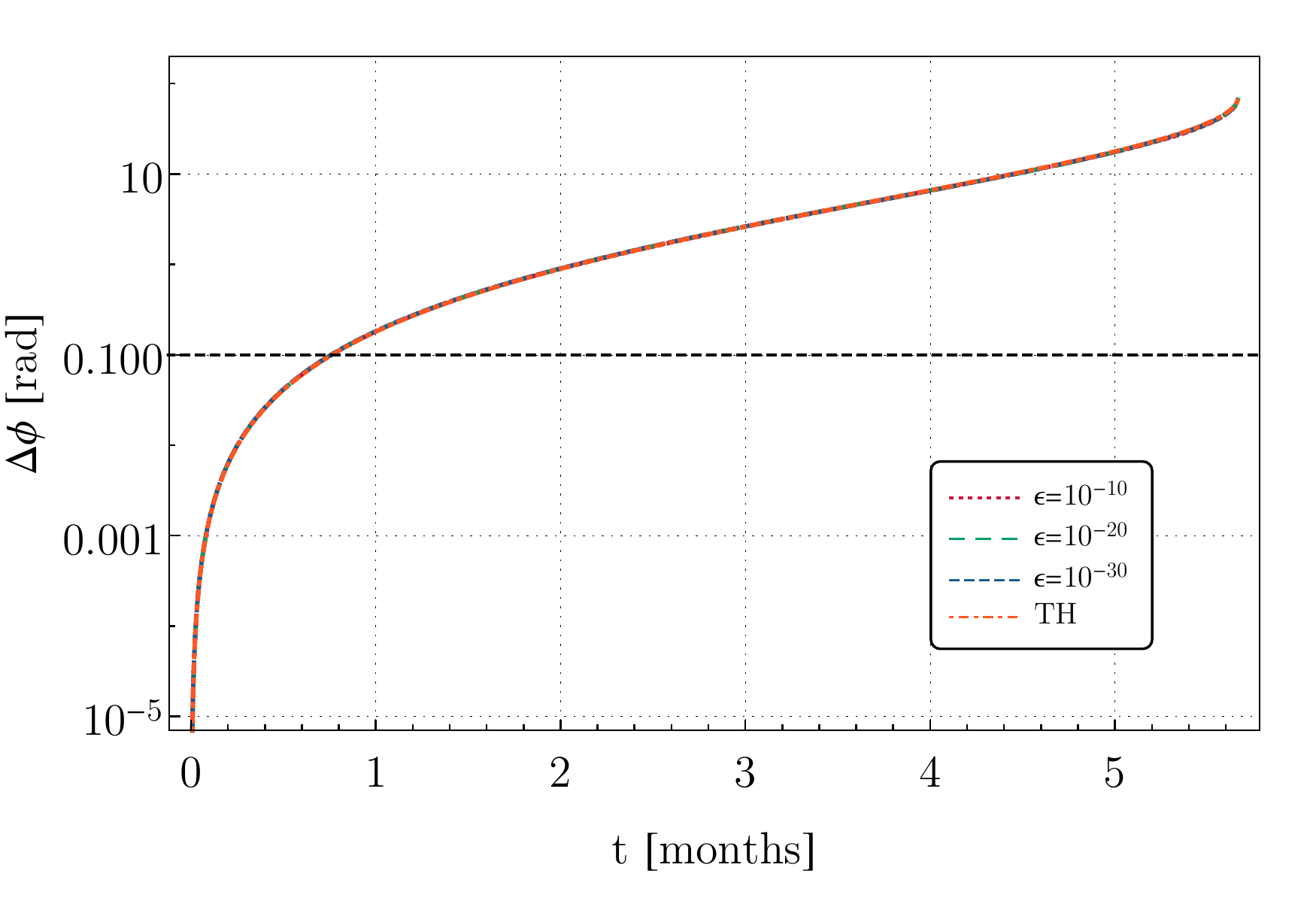}
\caption{Dephasing as a function of time in the case of a non-spinning and perfectly reflecting ECO relative to the Schwarzschild BH case for different values of the compactness parameter $\epsilon$ and $q=3 \times 10^{-5}$. The resonances in the energy flux do not contribute to the dephasing which is well approximated by the phenomenological model in Ref.~\cite{Datta:2019epe} marked as TH.~\cite{Maggio:2021uge}}
\label{fig:dephasing_spin0}
\end{figure}
Let us analyze the case of a central compact object which is non-spinning.
Fig.~\ref{fig:dephasing_spin0} shows the dephasing in the case of a perfectly reflecting ECO relative to the Schwarzschild BH, for different values of the compactness parameter $\epsilon$. As shown in Fig.~\ref{fig:dephasing_spin0}, the dephasing does not depend on $\epsilon$ and is not affected by the resonances, that are too narrow to be efficiently excited in the non-spinning case. For this reason, the dephasing is well approximated by the model adopted in Ref.~\cite{Datta:2019epe} that removes a fraction of TH from the BH energy flux.
Overall, the results in Fig.~\ref{fig:dephasing_spin0} are compatible with the analytical estimates in Ref.~\cite{Cardoso:2019nis} which assessed that the impact of resonances is negligible in the case of a non-spinning and perfectly reflecting compact object.

\subsection{Spinning central object} \label{sec:dephasing_spin}

%
\begin{figure}[t]
\centering
\includegraphics[width=0.69\textwidth]{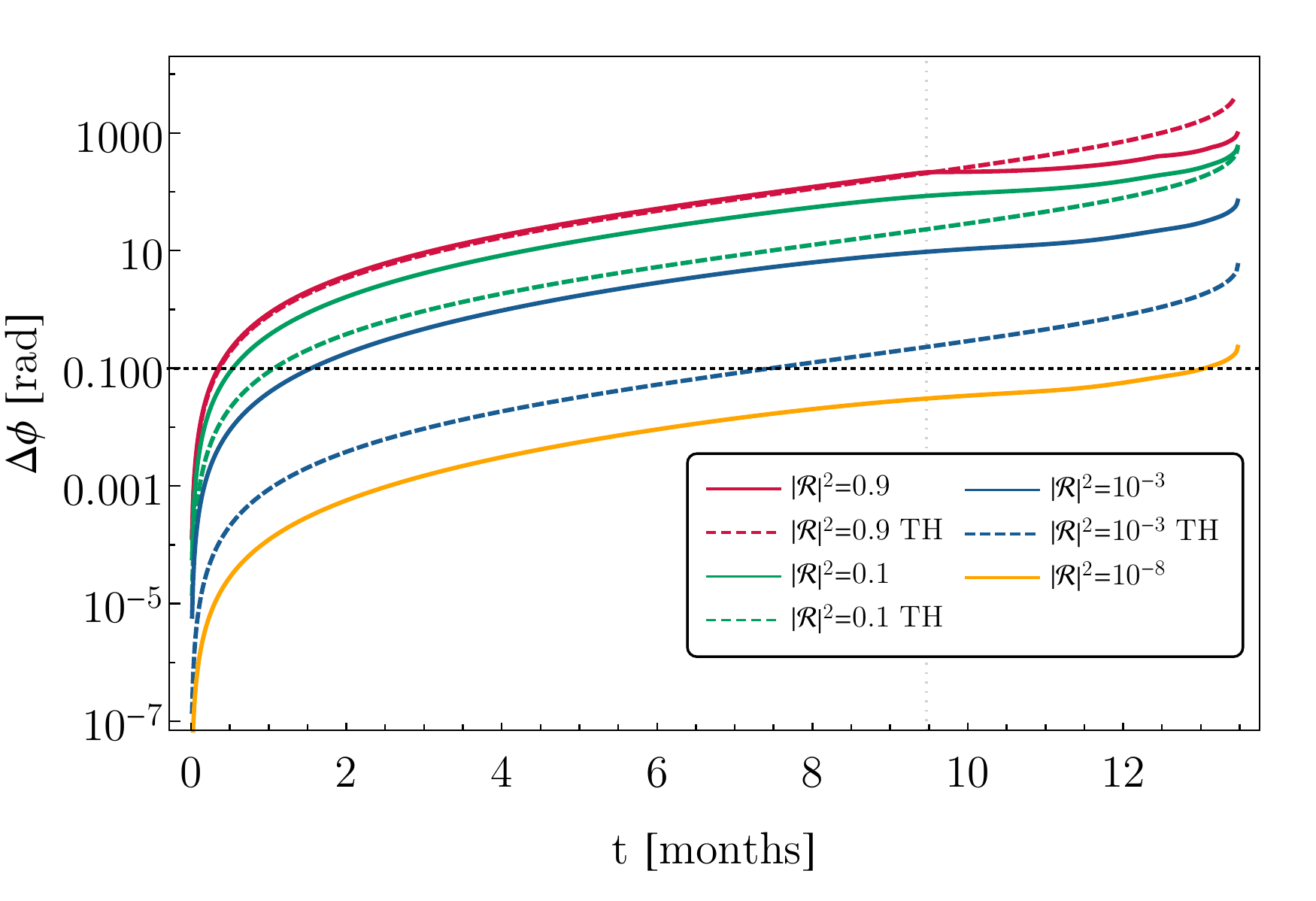}
\caption{GW dephasing between the BH and the ECO case as a  function of time for a binary with mass ratio $q=3 \times 10^{-5}$ and a central object with spin $\chi=0.8$, $\epsilon=10^{-10}$ and several values of the reflectivity. The dashed lines show the dephasing due to the absence of TH relative to the BH case as in Ref.~\cite{Datta:2019epe}.
The vertical dashed line corresponds to the time in which a resonant orbital frequency is excited. The horizontal line is a reference value $\Delta \phi=0.1\,{\rm rad}$ for the resolvability of the dephasing by LISA.~\cite{Maggio:2021uge}} 
\label{fig:dephasing}
\end{figure}
Let us analyze the dephasing between a spinning horizonless compact object and the standard Kerr case. 
This is shown in Fig.~\ref{fig:dephasing} for a fiducial binary with primary mass $M=10^6 M_\odot$, secondary mass $\mu=30M_\odot$, and a central object with spin $\chi=0.8$ and $\epsilon=10^{-10}$. 
We  analyze different values of the reflectivity and for each of them, we  compare our exact result with the one of the model in Ref.~\cite{Datta:2019epe}.

The dephasing increases monotonically in time and also as a function of the reflectivity.
When $|{\cal R}|^2\approx 1$, the difference to the model in Ref.~\cite{Datta:2019epe} is small 
until the inspiral moves across a resonance. In particular, for $|\mathcal{R}|^2=0.9$, the dephasing in the horizonless case deviates from the dephasing due to the absence of TH at $t=9.47\,{\rm months}$ (marked in Fig.~\ref{fig:dephasing} by a dashed vertical line) due to the presence of a $\ell=m=2$ resonance with $M\Omega=0.0473$ and $M\omega_I = -4.22\times 10^{-5}$. Subsequent resonances are excited at later times and are responsible for the deviations of the dephasing from the model adopted in Ref.~\cite{Datta:2019epe}.

Interestingly, the phenomenological model of Ref.~\cite{Datta:2019epe} and the exact result differ significantly for small reflectivities even if the resonances are less evident.
This is due to several factors: the energy fluxes at infinity and on the ECO side display some differences in the two models since a fraction of the energy is reflected by the object and leaves the system; moreover, both fluxes (at infinity and on the ECO side) can be resonantly excited only in our model and these resonances contribute significantly to the GW phase even for intermediate values of ${\cal R}$.
The dephasing in the consistent model is always larger than the dephasing with TH only. For small values of ${\cal R}$, the two models differ from each other but both produce a small dephasing to the Kerr case.

For a signal with $\text{SNR} \sim 30$, a dephasing $\Delta \phi = 0.1     \ \text{rad}$ is considered to be resolvable by LISA~\cite{Lindblom:2008cm,Bonga:2019ycj}. Fig.~\ref{fig:dephasing} shows that the phase difference is above this threshold only after half a month of observation for a horizonless compact object with $|\mathcal{R}|^2=0.9$. 
After twelve months of inspiral, values of the reflectivity as small as $|\mathcal{R}|^2=10^{-8}$ would be detectable by LISA. In Sec.~\ref{sec:overlap}, we shall assess the measurability of the reflectivity with a more robust method based on the computation of the overlap between the waveforms.

\subsection{The role of the compactness}

%
\begin{figure}[t]
\centering
\includegraphics[width=0.73\textwidth]{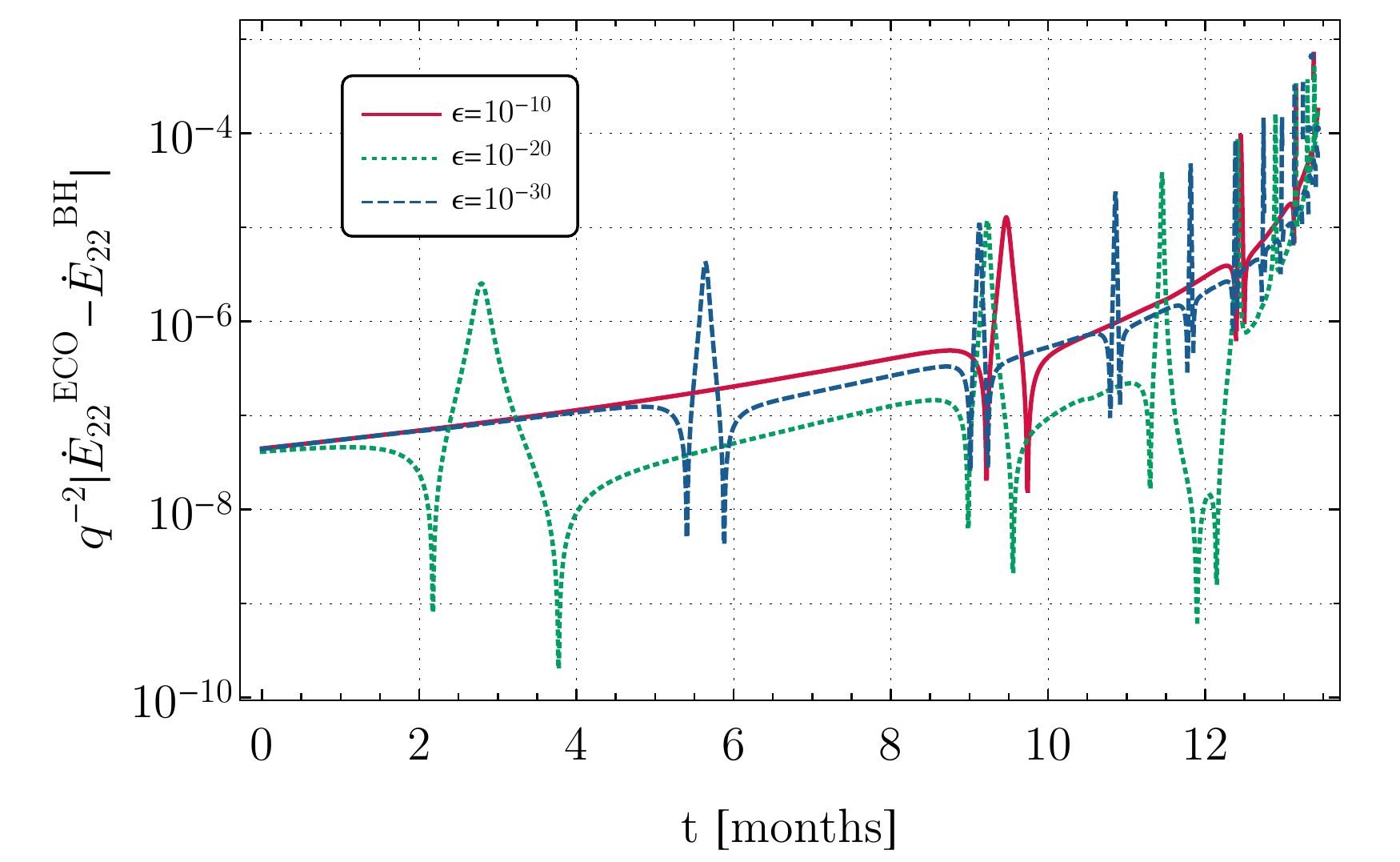}
\includegraphics[width=0.69\textwidth]{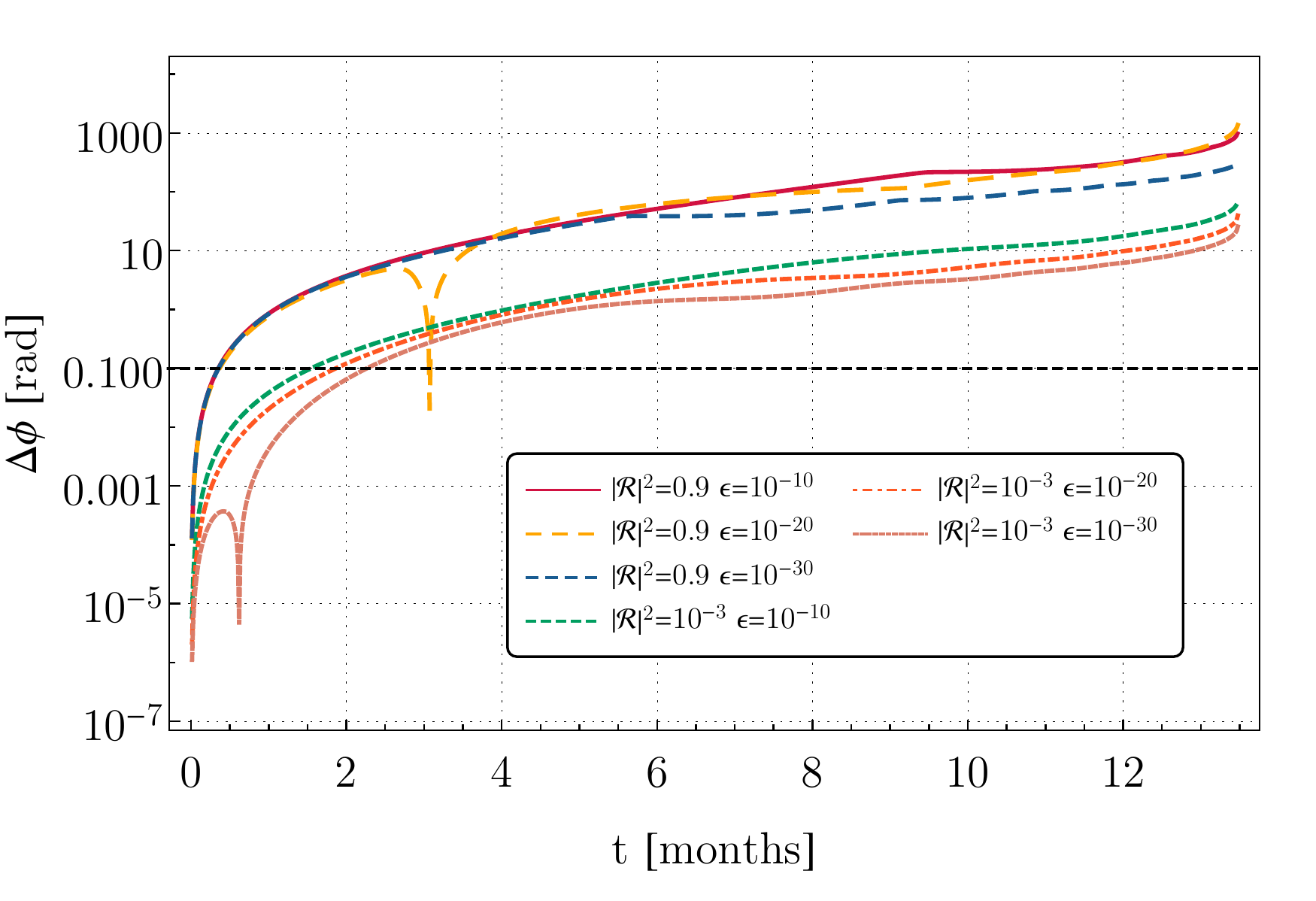}
\caption{Top panel: Resonances in the $\ell=m=2$ energy flux for an ECO with $\chi=0.8$, $|\mathcal{R}|^2=0.9$ and several values of $\epsilon$, as a function of time. Bottom panel: GW dephasing between the BH and the ECO case as a  function of time for $q=3 \times 10^{-5}$, $\chi=0.8$ and several values of $\epsilon$.~\cite{Maggio:2021uge}}
\label{fig:compactness}
\end{figure}
In this section, we analyze the role of the compactness in the energy fluxes emitted by an EMRI with a central horizonless compact object.
The top panel of Fig.~\ref{fig:compactness} shows the difference between the ECO and BH energy fluxes for several values of $\epsilon$ as a function of time. We  note that for $\epsilon \to 0$, more resonances appear and they also appear at lower frequencies. The first low-frequency resonances could give a large contribution to the GW phase since the orbital evolution is slower at low frequency and the particle can spend more time across the resonance. On the other hand, the width of each resonance is proportional to the imaginary part of the QNMs where $\omega_I \sim \omega_R^{2\ell+2}$, therefore low-frequency resonances are also more narrow. The two effects are competitive and the actual contribution of a resonance on the GW phase depends on the specific parameters of the configuration.

The bottom panel of Fig.~\ref{fig:compactness} shows the dephasing between the ECO and the BH case for several values of the reflectivity and compactness. The dependence on $\epsilon$ is mild, except for the excitation of the resonances whose impact depends on the specific values of $\chi$, $\epsilon$ and ${\cal R}$.

\section{Waveform and overlap} \label{sec:overlap}

The waveform emitted by an EMRI is computed from the Weyl scalar in Eq.~\eqref{weyl} at infinity and reads~\cite{Hughes:2001jr,Piovano:2020zin}
\begin{equation}
 h_+- i h_\times = -\frac{2}{\sqrt{2 \pi}} \frac{\mu}{D}  \sum_{\ell m} \frac{Z_{\ell m \omega}^{\infty}(t)}{\left[m \Omega(t)\right]^2} e^{i m\left( \Omega(t) r_*^D-\phi(t)\right)} ~_{-2}S_{\ell m \omega}(\theta,t) e^{i m \varphi} \,, \label{waveform}
\end{equation}
where $D$ is the luminosity distance of the source from the detector, $r_*^D \equiv r_*(D)$, and $(\theta,\varphi)$ identify the direction
of the detector in a reference frame centered at the source in Boyer-Lindquist coordinates. 
Since the initial phase is degenerate with the azimuthal direction, we  rescale the initial phase as $\varphi \equiv \phi(t=0)$.

Although the dephasing $\Delta\phi$ between two waveforms ($h_1$ and $h_2$) is a useful and quick measure to estimate the measurability of any deviation from a reference signal, a more reliable and robust measure is given by the overlap:
\begin{equation}\label{overlapw}
\mathit{O}(h_1|h_2) = \frac{\left\langle h_1|h_2\right\rangle}{\sqrt{\left\langle h_1|h_1\right\rangle \left\langle h_2|h_2\right\rangle}}\,,
\end{equation}
where the inner product is defined in Eq.~\eqref{innerproduct} od Appendix~\ref{app:fisher}. For the power spectral density, we adopt the LISA curve of Ref.~\cite{Cornish:2018dyw} adding the contribution of the confusion noise from the unresolved Galactic binaries for a one-year mission lifetime.
Since the waveforms are defined up to an arbitrary time and phase shift, it is also necessary to maximize the overlap in Eq.~\eqref{overlapw} over 
these quantities. This can be done by computing~\cite{Allen:2005fk} 
\begin{equation}\label{overlap2}
\mathcal{O}(h_1|h_2) = \frac{4}{\sqrt{\left\langle h_1|h_1\right\rangle \left\langle h_2|h_2\right\rangle}}\max_{t_0} 
\left|\mathcal{F}^{-1}\left[\frac{\tilde{h}_1 \tilde{h}^*_2}{S_n(f)}\right](t_0)\right|\,,
\end{equation}
where $\mathcal{F}^{-1}[g(f)](t) =\int_{-\infty}^{+\infty} g(f) e^{-2\pi i f t}df$ is the inverse Fourier 
transform. The overlap is defined such that $\mathcal{O}=1$ indicates a perfect agreement between two waveforms. 
It is also customary to define the mismatch as $\mathcal{M}\equiv 1-{\cal O}$.

\begin{figure}[t]
\centering
\includegraphics[width=0.69\textwidth]{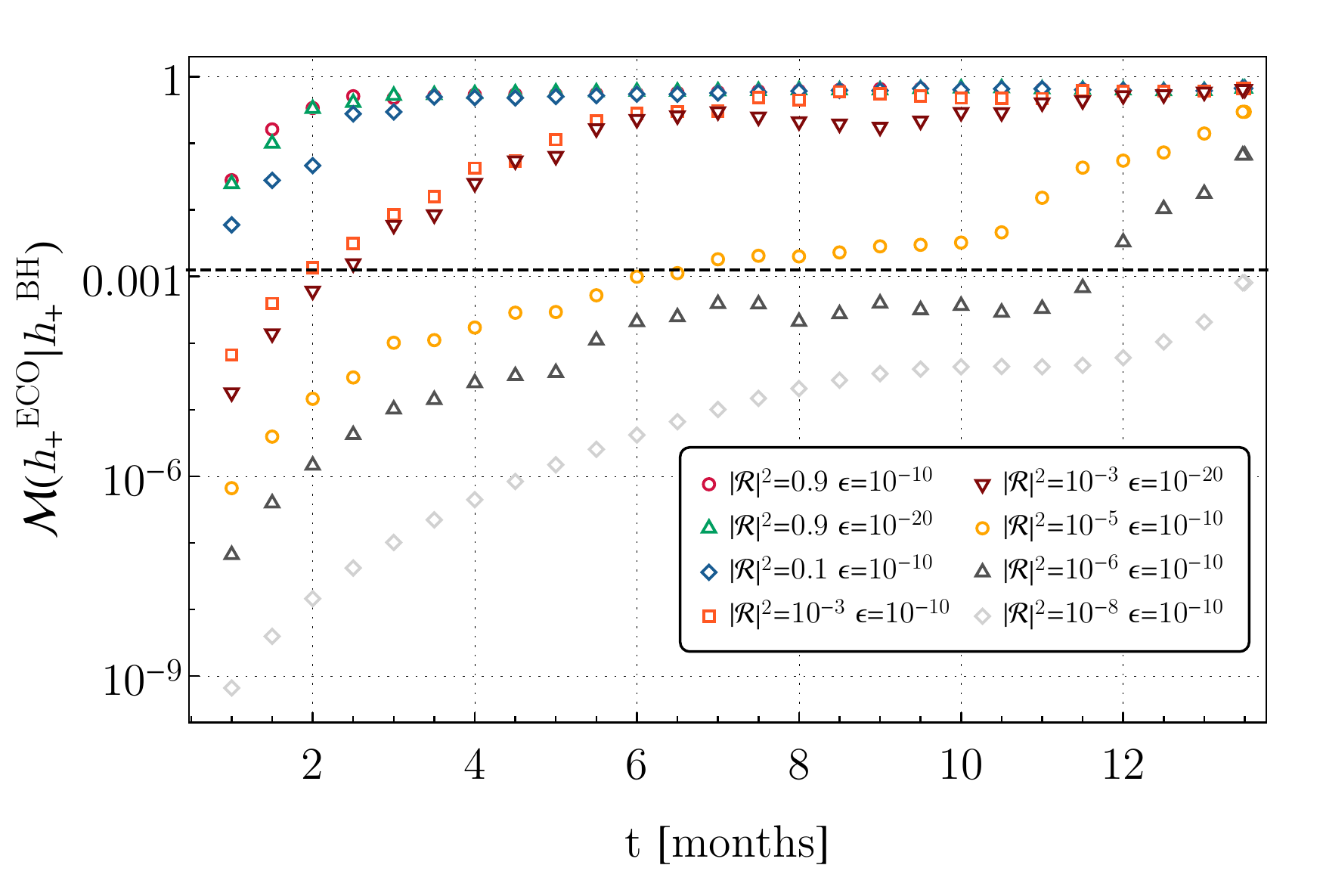}
\caption{Mismatch between the plus polarization of the waveforms with a central ECO and a central BH as a function of time, for a binary with mass ratio $q=3 \times 10^{-5}$ and a central object with spin $\chi=0.8$ and several values of the reflectivity and the compactness. The horizontal line is a reference value $\mathcal{M} \approx 10^{-3}$ for the resolvability by LISA.~\cite{Maggio:2021uge}} 
\label{fig:overlap}
\end{figure}

In Fig.~\ref{fig:overlap} we  show the mismatch between the waveforms in the ECO case and in the Kerr case with the same mass and spin for various values of the reflectivity and two choices of $\epsilon$. 
Let us notice that the compactness does not affect the mismatch significantly when $\epsilon\ll1$. Consistently with dephasing presented in Sec.~\ref{sec:dephasing}, the mismatch is larger in the consistent model than in the phenomenological approach of Ref.~\cite{Datta:2019epe}, especially at small reflectivity.

As a useful rule of thumb, two waveforms are indistinguishable for parameter estimation purposes if~\cite{Flanagan:1997kp,Lindblom:2008cm}
\begin{equation}
    \mathcal{M}\lesssim 1/(2\rho^2) \,,
\end{equation}
where $\rho$ is the SNR of the true signal. For an EMRI with $\rho\approx 20$ ($\rho\approx 100$) one has $\mathcal{M}\lesssim 10^{-3}$ ($\mathcal{M}\lesssim 
5\times 10^{-5}$). In Fig.~\ref{fig:overlap}, the more conservative threshold $\mathcal{M}= 10^{-3}$ is denoted with a 
dashed horizontal line. Exceeding this threshold is a necessary but not  sufficient condition  for a deviation to be detectable.
This level of mismatch is quickly exceeded after less than one year of data even for small values of the reflectivity. For example, for the fiducial case considered in Fig.~\ref{fig:overlap} ($\chi=0.8$, $M=10^6M_\odot$ and $\mu=30 M_\odot$), and assuming $\rho=20$, the threshold is exceeded after roughly one year unless
\begin{equation}
 |{\cal R}|^2 \lesssim 10^{-8}\,. \label{bound}
\end{equation}
This result is in agreement with the estimation based on the dephasing in Sec.~\ref{sec:dephasing}.
It shows that EMRIs could place stringent constraints on the reflectivity of supermassive compact objects at the remarkable level of $\mathcal{O}(10^{-6})\%$.
Let us notice that the bound in Eq.~\eqref{bound} is solely based on the mismatch calculation and does not take into account, e.g., correlations with the waveform parameters. Rigorous parameter estimation is necessary to derive an accurate projected upper bound in the case of no detection. 

\section{A case study: Boltzmann reflectivity} \label{sec:boltzmann}

%
\begin{figure*}[t]
\centering
\includegraphics[width=0.69\textwidth]{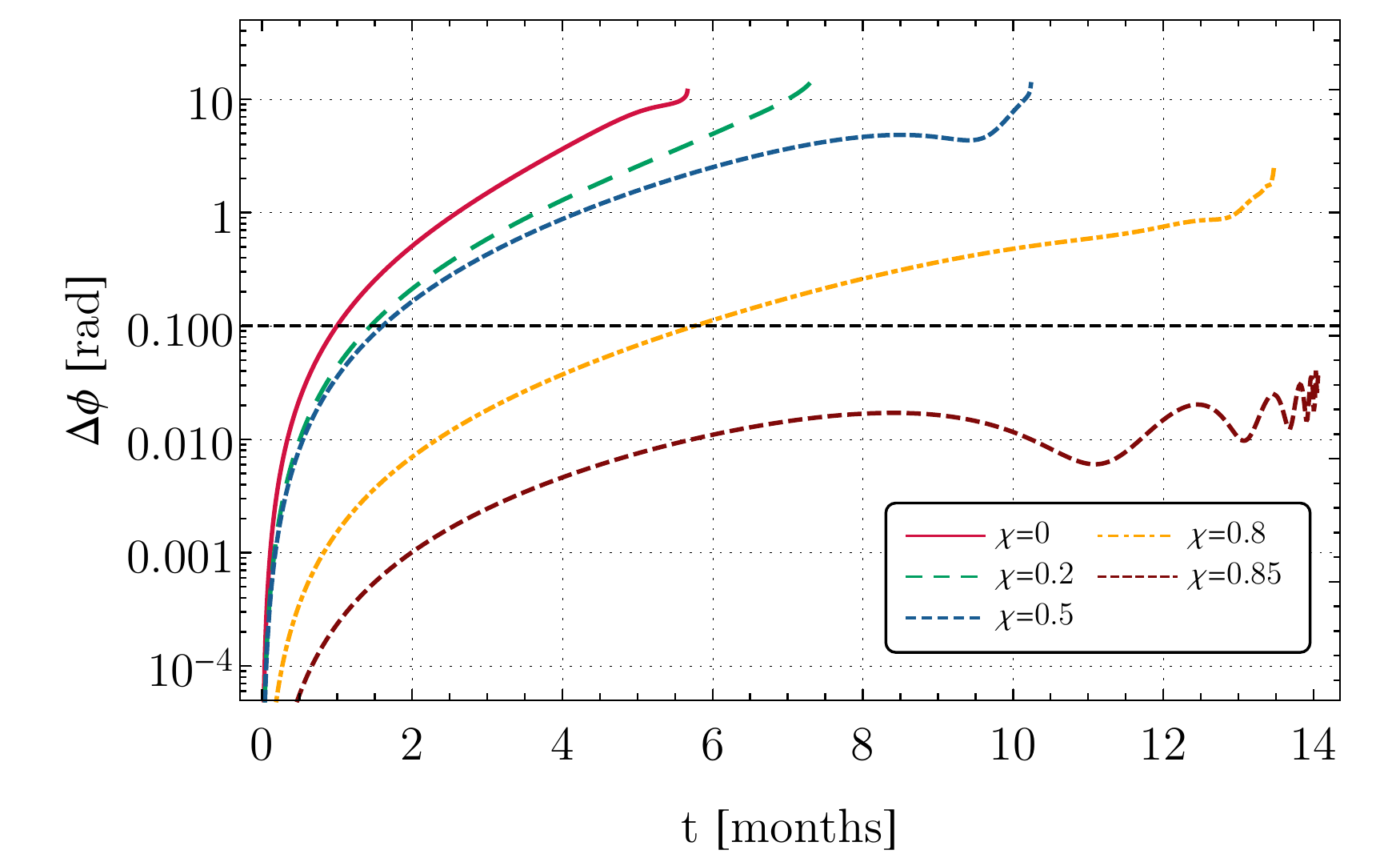}
\includegraphics[width=0.66\textwidth]{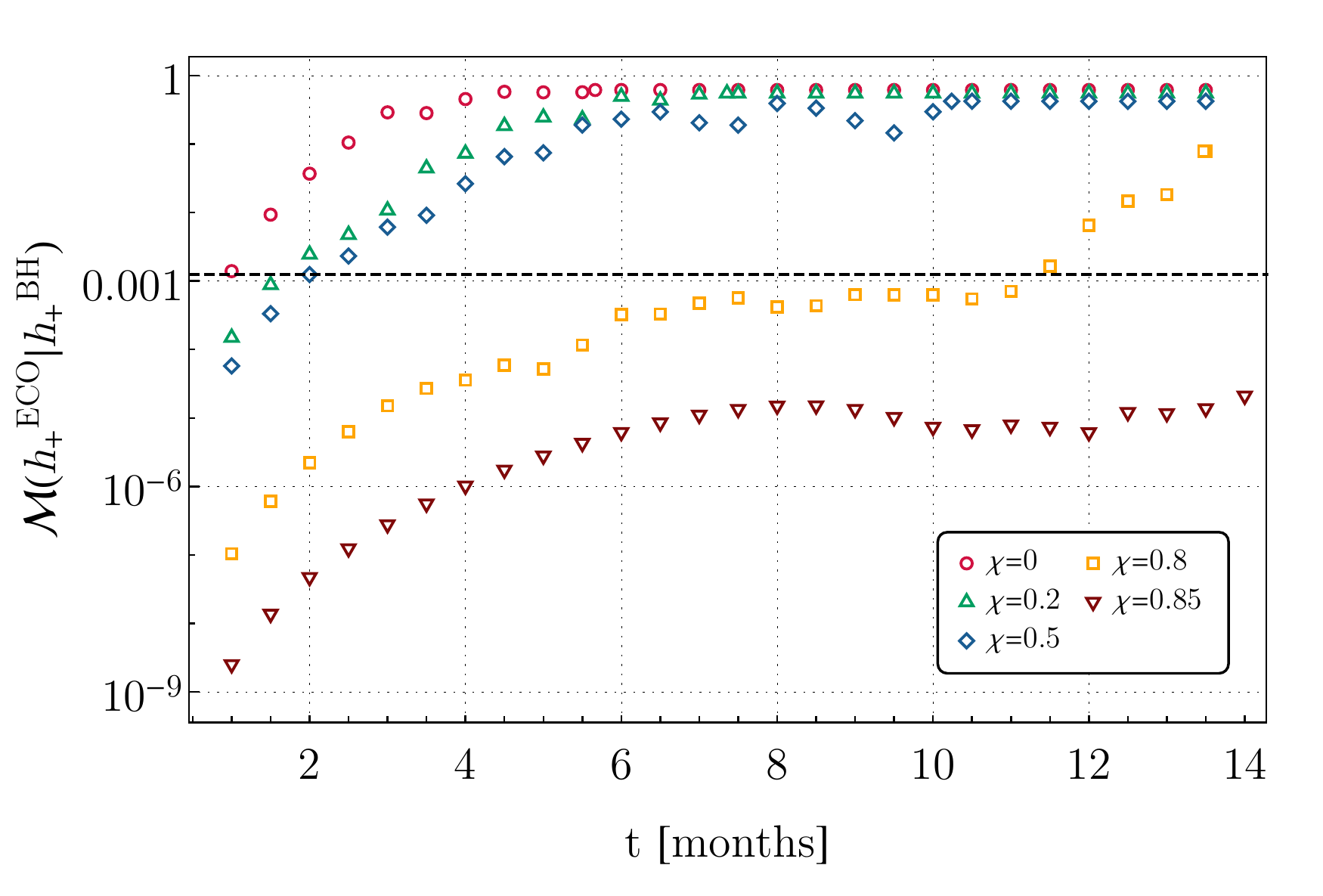}
\caption{Top panel: GW dephasing between the Kerr case and a quantum BH horizon with Boltzmann reflectivity [in Eq.~\eqref{Boltzmann}], where $q=3 \times 10^{-5}$, $\epsilon=10^{-10}$ as a function of time. Bottom panel: Mismatch between the plus polarization of the waveform with a quantum BH horizon with Boltzmann reflectivity and a standard BH as a function of time for several values of the spin.~\cite{Maggio:2021uge}}
\label{fig:Boltzmann}
\end{figure*}

The surface reflectivity of a horizonless compact object can be a complex function of the parameters of the model and the frequency. In this section, we consider a case study for the ECO reflectivity. In particular, we  analyze a model recently proposed to describe quantum BH horizons that gives rise to the ``Boltzmann'' reflectivity~\cite{Oshita:2019sat,Wang:2019rcf}
\begin{equation}
 \mathcal{R}(\omega) = e^{-\frac{|k|}{2 T_{\rm H}}} \,, \label{Boltzmann}
\end{equation}
where $T_{\rm H} = \frac{r_+ - r_-}{4 \pi (r_+^2 + a^2)}$ is the Hawking temperature of a Kerr BH. In this model, the reflectivity depends explicitly on the frequency and the spin of the compact object. The reflectivity in Eq.~\eqref{Boltzmann} provides sufficient absorption to quench the ergoregion instability and have a stable horizonless compact object with any spin~\cite{Oshita:2019sat}.
Let us notice that Eq.~\eqref{Boltzmann} can also contain a phase term, that depends on the specific model and the perturbation function on which the corresponding boundary condition is imposed~\cite{Oshita:2019sat,Wang:2019rcf,Xin:2021zir}. Recently, Refs.~\cite{Chen:2020htz,Xin:2021zir} proposed a model of the reflectivity that is related to the tidal response of the ECO to external curvature perturbations. In this model, the reflectivity contains extra terms that multiply the Boltzmann factor.
For simplicity, we neglect such phase terms, which would not affect our analysis significantly.

Figure~\ref{fig:Boltzmann} shows the dephasing (top panel) and the overlap (bottom panel) obtained in the Boltzmann reflectivity model as compared to the classical BH case.
An interesting feature of this model is that there is no free parameter that continuously connects the model to the classical Kerr case, so there is a concrete chance to rule it out with observations or to provide evidence for it. 
Interestingly, owing to its spin dependence, the Boltzmann reflectivity is much smaller at the relevant orbital frequencies when the central object is highly spinning. 
Consequently, the dephasing and the mismatch to the standard Kerr BH case are very small when $\chi\gtrsim0.8$ as shown in Fig.~\ref{fig:Boltzmann}. The oscillatory trend in the dephasing in the top panel is due to the contribution of high-frequency resonances appearing at late times.

\backmatter

\chapter{Conclusions and future prospects}

According to the BH paradigm, any compact object heavier than a few solar masses is described by the Kerr metric. 
In this thesis, we considered the possibility that the compact objects in the Universe are horizonless and singularity-free.
This hypothesis is supported by several models of ECOs that have been conceived in extensions of GR~\cite{Giudice:2016zpa,Cardoso:2019rvt}. ECOs are a tool that allows us to quantify the existence of horizons in astrophysical sources. 

In this thesis, we derived the characteristic frequencies of horizonless compact objects in the ringdown~\cite{Maggio:2017ivp,Maggio:2020jml}. We developed a model-independent framework relying on the membrane paradigm to quantify the deviations from the BH spectrum~\cite{Maggio:2020jml}. We assessed that current measurement accuracies impose a strong lower bound on the compactness of the merger remnant of $90\%$ of the BH compactness.

Spinning horizonless compact objects might be affected by an ergoregion instability when rotating sufficiently fast~\cite{Friedman:1978wla}.
In this thesis, we assessed the astrophysical viability of spinning horizonless compact objects under linear perturbations. From the analysis of the QNM frequencies, we determined the conditions for which horizonless compact objects are unstable~\cite{Maggio:2017ivp,Maggio:2018ivz}. Finally, we found a mechanism that allows for stable solutions, i.e., energy absorption within the object~\cite{Maggio:2017ivp}. We showed that a surface absorption of $60\%$ ensures the stability of horizonless compact objects for any spin~\cite{Maggio:2018ivz}.

In this thesis, we also explored the fingerprints of horizonless compact sources in the appearance of a modulated train of GW echoes at late times in the ringdown~\cite{Cardoso:2016rao}.
We provided an analytical template for the GW echoes that relates the key parameters of ECOs with the gravitational waveform~\cite{Maggio:2019zyv}. This template would allow us to constrain the parameters of exotic sources or perform model selection in the case of detection of GW echoes.

We also assessed how the future space-based interferometer LISA will allow us to perform tests of the BH paradigm, especially with new sources like EMRIs~\cite{Berti:2005ys}. 
We derived the gravitational waveform emitted by a stellar-mass object in orbital motion around a supermassive horizonless compact object. During the inspiral, we assessed the impact of extra resonances that would be excited when the orbital frequency matches the characteristic frequencies of the central ECO~\cite{Cardoso:2019nis,Maggio:2021uge}.
Finally, we estimated that EMRIs could potentially place the most stringent constraint on the reflectivity of supermassive compact objects at the remarkable level of $\mathcal{O}(10^{-6})\% $~\cite{Maggio:2021uge}.

A possible future research direction would be the development of data analysis techniques to infer the properties of compact sources from GW observations. Current constraints on the deviations from GR can be converted into constraints on the parameters of astrophysical sources. The mapping between the two descriptions needs to be developed and would be relevant to understand the nature of compact sources.

Moreover, full inspiral-merger-ringdown waveforms in various modified theories of gravity and alternative sources need to be developed. The accurate modeling of the gravitational waveform in alternative scenarios is crucial to look for new physics with current and future GW detections. 
The comparison of the echo templates obtained within perturbation theory with the postmerger signal of an ECO coalescence is an open problem. Numerical simulations of these systems are currently unavailable.

Furthermore, the extension of the membrane paradigm to spinning horizonless compact objects is left for future work. The membrane paradigm would allow us to describe several models of spinning ECOs with different interior solutions. The ECO phenomenology would be parametrized in terms of the properties of a fictitious rotating membrane located at the effective radius of the object.  

Concerning EMRIs detectable by LISA, a natural extension concerns the generalization to eccentric and inclined orbits. The bounds on the reflectivity of compact objects estimated in our work are based on the overlap calculation, and therefore neglect possible correlations among the waveform parameters. 
An interesting research line would be to perform accurate data analyses with exact waveforms either using the Fisher-information matrix or Bayesian inference.

\cleardoublepage
\phantomsection
\bibliographystyle{utphys}
\addcontentsline{toc}{chapter}{Bibliography}
\bibliography{bibliography}

\providecommand{\href}[2]{#2}\begingroup\raggedright\begin{thebibliography}{100}

\bibitem{LIGOScientific:2018mvr}
{\bfseries LIGO Scientific, Virgo} Collaboration, B.~P. Abbott {\em et~al.},
  ``{GWTC-1: A Gravitational-Wave Transient Catalog of Compact Binary Mergers
  Observed by LIGO and Virgo during the First and Second Observing Runs},''
  \href{http://dx.doi.org/10.1103/PhysRevX.9.031040}{{\em Phys. Rev.}
  {\bfseries X9} no.~3, (2019) 031040},
\href{http://arxiv.org/abs/1811.12907}{{\ttfamily arXiv:1811.12907
  [astro-ph.HE]}}.

\bibitem{LIGOScientific:2020ibl}
{\bfseries LIGO Scientific, Virgo} Collaboration, R.~Abbott {\em et~al.},
  ``{GWTC-2: Compact Binary Coalescences Observed by LIGO and Virgo During the
  First Half of the Third Observing Run},''
  \href{http://dx.doi.org/10.1103/PhysRevX.11.021053}{{\em Phys. Rev. X}
  {\bfseries 11} (2021) 021053},
  \href{http://arxiv.org/abs/2010.14527}{{\ttfamily arXiv:2010.14527 [gr-qc]}}.

\bibitem{LIGOScientific:2021usb}
{\bfseries LIGO Scientific, VIRGO} Collaboration, R.~Abbott {\em et~al.},
  ``{GWTC-2.1: Deep Extended Catalog of Compact Binary Coalescences Observed by
  LIGO and Virgo During the First Half of the Third Observing Run},''
  \href{http://arxiv.org/abs/2108.01045}{{\ttfamily arXiv:2108.01045 [gr-qc]}}.

\bibitem{LIGOScientific:2021djp}
{\bfseries LIGO Scientific, VIRGO, KAGRA} Collaboration, R.~Abbott {\em
  et~al.}, ``{GWTC-3: Compact Binary Coalescences Observed by LIGO and Virgo
  During the Second Part of the Third Observing Run},''
  \href{http://arxiv.org/abs/2111.03606}{{\ttfamily arXiv:2111.03606 [gr-qc]}}.

\bibitem{LIGOScientific:2016lio}
{\bfseries LIGO Scientific, Virgo} Collaboration, B.~P. Abbott {\em et~al.},
  ``{Tests of general relativity with GW150914},''
  \href{http://dx.doi.org/10.1103/PhysRevLett.116.221101}{{\em Phys. Rev.
  Lett.} {\bfseries 116} no.~22, (2016) 221101},
  \href{http://arxiv.org/abs/1602.03841}{{\ttfamily arXiv:1602.03841 [gr-qc]}}.
  [Erratum: Phys.Rev.Lett. 121, 129902 (2018)].

\bibitem{LIGOScientific:2017vwq}
{\bfseries LIGO Scientific, Virgo} Collaboration, B.~P. Abbott {\em et~al.},
  ``{GW170817: Observation of Gravitational Waves from a Binary Neutron Star
  Inspiral},'' \href{http://dx.doi.org/10.1103/PhysRevLett.119.161101}{{\em
  Phys. Rev. Lett.} {\bfseries 119} no.~16, (2017) 161101},
  \href{http://arxiv.org/abs/1710.05832}{{\ttfamily arXiv:1710.05832 [gr-qc]}}.

\bibitem{LIGOScientific:2017ync}
{\bfseries LIGO Scientific, Virgo, Fermi GBM, INTEGRAL, IceCube, AstroSat
  Cadmium Zinc Telluride Imager Team, IPN, Insight-Hxmt, ANTARES, Swift, AGILE
  Team, 1M2H Team, Dark Energy Camera GW-EM, DES, DLT40, GRAWITA, Fermi-LAT,
  ATCA, ASKAP, Las Cumbres Observatory Group, OzGrav, DWF (Deeper Wider Faster
  Program), AST3, CAASTRO, VINROUGE, MASTER, J-GEM, GROWTH, JAGWAR,
  CaltechNRAO, TTU-NRAO, NuSTAR, Pan-STARRS, MAXI Team, TZAC Consortium, KU,
  Nordic Optical Telescope, ePESSTO, GROND, Texas Tech University, SALT Group,
  TOROS, BOOTES, MWA, CALET, IKI-GW Follow-up, H.E.S.S., LOFAR, LWA, HAWC,
  Pierre Auger, ALMA, Euro VLBI Team, Pi of Sky, Chandra Team at McGill
  University, DFN, ATLAS Telescopes, High Time Resolution Universe Survey,
  RIMAS, RATIR, SKA South Africa/MeerKAT} Collaboration, B.~P. Abbott {\em
  et~al.}, ``{Multi-messenger Observations of a Binary Neutron Star Merger},''
  \href{http://dx.doi.org/10.3847/2041-8213/aa91c9}{{\em Astrophys. J. Lett.}
  {\bfseries 848} no.~2, (2017) L12},
  \href{http://arxiv.org/abs/1710.05833}{{\ttfamily arXiv:1710.05833
  [astro-ph.HE]}}.

\bibitem{LIGOScientific:2020iuh}
{\bfseries LIGO Scientific, Virgo} Collaboration, R.~Abbott {\em et~al.},
  ``{GW190521: A Binary Black Hole Merger with a Total Mass of $150
  M_{\odot}$},'' \href{http://dx.doi.org/10.1103/PhysRevLett.125.101102}{{\em
  Phys. Rev. Lett.} {\bfseries 125} no.~10, (2020) 101102},
  \href{http://arxiv.org/abs/2009.01075}{{\ttfamily arXiv:2009.01075 [gr-qc]}}.

\bibitem{LIGOScientific:2019fpa}
{\bfseries LIGO Scientific, Virgo} Collaboration, B.~Abbott {\em et~al.},
  ``{Tests of General Relativity with the Binary Black Hole Signals from the
  LIGO-Virgo Catalog GWTC-1},''
  \href{http://dx.doi.org/10.1103/PhysRevD.100.104036}{{\em Phys. Rev. D}
  {\bfseries 100} no.~10, (2019) 104036},
  \href{http://arxiv.org/abs/1903.04467}{{\ttfamily arXiv:1903.04467 [gr-qc]}}.

\bibitem{LIGOScientific:2020tif}
{\bfseries LIGO Scientific, Virgo} Collaboration, R.~Abbott {\em et~al.},
  ``{Tests of general relativity with binary black holes from the second
  LIGO-Virgo gravitational-wave transient catalog},''
  \href{http://dx.doi.org/10.1103/PhysRevD.103.122002}{{\em Phys. Rev. D}
  {\bfseries 103} no.~12, (2021) 122002},
  \href{http://arxiv.org/abs/2010.14529}{{\ttfamily arXiv:2010.14529 [gr-qc]}}.

\bibitem{LIGOScientific:2021sio}
{\bfseries LIGO Scientific, VIRGO, KAGRA} Collaboration, R.~Abbott {\em
  et~al.}, ``{Tests of General Relativity with GWTC-3},''
  \href{http://arxiv.org/abs/2112.06861}{{\ttfamily arXiv:2112.06861 [gr-qc]}}.

\bibitem{Giudice:2016zpa}
G.~F. Giudice, M.~McCullough, and A.~Urbano, ``{Hunting for Dark Particles with
  Gravitational Waves},''
  \href{http://dx.doi.org/10.1088/1475-7516/2016/10/001}{{\em JCAP} {\bfseries
  1610} no.~10, (2016) 001},
\href{http://arxiv.org/abs/1605.01209}{{\ttfamily arXiv:1605.01209 [hep-ph]}}.

\bibitem{Cardoso:2019rvt}
V.~Cardoso and P.~Pani, ``{Testing the nature of dark compact objects: a status
  report},'' \href{http://dx.doi.org/10.1007/s41114-019-0020-4}{{\em Living
  Rev. Rel.} {\bfseries 22} no.~1, (2019) 4},
\href{http://arxiv.org/abs/1904.05363}{{\ttfamily arXiv:1904.05363 [gr-qc]}}.

\bibitem{Mathur:2009hf}
S.~D. Mathur, ``{The Information paradox: A Pedagogical introduction},''
  \href{http://dx.doi.org/10.1088/0264-9381/26/22/224001}{{\em Class. Quant.
  Grav.} {\bfseries 26} (2009) 224001},
\href{http://arxiv.org/abs/0909.1038}{{\ttfamily arXiv:0909.1038 [hep-th]}}.

\bibitem{Abramowicz:2002vt}
M.~A. Abramowicz, W.~Kluzniak, and J.-P. Lasota, ``{No observational proof of
  the black hole event-horizon},''
  \href{http://dx.doi.org/10.1051/0004-6361:20021645}{{\em Astron.Astrophys.}
  {\bfseries 396} (2002) L31--L34},
\href{http://arxiv.org/abs/astro-ph/0207270}{{\ttfamily arXiv:astro-ph/0207270
  [astro-ph]}}.

\bibitem{EventHorizonTelescope:2019pgp}
{\bfseries Event Horizon Telescope} Collaboration, K.~Akiyama {\em et~al.},
  ``{First M87 Event Horizon Telescope Results. V. Physical Origin of the
  Asymmetric Ring},'' \href{http://dx.doi.org/10.3847/2041-8213/ab0f43}{{\em
  Astrophys. J. Lett.} {\bfseries 875} no.~1, (2019) L5},
  \href{http://arxiv.org/abs/1906.11242}{{\ttfamily arXiv:1906.11242
  [astro-ph.GA]}}.

\bibitem{Bustillo:2020syj}
J.~C. Bustillo, N.~Sanchis-Gual, A.~Torres-Forn\'e, J.~A. Font, A.~Vajpeyi,
  R.~Smith, C.~Herdeiro, E.~Radu, and S.~H.~W. Leong, ``{GW190521 as a Merger
  of Proca Stars: A Potential New Vector Boson of $8.7\times 10^{-13}$ eV},''
  \href{http://dx.doi.org/10.1103/PhysRevLett.126.081101}{{\em Phys. Rev.
  Lett.} {\bfseries 126} no.~8, (2021) 081101},
  \href{http://arxiv.org/abs/2009.05376}{{\ttfamily arXiv:2009.05376 [gr-qc]}}.

\bibitem{LIGOScientific:2020zkf}
{\bfseries LIGO Scientific, Virgo} Collaboration, R.~Abbott {\em et~al.},
  ``{GW190814: Gravitational Waves from the Coalescence of a 23 Solar Mass
  Black Hole with a 2.6 Solar Mass Compact Object},''
  \href{http://dx.doi.org/10.3847/2041-8213/ab960f}{{\em Astrophys. J. Lett.}
  {\bfseries 896} no.~2, (2020) L44},
  \href{http://arxiv.org/abs/2006.12611}{{\ttfamily arXiv:2006.12611
  [astro-ph.HE]}}.

\bibitem{Cardoso:2016rao}
V.~Cardoso, E.~Franzin, and P.~Pani, ``{Is the gravitational-wave ringdown a
  probe of the event horizon?},''
  \href{http://dx.doi.org/10.1103/PhysRevLett.116.171101}{{\em Phys. Rev.
  Lett.} {\bfseries 116} no.~17, (2016) 171101},
  \href{http://arxiv.org/abs/1602.07309}{{\ttfamily arXiv:1602.07309 [gr-qc]}}.
  [Erratum: Phys.Rev.Lett. 117, 089902 (2016)].

\bibitem{Carter:1971zc}
B.~Carter, ``{Axisymmetric Black Hole Has Only Two Degrees of Freedom},''
  \href{http://dx.doi.org/10.1103/PhysRevLett.26.331}{{\em Phys. Rev. Lett.}
  {\bfseries 26} (1971) 331--333}.

\bibitem{Robinson:1975bv}
D.~Robinson, ``{Uniqueness of the Kerr black hole},''
  \href{http://dx.doi.org/10.1103/PhysRevLett.34.905}{{\em Phys. Rev. Lett.}
  {\bfseries 34} (1975) 905--906}.

\bibitem{Maggio:2019zyv}
E.~Maggio, A.~Testa, S.~Bhagwat, and P.~Pani, ``{Analytical model for
  gravitational-wave echoes from spinning remnants},''
  \href{http://dx.doi.org/10.1103/PhysRevD.100.064056}{{\em Phys. Rev.}
  {\bfseries D100} no.~6, (2019) 064056},
\href{http://arxiv.org/abs/1907.03091}{{\ttfamily arXiv:1907.03091 [gr-qc]}}.

\bibitem{Abedi:2016hgu}
J.~Abedi, H.~Dykaar, and N.~Afshordi, ``{Echoes from the Abyss: Tentative
  evidence for Planck-scale structure at black hole horizons},''
  \href{http://dx.doi.org/10.1103/PhysRevD.96.082004}{{\em Phys. Rev.}
  {\bfseries D96} no.~8, (2017) 082004},
\href{http://arxiv.org/abs/1612.00266}{{\ttfamily arXiv:1612.00266 [gr-qc]}}.

\bibitem{Westerweck:2017hus}
J.~Westerweck, A.~Nielsen, O.~Fischer-Birnholtz, M.~Cabero, C.~Capano, T.~Dent,
  B.~Krishnan, G.~Meadors, and A.~H. Nitz, ``{Low significance of evidence for
  black hole echoes in gravitational wave data},''
  \href{http://dx.doi.org/10.1103/PhysRevD.97.124037}{{\em Phys. Rev.}
  {\bfseries D97} no.~12, (2018) 124037},
\href{http://arxiv.org/abs/1712.09966}{{\ttfamily arXiv:1712.09966 [gr-qc]}}.

\bibitem{Friedman:1978wla}
J.~L. Friedman, ``{Generic instability of rotating relativistic stars},''
\href{http://dx.doi.org/10.1007/BF01202527}{{\em Commun. Math. Phys.}
  {\bfseries 62} no.~3, (1978) 247--278}.

\bibitem{10.1093/mnras/282.2.580}
S.~Yoshida and Y.~Eriguchi, ``{Ergoregion instability revisited — a new and
  general method for numerical analysis of stability},''
  \href{http://dx.doi.org/10.1093/mnras/282.2.580}{{\em Monthly Notices of the
  Royal Astronomical Society} {\bfseries 282} no.~2, (09, 1996) 580--586}.

\bibitem{Kokkotas:2002sf}
K.~D. Kokkotas, J.~Ruoff, and N.~Andersson, ``{The w-mode instability of
  ultracompact relativistic stars},''
  \href{http://dx.doi.org/10.1103/PhysRevD.70.043003}{{\em Phys.Rev.}
  {\bfseries D70} (2004) 043003},
\href{http://arxiv.org/abs/astro-ph/0212429}{{\ttfamily arXiv:astro-ph/0212429
  [astro-ph]}}.

\bibitem{Cardoso:2014sna}
V.~Cardoso, L.~C.~B. Crispino, C.~F.~B. Macedo, H.~Okawa, and P.~Pani, ``{Light
  rings as observational evidence for event horizons: long-lived modes,
  ergoregions and nonlinear instabilities of ultracompact objects},''
  \href{http://dx.doi.org/10.1103/PhysRevD.90.044069}{{\em Phys.Rev.}
  {\bfseries D90} (2014) 044069},
\href{http://arxiv.org/abs/1406.5510}{{\ttfamily arXiv:1406.5510 [gr-qc]}}.

\bibitem{Brito:2015oca}
R.~Brito, V.~Cardoso, and P.~Pani,
  \href{http://dx.doi.org/10.1007/978-3-319-19000-6}{{\em {Superradiance}:
  {Energy Extraction, Black-Hole Bombs and Implications for Astrophysics and
  Particle Physics}}}, vol.~906.
\newblock Springer, 2015.
\newblock \href{http://arxiv.org/abs/1501.06570}{{\ttfamily arXiv:1501.06570
  [gr-qc]}}.

\bibitem{Maggio:2017ivp}
E.~Maggio, P.~Pani, and V.~Ferrari, ``{Exotic Compact Objects and How to Quench
  their Ergoregion Instability},''
  \href{http://dx.doi.org/10.1103/PhysRevD.96.104047}{{\em Phys. Rev.}
  {\bfseries D96} no.~10, (2017) 104047},
\href{http://arxiv.org/abs/1703.03696}{{\ttfamily arXiv:1703.03696 [gr-qc]}}.

\bibitem{Maggio:2018ivz}
E.~Maggio, V.~Cardoso, S.~R. Dolan, and P.~Pani, ``{Ergoregion instability of
  exotic compact objects: electromagnetic and gravitational perturbations and
  the role of absorption},''
  \href{http://dx.doi.org/10.1103/PhysRevD.99.064007}{{\em Phys. Rev.}
  {\bfseries D99} no.~6, (2019) 064007},
\href{http://arxiv.org/abs/1807.08840}{{\ttfamily arXiv:1807.08840 [gr-qc]}}.

\bibitem{Wang:2019rcf}
Q.~Wang, N.~Oshita, and N.~Afshordi, ``{Echoes from Quantum Black Holes},''
  \href{http://dx.doi.org/10.1103/PhysRevD.101.024031}{{\em Phys. Rev. D}
  {\bfseries 101} no.~2, (2020) 024031},
  \href{http://arxiv.org/abs/1905.00446}{{\ttfamily arXiv:1905.00446 [gr-qc]}}.

\bibitem{Punturo:2010zz}
M.~Punturo {\em et~al.}, ``{The Einstein Telescope: A third-generation
  gravitational wave observatory},''
\href{http://dx.doi.org/10.1088/0264-9381/27/19/194002}{{\em Class. Quant.
  Grav.} {\bfseries 27} (2010) 194002}.

\bibitem{Reitze:2019iox}
D.~Reitze {\em et~al.}, ``{Cosmic Explorer: The U.S. Contribution to
  Gravitational-Wave Astronomy beyond LIGO},'' {\em Bull. Am. Astron. Soc.}
  {\bfseries 51} no.~7, (2019) 035,
  \href{http://arxiv.org/abs/1907.04833}{{\ttfamily arXiv:1907.04833
  [astro-ph.IM]}}.

\bibitem{LISA:2017pwj}
{\bfseries LISA} Collaboration, P.~Amaro-Seoane {\em et~al.}, ``{Laser
  Interferometer Space Antenna},''
  \href{http://arxiv.org/abs/1702.00786}{{\ttfamily arXiv:1702.00786
  [astro-ph.IM]}}.

\bibitem{Gair:2017ynp}
J.~R. Gair, S.~Babak, A.~Sesana, P.~Amaro-Seoane, E.~Barausse, C.~P. Berry,
  E.~Berti, and C.~Sopuerta, ``{Prospects for observing extreme-mass-ratio
  inspirals with LISA},''
  \href{http://dx.doi.org/10.1088/1742-6596/840/1/012021}{{\em J. Phys. Conf.
  Ser.} {\bfseries 840} no.~1, (2017) 012021},
  \href{http://arxiv.org/abs/1704.00009}{{\ttfamily arXiv:1704.00009
  [astro-ph.GA]}}.

\bibitem{Cardoso:2019nis}
V.~Cardoso, A.~del Rio, and M.~Kimura, ``{Distinguishing black holes from
  horizonless objects through the excitation of resonances during inspiral},''
  \href{http://dx.doi.org/10.1103/PhysRevD.100.084046}{{\em Phys. Rev. D}
  {\bfseries 100} (2019) 084046},
  \href{http://arxiv.org/abs/1907.01561}{{\ttfamily arXiv:1907.01561 [gr-qc]}}.
  [Erratum: Phys.Rev.D 101, 069902 (2020)].

\bibitem{Maggio:2021uge}
E.~Maggio, M.~van~de Meent, and P.~Pani, ``{Extreme mass-ratio inspirals around
  a spinning horizonless compact object},''
  \href{http://arxiv.org/abs/2106.07195}{{\ttfamily arXiv:2106.07195 [gr-qc]}}.

\bibitem{Datta:2019epe}
S.~Datta, R.~Brito, S.~Bose, P.~Pani, and S.~A. Hughes, ``{Tidal heating as a
  discriminator for horizons in extreme mass ratio inspirals},''
  \href{http://dx.doi.org/10.1103/PhysRevD.101.044004}{{\em Phys. Rev.}
  {\bfseries D101} no.~4, (2020) 044004},
\href{http://arxiv.org/abs/1910.07841}{{\ttfamily arXiv:1910.07841 [gr-qc]}}.

\bibitem{LIGOScientific:2016aoc}
{\bfseries LIGO Scientific, Virgo} Collaboration, B.~P. Abbott {\em et~al.},
  ``{Observation of Gravitational Waves from a Binary Black Hole Merger},''
  \href{http://dx.doi.org/10.1103/PhysRevLett.116.061102}{{\em Phys. Rev.
  Lett.} {\bfseries 116} no.~6, (2016) 061102},
  \href{http://arxiv.org/abs/1602.03837}{{\ttfamily arXiv:1602.03837 [gr-qc]}}.

\bibitem{Abbott:2016xvh}
B.~P. Abbott {\em et~al.}, ``{Sensitivity of the Advanced LIGO detectors at the
  beginning of gravitational wave astronomy},''
  \href{http://dx.doi.org/10.1103/PhysRevD.93.112004}{{\em Phys. Rev. D}
  {\bfseries 93} no.~11, (2016) 112004},
  \href{http://arxiv.org/abs/1604.00439}{{\ttfamily arXiv:1604.00439
  [astro-ph.IM]}}. [Addendum: Phys.Rev.D 97, 059901 (2018)].

\bibitem{VIRGO:2014yos}
{\bfseries VIRGO} Collaboration, F.~Acernese {\em et~al.}, ``{Advanced Virgo: a
  second-generation interferometric gravitational wave detector},''
  \href{http://dx.doi.org/10.1088/0264-9381/32/2/024001}{{\em Class. Quant.
  Grav.} {\bfseries 32} no.~2, (2015) 024001},
  \href{http://arxiv.org/abs/1408.3978}{{\ttfamily arXiv:1408.3978 [gr-qc]}}.

\bibitem{LIGOScientific:2020stg}
{\bfseries LIGO Scientific, Virgo} Collaboration, R.~Abbott {\em et~al.},
  ``{GW190412: Observation of a Binary-Black-Hole Coalescence with Asymmetric
  Masses},'' \href{http://dx.doi.org/10.1103/PhysRevD.102.043015}{{\em Phys.
  Rev. D} {\bfseries 102} no.~4, (2020) 043015},
  \href{http://arxiv.org/abs/2004.08342}{{\ttfamily arXiv:2004.08342
  [astro-ph.HE]}}.

\bibitem{Barkat:1967zz}
Z.~Barkat, G.~Rakavy, and N.~Sack, ``{Dynamics of Supernova Explosion Resulting
  from Pair Formation},''
  \href{http://dx.doi.org/10.1103/PhysRevLett.18.379}{{\em Phys. Rev. Lett.}
  {\bfseries 18} (1967) 379--381}.

\bibitem{Ozel:2010su}
F.~Ozel, D.~Psaltis, R.~Narayan, and J.~E. McClintock, ``{The Black Hole Mass
  Distribution in the Galaxy},''
  \href{http://dx.doi.org/10.1088/0004-637X/725/2/1918}{{\em Astrophys. J.}
  {\bfseries 725} (2010) 1918--1927},
  \href{http://arxiv.org/abs/1006.2834}{{\ttfamily arXiv:1006.2834
  [astro-ph.GA]}}.

\bibitem{LIGOScientific:2021qlt}
{\bfseries LIGO Scientific, KAGRA, VIRGO} Collaboration, R.~Abbott {\em
  et~al.}, ``{Observation of Gravitational Waves from Two Neutron
  Star\textendash{}Black Hole Coalescences},''
  \href{http://dx.doi.org/10.3847/2041-8213/ac082e}{{\em Astrophys. J. Lett.}
  {\bfseries 915} no.~1, (2021) L5},
  \href{http://arxiv.org/abs/2106.15163}{{\ttfamily arXiv:2106.15163
  [astro-ph.HE]}}.

\bibitem{Blanchet:2013haa}
L.~Blanchet, ``{Gravitational Radiation from Post-Newtonian Sources and
  Inspiralling Compact Binaries},''
  \href{http://dx.doi.org/10.12942/lrr-2014-2}{{\em Living Rev. Rel.}
  {\bfseries 17} (2014) 2}, \href{http://arxiv.org/abs/1310.1528}{{\ttfamily
  arXiv:1310.1528 [gr-qc]}}.

\bibitem{Blanchet:1995ez}
L.~Blanchet, T.~Damour, B.~R. Iyer, C.~M. Will, and A.~G. Wiseman,
  ``{Gravitational radiation damping of compact binary systems to second
  postNewtonian order},''
  \href{http://dx.doi.org/10.1103/PhysRevLett.74.3515}{{\em Phys. Rev. Lett.}
  {\bfseries 74} (1995) 3515--3518},
  \href{http://arxiv.org/abs/gr-qc/9501027}{{\ttfamily arXiv:gr-qc/9501027}}.

\bibitem{Kidder:1995zr}
L.~E. Kidder, ``{Coalescing binary systems of compact objects to postNewtonian
  5/2 order. 5. Spin effects},''
  \href{http://dx.doi.org/10.1103/PhysRevD.52.821}{{\em Phys. Rev. D}
  {\bfseries 52} (1995) 821--847},
  \href{http://arxiv.org/abs/gr-qc/9506022}{{\ttfamily arXiv:gr-qc/9506022}}.

\bibitem{Damour:2001bu}
T.~Damour, P.~Jaranowski, and G.~Schaefer, ``{Dimensional regularization of the
  gravitational interaction of point masses},''
  \href{http://dx.doi.org/10.1016/S0370-2693(01)00642-6}{{\em Phys. Lett. B}
  {\bfseries 513} (2001) 147--155},
  \href{http://arxiv.org/abs/gr-qc/0105038}{{\ttfamily arXiv:gr-qc/0105038}}.

\bibitem{Arun:2008kb}
K.~G. Arun, A.~Buonanno, G.~Faye, and E.~Ochsner, ``{Higher-order spin effects
  in the amplitude and phase of gravitational waveforms emitted by inspiraling
  compact binaries: Ready-to-use gravitational waveforms},''
  \href{http://dx.doi.org/10.1103/PhysRevD.79.104023}{{\em Phys. Rev. D}
  {\bfseries 79} (2009) 104023},
  \href{http://arxiv.org/abs/0810.5336}{{\ttfamily arXiv:0810.5336 [gr-qc]}}.
  [Erratum: Phys.Rev.D 84, 049901 (2011)].

\bibitem{Teukolsky:1973ha}
S.~A. Teukolsky, ``{Perturbations of a rotating black hole. 1. Fundamental
  equations for gravitational electromagnetic and neutrino field
  perturbations},''
\href{http://dx.doi.org/10.1086/152444}{{\em Astrophysical Journal} {\bfseries
  185} (1973) 635--647}.

\bibitem{Press:1973zz}
W.~H. Press and S.~A. Teukolsky, ``{Perturbations of a Rotating Black Hole. II.
  Dynamical Stability of the Kerr Metric},''
\href{http://dx.doi.org/10.1086/152445}{{\em Astrophysical Journal} {\bfseries
  185} (1973) 649--674}.

\bibitem{Teukolsky:1974yv}
S.~Teukolsky and W.~Press, ``{Perturbations of a rotating black hole. III -
  Interaction of the hole with gravitational and electromagnet ic radiation},''
\href{http://dx.doi.org/10.1086/153180}{{\em Astrophysical Journal} {\bfseries
  193} (1974) 443--461}.

\bibitem{Chandrasekhar:1975zza}
S.~Chandrasekhar and S.~L. Detweiler, ``{The quasi-normal modes of the
  Schwarzschild black hole},''
{\em Proc.Roy.Soc.Lond.} {\bfseries A344} (1975) 441--452.

\bibitem{Leaver:1985ax}
E.~Leaver, ``{An Analytic representation for the quasi normal modes of Kerr
  black holes},''
{\em Proc.Roy.Soc.Lond.} {\bfseries A402} (1985) 285--298.

\bibitem{Kokkotas:1999bd}
K.~D. Kokkotas and B.~G. Schmidt, ``{Quasinormal modes of stars and black
  holes},'' {\em Living Rev.Rel.} {\bfseries 2} (1999) 2,
\href{http://arxiv.org/abs/gr-qc/9909058}{{\ttfamily arXiv:gr-qc/9909058
  [gr-qc]}}.

\bibitem{Berti:2009kk}
E.~Berti, V.~Cardoso, and A.~O. Starinets, ``{Quasinormal modes of black holes
  and black branes},''
  \href{http://dx.doi.org/10.1088/0264-9381/26/16/163001}{{\em
  Class.Quant.Grav.} {\bfseries 26} (2009) 163001},
\href{http://arxiv.org/abs/0905.2975}{{\ttfamily arXiv:0905.2975 [gr-qc]}}.

\bibitem{Ghosh:2016qgn}
A.~Ghosh {\em et~al.}, ``{Testing general relativity using golden black-hole
  binaries},'' \href{http://dx.doi.org/10.1103/PhysRevD.94.021101}{{\em Phys.
  Rev.} {\bfseries D94} no.~2, (2016) 021101},
\href{http://arxiv.org/abs/1602.02453}{{\ttfamily arXiv:1602.02453 [gr-qc]}}.

\bibitem{Li:2011cg}
T.~Li, W.~Del~Pozzo, S.~Vitale, C.~Van Den~Broeck, M.~Agathos, J.~Veitch,
  K.~Grover, T.~Sidery, R.~Sturani, and A.~Vecchio, ``{Towards a generic test
  of the strong field dynamics of general relativity using compact binary
  coalescence},'' \href{http://dx.doi.org/10.1103/PhysRevD.85.082003}{{\em
  Phys. Rev. D} {\bfseries 85} (2012) 082003},
  \href{http://arxiv.org/abs/1110.0530}{{\ttfamily arXiv:1110.0530 [gr-qc]}}.

\bibitem{Agathos:2013upa}
M.~Agathos, W.~Del~Pozzo, T.~G.~F. Li, C.~Van Den~Broeck, J.~Veitch, and
  S.~Vitale, ``{TIGER: A data analysis pipeline for testing the strong-field
  dynamics of general relativity with gravitational wave signals from
  coalescing compact binaries},''
  \href{http://dx.doi.org/10.1103/PhysRevD.89.082001}{{\em Phys. Rev. D}
  {\bfseries 89} no.~8, (2014) 082001},
  \href{http://arxiv.org/abs/1311.0420}{{\ttfamily arXiv:1311.0420 [gr-qc]}}.

\bibitem{Wex:2014nva}
N.~Wex, ``{Testing Relativistic Gravity with Radio Pulsars},''
  \href{http://arxiv.org/abs/1402.5594}{{\ttfamily arXiv:1402.5594 [gr-qc]}}.

\bibitem{Yunes:2009ke}
N.~Yunes and F.~Pretorius, ``{Fundamental Theoretical Bias in Gravitational
  Wave Astrophysics and the Parameterized Post-Einsteinian Framework},''
  \href{http://dx.doi.org/10.1103/PhysRevD.80.122003}{{\em Phys.Rev.}
  {\bfseries D80} (2009) 122003},
\href{http://arxiv.org/abs/0909.3328}{{\ttfamily arXiv:0909.3328 [gr-qc]}}.

\bibitem{Yunes:2016jcc}
N.~Yunes, K.~Yagi, and F.~Pretorius, ``{Theoretical Physics Implications of the
  Binary Black-Hole Mergers GW150914 and GW151226},''
  \href{http://dx.doi.org/10.1103/PhysRevD.94.084002}{{\em Phys. Rev.}
  {\bfseries D94} no.~8, (2016) 084002},
\href{http://arxiv.org/abs/1603.08955}{{\ttfamily arXiv:1603.08955 [gr-qc]}}.

\bibitem{Poisson:1997ha}
E.~Poisson, ``{Gravitational waves from inspiraling compact binaries: The
  Quadrupole moment term},''
  \href{http://dx.doi.org/10.1103/PhysRevD.57.5287}{{\em Phys. Rev. D}
  {\bfseries 57} (1998) 5287--5290},
  \href{http://arxiv.org/abs/gr-qc/9709032}{{\ttfamily arXiv:gr-qc/9709032}}.

\bibitem{Hansen:1974zz}
R.~O. Hansen, ``{Multipole moments of stationary space-times},''
  \href{http://dx.doi.org/10.1063/1.1666501}{{\em J. Math. Phys.} {\bfseries
  15} (1974) 46--52}.

\bibitem{Pappas:2012ns}
G.~Pappas and T.~A. Apostolatos, ``{Revising the multipole moments of numerical
  spacetimes, and its consequences},''
  \href{http://dx.doi.org/10.1103/PhysRevLett.108.231104}{{\em Phys. Rev.
  Lett.} {\bfseries 108} (2012) 231104},
  \href{http://arxiv.org/abs/1201.6067}{{\ttfamily arXiv:1201.6067 [gr-qc]}}.

\bibitem{Harry:2018hke}
I.~Harry and T.~Hinderer, ``{Observing and measuring the neutron-star
  equation-of-state in spinning binary neutron star systems},''
  \href{http://dx.doi.org/10.1088/1361-6382/aac7e3}{{\em Class. Quant. Grav.}
  {\bfseries 35} no.~14, (2018) 145010},
  \href{http://arxiv.org/abs/1801.09972}{{\ttfamily arXiv:1801.09972 [gr-qc]}}.

\bibitem{Ryan:1996nk}
F.~D. Ryan, ``{Spinning boson stars with large selfinteraction},''
\href{http://dx.doi.org/10.1103/PhysRevD.55.6081}{{\em Phys.Rev.} {\bfseries
  D55} (1997) 6081--6091}.

\bibitem{Herdeiro:2014goa}
C.~A.~R. Herdeiro and E.~Radu, ``{Kerr black holes with scalar hair},''
  \href{http://dx.doi.org/10.1103/PhysRevLett.112.221101}{{\em Phys.Rev.Lett.}
  {\bfseries 112} (2014) 221101},
\href{http://arxiv.org/abs/1403.2757}{{\ttfamily arXiv:1403.2757 [gr-qc]}}.

\bibitem{Krishnendu:2017shb}
N.~V. Krishnendu, K.~G. Arun, and C.~K. Mishra, ``{Testing the binary black
  hole nature of a compact binary coalescence},''
  \href{http://dx.doi.org/10.1103/PhysRevLett.119.091101}{{\em Phys. Rev.
  Lett.} {\bfseries 119} no.~9, (2017) 091101},
\href{http://arxiv.org/abs/1701.06318}{{\ttfamily arXiv:1701.06318 [gr-qc]}}.

\bibitem{Dreyer:2003bv}
O.~Dreyer, B.~J. Kelly, B.~Krishnan, L.~S. Finn, D.~Garrison, and
  R.~Lopez-Aleman, ``{Black hole spectroscopy: Testing general relativity
  through gravitational wave observations},''
  \href{http://dx.doi.org/10.1088/0264-9381/21/4/003}{{\em Class. Quant. Grav.}
  {\bfseries 21} (2004) 787--804},
  \href{http://arxiv.org/abs/gr-qc/0309007}{{\ttfamily arXiv:gr-qc/0309007}}.

\bibitem{Gossan:2011ha}
S.~Gossan, J.~Veitch, and B.~Sathyaprakash, ``{Bayesian model selection for
  testing the no-hair theorem with black hole ringdowns},''
  \href{http://dx.doi.org/10.1103/PhysRevD.85.124056}{{\em Phys. Rev. D}
  {\bfseries 85} (2012) 124056},
  \href{http://arxiv.org/abs/1111.5819}{{\ttfamily arXiv:1111.5819 [gr-qc]}}.

\bibitem{Brito:2018rfr}
R.~Brito, A.~Buonanno, and V.~Raymond, ``{Black-hole Spectroscopy by Making
  Full Use of Gravitational-Wave Modeling},''
  \href{http://dx.doi.org/10.1103/PhysRevD.98.084038}{{\em Phys. Rev. D}
  {\bfseries 98} no.~8, (2018) 084038},
  \href{http://arxiv.org/abs/1805.00293}{{\ttfamily arXiv:1805.00293 [gr-qc]}}.

\bibitem{Carullo:2019flw}
G.~Carullo, W.~Del~Pozzo, and J.~Veitch, ``{Observational Black Hole
  Spectroscopy: A time-domain multimode analysis of GW150914},''
  \href{http://dx.doi.org/10.1103/PhysRevD.99.123029}{{\em Phys. Rev. D}
  {\bfseries 99} no.~12, (2019) 123029},
  \href{http://arxiv.org/abs/1902.07527}{{\ttfamily arXiv:1902.07527 [gr-qc]}}.
  [Erratum: Phys.Rev.D 100, 089903 (2019)].

\bibitem{Isi:2019aib}
M.~Isi, M.~Giesler, W.~M. Farr, M.~A. Scheel, and S.~A. Teukolsky, ``{Testing
  the no-hair theorem with GW150914},''
  \href{http://dx.doi.org/10.1103/PhysRevLett.123.111102}{{\em Phys. Rev.
  Lett.} {\bfseries 123} no.~11, (2019) 111102},
  \href{http://arxiv.org/abs/1905.00869}{{\ttfamily arXiv:1905.00869 [gr-qc]}}.

\bibitem{Bhagwat:2019dtm}
S.~Bhagwat, X.~J. Forteza, P.~Pani, and V.~Ferrari, ``{Ringdown overtones,
  black hole spectroscopy, and no-hair theorem tests},''
  \href{http://dx.doi.org/10.1103/PhysRevD.101.044033}{{\em Phys. Rev. D}
  {\bfseries 101} no.~4, (2020) 044033},
  \href{http://arxiv.org/abs/1910.08708}{{\ttfamily arXiv:1910.08708 [gr-qc]}}.

\bibitem{Capano:2021etf}
C.~D. Capano, M.~Cabero, J.~Westerweck, J.~Abedi, S.~Kastha, A.~H. Nitz, A.~B.
  Nielsen, and B.~Krishnan, ``{Observation of a multimode quasi-normal spectrum
  from a perturbed black hole},''
  \href{http://arxiv.org/abs/2105.05238}{{\ttfamily arXiv:2105.05238 [gr-qc]}}.

\bibitem{Ghosh:2021mrv}
A.~Ghosh, R.~Brito, and A.~Buonanno, ``{Constraints on quasinormal-mode
  frequencies with LIGO-Virgo binary\textendash{}black-hole observations},''
  \href{http://dx.doi.org/10.1103/PhysRevD.103.124041}{{\em Phys. Rev. D}
  {\bfseries 103} no.~12, (2021) 124041},
  \href{http://arxiv.org/abs/2104.01906}{{\ttfamily arXiv:2104.01906 [gr-qc]}}.

\bibitem{Abedi:2020ujo}
J.~Abedi, N.~Afshordi, N.~Oshita, and Q.~Wang, ``{Quantum Black Holes in the
  Sky},'' \href{http://dx.doi.org/10.3390/universe6030043}{{\em Universe}
  {\bfseries 6} no.~3, (2020) 43},
  \href{http://arxiv.org/abs/2001.09553}{{\ttfamily arXiv:2001.09553 [gr-qc]}}.

\bibitem{Nakano:2017fvh}
H.~Nakano, N.~Sago, H.~Tagoshi, and T.~Tanaka, ``{Black hole ringdown echoes
  and howls},'' \href{http://dx.doi.org/10.1093/ptep/ptx093}{{\em PTEP}
  {\bfseries 2017} no.~7, (2017) 071E01},
\href{http://arxiv.org/abs/1704.07175}{{\ttfamily arXiv:1704.07175 [gr-qc]}}.

\bibitem{Wang:2018gin}
Q.~Wang and N.~Afshordi, ``{Black hole echology: The observer\textquoteright{}s
  manual},'' \href{http://dx.doi.org/10.1103/PhysRevD.97.124044}{{\em Phys.
  Rev. D} {\bfseries 97} no.~12, (2018) 124044},
  \href{http://arxiv.org/abs/1803.02845}{{\ttfamily arXiv:1803.02845 [gr-qc]}}.

\bibitem{Maselli:2017tfq}
A.~Maselli, S.~H. V\"olkel, and K.~D. Kokkotas, ``{Parameter estimation of
  gravitational wave echoes from exotic compact objects},''
  \href{http://dx.doi.org/10.1103/PhysRevD.96.064045}{{\em Phys. Rev. D}
  {\bfseries 96} no.~6, (2017) 064045},
  \href{http://arxiv.org/abs/1708.02217}{{\ttfamily arXiv:1708.02217 [gr-qc]}}.

\bibitem{Mark:2017dnq}
Z.~Mark, A.~Zimmerman, S.~M. Du, and Y.~Chen, ``{A recipe for echoes from
  exotic compact objects},''
  \href{http://dx.doi.org/10.1103/PhysRevD.96.084002}{{\em Phys. Rev.}
  {\bfseries D96} no.~8, (2017) 084002},
\href{http://arxiv.org/abs/1706.06155}{{\ttfamily arXiv:1706.06155 [gr-qc]}}.

\bibitem{Testa:2018bzd}
A.~Testa and P.~Pani, ``{Analytical template for gravitational-wave echoes:
  signal characterization and prospects of detection with current and future
  interferometers},'' \href{http://dx.doi.org/10.1103/PhysRevD.98.044018}{{\em
  Phys. Rev.} {\bfseries D98} no.~4, (2018) 044018},
\href{http://arxiv.org/abs/1806.04253}{{\ttfamily arXiv:1806.04253 [gr-qc]}}.

\bibitem{Tsang:2018uie}
K.~W. Tsang, M.~Rollier, A.~Ghosh, A.~Samajdar, M.~Agathos, K.~Chatziioannou,
  V.~Cardoso, G.~Khanna, and C.~Van Den~Broeck, ``{A morphology-independent
  data analysis method for detecting and characterizing gravitational wave
  echoes},'' \href{http://dx.doi.org/10.1103/PhysRevD.98.024023}{{\em Phys.
  Rev.} {\bfseries D98} no.~2, (2018) 024023},
\href{http://arxiv.org/abs/1804.04877}{{\ttfamily arXiv:1804.04877 [gr-qc]}}.

\bibitem{Tsang:2019zra}
K.~W. Tsang, A.~Ghosh, A.~Samajdar, K.~Chatziioannou, S.~Mastrogiovanni,
  M.~Agathos, and C.~Van Den~Broeck, ``{A morphology-independent search for
  gravitational wave echoes in data from the first and second observing runs of
  Advanced LIGO and Advanced Virgo},''
  \href{http://dx.doi.org/10.1103/PhysRevD.101.064012}{{\em Phys. Rev. D}
  {\bfseries 101} no.~6, (2020) 064012},
  \href{http://arxiv.org/abs/1906.11168}{{\ttfamily arXiv:1906.11168 [gr-qc]}}.

\bibitem{Conklin:2017lwb}
R.~S. Conklin, B.~Holdom, and J.~Ren, ``{Gravitational wave echoes through new
  windows},'' \href{http://dx.doi.org/10.1103/PhysRevD.98.044021}{{\em Phys.
  Rev.} {\bfseries D98} no.~4, (2018) 044021},
\href{http://arxiv.org/abs/1712.06517}{{\ttfamily arXiv:1712.06517 [gr-qc]}}.

\bibitem{Conklin:2019fcs}
R.~S. Conklin and B.~Holdom, ``{Gravitational wave echo spectra},''
  \href{http://dx.doi.org/10.1103/PhysRevD.100.124030}{{\em Phys. Rev.}
  {\bfseries D100} no.~12, (2019) 124030},
\href{http://arxiv.org/abs/1905.09370}{{\ttfamily arXiv:1905.09370 [gr-qc]}}.

\bibitem{Abedi:2018npz}
J.~Abedi and N.~Afshordi, ``{Echoes from the Abyss: A highly spinning black
  hole remnant for the binary neutron star merger GW170817},''
  \href{http://dx.doi.org/10.1088/1475-7516/2019/11/010}{{\em JCAP} {\bfseries
  11} (2019) 010}, \href{http://arxiv.org/abs/1803.10454}{{\ttfamily
  arXiv:1803.10454 [gr-qc]}}.

\bibitem{Nielsen:2018lkf}
A.~B. Nielsen, C.~D. Capano, O.~Birnholtz, and J.~Westerweck, ``{Parameter
  estimation and statistical significance of echoes following black hole
  signals in the first Advanced LIGO observing run},''
  \href{http://dx.doi.org/10.1103/PhysRevD.99.104012}{{\em Phys. Rev.}
  {\bfseries D99} no.~10, (2019) 104012},
\href{http://arxiv.org/abs/1811.04904}{{\ttfamily arXiv:1811.04904 [gr-qc]}}.

\bibitem{Uchikata:2019frs}
N.~Uchikata, H.~Nakano, T.~Narikawa, N.~Sago, H.~Tagoshi, and T.~Tanaka,
  ``{Searching for black hole echoes from the LIGO-Virgo Catalog GWTC-1},''
  \href{http://dx.doi.org/10.1103/PhysRevD.100.062006}{{\em Phys. Rev.}
  {\bfseries D100} no.~6, (2019) 062006},
\href{http://arxiv.org/abs/1906.00838}{{\ttfamily arXiv:1906.00838 [gr-qc]}}.

\bibitem{Lo:2018sep}
R.~K.~L. Lo, T.~G.~F. Li, and A.~J. Weinstein, ``{Template-based
  Gravitational-Wave Echoes Search Using Bayesian Model Selection},''
  \href{http://dx.doi.org/10.1103/PhysRevD.99.084052}{{\em Phys. Rev.}
  {\bfseries D99} no.~8, (2019) 084052},
\href{http://arxiv.org/abs/1811.07431}{{\ttfamily arXiv:1811.07431 [gr-qc]}}.

\bibitem{Hild:2010id}
S.~Hild {\em et~al.}, ``{Sensitivity Studies for Third-Generation Gravitational
  Wave Observatories},''
  \href{http://dx.doi.org/10.1088/0264-9381/28/9/094013}{{\em Class. Quant.
  Grav.} {\bfseries 28} (2011) 094013},
  \href{http://arxiv.org/abs/1012.0908}{{\ttfamily arXiv:1012.0908 [gr-qc]}}.

\bibitem{Klein:2015hvg}
A.~Klein {\em et~al.}, ``{Science with the space-based interferometer eLISA:
  Supermassive black hole binaries},''
  \href{http://dx.doi.org/10.1103/PhysRevD.93.024003}{{\em Phys. Rev.}
  {\bfseries D93} no.~2, (2016) 024003},
\href{http://arxiv.org/abs/1511.05581}{{\ttfamily arXiv:1511.05581 [gr-qc]}}.

\bibitem{Barack:2006pq}
L.~Barack and C.~Cutler, ``{Using LISA EMRI sources to test off-Kerr deviations
  in the geometry of massive black holes},''
  \href{http://dx.doi.org/10.1103/PhysRevD.75.042003}{{\em Phys. Rev. D}
  {\bfseries 75} (2007) 042003},
  \href{http://arxiv.org/abs/gr-qc/0612029}{{\ttfamily arXiv:gr-qc/0612029}}.

\bibitem{Glampedakis:2005cf}
K.~Glampedakis and S.~Babak, ``{Mapping spacetimes with LISA: Inspiral of a
  test-body in a `quasi-Kerr' field},''
  \href{http://dx.doi.org/10.1088/0264-9381/23/12/013}{{\em Class. Quant.
  Grav.} {\bfseries 23} (2006) 4167--4188},
  \href{http://arxiv.org/abs/gr-qc/0510057}{{\ttfamily arXiv:gr-qc/0510057}}.

\bibitem{Ghez:2008ms}
A.~M. Ghez {\em et~al.}, ``{Measuring Distance and Properties of the Milky
  Way's Central Supermassive Black Hole with Stellar Orbits},''
  \href{http://dx.doi.org/10.1086/592738}{{\em Astrophys. J.} {\bfseries 689}
  (2008) 1044--1062}, \href{http://arxiv.org/abs/0808.2870}{{\ttfamily
  arXiv:0808.2870 [astro-ph]}}.

\bibitem{EventHorizonTelescope:2019dse}
{\bfseries Event Horizon Telescope} Collaboration, K.~Akiyama {\em et~al.},
  ``{First M87 Event Horizon Telescope Results. I. The Shadow of the
  Supermassive Black Hole},''
  \href{http://dx.doi.org/10.3847/2041-8213/ab0ec7}{{\em Astrophys. J. Lett.}
  {\bfseries 875} (2019) L1}, \href{http://arxiv.org/abs/1906.11238}{{\ttfamily
  arXiv:1906.11238 [astro-ph.GA]}}.

\bibitem{Mazur:2001fv}
P.~O. Mazur and E.~Mottola, ``{Gravitational condensate stars: An alternative
  to black holes},''
\href{http://arxiv.org/abs/gr-qc/0109035}{{\ttfamily arXiv:gr-qc/0109035
  [gr-qc]}}.

\bibitem{Mathur:2005zp}
S.~D. Mathur, ``{The Fuzzball proposal for black holes: An Elementary
  review},'' \href{http://dx.doi.org/10.1002/prop.200410203}{{\em Fortsch.
  Phys.} {\bfseries 53} (2005) 793--827},
\href{http://arxiv.org/abs/hep-th/0502050}{{\ttfamily arXiv:hep-th/0502050
  [hep-th]}}.

\bibitem{Hawking:1974rv}
S.~W. Hawking, ``{Black hole explosions},''
  \href{http://dx.doi.org/10.1038/248030a0}{{\em Nature} {\bfseries 248} (1974)
  30--31}.

\bibitem{Hawking:1976de}
S.~W. Hawking, ``{Black Holes and Thermodynamics},''
  \href{http://dx.doi.org/10.1103/PhysRevD.13.191}{{\em Phys. Rev. D}
  {\bfseries 13} (1976) 191--197}.

\bibitem{Bekenstein:1973ur}
J.~D. Bekenstein, ``{Black holes and entropy},''
  \href{http://dx.doi.org/10.1103/PhysRevD.7.2333}{{\em Phys. Rev. D}
  {\bfseries 7} (1973) 2333--2346}.

\bibitem{Giddings:1992hh}
S.~B. Giddings, ``{Black holes and massive remnants},''
  \href{http://dx.doi.org/10.1103/PhysRevD.46.1347}{{\em Phys. Rev.} {\bfseries
  D46} (1992) 1347--1352},
\href{http://arxiv.org/abs/hep-th/9203059}{{\ttfamily arXiv:hep-th/9203059
  [hep-th]}}.

\bibitem{Giddings:2009ae}
S.~B. Giddings, ``{Nonlocality versus complementarity: A Conservative approach
  to the information problem},''
  \href{http://dx.doi.org/10.1088/0264-9381/28/2/025002}{{\em Class. Quant.
  Grav.} {\bfseries 28} (2011) 025002},
\href{http://arxiv.org/abs/0911.3395}{{\ttfamily arXiv:0911.3395 [hep-th]}}.

\bibitem{Giddings:2012bm}
S.~B. Giddings, ``{Black holes, quantum information, and unitary evolution},''
  \href{http://dx.doi.org/10.1103/PhysRevD.85.124063}{{\em Phys. Rev.}
  {\bfseries D85} (2012) 124063},
\href{http://arxiv.org/abs/1201.1037}{{\ttfamily arXiv:1201.1037 [hep-th]}}.

\bibitem{Lunin:2002qf}
O.~Lunin and S.~D. Mathur, ``{Statistical interpretation of Bekenstein entropy
  for systems with a stretched horizon},''
  \href{http://dx.doi.org/10.1103/PhysRevLett.88.211303}{{\em Phys. Rev. Lett.}
  {\bfseries 88} (2002) 211303},
\href{http://arxiv.org/abs/hep-th/0202072}{{\ttfamily arXiv:hep-th/0202072
  [hep-th]}}.

\bibitem{Mazur:2004fk}
P.~O. Mazur and E.~Mottola, ``{Gravitational vacuum condensate stars},''
  \href{http://dx.doi.org/10.1073/pnas.0402717101}{{\em Proc.Nat.Acad.Sci.}
  {\bfseries 101} (2004) 9545--9550},
\href{http://arxiv.org/abs/gr-qc/0407075}{{\ttfamily arXiv:gr-qc/0407075
  [gr-qc]}}.

\bibitem{Mathur:2008nj}
S.~D. Mathur, ``{Fuzzballs and the information paradox: A Summary and
  conjectures},''
\href{http://arxiv.org/abs/0810.4525}{{\ttfamily arXiv:0810.4525 [hep-th]}}.

\bibitem{Engelhardt:2014gca}
N.~Engelhardt and A.~C. Wall, ``{Quantum Extremal Surfaces: Holographic
  Entanglement Entropy beyond the Classical Regime},''
  \href{http://dx.doi.org/10.1007/JHEP01(2015)073}{{\em JHEP} {\bfseries 01}
  (2015) 073}, \href{http://arxiv.org/abs/1408.3203}{{\ttfamily arXiv:1408.3203
  [hep-th]}}.

\bibitem{Almheiri:2019psf}
A.~Almheiri, N.~Engelhardt, D.~Marolf, and H.~Maxfield, ``{The entropy of bulk
  quantum fields and the entanglement wedge of an evaporating black hole},''
  \href{http://dx.doi.org/10.1007/JHEP12(2019)063}{{\em JHEP} {\bfseries 12}
  (2019) 063}, \href{http://arxiv.org/abs/1905.08762}{{\ttfamily
  arXiv:1905.08762 [hep-th]}}.

\bibitem{Marolf:2020rpm}
D.~Marolf and H.~Maxfield, ``{Observations of Hawking radiation: the Page curve
  and baby universes},'' \href{http://dx.doi.org/10.1007/JHEP04(2021)272}{{\em
  JHEP} {\bfseries 04} (2021) 272},
  \href{http://arxiv.org/abs/2010.06602}{{\ttfamily arXiv:2010.06602
  [hep-th]}}.

\bibitem{Nicolini:2005vd}
P.~Nicolini, A.~Smailagic, and E.~Spallucci, ``{Noncommutative geometry
  inspired Schwarzschild black hole},''
  \href{http://dx.doi.org/10.1016/j.physletb.2005.11.004}{{\em Phys. Lett. B}
  {\bfseries 632} (2006) 547--551},
  \href{http://arxiv.org/abs/gr-qc/0510112}{{\ttfamily arXiv:gr-qc/0510112}}.

\bibitem{Bena:2007kg}
I.~Bena and N.~P. Warner, ``{Black holes, black rings and their microstates},''
  \href{http://dx.doi.org/10.1007/978-3-540-79523-0_1}{{\em Lect. Notes Phys.}
  {\bfseries 755} (2008) 1--92},
\href{http://arxiv.org/abs/hep-th/0701216}{{\ttfamily arXiv:hep-th/0701216
  [hep-th]}}.

\bibitem{Giddings:2014ova}
S.~B. Giddings, ``{Possible observational windows for quantum effects from
  black holes},'' \href{http://dx.doi.org/10.1103/PhysRevD.90.124033}{{\em
  Phys. Rev.} {\bfseries D90} no.~12, (2014) 124033},
\href{http://arxiv.org/abs/1406.7001}{{\ttfamily arXiv:1406.7001 [hep-th]}}.

\bibitem{Koshelev:2017bxd}
A.~S. Koshelev and A.~Mazumdar, ``{Do massive compact objects without event
  horizon exist in infinite derivative gravity?},''
  \href{http://dx.doi.org/10.1103/PhysRevD.96.084069}{{\em Phys. Rev.}
  {\bfseries D96} no.~8, (2017) 084069},
\href{http://arxiv.org/abs/1707.00273}{{\ttfamily arXiv:1707.00273 [gr-qc]}}.

\bibitem{Liebling:2012fv}
S.~L. Liebling and C.~Palenzuela, ``{Dynamical Boson Stars},'' {\em Living
  Rev.Rel.} {\bfseries 15} (2012) 6,
\href{http://arxiv.org/abs/1202.5809}{{\ttfamily arXiv:1202.5809 [gr-qc]}}.

\bibitem{Brito:2015pxa}
R.~Brito, V.~Cardoso, C.~A.~R. Herdeiro, and E.~Radu, ``{Proca stars:
  Gravitating Bose Einstein condensates of massive spin 1 particles},''
  \href{http://dx.doi.org/10.1016/j.physletb.2015.11.051}{{\em Phys. Lett.}
  {\bfseries B752} (2016) 291--295},
\href{http://arxiv.org/abs/1508.05395}{{\ttfamily arXiv:1508.05395 [gr-qc]}}.

\bibitem{Feinblum:1968nwc}
D.~A. Feinblum and W.~A. McKinley, ``{Stable States of a Scalar Particle in Its
  Own Gravational Field},''
  \href{http://dx.doi.org/10.1103/PhysRev.168.1445}{{\em Phys. Rev.} {\bfseries
  168} no.~5, (1968) 1445}.

\bibitem{Kaup:1968zz}
D.~J. Kaup, ``{Klein-Gordon Geon},''
  \href{http://dx.doi.org/10.1103/PhysRev.172.1331}{{\em Phys. Rev.} {\bfseries
  172} (1968) 1331--1342}.

\bibitem{Ruffini:1969qy}
R.~Ruffini and S.~Bonazzola, ``{Systems of selfgravitating particles in general
  relativity and the concept of an equation of state},''
  \href{http://dx.doi.org/10.1103/PhysRev.187.1767}{{\em Phys. Rev.} {\bfseries
  187} (1969) 1767--1783}.

\bibitem{Seidel:1991zh}
E.~Seidel and W.~Suen, ``{Oscillating soliton stars},''
\href{http://dx.doi.org/10.1103/PhysRevLett.66.1659}{{\em Phys.Rev.Lett.}
  {\bfseries 66} (1991) 1659--1662}.

\bibitem{Einstein:1935tc}
A.~Einstein and N.~Rosen, ``{The Particle Problem in the General Theory of
  Relativity},''
\href{http://dx.doi.org/10.1103/PhysRev.48.73}{{\em Phys. Rev.} {\bfseries 48}
  (1935) 73--77}.

\bibitem{Morris:1988cz}
M.~S. Morris and K.~S. Thorne, ``{Wormholes in space-time and their use for
  interstellar travel: A tool for teaching general relativity},''
\href{http://dx.doi.org/10.1119/1.15620}{{\em Am. J. Phys.} {\bfseries 56}
  (1988) 395--412}.

\bibitem{Damour:2007ap}
T.~Damour and S.~N. Solodukhin, ``{Wormholes as black hole foils},''
  \href{http://dx.doi.org/10.1103/PhysRevD.76.024016}{{\em Phys. Rev.}
  {\bfseries D76} (2007) 024016},
\href{http://arxiv.org/abs/0704.2667}{{\ttfamily arXiv:0704.2667 [gr-qc]}}.

\bibitem{Bowers:1974tgi}
R.~L. Bowers and E.~P.~T. Liang, ``{Anisotropic Spheres in General
  Relativity},'' \href{http://dx.doi.org/10.1086/152760}{{\em Astrophys. J.}
  {\bfseries 188} (1974) 657--665}.

\bibitem{Gimon:2007ur}
E.~G. Gimon and P.~Horava, ``{Astrophysical violations of the Kerr bound as a
  possible signature of string theory},''
  \href{http://dx.doi.org/10.1016/j.physletb.2009.01.026}{{\em Phys.Lett.}
  {\bfseries B672} (2009) 299--302},
\href{http://arxiv.org/abs/0706.2873}{{\ttfamily arXiv:0706.2873 [hep-th]}}.

\bibitem{Brustein:2016msz}
R.~Brustein and A.~J.~M. Medved, ``{Black holes as collapsed polymers},''
  \href{http://dx.doi.org/10.1002/prop.201600114}{{\em Fortsch. Phys.}
  {\bfseries 65} (2017) 0114},
\href{http://arxiv.org/abs/1602.07706}{{\ttfamily arXiv:1602.07706 [hep-th]}}.

\bibitem{Holdom:2016nek}
B.~Holdom and J.~Ren, ``{Not quite a black hole},''
  \href{http://dx.doi.org/10.1103/PhysRevD.95.084034}{{\em Phys. Rev.}
  {\bfseries D95} no.~8, (2017) 084034},
\href{http://arxiv.org/abs/1612.04889}{{\ttfamily arXiv:1612.04889 [gr-qc]}}.

\bibitem{Buoninfante:2019swn}
L.~Buoninfante and A.~Mazumdar, ``{Nonlocal star as a blackhole mimicker},''
  \href{http://dx.doi.org/10.1103/PhysRevD.100.024031}{{\em Phys. Rev. D}
  {\bfseries 100} no.~2, (2019) 024031},
  \href{http://arxiv.org/abs/1903.01542}{{\ttfamily arXiv:1903.01542 [gr-qc]}}.

\bibitem{Pani:2010jz}
P.~Pani, E.~Barausse, E.~Berti, and V.~Cardoso, ``{Gravitational instabilities
  of superspinars},'' \href{http://dx.doi.org/10.1103/PhysRevD.82.044009}{{\em
  Phys.Rev.} {\bfseries D82} (2010) 044009},
\href{http://arxiv.org/abs/1006.1863}{{\ttfamily arXiv:1006.1863 [gr-qc]}}.

\bibitem{Cardoso:2007az}
V.~Cardoso, P.~Pani, M.~Cadoni, and M.~Cavaglia, ``{Ergoregion instability of
  ultracompact astrophysical objects},''
  \href{http://dx.doi.org/10.1103/PhysRevD.77.124044}{{\em Phys.Rev.}
  {\bfseries D77} (2008) 124044},
\href{http://arxiv.org/abs/0709.0532}{{\ttfamily arXiv:0709.0532 [gr-qc]}}.

\bibitem{Lemos:2008cv}
J.~P.~S. Lemos and O.~B. Zaslavskii, ``{Black hole mimickers: Regular versus
  singular behavior},''
  \href{http://dx.doi.org/10.1103/PhysRevD.78.024040}{{\em Phys. Rev. D}
  {\bfseries 78} (2008) 024040},
  \href{http://arxiv.org/abs/0806.0845}{{\ttfamily arXiv:0806.0845 [gr-qc]}}.

\bibitem{Seidel:1993zk}
E.~Seidel and W.-M. Suen, ``{Formation of solitonic stars through gravitational
  cooling},'' \href{http://dx.doi.org/10.1103/PhysRevLett.72.2516}{{\em Phys.
  Rev. Lett.} {\bfseries 72} (1994) 2516--2519},
  \href{http://arxiv.org/abs/gr-qc/9309015}{{\ttfamily arXiv:gr-qc/9309015}}.

\bibitem{LIGOScientific:2020ufj}
{\bfseries LIGO Scientific, Virgo} Collaboration, R.~Abbott {\em et~al.},
  ``{Properties and Astrophysical Implications of the 150 M$_\odot$ Binary
  Black Hole Merger GW190521},''
  \href{http://dx.doi.org/10.3847/2041-8213/aba493}{{\em Astrophys. J. Lett.}
  {\bfseries 900} no.~1, (2020) L13},
  \href{http://arxiv.org/abs/2009.01190}{{\ttfamily arXiv:2009.01190
  [astro-ph.HE]}}.

\bibitem{Maggio:2021ans}
E.~Maggio, P.~Pani, and G.~Raposo, ``{Testing the nature of dark compact
  objects with gravitational waves},''
  \href{http://arxiv.org/abs/2105.06410}{{\ttfamily arXiv:2105.06410 [gr-qc]}}.

\bibitem{Cardoso:2017cqb}
V.~Cardoso and P.~Pani, ``{Tests for the existence of black holes through
  gravitational wave echoes},''
  \href{http://dx.doi.org/10.1038/s41550-017-0225-y}{{\em Nature Astron.}
  {\bfseries 1} no.~9, (2017) 586--591},
  \href{http://arxiv.org/abs/1709.01525}{{\ttfamily arXiv:1709.01525 [gr-qc]}}.

\bibitem{Raposo:2018xkf}
G.~Raposo, P.~Pani, and R.~Emparan, ``{Exotic compact objects with soft
  hair},'' \href{http://dx.doi.org/10.1103/PhysRevD.99.104050}{{\em Phys. Rev.
  D} {\bfseries 99} no.~10, (2019) 104050},
  \href{http://arxiv.org/abs/1812.07615}{{\ttfamily arXiv:1812.07615 [gr-qc]}}.

\bibitem{Buchdahl:1959zz}
H.~A. Buchdahl, ``{General Relativistic Fluid Spheres},''
\href{http://dx.doi.org/10.1103/PhysRev.116.1027}{{\em Phys. Rev.} {\bfseries
  116} (1959) 1027}.

\bibitem{Urbano:2018nrs}
A.~Urbano and H.~Veerm\"ae, ``{On gravitational echoes from ultracompact exotic
  stars},'' \href{http://dx.doi.org/10.1088/1475-7516/2019/04/011}{{\em JCAP}
  {\bfseries 04} (2019) 011}, \href{http://arxiv.org/abs/1810.07137}{{\ttfamily
  arXiv:1810.07137 [gr-qc]}}.

\bibitem{1974ApJ...188..657B}
R.~L. {Bowers} and E.~P.~T. {Liang}, ``{Anisotropic Spheres in General
  Relativity},'' \href{http://dx.doi.org/10.1086/152760}{{\em Astrophys. J.}
  {\bfseries 188} (Mar., 1974) 657}.

\bibitem{Raposo:2018rjn}
G.~Raposo, P.~Pani, M.~Bezares, C.~Palenzuela, and V.~Cardoso, ``{Anisotropic
  stars as ultracompact objects in General Relativity},''
  \href{http://dx.doi.org/10.1103/PhysRevD.99.104072}{{\em Phys. Rev.}
  {\bfseries D99} no.~10, (2019) 104072},
\href{http://arxiv.org/abs/1811.07917}{{\ttfamily arXiv:1811.07917 [gr-qc]}}.

\bibitem{Mottola:2006ew}
E.~Mottola and R.~Vaulin, ``{Macroscopic Effects of the Quantum Trace
  Anomaly},'' \href{http://dx.doi.org/10.1103/PhysRevD.74.064004}{{\em Phys.
  Rev.} {\bfseries D74} (2006) 064004},
\href{http://arxiv.org/abs/gr-qc/0604051}{{\ttfamily arXiv:gr-qc/0604051
  [gr-qc]}}.

\bibitem{Visserbook}
M.~Visser, {\em {Lorentzian wormholes: From Einstein to Hawking}}.
\newblock AIP Press, American Institute of Physics,
1995.
\newblock

\bibitem{Jetzer:1991jr}
P.~Jetzer, ``{Boson stars},''
\href{http://dx.doi.org/10.1016/0370-1573(92)90123-H}{{\em Phys. Rep.}
  {\bfseries 220} (1992) 163--227}.

\bibitem{Schunck:2003kk}
F.~Schunck and E.~Mielke, ``{General relativistic boson stars},'' {\em Class.
  Quant. Grav.} {\bfseries 20} (2003) R301--R356,
\href{http://arxiv.org/abs/0801.0307}{{\ttfamily arXiv:0801.0307 [astro-ph]}}.

\bibitem{Wald:106274}
R.~M. Wald, {\em {General relativity}}.
\newblock Chicago Univ. Press, Chicago, IL, 1984.
\newblock \url{https://cds.cern.ch/record/106274}.

\bibitem{Okawa:2013jba}
H.~Okawa, V.~Cardoso, and P.~Pani, ``{Collapse of self-interacting fields in
  asymptotically flat spacetimes: do self-interactions render Minkowski
  spacetime unstable?},''
  \href{http://dx.doi.org/10.1103/PhysRevD.89.041502}{{\em Phys. Rev.}
  {\bfseries D89} no.~4, (2014) 041502},
\href{http://arxiv.org/abs/1311.1235}{{\ttfamily arXiv:1311.1235 [gr-qc]}}.

\bibitem{Palenzuela:2007dm}
C.~Palenzuela, L.~Lehner, and S.~L. Liebling, ``{Orbital Dynamics of Binary
  Boson Star Systems},''
  \href{http://dx.doi.org/10.1103/PhysRevD.77.044036}{{\em Phys. Rev. D}
  {\bfseries 77} (2008) 044036},
  \href{http://arxiv.org/abs/0706.2435}{{\ttfamily arXiv:0706.2435 [gr-qc]}}.

\bibitem{Palenzuela:2017kcg}
C.~Palenzuela, P.~Pani, M.~Bezares, V.~Cardoso, L.~Lehner, and S.~Liebling,
  ``{Gravitational Wave Signatures of Highly Compact Boson Star Binaries},''
  \href{http://dx.doi.org/10.1103/PhysRevD.96.104058}{{\em Phys. Rev. D}
  {\bfseries 96} no.~10, (2017) 104058},
  \href{http://arxiv.org/abs/1710.09432}{{\ttfamily arXiv:1710.09432 [gr-qc]}}.

\bibitem{Sanchis-Gual:2018oui}
N.~Sanchis-Gual, C.~Herdeiro, J.~A. Font, E.~Radu, and F.~Di~Giovanni,
  ``{Head-on collisions and orbital mergers of Proca stars},''
  \href{http://dx.doi.org/10.1103/PhysRevD.99.024017}{{\em Phys. Rev. D}
  {\bfseries 99} no.~2, (2019) 024017},
  \href{http://arxiv.org/abs/1806.07779}{{\ttfamily arXiv:1806.07779 [gr-qc]}}.

\bibitem{Lee:1988av}
T.~D. Lee and Y.~Pang, ``{Stability of Mini - Boson Stars},''
\href{http://dx.doi.org/10.1016/0550-3213(89)90365-9}{{\em Nucl. Phys.}
  {\bfseries B315} (1989) 477}.

\bibitem{Colpi:1986ye}
M.~Colpi, S.~Shapiro, and I.~Wasserman, ``{Boson Stars: Gravitational
  Equilibria of Selfinteracting Scalar Fields},''
  \href{http://dx.doi.org/10.1103/PhysRevLett.57.2485}{{\em Phys. Rev. Lett.}
  {\bfseries 57} (1986) 2485--2488}.

\bibitem{Friedberg:1986tq}
R.~Friedberg, T.~D. Lee, and Y.~Pang, ``{Scalar Soliton Stars and Black
  Holes},'' \href{http://dx.doi.org/10.1103/PhysRevD.35.3658}{{\em Phys. Rev.}
  {\bfseries D35} (1987) 3658}.
[,73(1986)].

\bibitem{Lee:1986ts}
T.~D. Lee, ``{Soliton Stars and the Critical Masses of Black Holes},''
  \href{http://dx.doi.org/10.1103/PhysRevD.35.3637}{{\em Phys. Rev. D}
  {\bfseries 35} (1987) 3637}.

\bibitem{Visser:2003ge}
M.~Visser and D.~L. Wiltshire, ``{Stable gravastars: An Alternative to black
  holes?},'' \href{http://dx.doi.org/10.1088/0264-9381/21/4/027}{{\em Class.
  Quant. Grav.} {\bfseries 21} (2004) 1135--1152},
\href{http://arxiv.org/abs/gr-qc/0310107}{{\ttfamily arXiv:gr-qc/0310107
  [gr-qc]}}.

\bibitem{Mazur:2015kia}
P.~O. Mazur and E.~Mottola, ``{Surface tension and negative pressure interior
  of a non-singular "black hole"},''
  \href{http://dx.doi.org/10.1088/0264-9381/32/21/215024}{{\em Class. Quant.
  Grav.} {\bfseries 32} no.~21, (2015) 215024},
\href{http://arxiv.org/abs/1501.03806}{{\ttfamily arXiv:1501.03806 [gr-qc]}}.

\bibitem{Kanti:2011jz}
P.~Kanti, B.~Kleihaus, and J.~Kunz, ``{Wormholes in Dilatonic
  Einstein-Gauss-Bonnet Theory},''
  \href{http://dx.doi.org/10.1103/PhysRevLett.107.271101}{{\em Phys. Rev.
  Lett.} {\bfseries 107} (2011) 271101},
\href{http://arxiv.org/abs/1108.3003}{{\ttfamily arXiv:1108.3003 [gr-qc]}}.

\bibitem{Gonzalez:2008wd}
J.~A. Gonzalez, F.~S. Guzman, and O.~Sarbach, ``{Instability of wormholes
  supported by a ghost scalar field. I. Linear stability analysis},''
  \href{http://dx.doi.org/10.1088/0264-9381/26/1/015010}{{\em Class. Quant.
  Grav.} {\bfseries 26} (2009) 015010},
\href{http://arxiv.org/abs/0806.0608}{{\ttfamily arXiv:0806.0608 [gr-qc]}}.

\bibitem{Bronnikov:2012ch}
K.~A. Bronnikov, R.~A. Konoplya, and A.~Zhidenko, ``{Instabilities of wormholes
  and regular black holes supported by a phantom scalar field},''
  \href{http://dx.doi.org/10.1103/PhysRevD.86.024028}{{\em Phys. Rev.}
  {\bfseries D86} (2012) 024028},
\href{http://arxiv.org/abs/1205.2224}{{\ttfamily arXiv:1205.2224 [gr-qc]}}.

\bibitem{Lunin:2001jy}
O.~Lunin and S.~D. Mathur, ``{AdS / CFT duality and the black hole information
  paradox},'' \href{http://dx.doi.org/10.1016/S0550-3213(01)00620-4}{{\em Nucl.
  Phys.} {\bfseries B623} (2002) 342--394},
\href{http://arxiv.org/abs/hep-th/0109154}{{\ttfamily arXiv:hep-th/0109154
  [hep-th]}}.

\bibitem{Balasubramanian:2008da}
V.~Balasubramanian, J.~de~Boer, S.~El-Showk, and I.~Messamah, ``{Black Holes as
  Effective Geometries},''
  \href{http://dx.doi.org/10.1088/0264-9381/25/21/214004}{{\em Class. Quant.
  Grav.} {\bfseries 25} (2008) 214004},
\href{http://arxiv.org/abs/0811.0263}{{\ttfamily arXiv:0811.0263 [hep-th]}}.

\bibitem{Bena:2013dka}
I.~Bena and N.~P. Warner, ``{Resolving the Structure of Black Holes:
  Philosophizing with a Hammer},''
\href{http://arxiv.org/abs/1311.4538}{{\ttfamily arXiv:1311.4538 [hep-th]}}.

\bibitem{Gibbons:2013tqa}
G.~W. Gibbons and N.~P. Warner, ``{Global structure of five-dimensional
  fuzzballs},'' \href{http://dx.doi.org/10.1088/0264-9381/31/2/025016}{{\em
  Class. Quant. Grav.} {\bfseries 31} (2014) 025016},
  \href{http://arxiv.org/abs/1305.0957}{{\ttfamily arXiv:1305.0957 [hep-th]}}.

\bibitem{Bates:2003vx}
B.~Bates and F.~Denef, ``{Exact solutions for supersymmetric stationary black
  hole composites},'' \href{http://dx.doi.org/10.1007/JHEP11(2011)127}{{\em
  JHEP} {\bfseries 11} (2011) 127},
  \href{http://arxiv.org/abs/hep-th/0304094}{{\ttfamily arXiv:hep-th/0304094}}.

\bibitem{Bianchi:2020bxa}
M.~Bianchi, D.~Consoli, A.~Grillo, J.~F. Morales, P.~Pani, and G.~Raposo,
  ``{Distinguishing fuzzballs from black holes through their multipolar
  structure},'' \href{http://dx.doi.org/10.1103/PhysRevLett.125.221601}{{\em
  Phys. Rev. Lett.} {\bfseries 125} no.~22, (2020) 221601},
  \href{http://arxiv.org/abs/2007.01743}{{\ttfamily arXiv:2007.01743
  [hep-th]}}.

\bibitem{Bianchi:2020miz}
M.~Bianchi, D.~Consoli, A.~Grillo, J.~F. Morales, P.~Pani, and G.~Raposo,
  ``{The multipolar structure of fuzzballs},''
  \href{http://dx.doi.org/10.1007/JHEP01(2021)003}{{\em JHEP} {\bfseries 01}
  (2021) 003}, \href{http://arxiv.org/abs/2008.01445}{{\ttfamily
  arXiv:2008.01445 [hep-th]}}.

\bibitem{Bayin:1982vw}
S.~S. Bayin, ``{Anisotropic Fluid Spheres in General Relativity},''
\href{http://dx.doi.org/10.1103/PhysRevD.26.1262}{{\em Phys. Rev.} {\bfseries
  D26} (1982) 1262}.

\bibitem{Dev:2000gt}
K.~Dev and M.~Gleiser, ``{Anisotropic stars: Exact solutions},''
  \href{http://dx.doi.org/10.1023/A:1020707906543}{{\em Gen. Rel. Grav.}
  {\bfseries 34} (2002) 1793--1818},
\href{http://arxiv.org/abs/astro-ph/0012265}{{\ttfamily arXiv:astro-ph/0012265
  [astro-ph]}}.

\bibitem{Mak:2001eb}
M.~K. Mak and T.~Harko, ``{Anisotropic stars in general relativity},''
  \href{http://dx.doi.org/10.1098/rspa.2002.1014}{{\em Proc. Roy. Soc. Lond.}
  {\bfseries A459} (2003) 393--408},
\href{http://arxiv.org/abs/gr-qc/0110103}{{\ttfamily arXiv:gr-qc/0110103
  [gr-qc]}}.

\bibitem{Andreasson:2007ck}
H.~Andreasson, ``{Sharp bounds on 2m/r of general spherically symmetric static
  objects},'' \href{http://dx.doi.org/10.1016/j.jde.2008.05.010}{{\em J. Diff.
  Eq.} {\bfseries 245} (2008) 2243--2266},
\href{http://arxiv.org/abs/gr-qc/0702137}{{\ttfamily arXiv:gr-qc/0702137
  [gr-qc]}}.

\bibitem{Dev:2003qd}
K.~Dev and M.~Gleiser, ``{Anisotropic stars. 2. Stability},''
  \href{http://dx.doi.org/10.1023/A:1024534702166}{{\em Gen. Rel. Grav.}
  {\bfseries 35} (2003) 1435--1457},
\href{http://arxiv.org/abs/gr-qc/0303077}{{\ttfamily arXiv:gr-qc/0303077
  [gr-qc]}}.

\bibitem{Herrera:2004xc}
L.~Herrera, A.~Di~Prisco, J.~Martin, J.~Ospino, N.~O. Santos, and O.~Troconis,
  ``{Spherically symmetric dissipative anisotropic fluids: A General study},''
  \href{http://dx.doi.org/10.1103/PhysRevD.69.084026}{{\em Phys. Rev.}
  {\bfseries D69} (2004) 084026},
\href{http://arxiv.org/abs/gr-qc/0403006}{{\ttfamily arXiv:gr-qc/0403006
  [gr-qc]}}.

\bibitem{Doneva:2012rd}
D.~D. Doneva and S.~S. Yazadjiev, ``{Gravitational wave spectrum of anisotropic
  neutron stars in Cowling approximation},''
  \href{http://dx.doi.org/10.1103/PhysRevD.85.124023}{{\em Phys. Rev.}
  {\bfseries D85} (2012) 124023},
\href{http://arxiv.org/abs/1203.3963}{{\ttfamily arXiv:1203.3963 [gr-qc]}}.

\bibitem{Silva:2014fca}
H.~O. Silva, C.~F.~B. Macedo, E.~Berti, and L.~C.~B. Crispino, ``{Slowly
  rotating anisotropic neutron stars in general relativity and scalar-tensor
  theory},'' \href{http://dx.doi.org/10.1088/0264-9381/32/14/145008}{{\em
  Class. Quant. Grav.} {\bfseries 32} (2015) 145008},
\href{http://arxiv.org/abs/1411.6286}{{\ttfamily arXiv:1411.6286 [gr-qc]}}.

\bibitem{Yagi:2015hda}
K.~Yagi and N.~Yunes, ``{I-Love-Q anisotropically: Universal relations for
  compact stars with scalar pressure anisotropy},''
  \href{http://dx.doi.org/10.1103/PhysRevD.91.123008}{{\em Phys. Rev.}
  {\bfseries D91} no.~12, (2015) 123008},
\href{http://arxiv.org/abs/1503.02726}{{\ttfamily arXiv:1503.02726 [gr-qc]}}.

\bibitem{Almheiri:2012rt}
A.~Almheiri, D.~Marolf, J.~Polchinski, and J.~Sully, ``{Black Holes:
  Complementarity or Firewalls?},''
  \href{http://dx.doi.org/10.1007/JHEP02(2013)062}{{\em JHEP} {\bfseries 02}
  (2013) 062}, \href{http://arxiv.org/abs/1207.3123}{{\ttfamily arXiv:1207.3123
  [hep-th]}}.

\bibitem{Kaplan:2018dqx}
D.~E. Kaplan and S.~Rajendran, ``{Firewalls in General Relativity},''
  \href{http://dx.doi.org/10.1103/PhysRevD.99.044033}{{\em Phys. Rev.}
  {\bfseries D99} no.~4, (2019) 044033},
\href{http://arxiv.org/abs/1812.00536}{{\ttfamily arXiv:1812.00536 [hep-th]}}.

\bibitem{Zhang:2017jze}
J.~Zhang and S.-Y. Zhou, ``{Can the graviton have a large mass near black
  holes?},'' \href{http://dx.doi.org/10.1103/PhysRevD.97.081501}{{\em Phys.
  Rev. D} {\bfseries 97} no.~8, (2018) 081501},
  \href{http://arxiv.org/abs/1709.07503}{{\ttfamily arXiv:1709.07503 [gr-qc]}}.

\bibitem{Oshita:2018fqu}
N.~Oshita and N.~Afshordi, ``{Probing microstructure of black hole spacetimes
  with gravitational wave echoes},''
  \href{http://dx.doi.org/10.1103/PhysRevD.99.044002}{{\em Phys. Rev. D}
  {\bfseries 99} no.~4, (2019) 044002},
  \href{http://arxiv.org/abs/1807.10287}{{\ttfamily arXiv:1807.10287 [gr-qc]}}.

\bibitem{Frolov:2015bta}
V.~P. Frolov, ``{Mass-gap for black hole formation in higher derivative and
  ghost free gravity},''
  \href{http://dx.doi.org/10.1103/PhysRevLett.115.051102}{{\em Phys. Rev.
  Lett.} {\bfseries 115} no.~5, (2015) 051102},
  \href{http://arxiv.org/abs/1505.00492}{{\ttfamily arXiv:1505.00492
  [hep-th]}}.

\bibitem{Buoninfante:2018xif}
L.~Buoninfante, A.~S. Cornell, G.~Harmsen, A.~S. Koshelev, G.~Lambiase, J.~a.
  Marto, and A.~Mazumdar, ``{Towards nonsingular rotating compact object in
  ghost-free infinite derivative gravity},''
  \href{http://dx.doi.org/10.1103/PhysRevD.98.084041}{{\em Phys. Rev. D}
  {\bfseries 98} no.~8, (2018) 084041},
  \href{http://arxiv.org/abs/1807.08896}{{\ttfamily arXiv:1807.08896 [gr-qc]}}.

\bibitem{Piovano:2020ooe}
G.~A. Piovano, A.~Maselli, and P.~Pani, ``{Model independent tests of the Kerr
  bound with extreme mass ratio inspirals},''
  \href{http://dx.doi.org/10.1016/j.physletb.2020.135860}{{\em Phys. Lett. B}
  {\bfseries 811} (2020) 135860},
  \href{http://arxiv.org/abs/2003.08448}{{\ttfamily arXiv:2003.08448 [gr-qc]}}.

\bibitem{Herdeiro:2020kvf}
C.~A.~R. Herdeiro, J.~Kunz, I.~Perapechka, E.~Radu, and Y.~Shnir, ``{Multipolar
  boson stars: macroscopic Bose-Einstein condensates akin to hydrogen
  orbitals},'' \href{http://dx.doi.org/10.1016/j.physletb.2020.136027}{{\em
  Phys. Lett. B} {\bfseries 812} (2021) 136027},
  \href{http://arxiv.org/abs/2008.10608}{{\ttfamily arXiv:2008.10608 [gr-qc]}}.

\bibitem{Bena:2020see}
I.~Bena and D.~R. Mayerson, ``{Multipole Ratios: A New Window into Black
  Holes},'' \href{http://dx.doi.org/10.1103/PhysRevLett.125.221602}{{\em Phys.
  Rev. Lett.} {\bfseries 125} no.~22, (2020) 22},
  \href{http://arxiv.org/abs/2006.10750}{{\ttfamily arXiv:2006.10750
  [hep-th]}}.

\bibitem{Bena:2020uup}
I.~Bena and D.~R. Mayerson, ``{Black Holes Lessons from Multipole Ratios},''
  \href{http://dx.doi.org/10.1007/JHEP03(2021)114}{{\em JHEP} {\bfseries 03}
  (2021) 114}, \href{http://arxiv.org/abs/2007.09152}{{\ttfamily
  arXiv:2007.09152 [hep-th]}}.

\bibitem{Pani:2015tga}
P.~Pani, ``{I-Love-Q relations for gravastars and the approach to the
  black-hole limit},'' \href{http://dx.doi.org/10.1103/PhysRevD.92.124030}{{\em
  Phys. Rev.} {\bfseries D92} no.~12, (2015) 124030},
\href{http://arxiv.org/abs/1506.06050}{{\ttfamily arXiv:1506.06050 [gr-qc]}}.

\bibitem{Raposo:2020yjy}
G.~Raposo and P.~Pani, ``{Axisymmetric deformations of neutron stars and
  gravitational-wave astronomy},''
  \href{http://dx.doi.org/10.1103/PhysRevD.102.044045}{{\em Phys. Rev. D}
  {\bfseries 102} no.~4, (2020) 044045},
  \href{http://arxiv.org/abs/2002.02555}{{\ttfamily arXiv:2002.02555 [gr-qc]}}.

\bibitem{Kastha:2018bcr}
S.~Kastha, A.~Gupta, K.~Arun, B.~Sathyaprakash, and C.~Van Den~Broeck,
  ``{Testing the multipole structure of compact binaries using gravitational
  wave observations},''
  \href{http://dx.doi.org/10.1103/PhysRevD.98.124033}{{\em Phys. Rev. D}
  {\bfseries 98} no.~12, (2018) 124033},
  \href{http://arxiv.org/abs/1809.10465}{{\ttfamily arXiv:1809.10465 [gr-qc]}}.

\bibitem{Babak:2017tow}
S.~Babak, J.~Gair, A.~Sesana, E.~Barausse, C.~F. Sopuerta, C.~P.~L. Berry,
  E.~Berti, P.~Amaro-Seoane, A.~Petiteau, and A.~Klein, ``{Science with the
  space-based interferometer LISA. V: Extreme mass-ratio inspirals},''
  \href{http://dx.doi.org/10.1103/PhysRevD.95.103012}{{\em Phys. Rev.}
  {\bfseries D95} no.~10, (2017) 103012},
\href{http://arxiv.org/abs/1703.09722}{{\ttfamily arXiv:1703.09722 [gr-qc]}}.

\bibitem{Kastha:2019brk}
S.~Kastha, A.~Gupta, K.~Arun, B.~Sathyaprakash, and C.~Van Den~Broeck,
  ``{Testing the multipole structure and conservative dynamics of compact
  binaries using gravitational wave observations: The spinning case},''
  \href{http://dx.doi.org/10.1103/PhysRevD.100.044007}{{\em Phys. Rev. D}
  {\bfseries 100} no.~4, (2019) 044007},
  \href{http://arxiv.org/abs/1905.07277}{{\ttfamily arXiv:1905.07277 [gr-qc]}}.

\bibitem{Hartle:1973zz}
J.~B. Hartle, ``{Tidal Friction in Slowly Rotating Black Holes},''
\href{http://dx.doi.org/10.1103/PhysRevD.8.1010}{{\em Phys.Rev.} {\bfseries D8}
  (1973) 1010--1024}.

\bibitem{Hughes:1999bq}
S.~A. Hughes, ``{The Evolution of circular, nonequatorial orbits of Kerr black
  holes due to gravitational wave emission},''
  \href{http://dx.doi.org/10.1103/PhysRevD.65.069902}{{\em Phys. Rev. D}
  {\bfseries 61} no.~8, (2000) 084004},
  \href{http://arxiv.org/abs/gr-qc/9910091}{{\ttfamily arXiv:gr-qc/9910091}}.
  [Erratum: Phys.Rev.D 63, 049902 (2001), Erratum: Phys.Rev.D 65, 069902
  (2002), Erratum: Phys.Rev.D 67, 089901 (2003), Erratum: Phys.Rev.D 78, 109902
  (2008), Erratum: Phys.Rev.D 90, 109904 (2014)].

\bibitem{Hughes:2001jr}
S.~A. Hughes, ``{Evolution of circular, nonequatorial orbits of Kerr black
  holes due to gravitational wave emission. 2. Inspiral trajectories and
  gravitational wave forms},''
  \href{http://dx.doi.org/10.1103/PhysRevD.64.064004}{{\em Phys.Rev.}
  {\bfseries D64} (2001) 064004},
\href{http://arxiv.org/abs/gr-qc/0104041}{{\ttfamily arXiv:gr-qc/0104041
  [gr-qc]}}.

\bibitem{Maselli:2017cmm}
A.~Maselli, P.~Pani, V.~Cardoso, T.~Abdelsalhin, L.~Gualtieri, and V.~Ferrari,
  ``{Probing Planckian corrections at the horizon scale with LISA binaries},''
  \href{http://dx.doi.org/10.1103/PhysRevLett.120.081101}{{\em Phys. Rev.
  Lett.} {\bfseries 120} no.~8, (2018) 081101},
\href{http://arxiv.org/abs/1703.10612}{{\ttfamily arXiv:1703.10612 [gr-qc]}}.

\bibitem{poisson2014gravity}
E.~Poisson and C.~Will, {\em Gravity: Newtonian, Post-Newtonian, Relativistic}.
\newblock Cambridge University Press, 2014.
\newblock \url{https://books.google.it/books?id=PZ5cAwAAQBAJ}.

\bibitem{Binnington:2009bb}
T.~Binnington and E.~Poisson, ``{Relativistic theory of tidal Love numbers},''
  \href{http://dx.doi.org/10.1103/PhysRevD.80.084018}{{\em Phys. Rev. D}
  {\bfseries 80} (2009) 084018},
  \href{http://arxiv.org/abs/0906.1366}{{\ttfamily arXiv:0906.1366 [gr-qc]}}.

\bibitem{Damour:2009vw}
T.~Damour and A.~Nagar, ``{Relativistic tidal properties of neutron stars},''
  \href{http://dx.doi.org/10.1103/PhysRevD.80.084035}{{\em Phys. Rev. D}
  {\bfseries 80} (2009) 084035},
  \href{http://arxiv.org/abs/0906.0096}{{\ttfamily arXiv:0906.0096 [gr-qc]}}.

\bibitem{Poisson:2014gka}
E.~Poisson, ``{Tidal deformation of a slowly rotating black hole},''
  \href{http://dx.doi.org/10.1103/PhysRevD.91.044004}{{\em Phys. Rev. D}
  {\bfseries 91} no.~4, (2015) 044004},
  \href{http://arxiv.org/abs/1411.4711}{{\ttfamily arXiv:1411.4711 [gr-qc]}}.

\bibitem{Landry:2015zfa}
P.~Landry and E.~Poisson, ``{Tidal deformation of a slowly rotating material
  body. External metric},''
  \href{http://dx.doi.org/10.1103/PhysRevD.91.104018}{{\em Phys. Rev. D}
  {\bfseries 91} (2015) 104018},
  \href{http://arxiv.org/abs/1503.07366}{{\ttfamily arXiv:1503.07366 [gr-qc]}}.

\bibitem{Pani:2015hfa}
P.~Pani, L.~Gualtieri, A.~Maselli, and V.~Ferrari, ``{Tidal deformations of a
  spinning compact object},''
  \href{http://dx.doi.org/10.1103/PhysRevD.92.024010}{{\em Phys. Rev. D}
  {\bfseries 92} no.~2, (2015) 024010},
  \href{http://arxiv.org/abs/1503.07365}{{\ttfamily arXiv:1503.07365 [gr-qc]}}.

\bibitem{Chia:2020yla}
H.~S. Chia, ``{Tidal deformation and dissipation of rotating black holes},''
  \href{http://dx.doi.org/10.1103/PhysRevD.104.024013}{{\em Phys. Rev. D}
  {\bfseries 104} no.~2, (2021) 024013},
  \href{http://arxiv.org/abs/2010.07300}{{\ttfamily arXiv:2010.07300 [gr-qc]}}.

\bibitem{LeTiec:2020spy}
A.~Le~Tiec and M.~Casals, ``{Spinning Black Holes Fall in Love},''
  \href{http://dx.doi.org/10.1103/PhysRevLett.126.131102}{{\em Phys. Rev.
  Lett.} {\bfseries 126} no.~13, (2021) 131102},
  \href{http://arxiv.org/abs/2007.00214}{{\ttfamily arXiv:2007.00214 [gr-qc]}}.

\bibitem{LeTiec:2020bos}
A.~Le~Tiec, M.~Casals, and E.~Franzin, ``{Tidal Love Numbers of Kerr Black
  Holes},'' \href{http://dx.doi.org/10.1103/PhysRevD.103.084021}{{\em Phys.
  Rev. D} {\bfseries 103} no.~8, (2021) 084021},
  \href{http://arxiv.org/abs/2010.15795}{{\ttfamily arXiv:2010.15795 [gr-qc]}}.

\bibitem{Cardoso:2017cfl}
V.~Cardoso, E.~Franzin, A.~Maselli, P.~Pani, and G.~Raposo, ``{Testing
  strong-field gravity with tidal Love numbers},''
  \href{http://dx.doi.org/10.1103/PhysRevD.95.089901,
  10.1103/PhysRevD.95.084014}{{\em Phys. Rev.} {\bfseries D95} no.~8, (2017)
  084014}, \href{http://arxiv.org/abs/1701.01116}{{\ttfamily arXiv:1701.01116
  [gr-qc]}}.
[Addendum: Phys. Rev.D95,no.8,089901(2017)].

\bibitem{Sennett:2017etc}
N.~Sennett, T.~Hinderer, J.~Steinhoff, A.~Buonanno, and S.~Ossokine,
  ``{Distinguishing Boson Stars from Black Holes and Neutron Stars from Tidal
  Interactions in Inspiraling Binary Systems},''
  \href{http://dx.doi.org/10.1103/PhysRevD.96.024002}{{\em Phys. Rev. D}
  {\bfseries 96} no.~2, (2017) 024002},
  \href{http://arxiv.org/abs/1704.08651}{{\ttfamily arXiv:1704.08651 [gr-qc]}}.

\bibitem{Mendes:2016vdr}
R.~F.~P. Mendes and H.~Yang, ``{Tidal deformability of boson stars and dark
  matter clumps},'' \href{http://dx.doi.org/10.1088/1361-6382/aa842d}{{\em
  Class. Quant. Grav.} {\bfseries 34} no.~18, (2017) 185001},
  \href{http://arxiv.org/abs/1606.03035}{{\ttfamily arXiv:1606.03035
  [astro-ph.CO]}}.

\bibitem{Uchikata:2016qku}
N.~Uchikata, S.~Yoshida, and P.~Pani, ``{Tidal deformability and I-Love-Q
  relations for gravastars with polytropic thin shells},''
  \href{http://dx.doi.org/10.1103/PhysRevD.94.064015}{{\em Phys. Rev. D}
  {\bfseries 94} no.~6, (2016) 064015},
  \href{http://arxiv.org/abs/1607.03593}{{\ttfamily arXiv:1607.03593 [gr-qc]}}.

\bibitem{Berti:2006qt}
E.~Berti and V.~Cardoso, ``{Supermassive black holes or boson stars? Hair
  counting with gravitational wave detectors},''
  \href{http://dx.doi.org/10.1142/S0218271806009637}{{\em Int.J.Mod.Phys.}
  {\bfseries D15} (2006) 2209--2216},
  \href{http://arxiv.org/abs/gr-qc/0605101}{{\ttfamily arXiv:gr-qc/0605101
  [gr-qc]}}.

\bibitem{Macedo:2013jja}
C.~F. Macedo, P.~Pani, V.~Cardoso, and L.~C.~B. Crispino, ``{Astrophysical
  signatures of boson stars: quasinormal modes and inspiral resonances},''
  \href{http://dx.doi.org/10.1103/PhysRevD.88.064046}{{\em Phys.Rev.}
  {\bfseries D88} no.~6, (2013) 064046},
\href{http://arxiv.org/abs/1307.4812}{{\ttfamily arXiv:1307.4812 [gr-qc]}}.

\bibitem{Chirenti:2007mk}
C.~B. Chirenti and L.~Rezzolla, ``{How to tell a gravastar from a black
  hole},'' \href{http://dx.doi.org/10.1088/0264-9381/24/16/013}{{\em
  Class.Quant.Grav.} {\bfseries 24} (2007) 4191--4206},
\href{http://arxiv.org/abs/0706.1513}{{\ttfamily arXiv:0706.1513 [gr-qc]}}.

\bibitem{Pani:2009ss}
P.~Pani, E.~Berti, V.~Cardoso, Y.~Chen, and R.~Norte, ``{Gravitational wave
  signatures of the absence of an event horizon. I. Nonradial oscillations of a
  thin-shell gravastar},''
  \href{http://dx.doi.org/10.1103/PhysRevD.80.124047}{{\em Phys.Rev.}
  {\bfseries D80} (2009) 124047},
\href{http://arxiv.org/abs/0909.0287}{{\ttfamily arXiv:0909.0287 [gr-qc]}}.

\bibitem{Chirenti:2016hzd}
C.~Chirenti and L.~Rezzolla, ``{Did GW150914 produce a rotating gravastar?},''
  \href{http://dx.doi.org/10.1103/PhysRevD.94.084016}{{\em Phys. Rev.}
  {\bfseries D94} no.~8, (2016) 084016},
\href{http://arxiv.org/abs/1602.08759}{{\ttfamily arXiv:1602.08759 [gr-qc]}}.

\bibitem{Volkel:2017ofl}
S.~H. Völkel and K.~D. Kokkotas, ``{A Semi-analytic Study of Axial
  Perturbations of Ultra Compact Stars},''
  \href{http://dx.doi.org/10.1088/1361-6382/aa68cc}{{\em Class. Quant. Grav.}
  {\bfseries 34} no.~12, (2017) 125006},
\href{http://arxiv.org/abs/1703.08156}{{\ttfamily arXiv:1703.08156 [gr-qc]}}.

\bibitem{Konoplya:2016hmd}
R.~A. Konoplya and A.~Zhidenko, ``{Wormholes versus black holes: quasinormal
  ringing at early and late times},''
  \href{http://dx.doi.org/10.1088/1475-7516/2016/12/043}{{\em JCAP} {\bfseries
  1612} no.~12, (2016) 043},
\href{http://arxiv.org/abs/1606.00517}{{\ttfamily arXiv:1606.00517 [gr-qc]}}.

\bibitem{Bueno:2017hyj}
P.~Bueno, P.~A. Cano, F.~Goelen, T.~Hertog, and B.~Vercnocke, ``{Echoes of
  Kerr-like wormholes},''
  \href{http://dx.doi.org/10.1103/PhysRevD.97.024040}{{\em Phys. Rev.}
  {\bfseries D97} no.~2, (2018) 024040},
\href{http://arxiv.org/abs/1711.00391}{{\ttfamily arXiv:1711.00391 [gr-qc]}}.

\bibitem{Cardoso:2005gj}
V.~Cardoso, O.~J. Dias, J.~L. Hovdebo, and R.~C. Myers, ``{Instability of
  non-supersymmetric smooth geometries},''
  \href{http://dx.doi.org/10.1103/PhysRevD.73.064031}{{\em Phys.Rev.}
  {\bfseries D73} (2006) 064031},
\href{http://arxiv.org/abs/hep-th/0512277}{{\ttfamily arXiv:hep-th/0512277
  [hep-th]}}.

\bibitem{Eperon:2016cdd}
F.~C. Eperon, H.~S. Reall, and J.~E. Santos, ``{Instability of supersymmetric
  microstate geometries},''
  \href{http://dx.doi.org/10.1007/JHEP10(2016)031}{{\em JHEP} {\bfseries 10}
  (2016) 031},
\href{http://arxiv.org/abs/1607.06828}{{\ttfamily arXiv:1607.06828 [hep-th]}}.

\bibitem{Cardoso:2016oxy}
V.~Cardoso, S.~Hopper, C.~F.~B. Macedo, C.~Palenzuela, and P.~Pani,
  ``{Gravitational-wave signatures of exotic compact objects and of quantum
  corrections at the horizon scale},''
  \href{http://dx.doi.org/10.1103/PhysRevD.94.084031}{{\em Phys. Rev. D}
  {\bfseries 94} no.~8, (2016) 084031},
  \href{http://arxiv.org/abs/1608.08637}{{\ttfamily arXiv:1608.08637 [gr-qc]}}.

\bibitem{Barcelo:2017lnx}
C.~Barceló, R.~Carballo-Rubio, and L.~J. Garay, ``{Gravitational wave echoes
  from macroscopic quantum gravity effects},''
  \href{http://dx.doi.org/10.1007/JHEP05(2017)054}{{\em JHEP} {\bfseries 05}
  (2017) 054},
\href{http://arxiv.org/abs/1701.09156}{{\ttfamily arXiv:1701.09156 [gr-qc]}}.

\bibitem{Brustein:2017koc}
R.~Brustein, A.~J.~M. Medved, and K.~Yagi, ``{When black holes collide: Probing
  the interior composition by the spectrum of ringdown modes and emitted
  gravitational waves},''
  \href{http://dx.doi.org/10.1103/PhysRevD.96.064033}{{\em Phys. Rev.}
  {\bfseries D96} no.~6, (2017) 064033},
\href{http://arxiv.org/abs/1704.05789}{{\ttfamily arXiv:1704.05789 [gr-qc]}}.

\bibitem{Ferrari:2000sr}
V.~Ferrari and K.~Kokkotas, ``{Scattering of particles by neutron stars: Time
  evolutions for axial perturbations},''
  \href{http://dx.doi.org/10.1103/PhysRevD.62.107504}{{\em Phys. Rev. D}
  {\bfseries 62} (2000) 107504},
  \href{http://arxiv.org/abs/gr-qc/0008057}{{\ttfamily arXiv:gr-qc/0008057}}.

\bibitem{Pani:2018flj}
P.~Pani and V.~Ferrari, ``{On gravitational-wave echoes from neutron-star
  binary coalescences},''
  \href{http://dx.doi.org/10.1088/1361-6382/aacb8f}{{\em Class. Quant. Grav.}
  {\bfseries 35} no.~15, (2018) 15LT01},
  \href{http://arxiv.org/abs/1804.01444}{{\ttfamily arXiv:1804.01444 [gr-qc]}}.

\bibitem{Glampedakis:2017cgd}
K.~Glampedakis and G.~Pappas, ``{How well can ultracompact bodies imitate black
  hole ringdowns?},'' \href{http://dx.doi.org/10.1103/PhysRevD.97.041502}{{\em
  Phys. Rev. D} {\bfseries 97} no.~4, (2018) 041502},
  \href{http://arxiv.org/abs/1710.02136}{{\ttfamily arXiv:1710.02136 [gr-qc]}}.

\bibitem{Okounkova:2017yby}
M.~Okounkova, L.~C. Stein, M.~A. Scheel, and D.~A. Hemberger, ``{Numerical
  binary black hole mergers in dynamical Chern-Simons gravity: Scalar field},''
  \href{http://dx.doi.org/10.1103/PhysRevD.96.044020}{{\em Phys. Rev. D}
  {\bfseries 96} no.~4, (2017) 044020},
  \href{http://arxiv.org/abs/1705.07924}{{\ttfamily arXiv:1705.07924 [gr-qc]}}.

\bibitem{Blazquez-Salcedo:2016enn}
J.~L. Blázquez-Salcedo, C.~F.~B. Macedo, V.~Cardoso, V.~Ferrari, L.~Gualtieri,
  F.~S. Khoo, J.~Kunz, and P.~Pani, ``{Perturbed black holes in
  Einstein-dilaton-Gauss-Bonnet gravity: Stability, ringdown, and
  gravitational-wave emission},''
  \href{http://dx.doi.org/10.1103/PhysRevD.94.104024}{{\em Phys. Rev. D}
  {\bfseries 94} no.~10, (2016) 104024},
  \href{http://arxiv.org/abs/1609.01286}{{\ttfamily arXiv:1609.01286 [gr-qc]}}.

\bibitem{Tattersall:2018nve}
O.~J. Tattersall and P.~G. Ferreira, ``{Quasinormal modes of black holes in
  Horndeski gravity},''
  \href{http://dx.doi.org/10.1103/PhysRevD.97.104047}{{\em Phys. Rev. D}
  {\bfseries 97} no.~10, (2018) 104047},
  \href{http://arxiv.org/abs/1804.08950}{{\ttfamily arXiv:1804.08950 [gr-qc]}}.

\bibitem{Regge:1957td}
T.~Regge and J.~A. Wheeler, ``{Stability of a Schwarzschild singularity},''
\href{http://dx.doi.org/10.1103/PhysRev.108.1063}{{\em Phys.Rev.} {\bfseries
  108} (1957) 1063--1069}.

\bibitem{Zerilli:1970se}
F.~J. Zerilli, ``{Effective potential for even parity Regge-Wheeler
  gravitational perturbation equations},''
\href{http://dx.doi.org/10.1103/PhysRevLett.24.737}{{\em Phys.Rev.Lett.}
  {\bfseries 24} (1970) 737--738}.

\bibitem{Pani:2013pma}
P.~Pani, ``{Advanced Methods in Black-Hole Perturbation Theory},''
  \href{http://dx.doi.org/10.1142/S0217751X13400186}{{\em Int.J.Mod.Phys.}
  {\bfseries A28} (2013) 1340018},
\href{http://arxiv.org/abs/1305.6759}{{\ttfamily arXiv:1305.6759 [gr-qc]}}.

\bibitem{Chandrasekhar:1975nkd}
S.~Chandrasekhar, ``{On the equations governing the perturbations of the
  Schwarzschild black hole},''
  \href{http://dx.doi.org/10.1098/rspa.1975.0066}{{\em Proc. Roy. Soc. Lond. A}
  {\bfseries 343} no.~1634, (1975) 289--298}.

\bibitem{Chandra}
S.~Chandrasekhar, {\em {The mathematical theory of black holes}}.
\newblock Oxford University Press,
1983.
\newblock

\bibitem{Glampedakis:2017rar}
K.~Glampedakis, A.~D. Johnson, and D.~Kennefick, ``{Darboux transformation in
  black hole perturbation theory},''
  \href{http://dx.doi.org/10.1103/PhysRevD.96.024036}{{\em Phys. Rev. D}
  {\bfseries 96} no.~2, (2017) 024036},
  \href{http://arxiv.org/abs/1702.06459}{{\ttfamily arXiv:1702.06459 [gr-qc]}}.

\bibitem{Damour:1982}
T.~{Damour}, ``{Surface Effects in Black-Hole Physics},'' in {\em Marcel
  Grossmann Meeting: General Relativity}, p.~587.
\newblock Jan., 1982.

\bibitem{MembraneParadigm}
K.~S. Thorne, R.~Price, and D.~Macdonald, {\em {Black holes: the membrane
  paradigm}}.
\newblock Yale University Press, 1986.

\bibitem{Maggio:2020jml}
E.~Maggio, L.~Buoninfante, A.~Mazumdar, and P.~Pani, ``{How does a dark compact
  object ringdown?},''
  \href{http://dx.doi.org/10.1103/PhysRevD.102.064053}{{\em Phys. Rev. D}
  {\bfseries 102} no.~6, (2020) 064053},
  \href{http://arxiv.org/abs/2006.14628}{{\ttfamily arXiv:2006.14628 [gr-qc]}}.

\bibitem{Price:1986yy}
R.~Price and K.~Thorne, ``{Membrane Viewpoint on Black Holes: Properties and
  Evolution of the Stretched Horizon},''
  \href{http://dx.doi.org/10.1103/PhysRevD.33.915}{{\em Phys. Rev. D}
  {\bfseries 33} (1986) 915--941}.

\bibitem{Darmois:1927}
G.~Darmois, ``{Les \'equations de la gravitation einsteinienne},'' {\em
  M\'emorial de Sciences Math\'ematiques} {\bfseries fascicule 25} (1927)
  1--48.

\bibitem{Israel:1966rt}
W.~Israel, ``{Singular hypersurfaces and thin shells in general relativity},''
  \href{http://dx.doi.org/10.1007/BF02710419, 10.1007/BF02712210}{{\em Nuovo
  Cim.} {\bfseries B44S10} (1966) 1}.
[Nuovo Cim.B44,1(1966)].

\bibitem{Jaramillo:2020tuu}
J.~L. Jaramillo, R.~Panosso~Macedo, and L.~Al~Sheikh, ``{Pseudospectrum and
  Black Hole Quasinormal Mode Instability},''
  \href{http://dx.doi.org/10.1103/PhysRevX.11.031003}{{\em Phys. Rev. X}
  {\bfseries 11} no.~3, (2021) 031003},
  \href{http://arxiv.org/abs/2004.06434}{{\ttfamily arXiv:2004.06434 [gr-qc]}}.

\bibitem{Berti:2005ys}
E.~Berti, V.~Cardoso, and C.~M. Will, ``{On gravitational-wave spectroscopy of
  massive black holes with the space interferometer LISA},''
  \href{http://dx.doi.org/10.1103/PhysRevD.73.064030}{{\em Phys.Rev.}
  {\bfseries D73} (2006) 064030},
\href{http://arxiv.org/abs/gr-qc/0512160}{{\ttfamily arXiv:gr-qc/0512160
  [gr-qc]}}.

\bibitem{Pani:2009hk}
P.~Pani, E.~Berti, V.~Cardoso, Y.~Chen, and R.~Norte, ``{Gravitational-wave
  signature of a thin-shell gravastar},''
\href{http://dx.doi.org/10.1088/1742-6596/222/1/012032}{{\em J. Phys. Conf.
  Ser.} {\bfseries 222} (2010) 012032}.

\bibitem{Kojima:1992ie}
Y.~Kojima, ``{Equations governing the nonradial oscillations of a slowly
  rotating relativistic star},''
\href{http://dx.doi.org/10.1103/PhysRevD.46.4289}{{\em Phys.Rev.} {\bfseries
  D46} (1992) 4289--4303}.

\bibitem{Zerilli:1971wd}
F.~Zerilli, ``{Gravitational field of a particle falling in a schwarzschild
  geometry analyzed in tensor harmonics},''
\href{http://dx.doi.org/10.1103/PhysRevD.2.2141}{{\em Phys.Rev.} {\bfseries D2}
  (1970) 2141--2160}.

\bibitem{Bah:2021jno}
I.~Bah, I.~Bena, P.~Heidmann, Y.~Li, and D.~R. Mayerson, ``{Gravitational
  footprints of black holes and their microstate geometries},''
  \href{http://dx.doi.org/10.1007/JHEP10(2021)138}{{\em JHEP} {\bfseries 10}
  (2021) 138}, \href{http://arxiv.org/abs/2104.10686}{{\ttfamily
  arXiv:2104.10686 [hep-th]}}.

\bibitem{Penrose:1969pc}
R.~Penrose, ``{Gravitational collapse: The role of general relativity},''
  \href{http://dx.doi.org/10.1023/A:1016578408204}{{\em Riv. Nuovo Cim.}
  {\bfseries 1} (1969) 252--276}.

\bibitem{CominsSchutz}
N.~Comins and B.~F. Schutz, ``On the ergoregion instability,'' {\em Proceedings
  of the Royal Society of London. Series A, Mathematical and Physical Sciences}
  {\bfseries 364} no.~1717, (1978) pp. 211--226.
  \url{http://www.jstor.org/stable/79759}.

\bibitem{Moschidis:2016zjy}
G.~Moschidis, ``{A Proof of Friedman's Ergosphere Instability for Scalar
  Waves},'' \href{http://dx.doi.org/10.1007/s00220-017-3010-y}{{\em Commun.
  Math. Phys.} {\bfseries 358} no.~2, (2018) 437--520},
\href{http://arxiv.org/abs/1608.02035}{{\ttfamily arXiv:1608.02035 [math.AP]}}.

\bibitem{Chirenti:2008pf}
C.~B. Chirenti and L.~Rezzolla, ``{On the ergoregion instability in rotating
  gravastars},'' \href{http://dx.doi.org/10.1103/PhysRevD.78.084011}{{\em
  Phys.Rev.} {\bfseries D78} (2008) 084011},
\href{http://arxiv.org/abs/0808.4080}{{\ttfamily arXiv:0808.4080 [gr-qc]}}.

\bibitem{Pani:2012bp}
P.~Pani, V.~Cardoso, L.~Gualtieri, E.~Berti, and A.~Ishibashi, ``{Perturbations
  of slowly rotating black holes: massive vector fields in the Kerr metric},''
  \href{http://dx.doi.org/10.1103/PhysRevD.86.104017}{{\em Phys.Rev.}
  {\bfseries D86} (2012) 104017},
\href{http://arxiv.org/abs/1209.0773}{{\ttfamily arXiv:1209.0773 [gr-qc]}}.

\bibitem{PhysRevLett.29.1114}
S.~A. Teukolsky, ``Rotating black holes: Separable wave equations for
  gravitational and electromagnetic perturbations,''
  \href{http://dx.doi.org/10.1103/PhysRevLett.29.1114}{{\em Phys. Rev. Lett.}
  {\bfseries 29} (Oct, 1972) 1114--1118}.
  \url{https://link.aps.org/doi/10.1103/PhysRevLett.29.1114}.

\bibitem{1973ApJ...185..635T}
S.~A. {Teukolsky}, ``{Perturbations of a Rotating Black Hole. I. Fundamental
  Equations for Gravitational, Electromagnetic, and Neutrino-Field
  Perturbations},'' \href{http://dx.doi.org/10.1086/152444}{{\em The Astroph.
  J.} {\bfseries 185} (Oct., 1973) 635--648}.

\bibitem{1977RSPSA.352..381D}
S.~{Detweiler}, ``{On resonant oscillations of a rapidly rotating black
  hole},'' \href{http://dx.doi.org/10.1098/rspa.1977.0005}{{\em Proceedings of
  the Royal Society of London Series A} {\bfseries 352} (July, 1977) 381--395}.

\bibitem{Casals:2005kr}
M.~Casals and A.~C. Ottewill, ``{Canonical quantization of the electromagnetic
  field on the Kerr background},''
  \href{http://dx.doi.org/10.1103/PhysRevD.71.124016}{{\em Phys. Rev. D}
  {\bfseries 71} (2005) 124016},
  \href{http://arxiv.org/abs/gr-qc/0501005}{{\ttfamily arXiv:gr-qc/0501005}}.

\bibitem{Vilenkin:1978uc}
A.~Vilenkin, ``{Exponential Amplification of Waves in the Gravitational Field
  of Ultrarelativistic Rotating Body},''
\href{http://dx.doi.org/10.1016/0370-2693(78)90027-8}{{\em Phys.Lett.}
  {\bfseries B78} (1978) 301--303}.

\bibitem{Berti:2005gp}
E.~Berti, V.~Cardoso, and M.~Casals, ``{Eigenvalues and eigenfunctions of
  spin-weighted spheroidal harmonics in four and higher dimensions},''
  \href{http://dx.doi.org/10.1103/PhysRevD.73.109902,
  10.1103/PhysRevD.73.024013}{{\em Phys.Rev.} {\bfseries D73} (2006) 024013},
\href{http://arxiv.org/abs/gr-qc/0511111}{{\ttfamily arXiv:gr-qc/0511111
  [gr-qc]}}.

\bibitem{Fan:2017cfw}
X.-L. Fan and Y.-B. Chen, ``{Stochastic gravitational-wave background from spin
  loss of black holes},''
  \href{http://dx.doi.org/10.1103/PhysRevD.98.044020}{{\em Phys. Rev. D}
  {\bfseries 98} no.~4, (2018) 044020},
  \href{http://arxiv.org/abs/1712.00784}{{\ttfamily arXiv:1712.00784 [gr-qc]}}.

\bibitem{Du:2018cmp}
S.~M. Du and Y.~Chen, ``{Searching for near-horizon quantum structures in the
  binary black-hole stochastic gravitational-wave background},''
  \href{http://dx.doi.org/10.1103/PhysRevLett.121.051105}{{\em Phys. Rev.
  Lett.} {\bfseries 121} no.~5, (2018) 051105},
\href{http://arxiv.org/abs/1803.10947}{{\ttfamily arXiv:1803.10947 [gr-qc]}}.

\bibitem{Barausse:2018vdb}
E.~Barausse, R.~Brito, V.~Cardoso, I.~Dvorkin, and P.~Pani, ``{The stochastic
  gravitational-wave background in the absence of horizons},''
  \href{http://dx.doi.org/10.1088/1361-6382/aae1de}{{\em Class. Quant. Grav.}
  {\bfseries 35} no.~20, (2018) 20LT01},
\href{http://arxiv.org/abs/1805.08229}{{\ttfamily arXiv:1805.08229 [gr-qc]}}.

\bibitem{1987ApJ...314..234C}
C.~{Cutler} and L.~{Lindblom}, ``{The Effect of Viscosity on Neutron Star
  Oscillations},'' \href{http://dx.doi.org/10.1086/165052}{{\em The
  Astrophysical Journal} {\bfseries 314} (Mar., 1987) 234}.

\bibitem{1971ApJ...165..165E}
F.~P. {Esposito}, ``{Absorption of Gravitational Energy by a Viscous
  Compressible Fluid},'' \href{http://dx.doi.org/10.1086/150884}{{\em The
  Astrophysical Journal} {\bfseries 165} (Apr., 1971) 165}.

\bibitem{1985ApJ...292..330P}
D.~{Papadopoulos} and F.~P. {Esposito}, ``{Absorption of gravitational energy
  by a viscous compressible fluid in a curved spacetime},''
  \href{http://dx.doi.org/10.1086/163163}{{\em The Astrophysical Journal}
  {\bfseries 292} (May, 1985) 330--338}.

\bibitem{Oshita:2019sat}
N.~Oshita, Q.~Wang, and N.~Afshordi, ``{On Reflectivity of Quantum Black Hole
  Horizons},'' \href{http://dx.doi.org/10.1088/1475-7516/2020/04/016}{{\em
  JCAP} {\bfseries 04} (2020) 016},
  \href{http://arxiv.org/abs/1905.00464}{{\ttfamily arXiv:1905.00464
  [hep-th]}}.

\bibitem{10.2307/79029}
S.~Detweiler, ``On the equations governing the electromagnetic perturbations of
  the kerr black hole,'' {\em Proceedings of the Royal Society of London.
  Series A, Mathematical and Physical Sciences} {\bfseries 349} no.~1657,
  (1976) 217--230. \url{http://www.jstor.org/stable/79029}.

\bibitem{King:1977}
A.~R. King, ``{Black-hole magnetostatics},'' {\em Math. Proc. Camb. Phil. Soc.}
  {\bfseries 81} (1977) 149.

\bibitem{Starobinskij2}
A.~A. {Starobinskij} and S.~M. {Churilov}, ``{Amplification of electromagnetic
  and gravitational waves scattered by a rotating black hole.},'' {\em Zhurnal
  Eksperimentalnoi i Teoreticheskoi Fiziki} {\bfseries 65} (1973) 3--11.

\bibitem{Cardoso:2008kj}
V.~Cardoso, P.~Pani, M.~Cadoni, and M.~Cavaglia, ``{Instability of
  hyper-compact Kerr-like objects},''
  \href{http://dx.doi.org/10.1088/0264-9381/25/19/195010}{{\em
  Class.Quant.Grav.} {\bfseries 25} (2008) 195010},
\href{http://arxiv.org/abs/0808.1615}{{\ttfamily arXiv:0808.1615 [gr-qc]}}.

\bibitem{Correia:2018apm}
M.~R. Correia and V.~Cardoso, ``{Characterization of echoes: A Dyson-series
  representation of individual pulses},''
  \href{http://dx.doi.org/10.1103/PhysRevD.97.084030}{{\em Phys. Rev.}
  {\bfseries D97} no.~8, (2018) 084030},
\href{http://arxiv.org/abs/1802.07735}{{\ttfamily arXiv:1802.07735 [gr-qc]}}.

\bibitem{webpage}
 \noindent \url{http://www.darkgra.org}.

\bibitem{Harmark:2007jy}
T.~Harmark, J.~Natario, and R.~Schiappa, ``{Greybody Factors for d-Dimensional
  Black Holes},'' \href{http://dx.doi.org/10.4310/ATMP.2010.v14.n3.a1}{{\em
  Adv. Theor. Math. Phys.} {\bfseries 14} no.~3, (2010) 727--794},
  \href{http://arxiv.org/abs/0708.0017}{{\ttfamily arXiv:0708.0017 [hep-th]}}.

\bibitem{Buonanno:2006ui}
A.~Buonanno, G.~B. Cook, and F.~Pretorius, ``{Inspiral, merger and ring-down of
  equal-mass black-hole binaries},''
  \href{http://dx.doi.org/10.1103/PhysRevD.75.124018}{{\em Phys. Rev. D}
  {\bfseries 75} (2007) 124018},
  \href{http://arxiv.org/abs/gr-qc/0610122}{{\ttfamily arXiv:gr-qc/0610122}}.

\bibitem{zerodet}
{\bfseries LIGO} Collaboration, D.~Shoemaker, ``Advanced ligo anticipated
  sensitivity curves,'' Tech. Rep. T0900288-v3, 2010.
\newblock \url{https://dcc.ligo.org/LIGO-T0900288/public}.

\bibitem{Cardoso:2017njb}
V.~Cardoso and P.~Pani, ``{The observational evidence for horizons: from echoes
  to precision gravitational-wave physics},''
\href{http://arxiv.org/abs/1707.03021}{{\ttfamily arXiv:1707.03021 [gr-qc]}}.

\bibitem{Flanagan:1997sx}
E.~E. Flanagan and S.~A. Hughes, ``{Measuring gravitational waves from binary
  black hole coalescences: 1. Signal-to-noise for inspiral, merger, and
  ringdown},'' \href{http://dx.doi.org/10.1103/PhysRevD.57.4535}{{\em Phys.
  Rev. D} {\bfseries 57} (1998) 4535--4565},
  \href{http://arxiv.org/abs/gr-qc/9701039}{{\ttfamily arXiv:gr-qc/9701039}}.

\bibitem{Vallisneri:2007ev}
M.~Vallisneri, ``{Use and abuse of the Fisher information matrix in the
  assessment of gravitational-wave parameter-estimation prospects},''
  \href{http://dx.doi.org/10.1103/PhysRevD.77.042001}{{\em Phys. Rev. D}
  {\bfseries 77} (2008) 042001},
  \href{http://arxiv.org/abs/gr-qc/0703086}{{\ttfamily arXiv:gr-qc/0703086}}.

\bibitem{Evans:2016mbw}
{\bfseries LIGO Scientific} Collaboration, B.~P. Abbott {\em et~al.},
  ``{Exploring the Sensitivity of Next Generation Gravitational Wave
  Detectors},'' \href{http://dx.doi.org/10.1088/1361-6382/aa51f4}{{\em Class.
  Quant. Grav.} {\bfseries 34} no.~4, (2017) 044001},
\href{http://arxiv.org/abs/1607.08697}{{\ttfamily arXiv:1607.08697
  [astro-ph.IM]}}.

\bibitem{Essick:2017wyl}
R.~Essick, S.~Vitale, and M.~Evans, ``{Frequency-dependent responses in third
  generation gravitational-wave detectors},''
  \href{http://dx.doi.org/10.1103/PhysRevD.96.084004}{{\em Phys. Rev.}
  {\bfseries D96} no.~8, (2017) 084004},
\href{http://arxiv.org/abs/1708.06843}{{\ttfamily arXiv:1708.06843 [gr-qc]}}.

\bibitem{Teukolsky:1972my}
S.~A. Teukolsky, ``{Rotating black holes - separable wave equations for
  gravitational and electromagnetic perturbations},''
\href{http://dx.doi.org/10.1103/PhysRevLett.29.1114}{{\em Phys. Rev. Lett.}
  {\bfseries 29} (1972) 1114--1118}.

\bibitem{Fujita:2004rb}
R.~Fujita and H.~Tagoshi, ``{New numerical methods to evaluate homogeneous
  solutions of the Teukolsky equation},''
  \href{http://dx.doi.org/10.1143/PTP.112.415}{{\em Prog. Theor. Phys.}
  {\bfseries 112} (2004) 415--450},
  \href{http://arxiv.org/abs/gr-qc/0410018}{{\ttfamily arXiv:gr-qc/0410018}}.

\bibitem{vandeMeent:2014raa}
M.~van~de Meent, ``{Resonantly enhanced kicks from equatorial small mass-ratio
  inspirals},'' \href{http://dx.doi.org/10.1103/PhysRevD.90.044027}{{\em Phys.
  Rev. D} {\bfseries 90} no.~4, (2014) 044027},
  \href{http://arxiv.org/abs/1406.2594}{{\ttfamily arXiv:1406.2594 [gr-qc]}}.

\bibitem{vandeMeent:2015lxa}
M.~van~de Meent and A.~G. Shah, ``{Metric perturbations produced by eccentric
  equatorial orbits around a Kerr black hole},''
  \href{http://dx.doi.org/10.1103/PhysRevD.92.064025}{{\em Phys. Rev. D}
  {\bfseries 92} no.~6, (2015) 064025},
  \href{http://arxiv.org/abs/1506.04755}{{\ttfamily arXiv:1506.04755 [gr-qc]}}.

\bibitem{vandeMeent:2016pee}
M.~van~de Meent, ``{Gravitational self-force on eccentric equatorial orbits
  around a Kerr black hole},''
  \href{http://dx.doi.org/10.1103/PhysRevD.94.044034}{{\em Phys. Rev. D}
  {\bfseries 94} no.~4, (2016) 044034},
  \href{http://arxiv.org/abs/1606.06297}{{\ttfamily arXiv:1606.06297 [gr-qc]}}.

\bibitem{vandeMeent:2017bcc}
M.~van~de Meent, ``{Gravitational self-force on generic bound geodesics in Kerr
  spacetime},'' \href{http://dx.doi.org/10.1103/PhysRevD.97.104033}{{\em Phys.
  Rev. D} {\bfseries 97} no.~10, (2018) 104033},
  \href{http://arxiv.org/abs/1711.09607}{{\ttfamily arXiv:1711.09607 [gr-qc]}}.

\bibitem{Mano:1996vt}
S.~Mano, H.~Suzuki, and E.~Takasugi, ``{Analytic solutions of the Teukolsky
  equation and their low frequency expansions},''
  \href{http://dx.doi.org/10.1143/PTP.95.1079}{{\em Prog. Theor. Phys.}
  {\bfseries 95} (1996) 1079--1096},
  \href{http://arxiv.org/abs/gr-qc/9603020}{{\ttfamily arXiv:gr-qc/9603020}}.

\bibitem{Mano:1996gn}
S.~Mano and E.~Takasugi, ``{Analytic solutions of the Teukolsky equation and
  their properties},'' \href{http://dx.doi.org/10.1143/PTP.97.213}{{\em Prog.
  Theor. Phys.} {\bfseries 97} (1997) 213--232},
  \href{http://arxiv.org/abs/gr-qc/9611014}{{\ttfamily arXiv:gr-qc/9611014}}.

\bibitem{Fujita:2009us}
R.~Fujita, W.~Hikida, and H.~Tagoshi, ``{An Efficient Numerical Method for
  Computing Gravitational Waves Induced by a Particle Moving on Eccentric
  Inclined Orbits around a Kerr Black Hole},''
  \href{http://dx.doi.org/10.1143/PTP.121.843}{{\em Prog. Theor. Phys.}
  {\bfseries 121} (2009) 843--874},
  \href{http://arxiv.org/abs/0904.3810}{{\ttfamily arXiv:0904.3810 [gr-qc]}}.

\bibitem{Ori:2000zn}
A.~Ori and K.~S. Thorne, ``{The Transition from inspiral to plunge for a
  compact body in a circular equatorial orbit around a massive, spinning black
  hole},'' \href{http://dx.doi.org/10.1103/PhysRevD.62.124022}{{\em Phys. Rev.
  D} {\bfseries 62} (2000) 124022},
  \href{http://arxiv.org/abs/gr-qc/0003032}{{\ttfamily arXiv:gr-qc/0003032}}.

\bibitem{Bernuzzi:2012ku}
S.~Bernuzzi, A.~Nagar, and A.~Zenginoglu, ``{Horizon-absorption effects in
  coalescing black-hole binaries: An effective-one-body study of the
  non-spinning case},''
  \href{http://dx.doi.org/10.1103/PhysRevD.86.104038}{{\em Phys. Rev. D}
  {\bfseries 86} (2012) 104038},
  \href{http://arxiv.org/abs/1207.0769}{{\ttfamily arXiv:1207.0769 [gr-qc]}}.

\bibitem{Taracchini:2013wfa}
A.~Taracchini, A.~Buonanno, S.~A. Hughes, and G.~Khanna, ``{Modeling the
  horizon-absorbed gravitational flux for equatorial-circular orbits in Kerr
  spacetime},'' \href{http://dx.doi.org/10.1103/PhysRevD.88.044001}{{\em Phys.
  Rev. D} {\bfseries 88} (2013) 044001},
  \href{http://arxiv.org/abs/1305.2184}{{\ttfamily arXiv:1305.2184 [gr-qc]}}.
  [Erratum: Phys.Rev.D 88, 109903 (2013)].

\bibitem{Harms:2014dqa}
E.~Harms, S.~Bernuzzi, A.~Nagar, and A.~Zenginoglu, ``{A new gravitational wave
  generation algorithm for particle perturbations of the Kerr spacetime},''
  \href{http://dx.doi.org/10.1088/0264-9381/31/24/245004}{{\em Class. Quant.
  Grav.} {\bfseries 31} no.~24, (2014) 245004},
  \href{http://arxiv.org/abs/1406.5983}{{\ttfamily arXiv:1406.5983 [gr-qc]}}.

\bibitem{Pani:2010em}
P.~Pani, E.~Berti, V.~Cardoso, Y.~Chen, and R.~Norte, ``{Gravitational-wave
  signatures of the absence of an event horizon. II. Extreme mass ratio
  inspirals in the spacetime of a thin-shell gravastar},''
  \href{http://dx.doi.org/10.1103/PhysRevD.81.084011}{{\em Phys. Rev.}
  {\bfseries D81} (2010) 084011},
\href{http://arxiv.org/abs/1001.3031}{{\ttfamily arXiv:1001.3031 [gr-qc]}}.

\bibitem{Fransen:2020prl}
K.~Fransen, G.~Koekoek, R.~Tielemans, and B.~Vercnocke, ``{Modeling and
  detecting resonant tides of exotic compact objects},''
  \href{http://dx.doi.org/10.1103/PhysRevD.104.044044}{{\em Phys. Rev. D}
  {\bfseries 104} no.~4, (2021) 044044},
  \href{http://arxiv.org/abs/2005.12286}{{\ttfamily arXiv:2005.12286 [gr-qc]}}.

\bibitem{Pons:2001xs}
J.~A. Pons, E.~Berti, L.~Gualtieri, G.~Miniutti, and V.~Ferrari,
  ``{Gravitational signals emitted by a point mass orbiting a neutron star:
  Effects of stellar structure},''
  \href{http://dx.doi.org/10.1103/PhysRevD.65.104021}{{\em Phys. Rev.}
  {\bfseries D65} (2002) 104021},
\href{http://arxiv.org/abs/gr-qc/0111104}{{\ttfamily arXiv:gr-qc/0111104
  [gr-qc]}}.

\bibitem{Kapadia:2013kf}
S.~J. Kapadia, D.~Kennefick, and K.~Glampedakis, ``{Do floating orbits in
  extreme mass ratio binary black holes exist?},''
  \href{http://dx.doi.org/10.1103/PhysRevD.87.044050}{{\em Phys. Rev.}
  {\bfseries D87} no.~4, (2013) 044050},
\href{http://arxiv.org/abs/1302.1016}{{\ttfamily arXiv:1302.1016 [gr-qc]}}.

\bibitem{Cardoso:2011xi}
V.~Cardoso, S.~Chakrabarti, P.~Pani, E.~Berti, and L.~Gualtieri, ``{Floating
  and sinking: The Imprint of massive scalars around rotating black holes},''
  \href{http://dx.doi.org/10.1103/PhysRevLett.107.241101}{{\em Phys.Rev.Lett.}
  {\bfseries 107} (2011) 241101},
\href{http://arxiv.org/abs/1109.6021}{{\ttfamily arXiv:1109.6021 [gr-qc]}}.

\bibitem{Hinderer:2008dm}
T.~Hinderer and E.~E. Flanagan, ``{Two timescale analysis of extreme mass ratio
  inspirals in Kerr. I. Orbital Motion},''
  \href{http://dx.doi.org/10.1103/PhysRevD.78.064028}{{\em Phys. Rev.}
  {\bfseries D78} (2008) 064028},
\href{http://arxiv.org/abs/0805.3337}{{\ttfamily arXiv:0805.3337 [gr-qc]}}.

\bibitem{Lindblom:2008cm}
L.~Lindblom, B.~J. Owen, and D.~A. Brown, ``{Model Waveform Accuracy Standards
  for Gravitational Wave Data Analysis},''
  \href{http://dx.doi.org/10.1103/PhysRevD.78.124020}{{\em Phys. Rev. D}
  {\bfseries 78} (2008) 124020},
  \href{http://arxiv.org/abs/0809.3844}{{\ttfamily arXiv:0809.3844 [gr-qc]}}.

\bibitem{Bonga:2019ycj}
B.~Bonga, H.~Yang, and S.~A. Hughes, ``{Tidal resonance in extreme mass-ratio
  inspirals},'' \href{http://dx.doi.org/10.1103/PhysRevLett.123.101103}{{\em
  Phys. Rev. Lett.} {\bfseries 123} no.~10, (2019) 101103},
  \href{http://arxiv.org/abs/1905.00030}{{\ttfamily arXiv:1905.00030 [gr-qc]}}.

\bibitem{Piovano:2020zin}
G.~A. Piovano, A.~Maselli, and P.~Pani, ``{Extreme mass ratio inspirals with
  spinning secondary: a detailed study of equatorial circular motion},''
  \href{http://dx.doi.org/10.1103/PhysRevD.102.024041}{{\em Phys. Rev. D}
  {\bfseries 102} no.~2, (2020) 024041},
  \href{http://arxiv.org/abs/2004.02654}{{\ttfamily arXiv:2004.02654 [gr-qc]}}.

\bibitem{Cornish:2018dyw}
T.~Robson, N.~J. Cornish, and C.~Liug, ``{The construction and use of LISA
  sensitivity curves},'' \href{http://dx.doi.org/10.1088/1361-6382/ab1101}{{\em
  Class. Quant. Grav.} {\bfseries 36} no.~10, (2019) 105011},
\href{http://arxiv.org/abs/1803.01944}{{\ttfamily arXiv:1803.01944
  [astro-ph.HE]}}.

\bibitem{Allen:2005fk}
B.~Allen, W.~G. Anderson, P.~R. Brady, D.~A. Brown, and J.~D.~E. Creighton,
  ``{FINDCHIRP: An Algorithm for detection of gravitational waves from
  inspiraling compact binaries},''
  \href{http://dx.doi.org/10.1103/PhysRevD.85.122006}{{\em Phys. Rev. D}
  {\bfseries 85} (2012) 122006},
  \href{http://arxiv.org/abs/gr-qc/0509116}{{\ttfamily arXiv:gr-qc/0509116}}.

\bibitem{Flanagan:1997kp}
E.~E. Flanagan and S.~A. Hughes, ``{Measuring gravitational waves from binary
  black hole coalescences: 2. The Waves' information and its extraction, with
  and without templates},''
  \href{http://dx.doi.org/10.1103/PhysRevD.57.4566}{{\em Phys. Rev. D}
  {\bfseries 57} (1998) 4566--4587},
  \href{http://arxiv.org/abs/gr-qc/9710129}{{\ttfamily arXiv:gr-qc/9710129}}.

\bibitem{Xin:2021zir}
S.~Xin, B.~Chen, R.~K.~L. Lo, L.~Sun, W.-B. Han, X.~Zhong, M.~Srivastava,
  S.~Ma, Q.~Wang, and Y.~Chen, ``{Gravitational-wave echoes from spinning
  exotic compact objects: numerical waveforms from the Teukolsky equation},''
  \href{http://arxiv.org/abs/2105.12313}{{\ttfamily arXiv:2105.12313 [gr-qc]}}.

\bibitem{Chen:2020htz}
B.~Chen, Q.~Wang, and Y.~Chen, ``{Tidal response and near-horizon boundary
  conditions for spinning exotic compact objects},''
  \href{http://dx.doi.org/10.1103/PhysRevD.103.104054}{{\em Phys. Rev. D}
  {\bfseries 103} no.~10, (2021) 104054},
  \href{http://arxiv.org/abs/2012.10842}{{\ttfamily arXiv:2012.10842 [gr-qc]}}.

\end{thebibliography}\endgroup

\end{document}